\ifdefined\pdfobjcompresslevel\pdfobjcompresslevel=0\fi
\documentclass[11pt]{article}
\usepackage[a4paper,margin=25mm]{geometry}
\usepackage[T1]{fontenc}
\usepackage{lmodern,amsmath,amssymb,mathtools,microtype}
\usepackage{graphicx,tikz,booktabs,placeins,hyperref,float}
\usepackage{needspace}
\usetikzlibrary{arrows.meta,calc,decorations.markings}
\tikzset{
  every picture/.append style={color=black},
  paneltitle/.style={font=\small\normalfont,text=black,inner sep=2pt}
}
\hypersetup{hidelinks,pdftitle={Random Euclidean CFT2 from Triangulated dS3 Gravity}}
\numberwithin{equation}{section}
\allowdisplaybreaks[2]
\definecolor{gs}{rgb}{0.75,0.1,0.1}
\newcommand{\nw}[1]{#1}

\newcommand{\revnew}[1]{#1}

\newcommand{\vnew}[1]{#1}
\newcommand{\hnew}[1]{#1}
\newcommand{\fresh}[1]{#1}

\newcommand{\newpassage}[1]{#1}

\newcommand{\vone}[1]{#1}
\newenvironment{voneblock}{}{}
\newcommand{\vtwo}[1]{#1}
\newenvironment{vtwoblock}{}{}
\newcommand{\finalpolish}[1]{#1}
\newenvironment{finalpolishblock}{}{}
\newcommand{\aidrev}[1]{#1}

\definecolor{pt}{rgb}{0.1,0.3,0.75}
\definecolor{faceblue}{RGB}{222,232,241}
\definecolor{caporange}{RGB}{246,208,155}
\definecolor{edgeblue}{RGB}{55,90,137}
\definecolor{paircolor}{RGB}{169,88,27}
\definecolor{lightink}{RGB}{90,119,151}
\definecolor{copper}{RGB}{173,91,37}
\newcommand{\figpants}{%
  \path[fill=gray!9] (-.988,1.062)
   .. controls (-.35,.83) and (.28,.82) .. (.90,.92)
   .. controls (1.12,.50) and (1.12,-.50) .. (.90,-.92)
   .. controls (.28,-.82) and (-.35,-.83) .. (-.988,-1.062)
   --(-1.412,-.638)
   .. controls (-1.04,-.32) and (-1.04,.32) .. (-1.412,.638) --cycle;
  \draw (-.988,1.062) .. controls (-.35,.83) and (.28,.82) .. (.90,.92);
  \draw (-.988,-1.062) .. controls (-.35,-.83) and (.28,-.82) .. (.90,-.92);
  \draw (-1.412,-.638) .. controls (-1.04,-.32) and (-1.04,.32) .. (-1.412,.638);}
\newcommand{\figleftcuffs}{%
  \draw[fill=white,rotate around={45:(-1.20,.85)}] (-1.20,.85) ellipse (.30 and .15);
  \draw[fill=white,rotate around={-45:(-1.20,-.85)}] (-1.20,-.85) ellipse (.30 and .15);}
\newcommand{\figfoursphere}{%
  \begin{scope}[shift={(-.90,0)}]\figpants\figleftcuffs\end{scope}
  \begin{scope}[shift={(.90,0)},xscale=-1]\figpants\figleftcuffs\end{scope}}
\newcommand{\figsewcircle}[2][pt]{%
  \draw[#1,densely dashed,line width=.85pt] (#2,.92) arc[start angle=90,end angle=270,x radius=.16,y radius=.92];
  \draw[#1,line width=.85pt] (#2,-.92) arc[start angle=270,end angle=450,x radius=.16,y radius=.92];}
\newcommand{\figfourlabels}{%
  \node[left] at (-2.21,1.02) {$P_1$};\node[left] at (-2.21,-1.02) {$P_2$};
  \node[right] at (2.21,1.02) {$P_4$};\node[right] at (2.21,-1.02) {$P_3$};}
\newcommand{\figcell}{%
 \fill[faceblue] (-.27,1.09)--(.27,1.09)
  .. controls (.21,.44) and (.79,.20) .. (1.37,.25)
  --(1.37,-.25).. controls (.79,-.20) and (.21,-.44)..(.27,-1.09)
  --(-.27,-1.09).. controls (-.21,-.44) and (-.79,-.20)..(-1.37,-.25)
  --(-1.37,.25).. controls (-.79,.20) and (-.21,.44)..(-.27,1.09)--cycle;
 \draw[black!78,line width=.8pt] (-.27,1.09)--(.27,1.09)
  .. controls (.21,.44) and (.79,.20) .. (1.37,.25)
  --(1.37,-.25).. controls (.79,-.20) and (.21,-.44)..(.27,-1.09)
  --(-.27,-1.09).. controls (-.21,-.44) and (-.79,-.20)..(-1.37,-.25)
  --(-1.37,.25).. controls (-.79,.20) and (-.21,.44)..(-.27,1.09);
 \draw[edgeblue,line width=1pt] (0,.89).. controls (.18,.34) and (.14,-.26)..(0,-.89);
 \draw[black!58,densely dashed,line width=.62pt] (-1.12,0)
   .. controls (-.36,.09) and (.46,-.05).. (1.12,0);
 \filldraw[fill=caporange,draw=black!78,line width=.72pt] (-.27,1.09)--(.27,1.09)--(0,.89)--cycle;
 \filldraw[fill=caporange,draw=black!78,line width=.72pt] (-.27,-1.09)--(.27,-1.09)--(0,-.89)--cycle;
 \filldraw[fill=caporange,draw=black!78,line width=.72pt] (-1.37,.25)--(-1.37,-.25)--(-1.12,0)--cycle;
 \filldraw[fill=caporange,draw=black!78,line width=.72pt] (1.37,.25)--(1.37,-.25)--(1.12,0)--cycle;}
\newcommand{\figgentet}{%
 \draw[draw=black!48,line width=.65pt] (0,0) circle (2.02);
 \draw[black!40,line width=.72pt] (0,2.5457) -- (0.1760,1.0974);
 \draw[black!40,line width=.72pt] (0,2.5457) -- (-0.7481,1.5173);
 \draw[black!40,line width=.72pt] (0,2.5457) -- (0.5736,1.6203);
 \draw[black!21,line width=.45pt] (0.5895,-2.3042) -- (0.4045,-0.7818);
 \draw[black!21,line width=.45pt] (0.5895,-2.3042) -- (-0.3921,-1.5782);
 \draw[black!21,line width=.45pt] (0.5895,-2.3042) -- (0.9513,-1.4006);
 \draw[black!21,line width=.45pt] (-2.0876,-0.3241) -- (-1.5477,0.4180);
 \draw[black!21,line width=.45pt] (-2.0876,-0.3241) -- (-1.4136,-0.8226);
 \draw[black!21,line width=.45pt] (-2.0876,-0.3241) -- (-0.9282,-0.1998);
 \draw[black!21,line width=.45pt] (1.5379,0.0647) -- (1.1585,0.6767);
 \draw[black!21,line width=.45pt] (1.5379,0.0647) -- (1.3102,-0.5040);
 \draw[black!21,line width=.45pt] (1.5379,0.0647) -- (0.4758,-0.0492);
 \draw[black!26,densely dashed,line width=.48pt] (-0.9282,-0.1998) -- (-1.5477,0.4180) -- (-1.4136,-0.8226) -- cycle;
 \draw[black!26,densely dashed,line width=.48pt] (1.3102,-0.5040) -- (1.1585,0.6767) -- (0.4758,-0.0492) -- cycle;
 \draw[black!26,densely dashed,line width=.48pt] (-0.9282,-0.1998) -- (-1.4136,-0.8226) -- (-0.3921,-1.5782) -- (0.9513,-1.4006) -- (1.3102,-0.5040) -- (0.4758,-0.0492) -- cycle;
 \draw[black!26,densely dashed,line width=.48pt] (0.4758,-0.0492) -- (1.1585,0.6767) -- (0.5736,1.6203) -- (-0.7481,1.5173) -- (-1.5477,0.4180) -- (-0.9282,-0.1998) -- cycle;
 \filldraw[fill=lightink!17,draw=black!82,line width=.9pt] (0.9513,-1.4006) -- (0.4045,-0.7818) -- (0.1760,1.0974) -- (0.5736,1.6203) -- (1.1585,0.6767) -- (1.3102,-0.5040) -- cycle;
 \filldraw[fill=lightink!24,draw=black!82,line width=.9pt] (-1.4136,-0.8226) -- (-1.5477,0.4180) -- (-0.7481,1.5173) -- (0.1760,1.0974) -- (0.4045,-0.7818) -- (-0.3921,-1.5782) -- cycle;
 \filldraw[fill=copper!27,draw=black!82,line width=.9pt] (0.5736,1.6203) -- (0.1760,1.0974) -- (-0.7481,1.5173) -- cycle;
 \filldraw[fill=copper!27,draw=black!82,line width=.9pt] (-0.3921,-1.5782) -- (0.4045,-0.7818) -- (0.9513,-1.4006) -- cycle;
 \filldraw[fill=black!85,draw=black!85] (0,2.5457) circle (1.45pt);}
\newcommand{\figfunnel}[1]{%
 \fill[gray!9] (0,0) .. controls (-.22,.30*#1) and (-1.12,.70*#1) .. (-1.55,#1)
   -- (1.55,#1) .. controls (1.12,.70*#1) and (.22,.30*#1) .. cycle;
 \draw (0,0) .. controls (-.22,.30*#1) and (-1.12,.70*#1) .. (-1.55,#1);
 \draw (0,0) .. controls (.22,.30*#1) and (1.12,.70*#1) .. (1.55,#1);
 \fill (0,0) circle (1.3pt);}
\newcommand{\figwormhole}[1]{%
 \fill[gray!9] (-1.55,0) .. controls (-1.05,.33*#1) and (-1.05,.67*#1) .. (-1.55,#1)
   -- (1.55,#1) .. controls (1.05,.67*#1) and (1.05,.33*#1) .. (1.55,0) -- cycle;
 \draw (-1.55,0) .. controls (-1.05,.33*#1) and (-1.05,.67*#1) .. (-1.55,#1);
 \draw (1.55,0) .. controls (1.05,.33*#1) and (1.05,.67*#1) .. (1.55,#1);}
\newcommand{\figboundary}[2][white]{%
 \draw[fill=#1,fill opacity=.82] (0,#2) ellipse (1.55 and .42);
 \foreach \x in {-1.05,-.35,.35,1.05}{\fill (\x,#2) circle (1.35pt);}}
\makeatletter\newcommand{\eqalias}[1]{\ltx@label{#1}}\makeatother

\newcommand{\dd}{\mathrm d}
\newcommand{\ii}{\mathrm i}
\DeclareMathOperator{\Disc}{Disc}
\DeclareMathOperator{\Res}{Res}
\renewcommand{\Re}{\operatorname{Re}}
\renewcommand{\Im}{\operatorname{Im}}
\DeclareMathOperator{\arsinh}{arsinh}
\DeclareMathOperator{\arcosh}{arcosh}
\DeclareMathOperator{\PV}{PV}
\DeclareMathOperator{\Li}{Li}
\newcommand{\Eds}{\mathbb E_{\rm dS}}
\newcommand{\sixj}{\begin{Bmatrix}P_0&P_0&P_t\\P_0&P_0&P\end{Bmatrix}_{\!\rm Vir}}
\newcommand{\blockS}{\begin{tikzpicture}[baseline=-2pt,scale=.35,line cap=round]
\draw (-1,1)--(0,.4)--(1,1);\draw (-1,-1)--(0,-.4)--(1,-1);
\draw[gs,thick](0,-.4)--(0,.4);\node[right,font=\scriptsize]at(0,0){$P_s$};\end{tikzpicture}}
\newcommand{\blockT}{\begin{tikzpicture}[baseline=-2pt,scale=.35,line cap=round]
\draw (-1,1)--(-.4,0)--(-1,-1);\draw (1,1)--(.4,0)--(1,-1);
\draw[pt,thick](-.4,0)--(.4,0);\node[above,font=\scriptsize]at(0,0){$P_t$};\end{tikzpicture}}
\newcommand{\thetagraph}{\begin{tikzpicture}[baseline=-3pt,scale=.33,line cap=round]
\draw (0,0) circle (1);\draw (-1,0)--(1,0);\fill(-1,0)circle(2.2pt);\fill(1,0)circle(2.2pt);\end{tikzpicture}}
\newcommand{\tetgraph}{\begin{tikzpicture}[baseline=-3pt,scale=.36,line cap=round]
\draw (0,0) circle (1);\draw (0,0)--(90:1);\draw (0,0)--(210:1);\draw (0,0)--(330:1);
\foreach \a in {90,210,330}{\fill(\a:1)circle(2.2pt);}\fill(0,0)circle(2.2pt);\end{tikzpicture}}
\newcommand{\handcuffgraph}{\begin{tikzpicture}[baseline=-5pt,scale=.42,line cap=round]
\draw(-.5,0)++(58.6:.76)arc(58.6:394.6:.76);\draw(.5,0)++(238.6:.76)arc(238.6:574.6:.76);
\draw(-.5,-.76)..controls(-.5,-1.2)and(.5,-1.2)..(.5,-.76);
\fill(-.5,-.76)circle(2.1pt);\fill(.5,-.76)circle(2.1pt);\end{tikzpicture}}
\ifdefined\ComparisonLayout
\fi
\AtBeginDocument{\pdfpagewidth=\paperwidth\pdfpageheight=\paperheight}
\title{Random Euclidean CFT$_2$ from Triangulated dS$_3$ Gravity}
\author{Ji-seong Chae}
\date{23 October 2026}
\begin{document}
\thispagestyle{empty}
\begin{flushright}
\end{flushright}

\bigskip

\begin{center}
\noindent{\Large \textbf{Random Euclidean CFT$_2$ from Triangulated dS$_3$ Gravity}}\\ 
\vspace{1cm} \noindent{Ji-seong Chae${}^{a}$\footnote{e-mail:jiseongchae17@gmail.com}}

\vspace{1cm}
  {\it
Department of Physics, Hanyang University, Seoul 04763, Korea${}^{a}$\\
 }
\end{center}
\begin{abstract}
{We propose a random Euclidean CFT description of three-dimensional de Sitter gravity with fixed central charge and primary spectrum but fluctuating OPE coefficients. A normalized four-point coefficient extracted from the complex Liouville cosmological wavefunction matches the analytically continued tetrahedral Virasoro graph amplitude. Its two-copy form gives a non-Gaussian OPE correlation, with a modular $S$ transform relating the corresponding amplitude to fixed-angle Conformal Turaev--Viro theory. The $K_4$ graph also determines a compact hyperbolic core assembled from two generalized tetrahedra, with a pleated genus-three boundary. \vtwo{We reconstruct this core directly from a marked family of real Lorentzian de Sitter spacetimes expanding from a tetrahedral initial graph. Inverse normal flow collapses six future-boundary collars into geodesic cuffs. Initial spacelike arc lengths fix the exterior dihedral angles, while boost holonomies and marked seams determine the edge lengths and face pairings.} Semiclassically, the OPE correlation acquires an oscillatory phase governed by hyperbolic volume, whose angle derivatives measure Lorentzian holonomies. As four necks of the future boundary pinch and the remaining bends vanish, its genus-three surface degenerates into two four-punctured spheres joined at four nodes. Sphere-tree sewing extends this relation between geometry and wavefunction coefficients to higher-point channels.}
\end{abstract}
\vspace{0.3cm}
\clearpage
\begingroup
\small
\setlength{\parskip}{0pt}
\linespread{1.0}\selectfont
\tableofcontents
\endgroup
\clearpage

\FloatBarrier
\section{Introduction}\label{sec:intro}

Three-dimensional gravity with negative cosmological constant has become a precise setting for the statistical interpretation of holography.  Wormhole amplitudes compute averages of products of boundary observables, and in AdS$_3$ they are reproduced by treating the data of heavy states in a two-dimensional CFT as random variables \cite{CotlerJensen,BelindeBoer,CCHM,PollackETAL,MarolfMaxfield,StanfordNoise}.  Exact free-boson ensembles provide related averaged holographic examples \cite{AfkhamiJeddiNarain,MaloneyWittenNarain}.  The heavy operator product expansion (OPE) coefficients have \nw{a universal second moment governed by $C_0$ from crossing and modular invariance, often used as the covariance of a leading Gaussian ansatz} \cite{CMMT}.  Non-Gaussian corrections are governed by Virasoro $6j$ symbols and multiboundary wormholes \cite{BdBL,dBLPS,dBLP,HigherPointOPE,ChandraKnots}, Virasoro TQFT computes the corresponding amplitudes of specified three-manifolds \cite{CEZ,CEZ2}, and the full statistics must be compatible with crossing and with sums over topologies \cite{UniversalTopologies,CrossingOPE,MaloneyWitten,BlockFareyTail,GaberdielGenusTwo}.

Spectral statistics and their OPE-sensitive extensions have complementary random-matrix and surgery descriptions \cite{DiUbaldoPerlmutterRMT,BoruchRMT,GeneralizedSFF,SurgeryStatistics}. Positive torus spectral densities also depend on the completion of the pure-gravity saddle sum \cite{MaxfieldTuriaci,DiUbaldoPerlmutterStringy,YanOffShell}.

Candidate microscopic ensembles have been formulated through approximate CFTs and random tensor models, random ensembles of three-dimensional gravity and measures on the space of CFTs \cite{BdBJNS,JRW,CFTMeasure}. The distinction from apparent averaging of black-hole observables is emphasized in \cite{SchlenkerWitten}. Ensembles of boundary CFT data reproduce Virasoro TQFT and assemble the gravitational path integral from boundary correlators, and wormholes ending on end-of-the-world branes compute their averages \cite{CHJL,HJL,WWW,OpenClosedEnsemble}.

Hartman localized this correspondence to compact regions of spacetime \cite{Hartman}.  A compact region is assembled from generalized hyperbolic tetrahedra, constructed in the Klein model by truncating hyperideal vertices outside the unit ball representing $\mathbb H_3$ \cite{BaoBonahon,Ushijima}.  Related triangulations arise in quantum Teichm\"uller TQFT \cite{AndersenKashaev}.  Conformal Turaev--Viro (CTV) theory evaluates its gravitational amplitude with fixed dihedral angles on the finite boundary \cite{CTV}, Virasoro TQFT evaluates it with fixed geodesic lengths, and the two are related by the modular $S$ transform acting on each boundary label.  The open sector of a CFT ensemble defines an open Virasoro TQFT whose triangulations compute compact regions with fixed-length or fixed-angle boundary conditions, and for manifolds with boundary Wilson loops an open--closed duality relates CTV theory to the diagonal sector of two copies of Virasoro TQFT \cite{JafferisWang}.  Inverting this transform turns a fixed-angle amplitude into a moment of normalized OPE coefficients, so that each compact filling contributes a definite term to the OPE statistics \cite{Hartman}.

De Sitter space poses the same questions with a spacelike boundary.  Its horizons carry temperature and entropy \cite{GibbonsHawking}, and the entropy of dS$_3$ has been related to a microscopic count \cite{MaldacenaStrominger}.  Euclidean CFT calculations also predict quantum corrections to de Sitter entropy in four dimensions \cite{BobevdSEntropy}.  The dS/CFT proposal associates the late-time wavefunction of the universe with Euclidean conformal data at future infinity \cite{Strominger,Witten01,Maldacena03}.  The asymptotic Chern--Simons dynamics reduce to Euclidean Liouville theory, and the conical defects created by bulk point particles correspond to Liouville vertex operators \cite{CacciatoriKlemm,KlemmVanzo}.  Its wavefunction formulation, gravitational inner product, its analytic relation to Euclidean AdS amplitudes and the sum over dS$_3$ geometries have been studied in \cite{CastroMaloney,HarlowStanford,CLM,dSNormGravity}.  A higher-spin realization identifies a concrete Euclidean CFT at future infinity \cite{AHS}, the cosmological bootstrap organizes late-time correlators by conformal invariance \cite{ABLP,dSBootstrap,dSHilbertQFT}, and recent work constructs de Sitter observables from Euclidean CFT data \cite{Choudhury,HYang}.  Proposed duals of dS$_3$ and of more general cosmologies include nonunitary CFTs and large-$N$ theories with complex couplings \cite{HNTT,ThavanesanWall}.  In these examples the boundary theory is nonunitary, often with complex central charge, and its relation to bulk unitarity is part of the problem.

Low-dimensional models make the relation between gravity and its dual sharper.  Jackiw--Teitelboim gravity is described by a double-scaled matrix integral \cite{SSS}, and matrix integrals also describe the topological expansion of two-dimensional de Sitter gravity \cite{MTY,CJM}.  For pure dS$_3$, Collier, Eberhardt and M\"uhlmann (CEM) proposed a complex Liouville cosmological state, a function of the conformal geometry of future infinity \cite{CEM}.  Its Hermitian norm, integrated over the moduli of future infinity, gives the amplitudes of the complex Liouville string \cite{CEM,CLS}, which are computed by a double-scaled two-matrix integral \cite{Matrix}.  CEM identify this matrix integral, rather than a Euclidean CFT, as the dual of the moduli-integrated cosmological correlators \cite{CEM}.  Ensemble averaging also shapes the semiclassical wavefunctions of closed universes \cite{UsatyukZhao}, and observer-based approaches construct de Sitter observables through operator algebras, observer-dependent path integrals and worldline quantization \cite{CLPW,WittenObs,AAIL,HUZ,BTV}. Complex Virasoro TQFT also provides proposals for fixed-topology partition functions on a class of spherical cone manifolds and massive-observer correlators \cite{ChandraLiskaPost}. In four-dimensional minisuperspace, cosmological wavefunctions have also been related to static-patch partition functions \cite{BlackerHartnoll}.

We propose an OPE-statistical interpretation of the marked conformal-block coefficients of the CEM cosmological wavefunction before integration over the moduli of future infinity.  At fixed conformal geometry of future infinity we consider a \emph{random Euclidean CFT}$_2$, a two-dimensional Euclidean CFT whose central charge, labelled primary spectrum and two-point pairing are fixed while its nonidentity OPE coefficients are random variables.  Its moments are assigned by continuing Hartman's compact amplitudes from AdS$_3$ to dS$_3$ parameters.  This description is complementary to the matrix integral.  It assigns contributions to moments of normalized OPE coefficients for specified index contractions and marked conformal-block channels at fixed boundary moduli, whereas the matrix integral computes the norm after integration over those moduli.  Two kinds of test apply.  The continued moments must fit together as CFT statistics, consistently with crossing on the sphere and with modular transformations on the torus.  They must also carry a geometric meaning in de Sitter space.

The geometric question begins with the generalized hyperbolic tetrahedra used to compute compact amplitudes.  The faces of these tetrahedra have unit spacelike normals in Minkowski embedding space, and these normals lie on the de Sitter hyperboloid \cite{Hartman}.  Hyperideal polyhedra are determined by their dihedral angles and Gram matrices \cite{BaoBonahon,SchlenkerHyperideal,LuoYangRigidity}.  The boundary of a convex hyperbolic core is a pleated surface bent along geodesic laminations, and grafting and the renormalized volume relate its intrinsic geometry to the conformal structure at infinity \cite{EpsteinMarden,McMullen,KrasnovSchlenker,SchlenkerRV}.  Mess and Scannell associated maximal globally hyperbolic de Sitter spacetimes with hyperbolic ends and complex projective structures \cite{Mess,ABBGLSS,BenedettiBonsante,Scannell}.  The future boundary of such a spacetime is spacelike and carries a Euclidean conformal structure, on which the gravitational constraints become conformal Ward identities for the cosmological wavefunction \cite{CEM,dSHilbertGravity}.  \vtwo{For the marked tetrahedral family, we recover the \aidrev{two-cell compact core} from the initial arcs, boost holonomies and seam identifications of a real Lorentzian spacetime.} The extracted OPE coefficient then measures the common geometric data through its continued tetrahedral phase. Because the continued amplitudes are evaluated on complexified geometries, criteria for allowable complex metrics \cite{KontsevichSegal,WittenComplex} bear on which of these geometries enter the de Sitter path integral.

{The closed future surface is related to the sphere state through the joining annuli. Pinching them exposes two punctured spheres. An internal small-bending limit then matches their marked conformal geometries \cite{BonahonOtal,LecuireBending,SeriesBending}. The component states are related by sewing, with their puncture labels specified.}

\subsubsection*{What changes in de Sitter}

Much of the algebra is inherited from AdS$_3$.  The $S$ transform and the CTV relation \cite{Hartman,CTV}, the Virasoro fusion and recoupling identities \cite{CEZ,MooreSeiberg,PTBootstrap,PT,Eberhardt} and the crossing constraints on OPE statistics \cite{UniversalTopologies,CrossingOPE} are first established for real $b$.  Prism and $K_{3,3}$ contractions of the sixth moment have also been evaluated there \cite{CEZ2}.  Virasoro recursion and the large-exchange analysis of conformal blocks give complementary control over the descendant expansion \cite{Zam,Cardona2021}.  Here $b$ parametrizes the Virasoro central charge as $c=1+6(b+b^{-1})^2$.  The complex Liouville prescription for dS uses $b=re^{\ii\pi/4}$ and scalar momenta $P\in e^{\ii\pi/4}\mathbb R_+$, with $r>0$ \cite{CEM}.  Known kernel formulas and the change of these parameters already indicate five differences that a dS construction must address.
\begin{enumerate}
\item \emph{Volumes become phases.}  For fixed admissible hyperideal angles, known tetrahedral asymptotics contain the exponential $e^{-2V_A/(\pi b^2)}$ in $\{6j\}^2$, where $V_A$ is the hyperbolic cell volume \cite{Hartman,TV,LMSWY,MY,BellettiYang}.  At $b=re^{\ii\pi/4}$ this factor becomes $e^{2\ii V_A/(\pi r^2)}$.  Geometry should therefore be read from an oscillatory phase.  The comparison follows the fixed-angle family $b(\theta)=re^{\ii\theta}$ and $P(\theta)=\ii\psi/[2\pi b(\theta)]$ for $0\le\theta\le\pi/4$.  Along this family $4\pi bP=2\ii\psi$ is fixed, so the real-$b$ endpoint has imaginary $P$ and the dS endpoint has $P=e^{\ii\pi/4}\psi/(2\pi r)$.  The corresponding volume exponent rotates continuously from real suppression to an oscillatory phase.
\item \emph{Separated spectral cuts.}  With the left and right descendant levels matched to $n$, propagation along a boundary collar gives $e^{-2\pi y(P^2+n)}$, where $y$ is the collar length in conformal coordinates.  A Laplace transform in $y$ has its level-$n$ singularities at $\zeta=P^2+n$.  Real momenta give overlapping rays $[n,\infty)$, whereas the dS momenta give separated rays $n+\ii\mathbb R_+$.  This suggests a way to isolate primary data, provided the transform and the complete descendant sum admit the required continuation.
\item \emph{Complex twist frequencies.}  The relative twist $u$ of two sewn boundaries couples to the difference $\omega$ of their chiral weights.  The projection $\frac12\int_{\mathbb R}\dd u\,e^{\ii\pi u\omega}=\delta(\omega)$ is a real-frequency distribution.  We implement this projection in its real-frequency distributional domain, isolate the primary-channel coefficient using the radial transform, and analytically continue the resulting meromorphic coefficient to the dS ray.  The continuation transports the Virasoro-kernel contours and includes residues from poles crossed by their deformation \cite{CMMT,PT,WittenPL}.
\item \emph{Oscillatory internal measure.}  The density $\rho_0(P)=4\sqrt2\sinh(2\pi bP)\sinh(2\pi b^{-1}P)$ has a dominant growing exponential for real $b$ and large positive $P$ \cite{CMMT,CTV,Cardy,HKS}.  On the dS ray, setting the internal length $L=4\pi bP$ real makes the second hyperbolic sine proportional to $\sin[L/(2r^2)]$.  The two exponentials now have equal modulus.  A geometric saddle therefore requires both an admissible integration contour and a choice of oscillatory branch.
\item \emph{Two bilinears.}  Let $Z_\Gamma(b,\boldsymbol P)$ denote the chiral Virasoro TQFT amplitude of a specified embedded, framed graph $\Gamma$.  Define $Z_\Gamma^\vee(b,\boldsymbol P)$ as the chiral amplitude of the mirror image of the framed graph, with its twists and crossings reversed and with $b$ and the momentum labels held fixed.  In the real-$b$ normalization, framing reversal gives the complex-conjugate chiral amplitude, $Z_\Gamma^\vee(b,\boldsymbol P)=\overline{Z_\Gamma(b,\boldsymbol P)}$ \cite{Hartman,CTV}.  Analytic continuation of the real-domain product and Hermitian conjugation of the continued coefficient therefore give two different bilinears,
\begin{equation*}
\begin{aligned}
\big[Z_\Gamma(b)Z_\Gamma^\vee(b)\big]_{\rm cont}
 &=Z_\Gamma(b_{\rm dS})Z_\Gamma^\vee(b_{\rm dS}),\\
 Z_\Gamma(b_{\rm dS})\overline{Z_\Gamma(b_{\rm dS})}
 &=\lvert Z_\Gamma(b_{\rm dS})\rvert^2,
 \qquad b_{\rm dS}=re^{\ii\pi/4}.
\end{aligned}
\end{equation*}
For the second pairing, complex conjugation maps $b_{\rm dS}$ to $b_{\rm dS}^{*}=-\ii b_{\rm dS}$, maps $c(b_{\rm dS})$ to $c(b_{\rm dS})^{*}=26-c(b_{\rm dS})$, and conjugates the momentum contours.  In the unbraided tetrahedral graph $Z_\Gamma^\vee=Z_\Gamma=\{6j\}_{\rm Vir}$, so the first bilinear is $\{6j\}_{\rm Vir}^2$ and carries the continued hyperbolic-volume phase, while the second is $\lvert\{6j\}_{\rm Vir}\rvert^2$.  The former supplies the OPE statistical coefficient; the latter type of Hermitian pairing underlies the moduli-integrated cosmological observables \cite{CEM,CLS,Matrix}.
\end{enumerate}
These changes motivate two complementary calculations.  The boundary channel expansion can test the proposed statistical interpretation, while the internal-label integrals test whether the continued amplitudes admit a geometric saddle.  The second question requires an analysis of integration contours and saddles beyond the algebraic recoupling identities.

\subsubsection*{From compact amplitudes to OPE moments}

The asymptotic Virasoro algebra and the flat Chern--Simons description organize the boundary data of three-dimensional gravity \cite{BrownHenneaux,AchucarroTownsend,Witten88}.  Related geometric crossing kernels occur for global conformal symmetry \cite{KrasnovRacah,LiuPerlmutter6j}.  The tetrahedral fourth-order OPE contribution is governed by a Virasoro $6j$ symbol \cite{BdBL,dBLPS}, and Virasoro TQFT evaluates the corresponding manifold with its embedded, framed graph \cite{CEZ,CEZ2}.  Removing the asymptotic regions strips off the three-point normalizations and conformal-block propagation, leaving the compact amplitude \cite{Hartman,CTV}.

Let $\Gamma$ be a trivalent graph embedded in a closed auxiliary manifold $M_E$.  A framing specifies the twisting of a ribbon along each edge.  Removing small tubes around the edges and balls around the vertices gives the compact region $M=M_E-N(\Gamma)$.  The curves $\gamma$ wind once around the removed edge tubes on its finite boundary.  For this specified filling, the fixed-length amplitude $Z_L$ and fixed-angle amplitude $Z_A$ obey \cite{Hartman,CTV}
\begin{equation}
 Z_L=Z_{\rm Vir}Z_{\rm Vir}^{\vee},
 \qquad Z_A=S\,Z_L=Z_{\rm CTV}.
 \label{eq:introforward}
\end{equation}
Here $S$ acts on every external graph label.  It squares to the identity on positive real momenta.  The framing reversal $\vee$ was defined above; in the continued fixed-length amplitude both graph factors retain the same complex $b$ and label contour, as required by the real-domain CTV relation.

Our use of this relation starts with the quantum fixed-angle input and computes
\begin{equation}
 Z_L^{\rm cont}(M,\gamma)
 =\big[S^{-1}Z_{\rm CTV}(M_E,\Gamma)\big]_{\rm cont},
 \qquad b=re^{\ii\pi/4}.
 \label{eq:introbulkinverse}
\end{equation}
Here $r>0$.  The brackets mean that the real-domain transform is performed first and its result continued with the specified contours and residues.

Let $\Eds$ denote the proposed average of OPE coefficients, normalized by $\Eds[1]=1$.  We specify its contributions by continuing the real-$b$ compact amplitudes.  At each graph vertex $v$, $C_v$ is the OPE coefficient for its three incident labels, divided by the square root of the leading real-$b$ variance.  The subscript $c$ denotes the connected part of the average.  For the selected graph contributions considered below, we assign
\begin{equation}
 \boxed{\left.\Eds\!\left[\prod_{v\in V(\Gamma)}C_v\right]_c
 \right|_{(M_E,\Gamma)}
 =\nu_\Gamma Z_L^{\rm cont}(M,\gamma)
 =\nu_\Gamma\big[Z_{\rm Vir}Z_{\rm Vir}^{\vee}\big]_{\rm cont}.}
 \label{eq:inversedictionary}
\end{equation}
The multiplicity $\nu_\Gamma$ counts inequivalent ways of reattaching the removed asymptotic three-point regions, called $C_0$ flares.  It multiplies the contribution of one compact filling.  It is two for the linked handcuff and one for the other generic, distinctly labelled elementary examples \cite{Hartman}.  To assemble a full connected moment, one must also specify relative filling weights and subtract products of lower moments.

\hnew{For the tetrahedral $K_4$ filling, the continued $6j^2$ coefficient carries an oscillatory phase given by the hyperbolic cell volume. Using the Schl\"afli identity and \vtwo{the Lorentzian data from which we recover this core,} we show that its angle derivatives recover the rapidities $\lambda_e=\ell(\gamma_e)=2\ell_e$ of the boosts around the initial arcs. The non-Gaussian Euclidean OPE statistics thereby determine conjugacy-invariant Lorentzian holonomy data \cite{Hartman,Mess,ABBGLSS,BenedettiBonsante}.} \vtwo{The geometric reconstruction starts from the marked $K_4$ spacetime. Its initial arcs recover the hyperideal Gram matrix, its boosts \aidrev{test the resulting edge lengths}, and its identified pants seams fix the four truncation-face pairings.} The resulting pleated surface is the boundary of the same doubled generalized hyperbolic tetrahedron that defines the compact CTV amplitude. The continuation supplies a complex-saddle cap, not a second tessellation of the real Lorentzian spacetime.

Sewing imposes a different test.  At sixth order, joining two tetrahedral contractions produces a triangular prism, and changing the channel on a joining edge relates it to the $K_{3,3}$ graph \cite{dBLP,CEZ2,UniversalTopologies,CrossingOPE,CTV}.  A consistent moment assignment must respect this relation, including the internal-label integral in the $K_{3,3}$ amplitude.  Repeated OPE labels also produce torus channels, for which exchanging the two torus cycles gives an additional constraint on the same statistics.

The internal integral makes this a test of the dS continuation as well as of sewing.  The known $6j$ asymptotics describe individual cells \cite{TV,LMSWY,MY}; applying them inside an integral additionally requires control of its contour and of the approximation near the saddle.  In the metric description, smoothly glued cells must fill a total angle $2\pi$ around an internal edge \cite{Hartman}.  We therefore examine whether an allowed dS integration cycle selects this zero-deficit condition, using the symmetric $K_{3,3}$ integral as a concrete example.

Positive-cosmological-constant state sums have older precedents.  The compact Turaev--Viro model has a finite root-of-unity representation set \cite{TuraevViro,OoguriSasakura,MizoguchiTada,Barrett95}, and its graph observables and Fourier duality were constructed in \cite{BGIM}.  Their interpretation as geometrical measurements in three-dimensional quantum gravity was explored in \cite{Barrett2003}.  A possible de Sitter interpretation of these observables was suggested in \cite{Hartman}.  Here the representation labels are continuous.  On the real-$b$ starting contour, their measure $\rho_0(P)\dd P$ has infinite total weight,
\begin{equation}
 \int_0^\infty\dd P\,\rho_0(P)=\infty.
 \label{eq:infiniteCTVstates}
\end{equation}
Its continuation requires integration-cycle data as well as local recoupling identities.

\subsubsection*{The cosmological state and the statistical interpretation}

We use the complex Liouville state prepared by the topological initial condition of \cite{CEM}.  Its channel density, built from the DOZZ three-point coefficients \cite{DO,ZZ,SpectralFunction}, provides the reference for the boundary extraction.

Consider four scalar insertions with equal external momenta $P_0$.  Let $P$ label the intermediate primary of the $s$ channel in which the state is expanded, and $P_t$ the intermediate primary of a fixed $t$-channel component.  The reference state sums over intermediate primaries with density $\rho_s(P)$.  A single unweighted $t$-channel block pair instead gives a coefficient $\rho_{st}(P;P_t)$ after projection onto equal left- and right-moving $s$-channel momenta.  At primary level and real positive momenta these coefficients are
\begin{equation*}
 \rho_s(P)=\rho_0(P)C_0(P_0,P_0,P)^2,\qquad
 \rho_{st}(P;P_t)=\frac{\mathbb F_{P_tP}^{\,2}}{2P}.
\end{equation*}
The first expression is the Liouville sewing density \cite{CEZ,CEM}.  In the second, $\mathbb F_{P_tP}$ changes a $t$-channel block to the $s$ channel, and the two chiral blocks supply its square.  The twist projection gives $\delta(P_L^2-P_R^2)$; integrating over positive $P_R$ gives $1/(2P)$ at $P_L=P$.  The known conversion from the fusion kernel to Racah normalization is \cite{CTV,PT,TV}
\begin{equation*}
 \mathbb F_{P_tP}
 =\rho_0(P)\frac{C_0(P_0,P_0,P)}{C_0(P_0,P_0,P_t)}
 \begin{Bmatrix}P_0&P_0&P_t\\P_0&P_0&P\end{Bmatrix}_{\!\rm Vir}.
\end{equation*}
Substituting these expressions gives
\begin{equation}
 \frac{2P\,C_0(P_0,P_0,P_t)^2}{\rho_0(P)}
 \frac{\rho_{st}(P;P_t)}{\rho_s(P)}
 =\begin{Bmatrix}P_0&P_0&P_t\\P_0&P_0&P\end{Bmatrix}_{\!\rm Vir}^{\!2}.
 \label{eq:introresult}
\end{equation}
This algebraic relation identifies the coefficient that a boundary extraction should recover.  The task is to obtain it from the moduli dependence of the wavefunction while retaining all descendants.  The OPE degeneration develops a long conformal collar \cite{Wolpert}.  We use a radial transform to separate primary propagation from descendant thresholds.  For the crossed preparation, the twist projection and primary threshold monodromy are defined at real momenta before the extracted coefficient is continued to dS.  The resulting ratio can then be compared with the tetrahedral compact amplitude.  The analogous torus question uses the punctured modular kernel \cite{HJS,PostTsiares}.

The Virasoro cylinder evaluated at $(z,\bar z)$ gives the Liouville cosmological state \cite{CEM}.  The Gaussian contraction of two CFT four-point functions uses the same cylinder with chiral moduli paired across the correlators \cite{CCHM}.  Changing one sewing channel inserts the Virasoro fusion kernel, whose Racah normalization gives the tetrahedral $6j$ \cite{PT,TV}.  The two chiral factors then supply the square.  This common cylinder gives a concrete reason to compare the cosmological state with random-CFT correlations.

A normalizable vacuum obstructs the use of a compact CFT partition function as a cosmological wavefunction \cite{CLM,CEM}. The collective relation above makes a different identification. A fixed vacuum exchange cancels from a connected correlation of two four-point functions, and the remaining paired contribution contains the Liouville cylinder. We follow this cancellation and its channel change explicitly, keeping the vacuum condition on the selected cosmological state distinct from the conditions on individual CFT samples.

The proposal and the established integrated-observable dictionary use different contractions of these data,
\begin{equation}
\begin{aligned}
 \text{channel coefficients at fixed boundary moduli}
 &\ \longrightarrow\ \text{moments of a random Euclidean CFT}_2,\\
 \text{Hermitian norm integrated over moduli}
 &\ \longrightarrow\ \text{a double-scaled matrix integral}.
\end{aligned}
\label{eq:randomcftconjecture}
\end{equation}
The second line is the proposed duality for moduli-integrated cosmological correlators \cite{CEM}.  The first is the statistical interpretation we investigate through channel extraction, sewing and Lorentzian reconstruction.  In this dictionary the moments built from two chiral amplitudes at the same $b$ are distinct from the mixed moments that contain complex-conjugate coefficients, and crossing and modular invariance constrain the relative weights of different fillings.  Statistical crossing and topology sums constrain these data \cite{UniversalTopologies,CrossingOPE}; related microscopic proposals use matrix-tensor models and CFT measures \cite{BdBJNS,JRW,CFTMeasure}.

Section~\ref{sec:review} reviews the gravity dictionary, OPE statistics and the cosmological state. Section~\ref{sec:boundaryamp} extracts marked channel coefficients, and section~\ref{sec:gravity} identifies their compact cores, semiclassical phases and associated Lorentzian spacetimes, including the marked inverse from the latter to the same pleated core boundary. Section~\ref{sec:bulkinverse} assigns the OPE moments and tests statistical sewing; appendix~\ref{sec:sixth} contains the six-vertex calculations. Section~\ref{sec:physicalinterpretation} develops the random Euclidean CFT interpretation, from the Euclidean data at future infinity to the Lorentzian readout of OPE phases and the vacuum compatibility of the collective correlation.

{Appendix~\ref{ssec:sphere-tree-extension} gives the higher-point coefficient normalization and its continuation to fixed-angle amplitudes.}

\FloatBarrier
\Needspace{7\baselineskip}
\section{Boundary data, OPE statistics and the de Sitter state}\label{sec:review}\label{ssec:review}

We begin with the AdS construction that relates a compact gravitational region to OPE statistics \cite{Hartman,CTV}.  On its finite boundary we fix either angles or lengths; on the CFT side we specify which OPE coefficients are averaged.  The tetrahedral example then shows how an index contraction determines a graph and, after choosing its embedding, a compact region.  For de Sitter space we also need data at future infinity.  The last subsection introduces the cosmological state whose conformal-block coefficients we will extract, and distinguishes those coefficients from its gravitational norm \cite{CEM}.
\subsection{Gravity on a compact region}\label{ssec:bc}

We consider pure gravity with negative cosmological constant on a compact three-manifold $M$.  Its boundary $\partial M$ consists of closed two-dimensional surfaces at finite distance.  These are the surfaces on which we impose the gravitational boundary conditions.  In units of the AdS radius, the Euclidean path integral is weighted by $e^{-I_0}$, with
\begin{equation}
I_0=-\frac{1}{16\pi G_N}\int_M\dd^3x\sqrt g\,(R+2)
 -\frac{1}{8\pi G_N}\int_{\partial M}\dd^2x\sqrt h\,K .
\label{eq:revI0}
\end{equation}
Here $g$ is the bulk metric, $R$ its Ricci scalar, and $G_N$ Newton's constant.  The metric induced on $\partial M$ is $h_{ab}$; its extrinsic curvature $K_{ab}$ measures how that surface bends in the bulk, and $K=h^{ab}K_{ab}$ is its trace, with the normal pointing out of $M$.  The Gibbons--Hawking--York (GHY) term cancels normal derivatives of the metric variation, leaving a boundary term proportional to $\delta h_{ab}$.  We will use that term to impose a free boundary metric away from specified curves.  No asymptotic AdS counterterm is included at this finite boundary.  To match the compact amplitude to CFT data, we use the Brown--Henneaux central charge $c$ and parameterize it by $b$,
\begin{equation}
c=\frac{3}{2G_N}=1+6Q^2,\qquad Q=b+b^{-1}.
\label{eq:revc}
\end{equation}
The semiclassical limit is $b\to0$, in which $c\simeq6/b^2$.  

The interior of $M$ is smooth, but distinct faces of this same boundary $\partial M$ can meet at an angle.  We denote their meeting curves by $\gamma_i$, $i=1,\ldots,n$.  Locally, two faces share $\gamma_i$, just as two sides of a box share an edge.  A cross-section perpendicular to the curve shows the faces as two rays.  Their angle measured through $M$ is the interior dihedral angle $\psi_i$; a smooth join has $\psi_i=\pi$.  

We choose the $\gamma_i$ to be disjoint simple closed curves, with no endpoints, self-intersections, or intersections with one another.  They are essential, meaning noncontractible on $\partial M$, and mutually non-isotopic, meaning that no two can be continuously deformed into one another there.  The genus $g$ counts the surface's handles.  For a closed component with $g\geq2$, at most $3g-3$ such curves can be chosen; cutting along a maximal collection gives $2g-2$ three-holed spheres, called pairs of pants.  Their boundary circles are called cuffs.

The example sketched in~\cite{Hartman} has a genus-two compact boundary.  It enters the punctured-torus calculation of section~\ref{sec:torus}; the torus there is the surface carrying the CFT one-point function, distinct from this compact boundary,
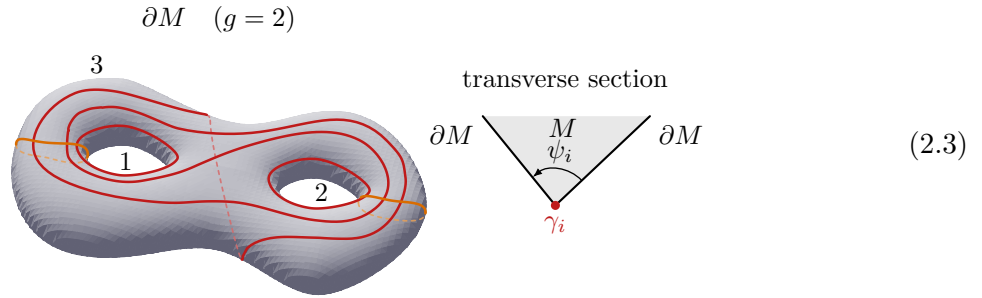
\begin{equation}
\begin{gathered}

\end{gathered}
\label{eq:boundarycurvesketch}
\end{equation}
 The three red curves labelled $1,2,3$ are $\gamma_1,\gamma_2,\gamma_3$; cutting along them gives two pairs of pants.  The orange curves bound disks inside $M$ and specify the bulk filling.  The right-hand sketch is a section perpendicular to a red curve, where $\gamma_i$ appears as a point, and $\psi_i$ is measured through the shaded interior of $M$.
 
Our first boundary condition fixes the interior dihedral angles $\psi_i$
along the curves $\gamma_i$, while leaving their lengths and the boundary
metric away from them free to vary. A classical saddle must make the action
stationary under all such allowed variations. For a metric variation
supported on a smooth boundary face, away from the curves, imposing the
bulk Einstein equation leaves
\begin{equation}
\left.\delta I_0\right|_{\rm face}
 =\frac{1}{16\pi G_N}\int_{\rm face}\dd^2x\sqrt h\,
 (K^{ab}-Kh^{ab})\,\delta h_{ab}.
\label{eq:s2freeboundaryvariation}
\end{equation}
Since $\delta h_{ab}$ is arbitrary there, stationarity requires
$K^{ab}-Kh^{ab}=0$. Contracting with $h_{ab}$ gives $K-2K=0$, so
$K=0$ and hence $K_{ab}=0$.

Each smooth face has vanishing extrinsic curvature and is therefore totally geodesic.  The bulk Einstein equation makes the three-dimensional saddle locally hyperbolic, so a totally geodesic face inherits intrinsic Gaussian curvature $-1$.  The equations at the joins also make each $\gamma_i$ a geodesic \cite{Hartman}.  Two adjacent faces of the same boundary meet along that curve at the prescribed interior angle $\psi_i$, as in the transverse sketch above.

Our second boundary condition fixes the geodesic lengths $\ell_i$ instead.  The smooth faces are again totally geodesic, but the angles at their joins are now determined by the classical solution.  We denote the two path integrals by $Z_A$ and $Z_L$, where $Z_A$ sums metrics with the prescribed angles, and $Z_L$ sums metrics with the prescribed lengths.  The topology of $M$ is held fixed in each path integral.  A choice of bulk manifold with the required boundary is called a filling; the boundary data alone need not select a unique filling.

The angle condition requires a distinction between two contributions at a
corner.  A non-smooth boundary already contributes to the GHY term in $I_0$, whether or not we add a boundary source.  To see why,
round a sharp join along $\gamma_i$ with a thin strip of transverse radius
$\epsilon$.  Let $t$ be proper length along the curve and $\alpha$ the
angle through which the outward normal turns across the strip.  Temporarily
write $\varphi_i$ for the geometric interior angle, to distinguish it from
the angle $\psi_i$ prescribed as boundary data.  On the rounded strip the
leading area element and extrinsic curvature are
\begin{align*}
\sqrt h\,\dd^2x&\simeq\epsilon\,\dd\alpha\,\dd t,
&K&\simeq\frac1\epsilon.
\end{align*}

Although the curvature diverges as the strip shrinks, its integral remains
finite,
\begin{align*}
\lim_{\epsilon\to0}\int_{\mathrm{round}(\gamma_i)}\dd^2x\sqrt h\,K
&=\lim_{\epsilon\to0}\int_{\gamma_i}\dd t
  \int_0^{\pi-\varphi_i}\dd\alpha\,
  \epsilon\frac1\epsilon\\
&=(\pi-\varphi_i)\ell_i,
\qquad \ell_i=\int_{\gamma_i}\dd t.
\end{align*}
The transverse integral measures the exterior turning angle; integrating
along the remaining direction gives the length of the corner.  Thus the
GHY action contains the geometric Hayward contribution
$-c(\pi-\varphi_i)\ell_i/(12\pi)$.  For a smooth join
$\varphi_i=\pi$, this corner contribution vanishes.

To select a prescribed angle while allowing the lengths $\ell_i$ to vary, a separate term supported on the curves is needed \cite{Hartman},
\begin{equation}
I_A=I_0+\frac{c}{12\pi}\sum_i(\pi-\psi_i)\int_{\gamma_i}\dd t .
\label{eq:revactions}
\end{equation}
Here $\psi_i$ is held fixed in the variation; the integral is the length
$\ell_i$. 
The variation of the boundary metric remains unrestricted away from the
curves, giving $K_{ab}=0$ on each smooth face.  At the curve the
boundary equation is, in the sharp-corner limit,
\begin{equation}
K=(\pi-\psi_i)\,\delta(s).
\label{eq:revcorner}
\end{equation}
Here $s$ measures proper distance across $\gamma_i$ within $\partial M$.
The delta function is the limit of the large curvature on the rounded
strip.  Indeed, before imposing the
boundary condition its coefficient is $\pi-\varphi_i$; equating it to
the source coefficient $\pi-\psi_i$ fixes $\varphi_i=\psi_i$.

The saddle condition can also be checked directly by varying the lengths
along a family of hyperbolic solutions with totally geodesic boundary
faces.  Writing $V=\operatorname{Vol}(M)$, the bulk action and the geometric GHY corner term give
$I_0^{\rm on\text{-}shell}=cV/(6\pi)-c\sum_i(\pi-\varphi_i)\ell_i/(12\pi)$.
Using the Schl\"afli identity
$\delta V=-\tfrac12\sum_i\ell_i\,\delta\varphi_i$, we obtain
\begin{align*}
\delta I_0^{\rm on\text{-}shell}
&=\frac{c}{6\pi}\delta V
  -\frac{c}{12\pi}\sum_i
     \big[(\pi-\varphi_i)\delta\ell_i-\ell_i\delta\varphi_i\big]\\
&=-\frac{c}{12\pi}\sum_i(\pi-\varphi_i)\delta\ell_i,\\
\delta I_A^{\rm on\text{-}shell}
&=\delta I_0^{\rm on\text{-}shell}
  +\frac{c}{12\pi}\sum_i(\pi-\psi_i)\delta\ell_i\\
&=\frac{c}{12\pi}\sum_i(\varphi_i-\psi_i)\delta\ell_i.
\end{align*}
Because the lengths are free to vary, stationarity requires
$\varphi_i=\psi_i$. 

For the length constraint, we restrict the metric variations by $\int_{\gamma_i}\dd t=\ell_i$ and require each curve to be extremal.  The action is
\begin{equation}
I_L=I_0+\frac{c}{12}\sum_i\ell_i .
\label{eq:revILaction}
\end{equation}
Since $\ell_i$ is fixed, the added term has zero variation.  It changes the weight $e^{-I_L}$ but not the classical equations.  Its usefulness for joining regions will follow from the on-shell action below.

On shell, $R_{\mu\nu}=-2g_{\mu\nu}$ gives $R=-6$.  At the
fixed-angle saddle $\varphi_i=\psi_i$, so the bulk and GHY pieces of
$I_0$ are
\begin{equation}
\begin{aligned}
I_{\rm bulk}&=-\frac{(-6+2)V}{16\pi G_N}=\frac{cV}{6\pi},\\
I_{\rm GHY}&=-\frac{1}{8\pi G_N}\sum_i(\pi-\psi_i)\ell_i
 =\frac{c}{12\pi}\sum_i(\psi_i-\pi)\ell_i.
\end{aligned}
\label{eq:s2actionpieces}
\end{equation}
The added fixed-angle term in \eqref{eq:revactions} is equal and
opposite to the GHY corner term.  Consequently,
\begin{equation}
I_A^{\rm on\text{-}shell}=\frac{c}{6\pi}\operatorname{Vol}(M).
\label{eq:revonshell}
\end{equation}
In $I_L$, the added constant cancels only the $-\pi\ell_i$ part of the corner term.  The result is
\begin{equation}
I_L^{\rm on\text{-}shell}=\frac{c}{6\pi}\Big(\operatorname{Vol}(M)+\frac12\sum_i\ell_i\psi_i\Big).
\label{eq:revonshellL}
\end{equation}
We can now explain gluing.  Choose a boundary face of $M_1$ and a boundary face of $M_2$ with the same intrinsic metric, and identify their corresponding points with opposite outward orientations.  This operation joins the regions into $M_1\cup M_2$; the identified faces cease to be part of its outer boundary.  For the totally geodesic faces considered here, the metrics join smoothly across the face.

Along the rim of the identified face, a transverse section consists of the two wedges placed side by side.  If the rim remains on the outer boundary, its angle is the sum of the two wedge angles.  Thus
\begin{equation}
\begin{aligned}
\operatorname{Vol}(M_1\cup M_2)&=\operatorname{Vol}(M_1)+\operatorname{Vol}(M_2),\\
\psi_i&=\psi_i^{(1)}+\psi_i^{(2)}.
\end{aligned}
\label{eq:s2gluingdata}
\end{equation}
If no edge becomes internal, and the remaining boundary curves retain their fixed-length markings, \eqref{eq:revonshellL} gives
\begin{equation}
I_L^{\rm on\text{-}shell}(M_1\cup M_2)
 =I_L^{\rm on\text{-}shell}(M_1)+I_L^{\rm on\text{-}shell}(M_2).
\label{eq:s2gluingadditive}
\end{equation}
An internal edge is a former boundary edge that lies in the interior after gluing, so that the wedges around it now form a full circle.  For a smooth join their angles sum to $2\pi$.  An edge of geometric length $\ell$ then contributes to the sum of the separate actions by
\begin{equation}
\frac{c\ell}{12\pi}\sum_{\rm pieces}\psi
 =\frac{c\ell}{6}=\frac{\ell}{4G_N}.
\label{eq:s2internalgluingterm}
\end{equation}
There is no boundary term on this edge in the joined region, so it must be subtracted,
\begin{equation}
\begin{aligned}
I_L^{\rm on\text{-}shell}(M)
 &=\sum_{\rm pieces} I_L^{\rm on\text{-}shell}
   -\frac{\ell}{4G_N},\\
e^{-I_L^{\rm on\text{-}shell}(M)}
 &=e^{\ell/(4G_N)}\prod_{\rm pieces}e^{-I_L^{\rm on\text{-}shell}}.
\end{aligned}
\label{eq:s2internalgluingweight}
\end{equation}
This is the compact-region gluing rule \cite{Hartman}; with several new internal edges there is one subtraction for each.  The exponential is the additional edge weight often called the entropy factor.  In the quantum state sum, its analogue is the measure for the internal label, explained in subsection~\ref{ssec:core}.

The difference between $I_L$ and $I_A$ is $(c/12\pi)\sum_i\ell_i\psi_i$.  This is a Legendre transform.  Adding the length--angle term changes which variable is held fixed.  The Schl\"afli identity shows this directly by varying a smooth family of hyperbolic solutions with the same face arrangement,
\begin{equation}
\dd\operatorname{Vol}(M)=-\frac12\sum_i\ell_i\,\dd\psi_i,
\qquad
\dd\!\left(\operatorname{Vol}(M)+\frac12\sum_i\ell_i\psi_i\right)
=\frac12\sum_i\psi_i\,\dd\ell_i.
\label{eq:reviewschlafli}
\end{equation}
For the hyperbolic cells these are the identities reviewed in \cite{Hartman}.  On a smooth gluing, internal-edge terms cancel because their total angle stays $2\pi$.  The conjugate variables are therefore
\begin{equation}
\frac{\partial I_A^{\rm on\text{-}shell}}{\partial\psi_i}
 =-\frac{c}{12\pi}\ell_i,
\qquad
\frac{\partial I_L^{\rm on\text{-}shell}}{\partial\ell_i}
 =\frac{c}{12\pi}\psi_i.
\label{eq:s2conjugatevariables}
\end{equation}
These derivatives compare neighboring classical solutions.  The path integral also includes off-shell metrics, which need not satisfy the Einstein equation.

The quantum dictionary encodes each curve's boundary datum by a complex number $P_i$, called the Liouville momentum.  Its CFT interpretation is given in the next subsection; here it is a boundary label.  In the fixed-length description its relation to the prescribed length is
\begin{equation}
\ell_i=4\pi bP_i .
\label{eq:revlabels}
\end{equation}
For real $b>0$, positive real $P_i$ therefore gives a positive real length.  The fixed-angle description uses a different relation,
\begin{equation}
\psi_i=2\pi\ii bP_i .
\label{eq:revangle}
\end{equation}
An angle $0<\psi_i<\pi$ corresponds to $P_i=-\ii\psi_i/(2\pi b)$ on this real-$b$ starting slice.  We use the same symbol $P_i$ for the representation parameter in the two descriptions, but real length data and real angle data occupy different momentum contours.  Section~\ref{ssec:dsproblem} will specify the continued contours and the reflection convention for positive angles in the de Sitter calculation.

At the quantum level, the change from fixed lengths to fixed angles is an integral transform \cite{Hartman,CTV}.  Its kernel is symmetric under momentum reflection and therefore includes both $P_i'$ and $-P_i'$.  Written in exponential form, it is
\begin{equation}
\begin{aligned}
Z_A(M,\gamma(P'))
 &=\int_{\mathbb R_+^n}\dd P\,
 \prod_{i=1}^n\sqrt2\big(e^{4\pi\ii P_i'P_i}
                         +e^{-4\pi\ii P_i'P_i}\big)\\
 &\hspace{3.5cm}\times Z_L(M,\gamma(P)).
\end{aligned}
\label{eq:revLaplace}
\end{equation}
Here $P_i$ labels the length in $Z_L$, and $P_i'$ the prescribed angle $\psi_i'$.  For $P_i'=-\ii\psi_i'/(2\pi b)$ with positive angles and lengths, the growing exponential is the leading branch at the interior semiclassical saddle.  Since $4\pi\ii P_i'P_i=\psi_i'\ell_i/(2\pi b^2)$ and $c\simeq6/b^2$, its classical saddle is
\begin{equation}
\begin{aligned}
Z_A(\psi')&\sim\int\dd^n\ell\,
 \exp\!\left[-I_L^{\rm on\text{-}shell}(\ell)
       +\frac{c}{12\pi}\sum_i\psi_i'\ell_i\right],\\
0&=-\frac{\partial I_L^{\rm on\text{-}shell}}{\partial\ell_i}
       +\frac{c}{12\pi}\psi_i'
 =\frac{c}{12\pi}\big(\psi_i'-\psi_i(\ell)\big).
\end{aligned}
\label{eq:s2anglesaddle}
\end{equation}
Thus the transform selects the lengths whose classical angles equal the prescribed ones, and its exponent becomes $-I_A^{\rm on\text{-}shell}$.  At finite $b$ both exponentials are retained, and each bracket in \eqref{eq:revLaplace} is exactly $2\sqrt2\cos(4\pi P_i'P_i)$, the kernel used in \eqref{eq:revZA}.  Their continuation follows the contour prescriptions of \cite{Hartman,CTV}.
\subsection{The CFT ensemble and OPE statistics}\label{ssec:cftensemble}

We begin with the real, unitary CFT conventions used in the AdS ensemble.  A primary operator $\mathcal O_i$ creates a state with left and right conformal weights $(h_i,\bar h_i)$.  Its descendants are obtained by applying the Virasoro generators of local conformal transformations.  In the left-moving sector,
\begin{equation}
\begin{aligned}
L_0|i\rangle&=h_i|i\rangle,\qquad L_{n>0}|i\rangle=0,\\
L_{-n_1}\cdots L_{-n_k}|i\rangle
&\ \hbox{has weight }h_i+\sum_{a=1}^k n_a,
\end{aligned}
\label{eq:s2primarystates}
\end{equation}
with $n_a>0$; the right-moving generators act in the same way.  A primary together with all these descendants is one conformal family.  Choose an orthonormal Hermitian basis of primaries.  The two-point normalization and the three-point coefficient are
\begin{equation}
\begin{aligned}
\langle\mathcal O_i(z,\bar z)\mathcal O_j(0)\rangle
 &=\frac{\delta_{ij}}{z^{2h_i}\bar z^{2\bar h_i}},\\
c_{ijk}&=\langle\mathcal O_k(\infty)\mathcal O_i(1)\mathcal O_j(0)\rangle.
\end{aligned}
\label{eq:s2cftcoefficients}
\end{equation}
The insertion at infinity includes the usual factors $z^{2h_k}\bar z^{2\bar h_k}$.  In this convention the operator product expansion (OPE) reads
\begin{equation}
\mathcal O_i(z,\bar z)\mathcal O_j(0)
 =\sum_k c_{ijk}\,z^{h_k-h_i-h_j}\bar z^{\bar h_k-\bar h_i-\bar h_j}
 \big[\mathcal O_k(0)+\hbox{Virasoro descendants}\big].
\label{eq:s2ope}
\end{equation}
The sum runs over primary operators; the descendant terms have coefficients fixed by Virasoro symmetry.  Thus $c_{ijk}$ is the dynamical coefficient multiplying the entire $k$th conformal family.  We parameterize the weights, dimension and spin by
\begin{equation}
h_i=\frac{Q^2}{4}+P_i^2,\qquad
\bar h_i=\frac{Q^2}{4}+\bar P_i^2,\qquad
\Delta_i=h_i+\bar h_i,\qquad \hbox{spin}=h_i-\bar h_i.
\label{eq:s2primaryweights}
\end{equation}
{We restrict to the above-threshold regime $P_i,\bar P_i\in\mathbb R_{>0}$ \cite{Hartman}.}
The two chiral sectors act on $z$ and $\bar z$ separately.  {The bar on $\bar P_i$ labels the right-moving sector.}  A star will denote complex conjugation, whereas an overline on a whole observable will denote an ensemble average.

The $c_{ijk}$ are fixed numbers in one CFT.  The ensemble of \cite{CCHM} instead assigns statistical weights to CFT data---the primary spectrum and OPE coefficients---at fixed central charge.  For an observable $X$ built from those data, the meaning of the average is
\begin{equation}
\overline X:=\int\dd\mu\,X,\qquad \int\dd\mu=1,
\qquad
\overline{|c_{ijk}|^2}=\int\dd\mu\,c_{ijk}c_{ijk}^*.
\label{eq:s2ensembleaverage}
\end{equation}
Here $\dd\mu$ denotes the normalized statistical weight {on CFT data}.  For example, $\overline{\rho}$ averages the spectra of the sampled data, whereas $\overline{G(z,\bar z)}$ averages correlators at fixed insertion positions.  The brackets $\langle\mathcal O\cdots\rangle$ are quantum expectation values within one CFT; they are evaluated before this ensemble average.  The leading model takes generic heavy coefficients to have $\overline{c_{ijk}}=0$, so their mean squared magnitude is their variance.  Whether these approximate moments extend to data obeying all CFT consistency conditions is the completion problem discussed after the statistical construction.

A different operation is to average nearby heavy states within one fixed CFT; this is the microcanonical average in \cite{CMMT}.  The ensemble prescription \cite{CCHM} adopts its universal asymptotic formula as a leading moment.  {The two operations share this leading formula.}  When discussing an OPE moment as a function of momenta, the external weights are held fixed and the indicated intermediate weights are conditioned on, as made explicit below.

We first need the number of primary states to sum over.  In a compact CFT the primary spectrum is discrete, and its heavy part is represented by the density
\begin{equation}
\rho(P,\bar P)=\sum_{i\,\mathrm{heavy}}\delta(P-P_i)\delta(\bar P-\bar P_i).
\label{eq:reviewprimarydensity}
\end{equation}
The delta functions turn a discrete sum into an integral.  For any test function $f$,
\begin{equation}
\sum_{i\,\mathrm{heavy}}f(P_i,\bar P_i)
 =\int_0^\infty\dd P\,\dd\bar P\,\rho(P,\bar P)f(P,\bar P).
\label{eq:s2densitysum}
\end{equation}
Only after averaging or coarse-graining over many levels do we replace $\rho$ by a smooth density.  To obtain its leading form, consider a torus with periods $1$ and $\tau$, where $\operatorname{Im}\tau>0$.  A character is the trace over one conformal family, counting the primary and its descendants with their propagation weights,
\begin{equation}
\chi_P(\tau)=\operatorname{Tr}_{P}q^{L_0-c/24},
\qquad q=e^{2\pi\ii\tau}.
\label{eq:s2charactertrace}
\end{equation}
The modular transformation $\tau\mapsto-1/\tau$ exchanges the torus's two cycles.  With $\eta(\tau)=q^{1/24}\prod_{n\geq1}(1-q^n)$, the non-vacuum and vacuum characters and their transform are \cite{CCHM,CMMT}
\begin{equation}
\begin{aligned}
\chi_P(\tau)
&=\frac{q^{P^2-1/24}}{\prod_{n\geq1}(1-q^n)}=\frac{q^{P^2}}{\eta(\tau)},
\\ \chi_{\mathbf1}(\tau)
&=q^{-c/24}\prod_{n=2}^{\infty}\frac{1}{1-q^n}
=\frac{q^{-Q^2/4}(1-q)}{\eta(\tau)},\\
\chi_{\mathbf{1}}(-1/\tau)&=\int_0^\infty\dd P\,\rho_0(P)\chi_P(\tau).
\end{aligned}
\label{eq:s2vacuumcharacter}
\end{equation}
Here $\tau$ is the torus modulus. The vacuum is invariant under translations, so \(L_{-1}|0\rangle=0\). Consequently, the vacuum character starts with descendants at level two, and the factor \(1-q\) removes the null submodule generated by \(L_{-1}|0\rangle\).
To obtain its transform, the Gaussian Fourier integral and $\eta(-1/\tau)=\sqrt{-\ii\tau}\,\eta(\tau)$ give the nondegenerate-character relation \cite{CMMT}
\begingroup
\[
 \chi_{P'}(-1/\tau)
 =\int_0^\infty\dd P\,2\sqrt2\cos(4\pi P'P)\,\chi_P(\tau),
 \qquad
 \chi_{\mathbf1}=\chi_{\ii Q/2}-\chi_{\ii(b-b^{-1})/2}.
\]
\endgroup
The second equality writes the vacuum's $1-q$ subtraction as two nondegenerate characters, using $Q^2-(b-b^{-1})^2=4$.  Substituting their imaginary momenta into the cosine kernel gives the two hyperbolic cosines below,
\begin{equation}
\begin{aligned}
\rho_0(P)&=2\sqrt2\big[\cosh(2\pi QP)-\cosh(2\pi(b-b^{-1})P)\big]\\
&=4\sqrt2\sinh(2\pi bP)\sinh(2\pi b^{-1}P).
\end{aligned}
\label{eq:s2vacuumdensitydifference}\eqalias{eq:revrho0}
\end{equation}
When the partition function after this exchange is dominated by the vacuum family, its inverse transform gives the leading heavy density $\overline{\rho(P,\bar P)}\approx\rho_0(P)\rho_0(\bar P)$, the Cardy density.  This counts primaries; descendants are already included in $\chi_P$.  Since $\dd h=2P\,\dd P$, the chiral density with respect to $\dd h$ is $\rho_0(P)/(2P)$.  The vacuum is separate, and light states and spin quantization contribute corrections.

To determine the leading variance, we compare two OPE expansions of the same four-point function.  A channel specifies which operator pairs are first replaced by their OPE.  For insertions at $(0,z,1,\infty)$, the $s$ channel pairs the first two and last two; the $t$ channel pairs the first with the last and the middle two.  The cross-ratio $z$ is the position left after fixing three insertions by conformal transformations.  Each conformal block $\mathcal F$ collects one exchanged primary and all its descendants.  For two distinct Hermitian scalar primaries, the $t$-channel expansion is
\begin{equation}
\begin{aligned}
G_{1122}(z,\bar z)
&=\langle\mathcal O_1(0)\mathcal O_1(z,\bar z)\mathcal O_2(1)\mathcal O_2(\infty)\rangle,\\
G_{1122}(z,\bar z)
&=\sum_k |c_{12k}|^2\,
\mathcal F_{P_k}^{(t)}(z)\,
\mathcal F_{\bar P_k}^{(t)}(\bar z).
\end{aligned}
\label{eq:reviewvariancecorrelator}
\end{equation}
For general external labels ordered as $(1,2,3,4)$ at $(0,z,1,\infty)$, the $s$-channel block starts as
\begin{equation}
\mathcal F_P^{(s)}(z)
 =z^{h(P)-h_1-h_2}\big[1+O(z)\big].
\label{eq:s2blockseries}
\end{equation}
The leading term propagates the primary.  The terms of order $z^n$ in the bracket collect level-$n$ descendants, with coefficients fixed by the central charge and the weights.  We write the right-moving block as $\mathcal F_{\bar P}^{(t)}(\bar z)$, with all external weights also right-moving, so no averaging over a block is implied.  The two-point normalization fixes $c_{11\mathbf{1}}=c_{22\mathbf{1}}=1$ and $c_{12\mathbf{1}}=0$.  Thus the $s$ channel of $G_{1122}$ contains the identity family with coefficient one, whereas the $t$ channel has no identity term.

Crossing symmetry requires the two channel expansions to describe the same correlator.  For the leading contribution considered in \cite{CCHM,Hartman}, the averaged correlator is approximated by the product of the $s$-channel identity blocks.  A fusion transform rewrites a block in the other channel; its integral kernel is the continuous-label change-of-basis coefficient.  For the identity block the transform is
\begin{equation}
\mathcal F_{\mathbf{1}}^{(s)}(z)
=\int_0^\infty\dd P\,\rho_0(P)C_0(P_1,P_2,P)\,
\mathcal F_P^{(t)}(z).
\label{eq:reviewidentityfusion}
\end{equation}
Thus $C_0$ is the identity fusion coefficient after removing $\rho_0$ \cite{CMMT,CTV}.  The vacuum block includes its descendants.

We now compare the two expansions at fixed intermediate weights.  In continuum notation $c_{12P}$ means $c_{12k}$ for an intermediate state with labels $(P,\bar P)$.  Its conditional mean squared coefficient is the mean OPE-weighted number of states divided by the mean number of states in the same small momentum window.  In density notation this is
\begin{equation}
\overline{|c_{12P}|^2}:=
\frac{\displaystyle\overline{\sum_{k\,\mathrm{heavy}}|c_{12k}|^2
 \delta(P-P_k)\delta(\bar P-\bar P_k)}}
 {\displaystyle\overline{\rho(P,\bar P)}}.
\label{eq:s2weightedensity}
\end{equation}
The numerator and denominator are understood after integration over that window, or against the same narrow smooth test function.  With this meaning of $\overline{|c_{12P}|^2}$, the two leading expansions give
\begin{equation}
\begin{aligned}
\overline{G_{1122}}
 &\approx\int_0^\infty\dd P\,\dd\bar P\,
 \overline{\rho(P,\bar P)}\,\overline{|c_{12P}|^2}
 \mathcal F_P^{(t)}(z)\mathcal F_{\bar P}^{(t)}(\bar z)\\
 &\approx\int_0^\infty\dd P\,\dd\bar P\,
 \rho_0(P)\rho_0(\bar P)C_0(P_1,P_2,P)C_0(\bar P_1,\bar P_2,\bar P)
 \mathcal F_P^{(t)}(z)\mathcal F_{\bar P}^{(t)}(\bar z).
\end{aligned}
\label{eq:reviewcontinuumsum}
\end{equation}
Matching the coefficients of the same blocks gives
\begin{equation}
\overline{|c_{12P}|^2}\approx
 \frac{\rho_0(P)\rho_0(\bar P)}{\overline{\rho(P,\bar P)}}
 C_0(P_1,P_2,P)C_0(\bar P_1,\bar P_2,\bar P).
\label{eq:s2variancequotient}
\end{equation}
Replacing the mean density by its leading Cardy form cancels the ratio. Thus identity-block dominance determines the variance in the same approximation. Relabelling the three primaries gives the ensemble inputs
\begin{align}
\overline{|c_{ijk}|^2}&\approx C_0(P_i,P_j,P_k)\,C_0(\bar P_i,\bar P_j,\bar P_k),\nonumber\\
 \overline{\rho(P,\bar P)}&\approx\rho_0(P)\rho_0(\bar P).\label{eq:revC0}
\end{align}
Thus $\rho_0(P)\rho_0(\bar P)$ counts the available states, while $C_0(P_i,P_j,P_k)C_0(\bar P_i,\bar P_j,\bar P_k)$ is their mean squared OPE coefficient \cite{Hartman}\cite{CCHM}.

For completeness, the exact function that fixes our normalization is
\begin{equation}
C_0(P_1,P_2,P_3)
 =\frac{\Gamma_b(2Q)}{\sqrt2\,\Gamma_b(Q)^3}
 \frac{\displaystyle\prod_{\sigma_1,\sigma_2,\sigma_3=\pm1}
 \Gamma_b\!\left(\frac Q2+\ii\sigma_1P_1+\ii\sigma_2P_2+\ii\sigma_3P_3\right)}
 {\displaystyle\prod_{j=1}^3\Gamma_b(Q+2\ii P_j)\Gamma_b(Q-2\ii P_j)}.
 \label{eq:Czero}
\end{equation}
Here $\Gamma_b$ is the Barnes double gamma function, and the numerator contains all eight independent choices of signs.  The formula makes reflection symmetry in each momentum explicit.

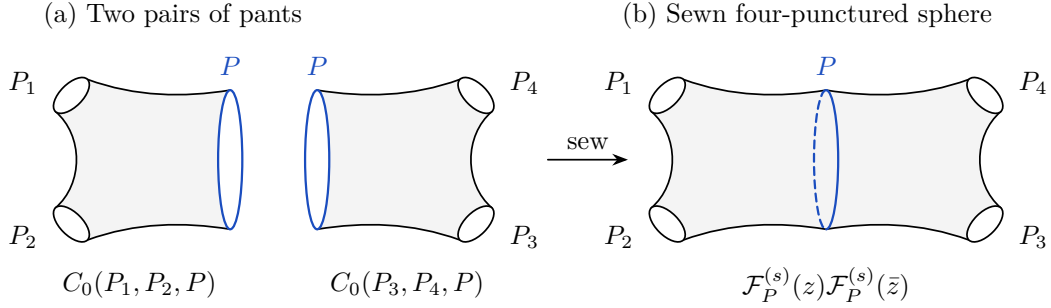
\begin{figure}[!ht]
\centering
\begin{tikzpicture}[x=1cm,y=1cm,line width=.65pt,line cap=round,line join=round,
    >=Stealth,font=\small]

\def\pantssurface{%
  \path[fill=gray!9] (-.988,1.062)
   .. controls (-.35,.83) and (.28,.82) .. (.90,.92)
   .. controls (1.12,.50) and (1.12,-.50) .. (.90,-.92)
   .. controls (.28,-.82) and (-.35,-.83) .. (-.988,-1.062)
   --(-1.412,-.638)
   .. controls (-1.04,-.32) and (-1.04,.32) .. (-1.412,.638) --cycle;
  \draw (-.988,1.062) .. controls (-.35,.83) and (.28,.82) .. (.90,.92);
  \draw (-.988,-1.062) .. controls (-.35,-.83) and (.28,-.82) .. (.90,-.92);
  \draw (-1.412,-.638) .. controls (-1.04,-.32) and (-1.04,.32) .. (-1.412,.638);
}
\def\leftcuffs{%
  \draw[fill=white,rotate around={45:(-1.20,.85)}] (-1.20,.85) ellipse (.30 and .15);
  \draw[fill=white,rotate around={-45:(-1.20,-.85)}] (-1.20,-.85) ellipse (.30 and .15);
}

\node[anchor=west,paneltitle] at (-1.65,1.91) {(a) Two pairs of pants};
\node[anchor=west,paneltitle] at (6.00,1.91) {(b) Sewn four-punctured sphere};
\begin{scope}
 \pantssurface\leftcuffs
 \draw[pt,fill=white,line width=.85pt] (.90,0) ellipse (.16 and .92);
 \node[left] at (-1.51,1.02) {$P_1$};
 \node[left] at (-1.51,-1.02) {$P_2$};
 \node[pt,above] at (.90,.98) {$P$};
 \node at (-.30,-1.65) {$C_0(P_1,P_2,P)$};
\end{scope}
\begin{scope}[shift={(2.95,0)},xscale=-1]
 \pantssurface\leftcuffs
 \draw[pt,fill=white,line width=.85pt] (.90,0) ellipse (.16 and .92);
\end{scope}
\node[right] at (4.46,1.02) {$P_4$};
\node[right] at (4.46,-1.02) {$P_3$};
\node[pt,above] at (2.05,.98) {$P$};
\node at (3.24,-1.65) {$C_0(P_3,P_4,P)$};
\draw[->] (5.10,0)--(6.16,0) node[midway,above] {sew};
\begin{scope}[shift={(7.88,0)}]
 \pantssurface\leftcuffs
\end{scope}
\begin{scope}[shift={(9.68,0)},xscale=-1]
 \pantssurface\leftcuffs
\end{scope}
\draw[pt,densely dashed,line width=.85pt] (8.78,.92) arc[start angle=90,end angle=270,x radius=.16,y radius=.92];
\draw[pt,line width=.85pt] (8.78,-.92) arc[start angle=270,end angle=450,x radius=.16,y radius=.92];
\node[left] at (6.37,1.02) {$P_1$};
\node[left] at (6.37,-1.02) {$P_2$};
\node[right] at (11.19,1.02) {$P_4$};
\node[right] at (11.19,-1.02) {$P_3$};
\node[pt,above] at (8.78,.98) {$P$};
\node[paneltitle] at (8.78,-1.65) {$\mathcal F_P^{(s)}(z)\mathcal F_P^{(s)}(\bar z)$};
\end{tikzpicture}
\caption{{Sewing two pairs of pants along the blue circles labelled $P$ produces the four-punctured sphere in the $(12)(34)$ channel.  The four remaining circles carry the external primary insertions.  The two pants supply the $C_0$ factors, and the complete-state sum on the common circle supplies $\rho_0(P)\dd P$ and the descendants in \eqref{eq:s2liouvillesewing}.}}
\label{fig:v72sewing}
\end{figure}

{The function $C_0$ in \eqref{eq:Czero} is fixed by the Virasoro vacuum fusion kernel in \eqref{eq:reviewidentityfusion}. In compact CFTs, it governs the universal asymptotic heavy-primary OPE variance \eqref{eq:revC0} \cite{CMMT}; the individual three-point coefficients remain the theory-dependent $c_{ijk}$ of \eqref{eq:s2cftcoefficients}. In Liouville CFT, the same function is obtained exactly from the DOZZ structure constant by a change of operator normalization \cite{CEZ}. To exhibit this normalization and its sewing measure, we use the Liouville primary vertex operators $V_P(z,\bar z)$, distinct from the compact-CFT primaries $\mathcal O_i$, with $h_P=\bar h_P=Q^2/4+P^2$, $P>0$, and reflection convention $V_P=V_{-P}$. Their two- and three-point functions are}
{
\begin{equation}
\begin{aligned}
\langle V_P(0)V_{P'}(\infty)\rangle
 &=\frac{\delta(P-P')}{\rho_0(P)},\\
\langle V_{P_1}(0)V_{P_2}(1)V_{P_3}(\infty)\rangle
 &=C_0(P_1,P_2,P_3).
\end{aligned}
\label{eq:twopoint}
\end{equation}
}
{The second line is the DOZZ three-point function after the normalization choice of \cite{CEZ}, while the first fixes the inverse two-point pairing.}
\begin{table}[!t]
\centering
\small

\begin{tabular}{@{}p{.25\linewidth}p{.31\linewidth}p{.34\linewidth}@{}}
\toprule
 & {Liouville CFT} & {Compact-CFT ensemble} \\
\midrule
{Primary operators} & {$V_P$} & {$\mathcal O_i$} \\[3pt]
{Spectrum and weights} & {Continuous $P>0$, $h_P=\bar h_P$} & {Discrete in each CFT; $h_i\ne\bar h_i$ allowed} \\[3pt]
{Three-point function on $S^2$} & {$C_0(P_1,P_2,P_3)$} & {$c_{ijk}$} \\[3pt]
{Role of $C_0$} & {Exact three-point coefficient in the chosen normalization} & {Leading variance in \eqref{eq:revC0}} \\
\bottomrule
\end{tabular}
\caption{{The same universal function $C_0$ appears in two settings, as a normalized Liouville three-point coefficient and in the averaged OPE data of compact CFTs \cite{CMMT}\cite{CEZ}.}}
\label{tab:v72liouvillecompact}
\end{table}
Writing $|P\rangle$ for the primary state, the resolution of the identity on the primary subspace is therefore
\begin{equation}
\begin{aligned}
\mathbf{1}_{\rm prim}&=\int_0^\infty\dd P\,\rho_0(P)|P\rangle\langle P|,\\
\mathbf{1}_{\rm prim}|P'\rangle
 &=\int_0^\infty\dd P\,\rho_0(P)|P\rangle
 \frac{\delta(P-P')}{\rho_0(P)}=|P'\rangle.
\end{aligned}
\label{eq:s2primaryresolution}
\end{equation}
{A four-point function is obtained by sewing two three-point functions.  Following the construction reviewed in \cite{CMMT}, replace each insertion by a small circular boundary, {so that each three-punctured sphere becomes a pair of pants,} and join one boundary circle of the first pair of pants to one of the second, as in figure~\ref{fig:v72sewing}.  On the joined circle one inserts a complete set of states.  The primary pairing in \eqref{eq:twopoint} supplies the measure $\rho_0(P)\dd P$, while summing over the descendants in each intermediate family produces the conformal blocks.}

{The resulting Liouville four-point function is \cite{CEZ}}
\begin{equation}
\begin{aligned}
&\langle V_{P_1}(0)V_{P_2}(z,\bar z)V_{P_3}(1)V_{P_4}(\infty)\rangle\\
&\quad=\int_0^\infty\dd P\,
 \rho_0(P)C_0(P_1,P_2,P)C_0(P_3,P_4,P)
 \mathcal F_P^{(s)}(z)\mathcal F_P^{(s)}(\bar z).
\end{aligned}
\label{eq:s2liouvillesewing}
\end{equation}
{Equation~\eqref{eq:s2liouvillesewing} is a correlator of the Liouville operators $V_P$.  Liouville sewing has one scalar intermediate momentum $P$, whereas compact CFTs can have separate left- and right-moving momenta. The connection is the chiral coefficient $\rho_0(P)C_0(P_1,P_2,P)C_0(P_3,P_4,P)$, which reappears in the averaged product of compact-CFT correlators in \eqref{eq:s2channelnormalizations}.  Table~\ref{tab:v72liouvillecompact} distinguishes the operators and the two uses of $C_0$.}

It is convenient to remove the known variance from each OPE coefficient,
\begin{equation}
C_{ijk}=\frac{c_{ijk}}{\big[C_0(P_i,P_j,P_k)C_0(\bar P_i,\bar P_j,\bar P_k)\big]^{1/2}} .
\label{eq:revnormalized}
\end{equation}
For a generic triple of distinct heavy primaries, the normalized coefficient then has
\begin{equation}
\overline{|C_{123}|^2}
 =\frac{\overline{|c_{123}|^2}}
 {C_0(P_1,P_2,P_3)C_0(\bar P_1,\bar P_2,\bar P_3)}
 =1+\cdots.
\label{eq:revgaussian}
\end{equation}
The dots denote corrections to the leading statistics.  We use lowercase $c_{ijk}$ for the original OPE coefficient, uppercase $C_{ijk}$ for this normalized random coefficient, and $C_0$ for the fixed universal function.

{Let $S_{ijk}$ denote the Riemann sphere with the three operator insertions defining $c_{ijk}$ in \eqref{eq:s2cftcoefficients}.  In the gravitational description it is an asymptotic boundary component.  The product $c_{ijk}c_{ijk}^*$ therefore specifies two such components.  Their leading connected bulk filling is the $C_0$ wormhole, whose amplitude reproduces the variance in \eqref{eq:revC0} \cite{Hartman}.}

\begin{figure}[!htbp]
\centering
\begin{tikzpicture}[x=1cm,y=1cm,line width=.65pt,line cap=round,line join=round,
    >=Stealth,font=\small]
\node[anchor=west,paneltitle] at (-1.72,2.95) {(a) The two $C_0$ flares};
\node[anchor=west,paneltitle] at (7.08,2.95) {(b) After stripping off both};
\def\oneflare{%
  \path[fill=gray!13] (-1.38,1.42)
    .. controls (-1.27,.88) and (-.95,.61) .. (-1.02,.03)
    arc[start angle=180,end angle=360,x radius=1.02,y radius=.29]
    .. controls (.95,.61) and (1.27,.88) .. (1.38,1.42) --cycle;
  \draw (-1.38,1.42) .. controls (-1.27,.88) and (-.95,.61) .. (-1.02,.03);
  \draw (1.38,1.42) .. controls (1.27,.88) and (.95,.61) .. (1.02,.03);
  \draw[fill=white] (0,1.42) ellipse (1.38 and .45);
  \draw[gray!65,densely dashed,line width=.45pt] (-1.37,1.42)
   .. controls (-.60,1.70) and (.60,1.70) .. (1.37,1.42);
  \draw[gray!65,line width=.45pt] (-1.37,1.42)
   .. controls (-.60,1.14) and (.60,1.14) .. (1.37,1.42);
  \foreach \x/\y/\lab in {-.62/1.41/i,0/1.24/j,.63/1.43/k}{
   \fill (\x,\y) circle (1.3pt);
   \node[above,font=\scriptsize,inner sep=2pt] at (\x,\y) {$\lab$};}
  \path[fill=pt!7,even odd rule] (0,.03) ellipse (1.02 and .29)
   (-.39,.03) ellipse (.23 and .11) (.39,.03) ellipse (.23 and .11);
  \draw[pt,line width=.8pt] (0,.03) ellipse (1.02 and .29);
  \draw[pt,fill=white,line width=.8pt] (-.39,.03) ellipse (.23 and .11);
  \draw[pt,fill=white,line width=.8pt] (.39,.03) ellipse (.23 and .11);
  \node at (0,.71) {$C_0$ flare};
}
\begin{scope}\oneflare\end{scope}
\begin{scope}[shift={(3.75,0)}]\oneflare\end{scope}
\node[above] at (0,1.94) {$S_{ijk}$};
\node[above] at (3.75,1.94) {$S_{ijk}^{*}$};
\node[pt,below] at (0,-.39) {$\Pi_+$};
\node[pt,below] at (3.75,-.39) {$\Pi_-$};
\draw[pt,<->] (1.19,.03)--(2.56,.03);
\node[pt,align=center,font=\scriptsize] at (1.875,-.48) {identify the\\whole surfaces};
\node[align=center] at (1.88,-1.25) {$\Pi_+\simeq\Pi_-$: one common finite pair of pants};
\draw[->] (5.36,.76)--(6.78,.76);
\node[font=\scriptsize,align=center] at (6.07,1.17) {strip off\\both};
\begin{scope}[shift={(9.10,.74)},rotate=-7]
 \path[fill=gray!19,even odd rule] (0,0) ellipse (1.85 and .52)
   (-.70,0) ellipse (.46 and .18) (.70,0) ellipse (.46 and .18);
 \draw[pt,line width=.9pt] (0,0) ellipse (1.85 and .52);
 \draw[pt,line width=.9pt] (-.70,0) ellipse (.46 and .18);
 \draw[pt,line width=.9pt] (.70,0) ellipse (.46 and .18);
\end{scope}
\node[above] at (9.08,1.50) {$M:\ \Pi\times I,\quad |I|\to0$};
\node[align=center] at (9.08,-.56) {$\operatorname{Vol}(M)=0$\\[4pt] $\partial M=\Pi_+\cup_{\partial\Pi}\Pi_-$};
\node at (9.08,-1.58) {genus-two boundary};
\end{tikzpicture}
\caption{{The $C_0$ wormhole and its leading compact core.  (a) The two halves of the wormhole are the two $C_0$ flares, each ending on one asymptotic three-point sphere.  The blue surfaces $\Pi_+$ and $\Pi_-$ are the two sides of $\Pi$ and are identified as whole surfaces.  (b) After both flares are stripped off, the interval thickness tends to zero.  The upper and lower faces of the flattened core are glued along their three cuffs, giving a genus-two boundary.}}
\label{fig:v72wormhole}
\end{figure}
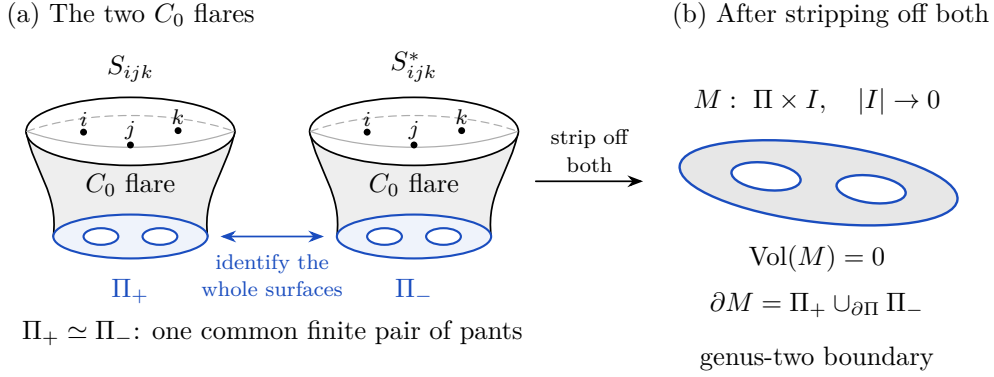

{For the geometric description we take real $b$ and scalar labels, $P_i=\bar P_i$.  A $C_0$ flare is one half of the heavy $C_0$ wormhole, connecting one asymptotic sphere to the totally geodesic pair of pants $\Pi$ in its middle \cite{Hartman}.  Each flare contributes the square root of the variance in \eqref{eq:revC0} \cite{Hartman}.  Dividing an OPE coefficient by this factor, as in \eqref{eq:revnormalized}, strips off one flare.  For the leading variance, stripping off both flares leaves the zero-thickness limit of $\Pi\times I$.  The compact core has zero volume, while its boundary is the double of $\Pi$, a genus-two surface \cite{Hartman}.  Figure~\ref{fig:v72wormhole} displays the two flares and this degenerate core.}

Equation~\eqref{eq:revgaussian} fixes the leading variance, but a variance alone does not fix a probability distribution.  The leading ensemble model of \cite{CCHM} additionally takes the independent heavy coefficients to be zero-mean Gaussian variables.  Dividing by the square root of their variance in \eqref{eq:revnormalized} then gives unit variance.  For three distinct Hermitian scalar operators, $C_{123}$ is real, and this assumption means
\begin{equation}
\begin{aligned}
 \dd\mu_{\rm G}(C_{123})&=\frac{e^{-C_{123}^2/2}}{\sqrt{2\pi}}\,\dd C_{123},\\
 \overline{C_{123}}\big|_{\rm G}&=\int\dd\mu_{\rm G}\,C_{123}=0,\qquad
 \overline{C_{123}^{\,2}}\big|_{\rm G}=\int\dd\mu_{\rm G}\,C_{123}^2=1.
\end{aligned}
\label{eq:s2gaussianaverage}
\end{equation}
Here G denotes this leading approximation.  For canonically ordered triples $\alpha=(i,j,k)$ and $\beta=(l,m,n)$ of distinct primaries, independence gives
\begin{equation}
\overline{C_\alpha}=0,\qquad
\overline{C_\alpha C_\beta^*}\approx\delta_{\alpha\beta}.
\label{eq:reviewgaussiancovariance}
\end{equation}
The Kronecker delta is one when the two ordered triples coincide and zero otherwise.  A contraction means replacing a pair of coefficients by this two-coefficient average.  Wick's rule sums all pairings; for four real scalar coefficients there are three,
\begin{equation}
\overline{C_\alpha C_\beta C_\gamma C_\delta}_{\rm G}
 =\delta_{\alpha\beta}\delta_{\gamma\delta}
 +\delta_{\alpha\gamma}\delta_{\beta\delta}
 +\delta_{\alpha\delta}\delta_{\beta\gamma}.
\label{eq:s2wickfour}
\end{equation}
For centered coefficients, the connected fourth moment is what remains after subtracting these three pairings,
\begin{equation}
\begin{aligned}
\overline{C_\alpha C_\beta C_\gamma C_\delta}_{c}
 ={}&\overline{C_\alpha C_\beta C_\gamma C_\delta}
 -\overline{C_\alpha C_\beta}\,\overline{C_\gamma C_\delta}\\
 &-\overline{C_\alpha C_\gamma}\,\overline{C_\beta C_\delta}
 -\overline{C_\alpha C_\delta}\,\overline{C_\beta C_\gamma}.
\end{aligned}
\label{eq:s2connectedfour}
\end{equation}
It vanishes for a zero-mean Gaussian ensemble.  Unordered triples require permutation terms and spin-dependent signs, and repeated labels can give extra multiplicities.  Identity coefficients are fixed, $c_{ij\mathbf{1}}=\delta_{ij}$.

Crossing and modular covariance require correlations beyond these Gaussian moments \cite{CCHM,Hartman}.  The next subsection derives the tetrahedral fourth moment and explains how the compact gravitational region encodes that contribution.
\subsection{From OPE moments to the tetrahedral core}\label{ssec:core}

Why does a tetrahedron enter OPE statistics?  Each four-point function contains two three-point coefficients, so averaging a product of two correlators probes a fourth OPE moment.  We first evaluate its Gaussian contribution with both correlators in the same channel, then re-expand one correlator in a different channel.  The resulting moment has the index pattern of a tetrahedral graph.  Only after identifying that graph do we construct its compact bulk region.

Take four distinct heavy external operators.  Applying the OPE twice gives the four-point function
\begin{equation}
G_{1234}(z,\bar z)=\sum_p c_{12p}c_{34p}\,
 \mathcal F_{P_p}^{(s)}(z)\mathcal F_{\bar P_p}^{(s)}(\bar z).
\label{eq:s2fourpointope}
\end{equation}
The two correlators may be evaluated at different cross-ratios.  Their product contains four OPE coefficients,
\begin{equation}
\begin{aligned}
\overline{G_{1234}(z,\bar z)G_{1234}(w,\bar w)^*}
 &=\sum_{p,q}\overline{c_{12p}c_{34p}c_{12q}^*c_{34q}^*}\\
 &\quad\times\mathcal F_{P_p}^{(s)}(z)\mathcal F_{\bar P_p}^{(s)}(\bar z)
 \big[\mathcal F_{P_q}^{(s)}(w)\mathcal F_{\bar P_q}^{(s)}(\bar w)\big]^*.
\end{aligned}
\label{eq:s2twocorrelators}
\end{equation}
For generic intermediate labels, the Gaussian pairing in \eqref{eq:reviewgaussiancovariance} matches $(12p)$ with $(12q)$ and $(34p)$ with $(34q)$.  Both matches require $p=q$, so $\delta_{pq}^2=\delta_{pq}$ gives
\begin{equation}
\overline{c_{12p}c_{34p}c_{12q}^*c_{34q}^*}
\approx\delta_{pq}\,
\overline{|c_{12p}|^2}\,\overline{|c_{34p}|^2}.
\label{eq:reviewsamechannelwick}
\end{equation}
Here $p,q$ are primary-state indices.  The Kronecker delta collapses the double sum to one sum; replacing it by the heavy continuum supplies one density $\rho_0(P)\rho_0(\bar P)$.  The two variances supply four $C_0$ factors.  The resulting chiral coefficient is the same sewing coefficient that appeared in \eqref{eq:s2liouvillesewing}; we denote it by
\begin{equation}
\rho_s(P)=\rho_0(P)C_0(P_1,P_2,P)C_0(P_3,P_4,P).
\label{eq:s2channelnormalizations}
\end{equation}
The other sector uses the corresponding expression with all labels barred.  Rewriting this averaged correlator in mixed channels yields the tetrahedral contribution \cite{Hartman}.

To obtain a mixed-channel expansion, re-expand the conjugated correlator at $w$ in the $t$ channel, with pairings $(14)$ and $(23)$.  We first display the transformation of one unconjugated chiral block at a generic coordinate $z$.  Its complex conjugate and right-moving counterpart act on the second correlator in \eqref{eq:s2twocorrelators}.  The change-of-basis coefficient is the fusion kernel $\mathbb F$,
\begin{equation}
\mathcal F_P^{(s)}(z)=\int_0^\infty\dd R\,
\mathbb F_{PR}\begin{bmatrix}P_3&P_2\\P_4&P_1\end{bmatrix}
\mathcal F_R^{(t)}(z).
\label{eq:reviewblockfusion}
\end{equation}
Its external labels are displayed in the bracket.  In the normalization of \cite{Hartman,CTV}, it is related to the tetrahedrally symmetric Virasoro $6j$ symbol by
\begin{equation}
\mathbb F_{PR}\begin{bmatrix}P_3&P_2\\P_4&P_1\end{bmatrix}
 =\rho_0(R)\left[\frac{C_0(P_1,P_4,R)C_0(P_2,P_3,R)}{C_0(P_1,P_2,P)C_0(P_3,P_4,P)}\right]^{1/2}
 \begin{Bmatrix}P_1&P_2&P\\P_3&P_4&R\end{Bmatrix}_{\!\rm Vir}.
 \label{eq:revfusion}
\end{equation}
The square root changes the three-point normalization from the $(12)(34)$ channel to $(14)(23)$, and $\rho_0(R)$ supplies the new state density.  Removing these factors from $\mathbb F$ defines the tetrahedrally symmetric Virasoro $6j$ symbol.  Multiplying the fusion kernel by the original sewing density $\rho_s(P)$ gives the coefficient in the mixed-channel expansion,
\begin{equation}
\begin{aligned}
\rho_s(P)\mathbb F_{PR}\begin{bmatrix}P_3&P_2\\P_4&P_1\end{bmatrix}
 &=\rho_0(P)\rho_0(R)\big[C_0(P_1,P_2,P)C_0(P_3,P_4,P)\big]^{1/2}\\
 &\quad\times\big[C_0(P_1,P_4,R)C_0(P_2,P_3,R)\big]^{1/2}
 \begin{Bmatrix}P_1&P_2&P\\P_3&P_4&R\end{Bmatrix}_{\!\rm Vir}.
\end{aligned}
\label{eq:reviewmixedweight}
\end{equation}

Expanding the product in mixed channels from the outset gives the moment $\overline{c_{12P}c_{34P}c^*_{14R}c^*_{23R}}$.  Here $P,R$ in an OPE subscript stand for the full intermediate states, including their right-moving labels.  Its double state sum contributes $\rho_0(P)\rho_0(R)$ in each chirality.  Dividing the transformed coefficient by these densities gives
\begin{equation}
\begin{aligned}
\left.\overline{c_{12P}c_{34P}c^*_{14R}c^*_{23R}}\right|_{\rm tet}
 ={}&\frac{C_0(P_1,P_2,P)C_0(P_3,P_4,P)\mathbb F_{PR}}{\rho_0(R)}\\
 &\times\frac{C_0(\bar P_1,\bar P_2,\bar P)C_0(\bar P_3,\bar P_4,\bar P)
 \mathbb F_{\bar P\bar R}}{\rho_0(\bar R)}.
\end{aligned}
\label{eq:s2rawtetmoment}
\end{equation}
The external labels of $\mathbb F_{PR}$ are those in \eqref{eq:revfusion}, and $\mathbb F_{\bar P\bar R}$ uses all right-moving labels.  Replacing each $c$ by $C$ divides out the square root of its left/right $C_0$ product, as specified in \eqref{eq:revnormalized}.  In the left-moving sector the cancellation is
\begin{equation}
\frac{\mathbb F_{PR}}{\rho_0(R)}
 \left[\frac{C_0(P_1,P_2,P)C_0(P_3,P_4,P)}
 {C_0(P_1,P_4,R)C_0(P_2,P_3,R)}\right]^{1/2}
 =\begin{Bmatrix}P_1&P_2&P\\P_3&P_4&R\end{Bmatrix}_{\!\rm Vir}.
\label{eq:s2tetnormalization}
\end{equation}
The right-moving factors cancel in the same way.  Only the two $6j$ symbols remain,
\begin{equation}
\overline{C_{12P}C_{34P}C^*_{14R}C^*_{23R}}\Big|_{\rm tet}
 =\left|\begin{Bmatrix}P_1&P_2&P\\P_3&P_4&R\end{Bmatrix}_{\!\rm Vir}\right|^2 .
 \label{eq:revnongauss}
\end{equation}
The notation $|_{\rm tet}$ selects this connected tetrahedral contribution.  The absolute-square notation is literal for scalar labels.  For spinning primaries, the second factor instead uses the right-moving labels $\bar P_i$.  For generic distinct labels, the Gaussian pairings alone do not produce this contribution.  The equation specifies the contribution required by this crossing comparison, with the associated bulk filling held fixed; a full ensemble moment can also contain other fillings and contractions.

The contraction graph $\Gamma$ can be read directly from the four coefficients.  Label its vertices $A=(12P)$, $B=(34P)$, $C=(14R)$ and $D=(23R)$.  Draw the two OPE decompositions as trivalent trees, with an edge for each primary label.  Joining external legs with equal labels produces the graph on the right,
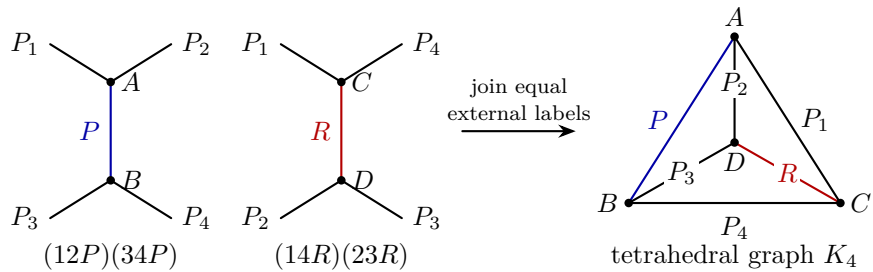
\begin{center}
\begin{tikzpicture}[x=1cm,y=1cm,line cap=round,line join=round,font=\small,>=Stealth]
\begin{scope}[shift={(0,0)}]
 \coordinate(A)at(0,.65);\coordinate(B)at(0,-.65);
 \draw[thick](A)--(-.8,1.15)node[left]{$P_1$};
 \draw[thick](A)--(.8,1.15)node[right]{$P_2$};
 \draw[thick](B)--(-.8,-1.15)node[left]{$P_3$};
 \draw[thick](B)--(.8,-1.15)node[right]{$P_4$};
 \draw[thick,blue!65!black](A)--(B)node[midway,left]{$P$};
 \foreach \p in {A,B}{\fill(\p)circle(1.6pt);}
 \node[right]at(A){$A$};\node[right]at(B){$B$};
 \node at(0,-1.65){$(12P)(34P)$};
\end{scope}
\begin{scope}[shift={(3.05,0)}]
 \coordinate(C)at(0,.65);\coordinate(D)at(0,-.65);
 \draw[thick](C)--(-.8,1.15)node[left]{$P_1$};
 \draw[thick](C)--(.8,1.15)node[right]{$P_4$};
 \draw[thick](D)--(-.8,-1.15)node[left]{$P_2$};
 \draw[thick](D)--(.8,-1.15)node[right]{$P_3$};
 \draw[thick,red!70!black](C)--(D)node[midway,left]{$R$};
 \foreach \p in {C,D}{\fill(\p)circle(1.6pt);}
 \node[right]at(C){$C$};\node[right]at(D){$D$};
 \node at(0,-1.65){$(14R)(23R)$};
\end{scope}
\draw[->,thick](4.65,0)--(6.15,0)node[midway,above,align=center,font=\scriptsize]{join equal\\external labels};
\begin{scope}[shift={(8.25,0)}]
 \coordinate(A)at(0,1.25);\coordinate(B)at(-1.4,-.95);\coordinate(C)at(1.4,-.95);\coordinate(D)at(0,-.15);
 \draw[thick,blue!65!black](A)--(B)node[midway,left=1pt]{$P$};
 \draw[thick](A)--(C)node[midway,right=1pt]{$P_1$};
 \draw[thick](B)--(C)node[midway,below=1pt]{$P_4$};
 \draw[thick](A)--(D)node[pos=.45,fill=white,inner sep=1pt]{$P_2$};
 \draw[thick](B)--(D)node[pos=.5,fill=white,inner sep=1pt]{$P_3$};
 \draw[thick,red!70!black](C)--(D)node[pos=.5,fill=white,inner sep=1pt]{$R$};
 \foreach \p in {A,B,C,D}{\fill(\p)circle(1.6pt);}
 \node[above]at(A){$A$};\node[left]at(B){$B$};\node[right]at(C){$C$};\node[below]at(D){$D$};
 \node at(0,-1.65){tetrahedral graph $K_4$};
\end{scope}
\end{tikzpicture}
\end{center}
For example, $P_1$ joins $A$ to $C$, and $P$ joins $A$ to $B$.  Every vertex is joined to the other three, so the graph is the complete four-vertex graph $K_4$.  Complex conjugation in the moment acts on the coefficients, while this diagram records their shared indices.  All six edge labels are fixed in the moment; the state sums belong to the correlator in \eqref{eq:s2twocorrelators}.

{For a higher OPE moment, the compact core can be nondegenerate.  One $C_0$ flare is attached to each boundary pair of pants by identifying the entire two-dimensional surfaces with matching cuff lengths \cite{Hartman}.  This bulk attachment removes or restores an asymptotic region; the sewing in figure~\ref{fig:v72sewing}, instead, joins two-dimensional CFT surfaces along a circle.  With one flare for each vertex $v$ of $\Gamma$, whose incident labels are $(i_v,j_v,k_v)$, the amplitude for a fixed attachment is}
\begin{equation}
\begin{aligned}
&Z_{\rm grav}(\hbox{core with flares})\\
&\quad=\prod_{v\in V(\Gamma)}
 \big[C_0(P_{i_v},P_{j_v},P_{k_v})C_0(\bar P_{i_v},\bar P_{j_v},\bar P_{k_v})\big]^{1/2}
 Z_L(M,\gamma).
\end{aligned}
\label{eq:s2flarefactorization}
\end{equation}
Here $V(\Gamma)$ is the vertex set.  Dividing by the product in \eqref{eq:s2flarefactorization}, exactly as in the definition of $C_{ijk}$, removes the flares and leaves the compact amplitude $Z_L$.  Each exposed pair of pants has three cuffs of fixed lengths.  This formula concerns one attachment.  Summing the inequivalent attachments that realize the same index contraction gives the multiplicity $\nu_\Gamma$ of \eqref{eq:inversedictionary}, which is two for the linked handcuff.  The relative weights of different fillings are separate data, defined in section~\ref{ssec:bulkstatisticalinput}.

{A filling of the contraction graph is specified by an embedding of $\Gamma$ in a closed three-manifold $M_E$ together with a framing, which records how a thin ribbon around each edge twists.}  The framing fixes phases in chiral amplitudes.  Virasoro topological quantum field theory (TQFT) assigns an amplitude to this labelled, framed pair $(M_E,\Gamma(P))$ \cite{CEZ,CTV}.
{The embedding and the graph complement give the compact region.  We take $N(\Gamma)$ to be an open regular neighbourhood, as in \cite{Hartman},}
\begin{equation}
 {\begin{aligned}
 \Gamma&\longrightarrow (M_E,\Gamma),\\
 M&=M_E\setminus N(\Gamma),\qquad \partial M=\partial N(\Gamma).
 \end{aligned}}
 \label{eq:revcomplement}
\end{equation}

\begin{figure}[!htbp]
\centering
\begin{tikzpicture}[x=1cm,y=1cm,line width=.65pt,line cap=round,line join=round,
    >=Stealth,font=\small]

\def\pantssurface{%
  \path[fill=gray!9] (-.988,1.062)
   .. controls (-.35,.83) and (.28,.82) .. (.90,.92)
   .. controls (1.12,.50) and (1.12,-.50) .. (.90,-.92)
   .. controls (.28,-.82) and (-.35,-.83) .. (-.988,-1.062)
   --(-1.412,-.638)
   .. controls (-1.04,-.32) and (-1.04,.32) .. (-1.412,.638) --cycle;
  \draw (-.988,1.062) .. controls (-.35,.83) and (.28,.82) .. (.90,.92);
  \draw (-.988,-1.062) .. controls (-.35,-.83) and (.28,-.82) .. (.90,-.92);
  \draw (-1.412,-.638) .. controls (-1.04,-.32) and (-1.04,.32) .. (-1.412,.638);
}
\def\leftcuffs{%
  \draw[fill=white,rotate around={45:(-1.20,.85)}] (-1.20,.85) ellipse (.30 and .15);
  \draw[fill=white,rotate around={-45:(-1.20,-.85)}] (-1.20,-.85) ellipse (.30 and .15);
}

\node[anchor=west,paneltitle] at (-3.24,2.93) {(a) $M_E=S^3=\mathbb R^3\cup\{\infty\}$};
\path[fill=gray!16] (-2.70,0)
 .. controls (-2.70,2.57) and (2.70,2.57) .. (2.70,0)
 .. controls (2.70,-2.57) and (-2.70,-2.57) ..cycle;
\draw (-2.70,0) .. controls (-2.70,2.57) and (2.70,2.57) .. (2.70,0)
 .. controls (2.70,-2.57) and (-2.70,-2.57) ..cycle;
\foreach \sgn in {1,-1}{
 \begin{scope}[yscale=\sgn]
  \path[fill=white] (-1.76,.68) .. controls (-.94,1.33) and (.94,1.33) .. (1.76,.68)
    .. controls (.86,.42) and (-.86,.42) ..cycle;
  \draw (-1.76,.68) .. controls (-.94,1.33) and (.94,1.33) .. (1.76,.68)
    .. controls (.86,.42) and (-.86,.42) ..cycle;
  \draw[gray!65,line width=.4pt] (-1.76,.68)
    .. controls (-.97,1.04) and (.97,1.04) .. (1.76,.68);
 \end{scope}
}
\draw[black!75,densely dashed,line width=.55pt] (-2.22,0)
 .. controls (-1.92,2.05) and (1.92,2.05) .. (2.22,0);
\draw[black!75,densely dashed,line width=.55pt] (-2.22,0)--(2.22,0);
\draw[black!75,densely dashed,line width=.55pt] (-2.22,0)
 .. controls (-1.92,-2.05) and (1.92,-2.05) .. (2.22,0);
\fill (-2.22,0) circle (1.6pt);\fill (2.22,0) circle (1.6pt);
\node[below left,inner sep=3pt] at (-2.22,0) {$A$};
\node[below right,inner sep=3pt] at (2.22,0) {$B$};
\foreach \yy/\rr/\nn in {1.5475/.38/1,0/.485/2,-1.5475/.38/3}{
 \draw[pt,densely dashed,line width=.8pt] (0,{\yy+\rr})
   arc[start angle=90,end angle=270,x radius=.17,y radius=\rr];
 \draw[pt,line width=.9pt] (0,{\yy-\rr})
   arc[start angle=270,end angle=450,x radius=.17,y radius=\rr];
 \node[pt,right,inner sep=3pt] at (.19,\yy) {$\gamma_{\nn}$};
}
\node at (1.09,.18) {$\Gamma$};
\node[anchor=east] at (-2.15,2.06) {$N(\Gamma)$};
\draw[->] (-2.12,1.98)--(-1.62,1.39);
\node[anchor=west] at (1.68,2.16) {$M\ni\infty$};
\node[anchor=east] at (-.74,-2.39) {$\partial M=\partial N(\Gamma)$};
\draw[->] (-.63,-2.31)--(-1.37,-1.72);
\node at (-.02,-3.14) {$M=M_E\setminus N(\Gamma)$};
\begin{scope}[shift={(7.22,1.77)}]
 \node[anchor=west,paneltitle] at (-2.88,1.16) {(b) A meridian bounds a disk in $N(\Gamma)$};
 \path[fill=gray!13] (-1.95,.48)--(1.86,.48)
   arc[start angle=90,end angle=-90,x radius=.19,y radius=.48]
   --(-1.95,-.48) --cycle;
 \draw (-1.95,.48)--(1.86,.48);
 \draw (-1.95,-.48)--(1.86,-.48);
 \draw[fill=gray!7] (-1.95,0) ellipse (.19 and .48);
 \draw (1.86,0) ellipse (.19 and .48);
 \path[fill=pt!13] (0,0) ellipse (.22 and .48);
 \draw[pt,densely dashed,line width=.8pt] (0,.48)
   arc[start angle=90,end angle=270,x radius=.22,y radius=.48];
 \draw[pt,line width=.9pt] (0,-.48)
   arc[start angle=270,end angle=450,x radius=.22,y radius=.48];
 \draw[black!70,densely dashed] (-2.19,0)--(2.1,0);
 \fill (0,0) circle (1.25pt);
 \node[above,inner sep=2pt] at (-1.08,.03) {$e\subset\Gamma$};
 \node at (1.13,-.25) {$N(\Gamma)$};
 \node[above] at (1.58,.66) {$M$};
 \node[pt,anchor=east] at (-.58,-.86) {$D_e$};
 \draw[pt,->] (-.49,-.79)--(-.065,-.23);
 \node[pt,anchor=west] at (.45,-.89) {$\gamma_e=\partial D_e$};
 \draw[pt,->] (.49,-.76)--(.18,-.27);
\end{scope}
\node[anchor=west,paneltitle] at (4.34,-.17) {(c) Cutting $\partial M$ along the three meridians};
\begin{scope}[shift={(5.83,-1.95)},scale=.76]
 \pantssurface
 \begin{scope}[pt,line width=.9pt]\leftcuffs\end{scope}
 \draw[pt,fill=white,line width=.9pt] (.90,0) ellipse (.16 and .92);
 \draw[black!65,densely dashed] (-1.16,.85)--(-.36,0)--(-1.16,-.85);
 \draw[black!65,densely dashed] (-.36,0)--(.86,0);
 \fill (-.36,0) circle (1.6pt);
 \node[above] at (-.34,.03) {$A$};
\end{scope}
\begin{scope}[shift={(9.20,-1.95)},scale=.76,xscale=-1]
 \pantssurface
 \begin{scope}[pt,line width=.9pt]\leftcuffs\end{scope}
 \draw[pt,fill=white,line width=.9pt] (.90,0) ellipse (.16 and .92);
 \draw[black!65,densely dashed] (-1.16,.85)--(-.36,0)--(-1.16,-.85);
 \draw[black!65,densely dashed] (-.36,0)--(.86,0);
 \fill (-.36,0) circle (1.6pt);
\end{scope}
\node[above] at (9.46,-1.925) {$B$};
\foreach \x/\y/\lab in {4.52/-1.14/1,4.52/-2.76/3,6.89/-1.95/2,10.51/-1.14/1,10.51/-2.76/3,8.13/-1.95/2}{
 \node[pt,font=\scriptsize] at (\x,\y) {$\gamma_{\lab}$};}
\node at (5.48,-3.14) {$\Pi_A$};
\node at (9.55,-3.14) {$\Pi_B$};
\node at (7.43,-3.80) {$\Pi_v\simeq S^2\setminus\bigcup_{a=1}^3\operatorname{int}D_a$};
\end{tikzpicture}
\caption{{Graph complement and its boundary pants.  (a) For the planar theta graph in $M_E=S^3$, the grey regular neighbourhood $N(\Gamma)$ is excised.  Its exterior, including the point at infinity, is the retained region $M$; their common surface is $\partial M$.  (b) A transverse disk $D_e$ in a removed edge tube has boundary $\gamma_e$ on that surface.  (c) Cutting along all three meridians separates the genus-two boundary into $\Pi_A$ and $\Pi_B$.  The matching blue cuffs recover the original surface upon gluing.}}
\label{fig:v72complement}
\end{figure}
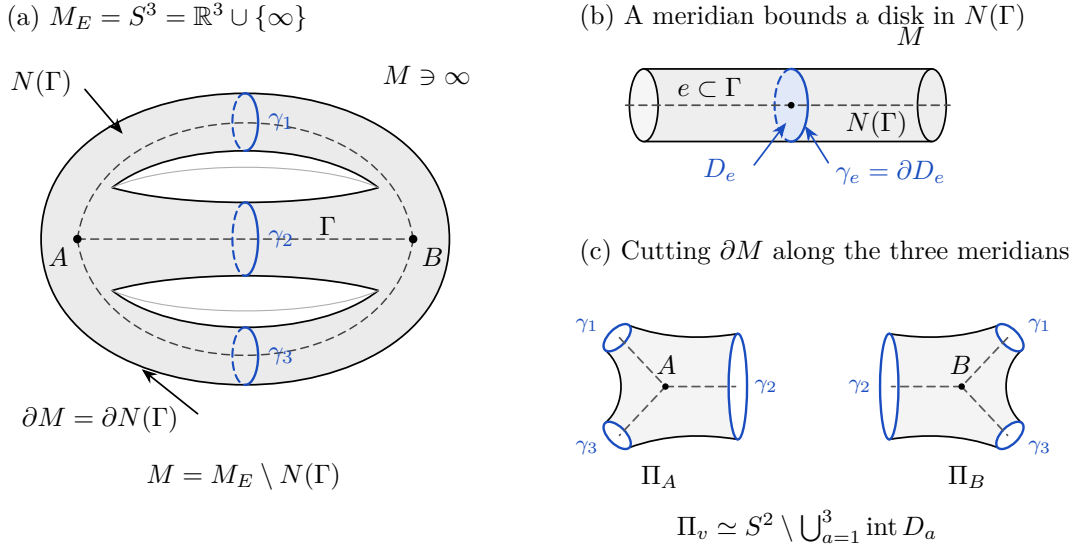

{Thicken each vertex of $\Gamma$ to a ball and each edge to a tube.  The boundary of this neighbourhood is $\partial M$; a transverse disk $D_e$ in an edge tube has boundary $\gamma_e=\partial D_e$ on $\partial M$.  The disk itself is inside the excised neighbourhood, as shown in figure~\ref{fig:v72complement}(b).}

{Cutting $\partial M$ along one meridian for each edge separates it into one pair of pants $\Pi_v$ for each trivalent vertex.  Indeed, the three incident tubes meet the vertex ball in three disks, so the exposed part of its boundary is $S^2$ with the interiors of those disks removed.  The half-tubes only extend its three circular cuffs.  Each edge therefore identifies a cuff of one vertex's pair of pants with a cuff at the other end.  Figure~\ref{fig:v72complement}(c) exhibits this decomposition for the theta graph, whose two vertices give two pairs of pants \cite{CEZ}.}

{The meridians of figure~\ref{fig:v72complement} are thus the cycles on which the fixed-length boundary data are imposed.  For a connected graph with $V$ vertices and $E$ edges, $N(\Gamma)$ is a handlebody of genus $E-V+1$; its boundary is also $\partial M$.  On the real-AdS length contour,}
\begin{equation}
{g(\partial M)=E-V+1,\qquad \ell(\gamma_e)=4\pi bP_e.}
\label{eq:revgraphboundary}
\end{equation}
{The theta graph, for example, has $E=3$, $V=2$ and a genus-two boundary, while the tetrahedral graph has $E=6$, $V=4$ and a genus-three boundary.  The graph therefore specifies the boundary topology and the labelled geodesics; the Virasoro TQFT amplitude supplies the gravitational path integral with precisely these fixed-length data.  With the fixed-length normalization \cite{Hartman},}
\begin{equation}
Z_L\big(M,\gamma(P)\big)=\big|Z_{\rm Vir}\big(M_E,\Gamma(P)\big)\big|^2 ,\qquad
Z_{\rm Vir}\big(S^3,\thetagraph\,\big)=1.
\label{eq:revZL}
\end{equation}
The right-hand side combines the two chiral factors.  For scalar real-AdS labels, it is the ordinary absolute square.  More generally, the CFT notation pairs the chiral amplitude evaluated at $P_i$ with its conjugate evaluated at the right-moving labels $\bar P_i$.  The length-and-angle geometry used here is the scalar case.  Equation~\eqref{eq:revZL} evaluates one specified filling.  Recovering an entire OPE moment also requires summing the appropriate contraction patterns and fillings.

We can now change its boundary data from lengths to angles.  The modular kernel $S_{P'P}=2\sqrt2\cos(4\pi P'P)$ combines the two reflected momentum contributions.  Applying it to every graph label gives the Conformal Turaev--Viro amplitude \cite{CTV},
\begin{equation}
Z_{\rm CTV}\big(M_E,\Gamma(P')\big)=\int_{\mathbb R_+^n}\dd P\,\prod_iS_{P_i'P_i}\,
 \big|Z_{\rm Vir}\big(M_E,\Gamma(P)\big)\big|^2 ,
\label{eq:revZA}
\end{equation}
The graph edges are called external because their labels prescribe boundary data on the compact region; they need not be open lines in the drawing.  The external labels are continued to the appropriate contours when they specify real angles.  This is the fixed-angle path integral,
\begin{equation}
Z_A\big(M,\gamma(P)\big)=Z_{\rm CTV}\big(M_E,\Gamma(P)\big).
\label{eq:revZAeq}
\end{equation}
The fusion kernel $\mathbb F$ above changes a four-point OPE channel, whereas the external modular transform here exchanges the geometric boundary data of the whole graph amplitude.

CTV computes this amplitude by triangulating $(M_E,\Gamma)$, so that the closed manifold is assembled from tetrahedra with specified identifications of their triangular faces, and $\Gamma$ is a subset of the resulting edges.  For each tetrahedron $\Delta$, its six edge labels determine the quantum weight $W_b(\Delta)$, a $6j$ symbol.  Labels on $\Gamma$ are prescribed, while labels on all other edges are summed over.  Denoting the latter edge set by $E_{\rm int}$, this continuous state sum is
\begin{equation}
Z_A(M,\gamma)=Z_{\rm CTV}(M_E,\Gamma)
 =\int\prod_{e\in E_{\rm int}}\dd P_e\,\rho_0(P_e)
 \prod_\Delta W_b(\Delta),
\label{eq:revgeneralCTV}
\end{equation}
The product supplies one weight per tetrahedron.  Cells meeting at an edge use the same label there; if that label is not boundary data, it is integrated with $\rho_0(P_e)\dd P_e$, the inverse pairing introduced in \eqref{eq:s2primaryresolution}.

{The tetrahedra in \eqref{eq:revgeneralCTV} initially triangulate the closed embedding manifold $M_E$.  To obtain cells of $M$, one truncates their vertices and then dualizes the edges lying on the graph $\Gamma$ \cite{Hartman}.  In the all-external-edge tetrahedral example this gives generalized cells $\widehat\Delta$ with triangular gluing faces and hexagonal faces on $\partial M$; the weight is inherited from the original CTV tetrahedron,}
\begin{equation}
{W_b(\widehat\Delta):=W_b(\Delta).}
\label{eq:revgeneralcellweight}
\end{equation}

{In the semiclassical chamber, the generalized cell $\widehat\Delta$ is represented by a generalized hyperbolic tetrahedron.  In the Klein ball model, hyperbolic space is the interior of the unit ball and a \emph{hyperideal} vertex lies outside it.  Such a point is not itself in $H_3$, so it cannot remain as a physical corner of the cell.  One therefore replaces it by its canonical polar truncation plane}
\begin{equation}
{\Pi_v=\{x\in H_3:v\cdot x=1\},}
\label{eq:revpolarplane}
\end{equation}
{retaining the half-space $v\cdot x\leq1$.  The tangents from $v$ to the boundary sphere touch it along a circle lying in $\Pi_v$; this determines the plane without choosing an arbitrary cut depth.  The plane is orthogonal, in the hyperbolic metric, to the three incident faces \cite{BaoBonahon}\cite{Hartman}.  {Its intersection with the tetrahedron is a smaller triangle, whose three corners are edge--plane intersections inside $H_3$.}  Figure~\ref{fig:gentet} shows how tangency fixes the cut and how the four truncations form the complete hyperbolic cell.}
\begin{figure}[!htbp]
\centering
\begingroup\color{black}
\begin{tikzpicture}[x=1cm,y=1cm,line cap=round,line join=round,>=Stealth,font=\small]
\definecolor{cutblue}{RGB}{32,89,133}
\definecolor{copper}{RGB}{173,91,37}
\definecolor{lightink}{RGB}{90,119,151}
\node[anchor=west,font=\small,paneltitle] at (0.2200,6.5400) {(a) Polar truncation in a Klein section};
\node[anchor=west,font=\small,paneltitle] at (8.2100,6.5400) {(b) Truncated hyperideal tetrahedron};
\filldraw[fill=black!1,draw=black!52,line width=.65pt] (2.3500,3.6200) circle (2.1000);
\fill[fill=lightink!19,draw=none] (0.9220,2.6614) -- (2.2338,1.9040) -- (3.8942,2.8626) -- (3.8942,4.3774) -- (2.2338,5.3360) -- (0.9220,4.5786) -- cycle;
\draw[black!34,densely dotted,line width=.6pt] (5.2059,3.6200) -- (3.8942,4.3774);
\draw[black!34,densely dotted,line width=.6pt] (2.2338,5.3360) -- (0.9220,6.0933);
\draw[black!80,line width=.92pt] (3.8942,4.3774) -- (2.2338,5.3360);
\draw[black!34,densely dotted,line width=.6pt] (0.9220,6.0933) -- (0.9220,4.5786);
\draw[black!34,densely dotted,line width=.6pt] (0.9220,2.6614) -- (0.9220,1.1467);
\draw[black!80,line width=.92pt] (0.9220,4.5786) -- (0.9220,2.6614);
\draw[black!34,densely dotted,line width=.6pt] (0.9220,1.1467) -- (2.2338,1.9040);
\draw[black!34,densely dotted,line width=.6pt] (3.8942,2.8626) -- (5.2059,3.6200);
\draw[black!80,line width=.92pt] (2.2338,1.9040) -- (3.8942,2.8626);
\draw[copper!55,line width=.75pt] (3.8942,4.3774) -- (3.8942,2.8626);
\draw[copper!55,line width=.75pt] (2.2338,5.3360) -- (0.9220,4.5786);
\draw[copper!55,line width=.75pt] (2.2338,1.9040) -- (0.9220,2.6614);
\draw[cutblue!70,densely dashed,line width=.85pt] (5.2059,3.6200) -- (3.8942,5.0432);
\draw[cutblue!70,densely dashed,line width=.85pt] (5.2059,3.6200) -- (3.8942,2.1968);
\draw[copper!65,densely dashed,line width=.85pt] (3.8942,5.0432) -- (3.8942,2.1968);
\draw[copper,line width=1.8pt] (3.8942,4.3774) -- (3.8942,2.8626);
\draw[black!43,dash pattern=on 2pt off 2pt,line width=.62pt] (4.3975,4.0868) -- (4.3975,3.1532);
\filldraw[fill=white,draw=cutblue,line width=.95pt] (3.8542,5.0032) rectangle (3.9342,5.0832);
\filldraw[fill=white,draw=cutblue,line width=.95pt] (3.8542,2.1568) rectangle (3.9342,2.2368);
\filldraw[fill=white,draw=black!63,line width=.8pt] (4.3975,4.0868) circle (1.75pt);
\filldraw[fill=white,draw=black!63,line width=.8pt] (4.3975,3.1532) circle (1.75pt);
\filldraw[fill=copper,draw=white,line width=.4pt] (3.8942,4.3774) circle (2.4pt);
\filldraw[fill=copper,draw=white,line width=.4pt] (3.8942,2.8626) circle (2.4pt);
\filldraw[fill=black,draw=black] (5.2059,3.6200) circle (1.5pt);
\node[anchor=west,font=\normalsize] at (5.3659,3.6400) {$v$};
\node[text=black,font=\normalsize] at (2.2400,3.6000) {$H_3$};
\node[text=black!65] at (1.9800,1.2200) {$S^2_\infty$};
\node[anchor=west,text=black,font=\footnotesize] at (5.3500,4.8500) {tangent};
\draw[black!65,-Stealth,line width=.5pt] (5.3000,4.7400) -- (4.5500,4.3316);
\node[anchor=west,text=black,font=\footnotesize] at (5.3500,2.4300) {polar line $\Pi_v$};
\draw[black!65,-Stealth,line width=.5pt] (5.2700,2.5100) -- (3.8942,2.3068);
\draw[draw=black!48,line width=.65pt] (11.2500,3.5000) circle (2.0200);
\draw[black!40,line width=.72pt] (11.2500,6.0457) -- (11.4260,4.5974);
\draw[black!40,line width=.72pt] (11.2500,6.0457) -- (10.5019,5.0173);
\draw[black!40,line width=.72pt] (11.2500,6.0457) -- (11.8236,5.1203);
\draw[black!21,line width=.45pt] (11.8395,1.1958) -- (11.6545,2.7182);
\draw[black!21,line width=.45pt] (11.8395,1.1958) -- (10.8579,1.9218);
\draw[black!21,line width=.45pt] (11.8395,1.1958) -- (12.2013,2.0994);
\draw[black!21,line width=.45pt] (9.1624,3.1759) -- (9.7023,3.9180);
\draw[black!21,line width=.45pt] (9.1624,3.1759) -- (9.8364,2.6774);
\draw[black!21,line width=.45pt] (9.1624,3.1759) -- (10.3218,3.3002);
\draw[black!21,line width=.45pt] (12.7879,3.5647) -- (12.4085,4.1767);
\draw[black!21,line width=.45pt] (12.7879,3.5647) -- (12.5602,2.9960);
\draw[black!21,line width=.45pt] (12.7879,3.5647) -- (11.7258,3.4508);
\draw[black!26,densely dashed,line width=.48pt] (10.3218,3.3002) -- (9.7023,3.9180) -- (9.8364,2.6774) -- cycle;
\draw[black!26,densely dashed,line width=.48pt] (12.5602,2.9960) -- (12.4085,4.1767) -- (11.7258,3.4508) -- cycle;
\draw[black!26,densely dashed,line width=.48pt] (10.3218,3.3002) -- (9.8364,2.6774) -- (10.8579,1.9218) -- (12.2013,2.0994) -- (12.5602,2.9960) -- (11.7258,3.4508) -- cycle;
\draw[black!26,densely dashed,line width=.48pt] (11.7258,3.4508) -- (12.4085,4.1767) -- (11.8236,5.1203) -- (10.5019,5.0173) -- (9.7023,3.9180) -- (10.3218,3.3002) -- cycle;
\filldraw[fill=lightink!17,draw=black!82,line width=.9pt] (12.2013,2.0994) -- (11.6545,2.7182) -- (11.4260,4.5974) -- (11.8236,5.1203) -- (12.4085,4.1767) -- (12.5602,2.9960) -- cycle;
\filldraw[fill=lightink!24,draw=black!82,line width=.9pt] (9.8364,2.6774) -- (9.7023,3.9180) -- (10.5019,5.0173) -- (11.4260,4.5974) -- (11.6545,2.7182) -- (10.8579,1.9218) -- cycle;
\filldraw[fill=copper!27,draw=black!82,line width=.9pt] (11.8236,5.1203) -- (11.4260,4.5974) -- (10.5019,5.0173) -- cycle;
\filldraw[fill=copper!27,draw=black!82,line width=.9pt] (10.8579,1.9218) -- (11.6545,2.7182) -- (12.2013,2.0994) -- cycle;
\filldraw[fill=black!85,draw=black!85] (11.2500,6.0457) circle (1.45pt);
\node[anchor=west,font=\normalsize] at (11.4000,6.1057) {$v$};
\node[text=black,anchor=west,font=\footnotesize] at (13.2200,5.2400) {truncation triangle};
\draw[black!65,-Stealth,line width=.5pt] (13.1900,5.1300) -- (12.9500,5.1200) -- (11.2500,4.9224);
\node[text=black,anchor=west,font=\footnotesize] at (13.2200,2.2700) {hexagonal face};
\draw[black!65,-Stealth,line width=.5pt] (13.1900,2.3700) -- (12.9400,2.4500) -- (10.6083,3.4730);
\end{tikzpicture}
\endgroup
\caption{{Polar truncation in the Klein model.  (a) Tangents from the hyperideal vertex $v$ to the sphere at infinity determine the polar plane $\Pi_v$; the cut points are its intersections with the original tetrahedron edges.  (b) Truncating all four hyperideal vertices gives a compact generalized tetrahedron with four triangular truncation faces and four original hexagonal faces.  Squares mark tangency points, open circles mark the intersections of the original edges with the sphere, and filled dots mark their intersections with the polar plane.  Only the visible faces of the truncated tetrahedron are shaded \cite{Hartman}.}}
\label{fig:gentet}
\end{figure}

{When the four truncations are mild, the cell has four triangular truncation faces and four original hexagonal faces.  The six surviving original edges run between pairs of truncation planes.  If their lengths are $\ell_e^{\rm cell}$ and the dihedral angles between the adjacent hexagonal faces are $\psi_e$, then}
\begin{equation}
{\dd V_A=-\frac12\sum_{e=1}^{6}\ell_e^{\rm cell}\,\dd\psi_e,
\qquad \psi(\text{triangle},\text{hexagon})=\frac{\pi}{2}.}
\label{eq:revcellschlafli}
\end{equation}
{This is the cell version of \eqref{eq:reviewschlafli}.  The quantum weight $W_b(\widehat\Delta)$ is defined by the representation labels whether or not a real hyperbolic saddle exists, while the geometric chamber is the one in which such a truncated tetrahedron realizes the cell.}

{For the tetrahedral OPE contraction, the dualization of the six graph edges exchanges graph vertices with the original hexagonal faces of the generalized cell \cite{Hartman}.  Thus the graph vertex $A=(12P)$ corresponds to the hexagonal face $F_A$, bounded by the three labelled edges $(P_1,P_2,P)$.  The following incidence diagram records this vertex--face correspondence.  The identification of the graph labels with the cell dihedral angles uses the external modular transform.}
\begin{center}
\begin{tikzpicture}[line cap=round,line join=round,font=\small,>=Stealth]
\begin{scope}[shift={(0,0)}]
 \coordinate(A)at(0,1.35);\coordinate(B)at(-1.5,-1);\coordinate(C)at(1.5,-1);\coordinate(D)at(0,-.1);
 \draw[thick,blue!65!black](A)--(B)node[midway,left=1pt]{$P$};
 \draw[thick](A)--(C)node[midway,right=1pt]{$P_1$};
 \draw[thick](B)--(C)node[midway,below=1pt]{$P_4$};
 \draw[thick](A)--(D)node[pos=.45,fill=white,inner sep=1pt]{$P_2$};
 \draw[thick](B)--(D)node[pos=.5,fill=white,inner sep=1pt]{$P_3$};
 \draw[thick,red!70!black](C)--(D)node[pos=.5,fill=white,inner sep=1pt]{$R$};
 \foreach \p in {B,C,D}{\fill(\p)circle(1.6pt);}
 \fill[pt!85!black](A)circle(3pt);
 \node[above]at(A){$A=(12P)$};\node[left]at(B){$B$};\node[right]at(C){$C$};\node[below]at(D){$D$};
 \node at(0,-1.7){graph vertex $A$};
\end{scope}
\draw[->,thick](2.1,.15)--(4.15,.15)node[midway,above,align=center,font=\scriptsize]{vertex $\leftrightarrow$ face\\edge $\leftrightarrow$ edge};
\begin{scope}[shift={(6.45,0)}]
 \coordinate(a)at(0,1.55);\coordinate(b)at(-1.7,-1);\coordinate(c)at(1.7,-1);\coordinate(d)at(.05,-.05);
 \foreach \i/\j in {a/b,a/c,a/d,b/a,b/c,b/d,c/a,c/b,c/d,d/a,d/b,d/c}{\coordinate(T\i\j)at($(\i)!.22!(\j)$);}
 \fill[pt!15](Tbc)--(Tcb)--(Tcd)--(Tdc)--(Tdb)--(Tbd)--cycle;
 \foreach \i/\j/\k/\l in {a/b/c/d,b/a/c/d,c/a/b/d,d/a/b/c}{\fill[gray!15](T\i\j)--(T\i\k)--(T\i\l)--cycle;\draw[gray!65](T\i\j)--(T\i\k)--(T\i\l)--cycle;}
 \draw[thick,red!70!black](Tab)--(Tba)node[midway,left=1pt]{$R$};
 \draw[thick](Tac)--(Tca)node[midway,right=1pt]{$P_3$};
 \draw[thick](Tad)--(Tda)node[pos=.5,fill=white,inner sep=1pt]{$P_4$};
 \draw[thick](Tbc)--(Tcb)node[midway,below=1pt]{$P_2$};
 \draw[thick](Tbd)--(Tdb)node[pos=.5,fill=pt!15,inner sep=1pt]{$P_1$};
 \draw[thick,blue!65!black](Tcd)--(Tdc)node[pos=.5,fill=pt!15,inner sep=1pt]{$P$};
 \node[font=\scriptsize]at(.05,-.68){$F_A$};
 \node at(0,-1.7){original face $F_A$ after truncation};
\end{scope}
\end{tikzpicture}
\end{center}
{Likewise the other original faces carry $(P_3,P_4,P)$, $(P_1,P_4,R)$ and $(P_2,P_3,R)$.  The four triangular truncation faces, by contrast, are paired when the two generalized tetrahedra are glued.  Their unpaired hexagonal faces make the finite boundary $\partial M$; the six graph labels become meridian labels on this boundary \cite{Hartman}.}

{Figure~\ref{fig:reviewcores} depicts each generalized tetrahedral cell by a topological face-pairing schematic.  The four orange tips represent its triangular truncation faces, while the blue interior stands collectively for its four original hexagonal faces.  The letters $A,B,C,D$ label truncation faces, and only the face identifications are shown.}

\begin{figure}[!htbp]
\centering
\begingroup\color{black}
\begin{tikzpicture}[x=1cm,y=1cm,>=Stealth,line cap=round,line join=round,font=\small]
\definecolor{faceblue}{RGB}{222,232,241}
\definecolor{caporange}{RGB}{246,208,155}
\definecolor{edgeblue}{RGB}{55,90,137}
\definecolor{paircolor}{RGB}{169,88,27}
\tikzset{pairline/.style={-{Stealth[length=2mm]},line width=.75pt,draw=paircolor}}
\newcommand{\fourcaps}{%
 \fill[faceblue] (-.27,1.09)--(.27,1.09)
  .. controls (.21,.44) and (.79,.20) .. (1.37,.25)
  --(1.37,-.25).. controls (.79,-.20) and (.21,-.44)..(.27,-1.09)
  --(-.27,-1.09).. controls (-.21,-.44) and (-.79,-.20)..(-1.37,-.25)
  --(-1.37,.25).. controls (-.79,.20) and (-.21,.44)..(-.27,1.09)--cycle;
 \draw[black!78,line width=.8pt] (-.27,1.09)--(.27,1.09)
  .. controls (.21,.44) and (.79,.20) .. (1.37,.25)
  --(1.37,-.25).. controls (.79,-.20) and (.21,-.44)..(.27,-1.09)
  --(-.27,-1.09).. controls (-.21,-.44) and (-.79,-.20)..(-1.37,-.25)
  --(-1.37,.25).. controls (-.79,.20) and (-.21,.44)..(-.27,1.09);
 \draw[edgeblue,line width=1pt] (0,.89).. controls (.18,.34) and (.14,-.26)..(0,-.89);
 \draw[black!58,densely dashed,line width=.62pt] (-1.12,0)
   .. controls (-.36,.09) and (.46,-.05).. (1.12,0);
 \filldraw[fill=caporange,draw=black!78,line width=.72pt]
       (-.27,1.09)--(.27,1.09)--(0,.89)--cycle;
 \filldraw[fill=caporange,draw=black!78,line width=.72pt]
       (-.27,-1.09)--(.27,-1.09)--(0,-.89)--cycle;
 \filldraw[fill=caporange,draw=black!78,line width=.72pt]
       (-1.37,.25)--(-1.37,-.25)--(-1.12,0)--cycle;
 \filldraw[fill=caporange,draw=black!78,line width=.72pt]
       (1.37,.25)--(1.37,-.25)--(1.12,0)--cycle;
}
\tikzset{facename/.style={font=\small,inner sep=.7pt}}
\node[anchor=west,font=\small\bfseries,paneltitle] at (-6.1,3.62)
  {(a) Tetrahedral core --- two cells};
\begin{scope}[shift={(-3.25,1.54)}]
 \fourcaps
 \node[facename] at (-1.70,0) {$A$};
 \node[facename] at (1.70,0) {$C$};
 \node[facename] at (0,1.42) {$D$};
 \node[facename] at (0,-1.42) {$B$};
\end{scope}
\begin{scope}[shift={(3.25,1.54)},xscale=-1]
 \fourcaps
\end{scope}
\node[facename] at (3.25-1.70,1.54) {$C$};
\node[facename] at (3.25+1.70,1.54) {$A$};
\node[facename] at (3.25,1.54+1.42) {$D$};
\node[facename] at (3.25,1.54-1.42) {$B$};
\node[font=\scriptsize,text=black!65,align=center] at (0,2.74) {glued faces};
\node[font=\small,align=center,inner sep=2pt] at (0,1.48)
  {$A\leftrightarrow A$\\[1pt]$B\leftrightarrow B$\\[1pt]$C\leftrightarrow C$\\[1pt]$D\leftrightarrow D$};
\node[font=\scriptsize,text=black!65] at (0,.02) {$g(\partial M)=3$};
\node[anchor=west,font=\small\bfseries,paneltitle] at (-6.1,-.88)
 {(b) Linked handcuff --- one cell};
\begin{scope}[shift={(-3.85,-2.67)}]
 \draw[black!83,line width=1pt] (-.52,0) circle (.76);
 \draw[black!83,line width=1pt] (.52,0) circle (.76);
 \draw[white,line width=4.6pt] (-.11,.455)--(.11,.655);
 \draw[black!83,line width=1pt] (-.11,.455)--(.11,.655);
 \draw[white,line width=4.6pt] (-.11,-.655)--(.11,-.455);
 \draw[black!83,line width=1pt] (-.11,-.655)--(.11,-.455);
 \draw[black!83,line width=1pt] (-.24,0)--(.24,0);
 \fill[black!85] (-.24,0) circle (1.4pt);
 \fill[black!85] (.24,0) circle (1.4pt);
 \node[font=\scriptsize,anchor=north,inner sep=1pt] at (0,-1.05) {linked graph};
\end{scope}
\draw[-{Stealth[length=2mm]},black!70,line width=.8pt]
 (-2.0,-2.66)--(-.74,-2.66);
\node[font=\scriptsize,align=center,anchor=south] at (-1.38,-2.55) {graph\\complement};
\begin{scope}[shift={(2.24,-2.67)}]
 \fourcaps
 \node[facename] at (-1.67,0) {$A$};
 \node[facename] at (1.67,0) {$A$};
 \node[facename] at (0,1.43) {$B$};
 \node[facename] at (0,-1.43) {$B$};
 \draw[pairline] (-1.43,-.27)..controls (-1.52,-2.60) and (1.53,-2.57)..(1.43,-.27);
 \draw[pairline] (.31,1.10)..controls (2.88,1.54) and (2.88,-1.54)..(.31,-1.10);
\end{scope}
\node[font=\scriptsize,anchor=west] at (5.22,-2.22) {$B\leftrightarrow B$};
\node[font=\scriptsize,anchor=west] at (5.22,-3.05) {$A\leftrightarrow A$};
\node[font=\scriptsize,text=black!65,anchor=west] at (5.22,-3.63)
 {$g(\partial M)=2$};
\end{tikzpicture}
\endgroup
\caption{{Face identifications for two compact cores.  (a) Two generalized tetrahedra are glued along the four pairs of triangular truncation faces $A$, $B$, $C$, $D$.  (b) In the linked handcuff the two $A$ faces and two $B$ faces of a single tetrahedron are identified.  The unpaired hexagonal faces form $\partial M$.}}
\label{fig:reviewcores}
\end{figure}

For the tetrahedral graph in $S^3$, the fixed-length chiral amplitude is one $6j$ symbol.  We now use the standard six-label order, replacing $(P_1,P_2,P;P_3,P_4,R)$ by $(P_1,P_2,P_3;P_4,P_5,P_6)$,
\begin{equation}
Z_{\rm Vir}\big(S^3,\tetgraph\,\big)
 =\begin{Bmatrix}P_1&P_2&P_3\\P_4&P_5&P_6\end{Bmatrix}_{\!\rm Vir}.
\label{eq:revtet}
\end{equation}
Its fixed-angle CTV evaluation instead uses two tetrahedral cell weights.  The triangulation of $S^3$ consists of two ordinary tetrahedra with their four faces paired.  After the conversion to the compact region, the two generalized hyperbolic tetrahedra are glued along their four truncation triangles, as shown in figure~\ref{fig:reviewcores}.  Their remaining faces form one connected genus-three boundary.  Its six labelled geodesics are the meridians of the six graph edges, in agreement with $3g-3=6$.

Each closed geodesic consists of one edge segment from each of the two identical cells.  Consequently its total length is $\ell_i=2\ell_i^{\rm cell}$.  In the fixed-length description, $\ell_i=4\pi bP_i$ labels this whole closed curve \cite{Hartman}.  The dihedral angle is constant along the joined curve.  Keeping total curve lengths separate from cell-edge lengths is essential when comparing the graph labels with a single tetrahedron.

In a geometric chamber where the corresponding hyperbolic cell exists and the geometric saddle gives the asymptotics, one cell weight has leading behaviour \cite{Hartman,TV}
\begin{equation}
\begin{Bmatrix}P_4&P_5&P_6\\P_1&P_2&P_3\end{Bmatrix}_{\!\rm Vir}
 \approx\exp\Big[-\frac{c}{6\pi}V_A(\psi)\Big]=e^{-I_A^{\rm tet}} ,
\label{eq:revsixj}
\end{equation}
Here $V_A(\psi)$ is the cell volume at the prescribed dihedral angles.  The same $6j$ function has two roles.  Its labels encode lengths in the graph amplitude \eqref{eq:revtet}, but angles in this cell weight.  Real lengths and real angles lie on the different momentum contours specified in subsection~\ref{ssec:bc}; the external transform relates the two evaluations.  For the two-cell triangulation, the exact identity and its semiclassical limit are
\begin{equation}
\boxed{\;\begin{aligned}
 Z_A\big(M,\gamma(P)\big)
 &=Z_{\rm CTV}\big(S^3,\Gamma_{\rm tet}\big)\\
 &=\begin{Bmatrix}P_4&P_5&P_6\\P_1&P_2&P_3\end{Bmatrix}_{\!\rm Vir}^{\!2}\\
 &\approx\exp\Big[-\frac{c}{3\pi}V_A(\psi)\Big]\\
 &=e^{-I_A^{\rm on\text{-}shell}(M)}
\end{aligned}\;}
\label{eq:revsemicl}
\end{equation}
The two logarithms add, and the total volume is $2V_A$.  This reproduces the gravitational action \eqref{eq:revonshell}.  The quantum $6j$ weights are defined before this semiclassical interpretation; their small-$b$ asymptotics identifies the metric and action of the compact core.

The tetrahedral fourth moment is the first non-Gaussian term required by the mixed-channel comparison above.  There are also corrections to the two-coefficient average.  To see their torus origin, sew two legs of a three-point function together.  This leaves one insertion and sums over the state entering and exiting the same circle.  The torus one-point function, before division by the partition function, has the expansion
\begin{equation}
\langle\mathcal O_0\rangle_{\tau}
 =\sum_p c_{pp0}\,
 \mathcal F_{P_p}^{(1)}[P_0](\tau)
 \mathcal F_{\bar P_p}^{(1)}[\bar P_0](-\bar\tau).
\label{eq:s2torusonepoint}
\end{equation}
The superscript $(1)$ denotes a genus-one block{, and $0$ labels the inserted primary $\mathcal O_0$}.  The argument $-\bar\tau$ implements the right-moving propagation variable $e^{-2\pi\ii\bar\tau}$.  A product of two one-point functions contains $\overline{c_{pp0}c_{qq0}^*}$.  Exchanging the torus cycles in one factor and comparing expansions requires an additional two-coefficient correlation \cite{Hartman}.  After normalization and relabelling, it correlates $C_{11p}$ with $C_{22p}^*$.  The repeated label at each vertex makes a loop, and the common label $p$ joins the vertices, giving the handcuff graph.  The relevant embedding links its two loops, as in figure~\ref{fig:reviewcores}.  Its chiral amplitude uses the punctured-torus modular kernel $\mathbb S$, the integral change of block basis under the cycle exchange; section~\ref{sec:crossed} fixes its normalization and framing phase.

The same correction can be obtained on the sphere by continuing an insertion around another before comparing channels \cite{Hartman}.  Such a winding changes the branch of an individual conformal block.  Re-expanding the wound identity block in the original channel gives the same linked-handcuff contribution.

When the cell decomposition has an internal edge, its label is integrated rather than prescribed.  The stationary equation of this integral imposes smoothness around the edge.  Write $L_e=4\pi bP_e$ for the integrated label length; the segment in one cell has length $L_e/2$, as explained in {appendix~\ref{ssec:generalinternal}}.  {Thus $L_e$ denotes the full geodesic length, whereas $\ell_e^{\rm cell}$ in \eqref{eq:revcellschlafli} is the length of one tetrahedral edge segment.}  For real $b$ and fixed positive $L_e$ as $b\to0$, \eqref{eq:revrho0} and Schl\"afli's identity give, on the geometric branch,
\begin{equation}
\begin{aligned}
\log\rho_0(P_e)&=\frac{cL_e}{12}+o(c),\\
I_\Delta&=\frac{c}{6\pi}\left(V_\Delta+
 \frac14\sum_{e\subset\Delta\,\mathrm{internal}}L_e\psi_e^{(\Delta)}\right),\\
\frac{\partial I_\Delta}{\partial L_e}
 &=\frac{c}{24\pi}\psi_e^{(\Delta)}.
\end{aligned}
\label{eq:s2mixedcellaction}
\end{equation}
Here $V_\Delta$ is the cell volume and $I_\Delta$ its action with external angles and internal lengths fixed.  The factor $1/4$ is $1/2$ from the Legendre term times $1/2$ from the segment length.  Moreover, $cL_e/12=(L_e/2)/(4G_N)$ is exactly the entropy exponent for that segment in \eqref{eq:s2internalgluingweight}.  At fixed external angles the full integrand therefore obeys
\begin{equation}
\frac{\partial}{\partial L_e}
 \log\!\left[\rho_0(P_e)\prod_\Delta e^{-I_\Delta}\right]
 =\frac{c}{24\pi}\left(2\pi-\sum_{\Delta\ni e}\psi_e^{(\Delta)}\right)+o(c).
\label{eq:s2internalstationarity}
\end{equation}
The sum counts each appearance of the edge in a cell, including repeated appearances in the same cell if faces are identified.  Setting the leading derivative to zero yields
\begin{equation}
\sum_{\Delta\ni e}\psi_e^{(\Delta)}=2\pi .
\label{eq:revinternal}
\end{equation}
This is the zero-deficit equation, stating that the wedges of the incident cells fill a complete angle around a bulk edge.  It is distinct from the prescribed angle between two faces on the boundary of $M$.  Whether the smooth internal saddle contributes to the integral also depends on its integration cycle, as studied in {appendix~\ref{ssec:internaledge}}.

We have now related three quantities, namely the normalized OPE moment for a specified contraction and filling, its fixed-length compact-region amplitude, and the fixed-angle state sum obtained by the external modular transform.  The de Sitter proposal uses this relation in the angle-to-length direction, beginning with a specified compact quantum amplitude.  We first review the separate cosmological-state input needed for its later boundary realization.
\subsection{The de Sitter state and its physical pairing}\label{ssec:dsproblem}\label{ssec:dsobservables}

{The compact amplitudes above depend on boundary data on a finite three-dimensional region $M$. The de Sitter input is a cosmological wavefunction, which assigns an amplitude to the conformal geometry at future infinity $\mathcal I^+$. We use the complex Liouville state \cite{CEM}. Section~\ref{sec:boundaryamp} extracts primary coefficients while keeping the boundary moduli fixed. Its gravitational norm instead pairs full states and integrates over moduli. This subsection fixes the continued parameters, the state and its two pairings. The continued OPE-moment functional is defined in section~\ref{ssec:ensemblesample}, after the real-$b$ sewing and boundary reconstruction.}

\paragraph{{Continued parameters.}}

Pure dS$_3$ gravity obeys the Einstein equation below.  Diffeomorphisms preserving its future-boundary conditions give two asymptotic Virasoro algebras, with central charge \cite{Strominger},
\begin{equation}
 R_{\mu\nu}=\frac{2}{\ell_{\rm dS}^2}g_{\mu\nu},
 \qquad c_{\rm asym}=\frac{3\ell_{\rm dS}}{2G_N}.
 \label{eq:s61puregravity}
\end{equation}
The parameter in the Ward identities {\eqref{eq:s61stateconstraints}} of the cosmological wavefunction instead comes from continuing the AdS Liouville description \cite{CEM}.  We denote this parameter by $c$, reserving $c_{\rm asym}$ for the real asymptotic charge.  All subsequent blocks and sewing kernels use $c\in13+\ii\mathbb R$, with $|\Im c|\simeq3\ell_{\rm dS}/(2G_N)$ semiclassically.  The continuation branch fixes its sign.

We continue the Virasoro parameter from real $b$ to
\begin{equation}
b=re^{\ii\pi/4}.
\label{eq:physicalray}
\end{equation}
Here $r>0$.  Substitution in $c=1+6(b+b^{-1})^2$ gives
\begin{equation}
c=13+6\ii(r^2-r^{-2}).
\label{eq:physicalc}
\end{equation}
Matching the magnitude of the leading imaginary term to the gravitational coupling gives $r^2\simeq4G_N/\ell_{\rm dS}$.  Thus $r\to0$ is the semiclassical limit on our branch.  The scalar representations of the selected state lie on the oriented ray
\begin{equation}
P\in e^{\ii\pi/4}\mathbb R_+,\qquad h(P)=\frac{Q^2}{4}+P^2 .
\label{eq:physicalP}
\end{equation}
Writing $P=e^{\ii\pi/4}k$ with $k>0$ makes their dimensions explicit,
\begin{equation}
Q^2=2+\ii(r^2-r^{-2}),\qquad
\Delta(P)=2h(P)
 =1+\ii\left[\frac{r^2-r^{-2}}2+2k^2\right].
\label{eq:s2principaldimension}
\end{equation}
Hence $h=\bar h$ and $\Re\Delta=1$, the scalar principal-series condition \cite{CEM}.  {As in section~\ref{ssec:cftensemble}, the bar on $h$ labels the right-moving representation.}  At fixed $b$, the condition $k^2\geq0$ selects only part of the principal-series line \cite{CEM}.

The same representation label also parameterizes the continued compact-boundary data.  To see which angle convention gives positive angles, note that on this ray
\begin{equation}
bP=\ii r|P| .
\label{eq:introangles}
\end{equation}
The AdS expression $2\pi\ii bP$ is therefore negative.  Reflection invariance under $P\mapsto-P$ lets us use the representative with positive physical angle,
\begin{equation}
\psi=-2\pi\ii bP=2\pi r|P| .
\label{eq:introanglespsi}
\end{equation}
Thus the former length parameter becomes $4\pi bP=2\ii\psi$.  A real angle is now specified by a momentum on the dS ray, whereas the corresponding real-$b$ angle data used imaginary momenta.  In the geometric limit we keep $\psi$ fixed as $r\to0$, so $|P|$ scales as $r^{-1}$.  Where the real hyperbolic particle geometry of \eqref{eq:cosmology_metric_v46} exists, the corresponding opening angle is $2\psi$, consistently with the deficits in \eqref{eq:cosmology_cone_domain_v46}; this is distinct from the compact-core interior dihedral angle $\psi$.

\paragraph{{Inflating geometry and its initial condition.}}
The state is prepared by an expanding spacetime with a specified initial condition at its zero-size end \cite{CEM}.  Its future boundary is distinct from the finite boundary of the compact region.  Let $h$ be a hyperbolic metric on the punctured spatial surface $\Sigma_{g,n}${, of genus $g$ with $n$ particle punctures}.  The inflating solution is
\begin{equation}
 ds^2=\ell_{\rm dS}^2\big[-\dd t^2+a(t)^2h_{ij}\dd x^i\dd x^j\big],
 \qquad a(t)=\sinh t,\quad t>0 .
 \label{eq:cosmology_metric_v46}
\end{equation}
The metric $h$ has Gaussian curvature $-1$ away from the particles, whose positions stay fixed as the scale factor grows; $t$ is dimensionless proper time.  At each particle, a small circle has circumference $2\psi_i$ times its proper radius, so the deficit angle is $2\pi-2\psi_i$.  On a sphere, Gauss--Bonnet requires
\begin{equation}
 \begin{aligned}
 -\operatorname{Area}(h)+\sum_{i=1}^n(2\pi-2\psi_i)&=4\pi,\\
 \operatorname{Area}(h)&=2\pi(n-2)-2\sum_i\psi_i>0 .
 \end{aligned}
 \label{eq:cosmology_cone_domain_v46}
\end{equation}
At genus $g$, replace $n-2$ by $2g-2+n$.  Positive area is necessary for this real hyperbolic metric.  Outside its geometric domain, we use the analytic continuation of the correlator.

To check the expansion, let a dot denote $\dd/\dd t$.  Away from the particle worldlines,
\begin{equation}
 \begin{aligned}
 R_{tt}&=-2\frac{\ddot a}{a},\\
 R_{ij}&=\big[-1+a\ddot a+\dot a^2\big]h_{ij} .
 \end{aligned}
 \label{eq:cosmology_ricci_v46}
\end{equation}
The Einstein equation $R_{\mu\nu}=2g_{\mu\nu}/\ell_{\rm dS}^2$ then gives
\begin{equation}
 \begin{aligned}
 \ddot a&=a,\\
 \dot a^2&=1+a^2 .
 \end{aligned}
 \label{eq:cosmology_friedmann_v46}
\end{equation}
The positive real scale factor has no turning point.  It expands from a zero-size, Milne-type singularity toward one connected $\mathcal I^+$, whose conformal geometry is that of $\Sigma_{g,n}$.  In particular, $\Sigma_{0,4}$ describes four particle worldlines in one universe.  The familiar two-ended wormhole is obtained on a different time contour \cite{CEM},
\begin{equation}
 t=\rho+\frac{\ii\pi}{2}
 \quad\Longrightarrow\quad
 ds^2=-\ell_{\rm dS}^2
 \big[\dd\rho^2+\cosh^2\rho\,h_{ij}\dd x^i\dd x^j\big] .
 \label{eq:cosmology_wormhole_contour_v46}
\end{equation}
Real $\rho\in\mathbb R$ gives two conformal ends, whereas the physical inflating branch has $t>0$ and the specified zero-size end.  This continuation relates different boundary-value problems.  The finite boundaries of the compact fillings are counted separately in section~\ref{ssec:labels}.

{A quantum state on this expanding geometry also needs a boundary condition at the zero-size end.  In the first-order formulation the frame $e^a$ and the spin connection $\omega^a=\frac12\epsilon^{abc}\omega_{bc}$ combine into two $\mathfrak{sl}(2,\mathbb C)$ Chern--Simons connections \cite{CEM},
\begin{equation}
 \mathcal A=\omega+\frac{\ii}{\ell_{\rm dS}}\,e,\qquad
 \bar{\mathcal A}=\omega-\frac{\ii}{\ell_{\rm dS}}\,e .
 \label{eq:csconnections}
\end{equation}
The topological initial condition is \cite{CEM}
\begin{equation}
 \left.\mathcal A\right|_{t=0}=\left.\bar{\mathcal A}\right|_{t=0}
 \quad\Longleftrightarrow\quad
 \left.e^a\right|_{t=0}=0 .
 \label{eq:s55topologicalstate}
\end{equation}
Here $|_{t=0}$ is the pull-back to the slice $t=0$.  The difference of the two connections is $2\ii e/\ell_{\rm dS}$, so the condition sets the pulled-back frame to zero and leaves $\omega$ unconstrained.  For \eqref{eq:cosmology_metric_v46} the pulled-back frame is $\ell_{\rm dS}\sinh t$ times a frame of $h$, which vanishes at $t=0$, while the pulled-back spin connection contains the connection of $h$ and the extrinsic term $\cosh t$ times a frame of $h$, which stays finite.  The condition cancels the boundary variation of the action without additional dynamical fields at that end, and is called a topological boundary condition.}

{The state prepared by \eqref{eq:s55topologicalstate} has one future boundary, unlike the two-ended AdS cylinder.  After the channel coefficients have been extracted, figure~\ref{fig:dsbranches} relates its two channels to the compact core of the TQFT coefficient.}

\paragraph{{The state and its Ward identities.}}
{At real $b$, in the AdS description, the condition \eqref{eq:s55topologicalstate}} joins two oppositely oriented Virasoro TQFT copies.  Unfolding them gives a single-copy cylinder.  The left- and right-moving conformal data sit at opposite ends.  Evaluating its amplitude at conjugate boundary moduli gives a Liouville correlator.  Continuing the central charge gives the proposed cosmological state \cite{CEM},
\begin{equation}
\Psi_{g,n}^{(b)}[m;P_1,\ldots,P_n]
 =\left\langle\prod_{i=1}^n V_{P_i}\right\rangle_{\Sigma_{g,n},b}[m],
 \qquad b=re^{\ii\pi/4}.
\label{eq:s55CEMstate}
\end{equation}
Here $m$ denotes conformal moduli, including relative puncture positions.  The correlator is defined by continuing the Liouville OPE data and sewing its conformal blocks.  We adopt this scalar state as our boundary input.

{The state is a functional of the late-time spatial metric $h$ on $\Sigma_{g,n}$ and of the particle labels.  The momentum constraints make it invariant under spatial diffeomorphisms, and the Hamiltonian constraint fixes its response to a change of the Weyl factor \cite{CEM}.  For \eqref{eq:s55CEMstate} the two conditions are the Ward identities of a two-dimensional CFT with central charge $c$,
\begin{equation}
\begin{aligned}
 \Psi[\varphi^*h]&=\Psi[h],\\
 \Psi[e^{2\sigma}h]&=\exp\!\Big[\frac{c}{24\pi}\int_{\Sigma_{g,n}}\!\dd^2x\,\sqrt h\,
 \big(h^{ab}\partial_a\sigma\,\partial_b\sigma+R_h\,\sigma\big)\Big]
 \prod_{i=1}^n e^{-\Delta_i\sigma(x_i)}\;\Psi[h].
\end{aligned}
\label{eq:s61stateconstraints}
\end{equation}
Here $\varphi$ is a diffeomorphism isotopic to the identity that fixes the punctures $x_i$, $\sigma$ is a smooth Weyl factor, $R_h$ is the scalar curvature of $h$, and $\Delta_i=2h(P_i)$ is given by \eqref{eq:s2principaldimension}.  For constant $\sigma$ on a sphere without punctures the second line gives $\Psi\propto\operatorname{Area}(h)^{c/6}$.  The first line reduces the dependence on $h$ to its conformal structure, and the second fixes the dependence on the Weyl factor through explicit anomaly and puncture factors.  Teichm\"uller space $\mathcal T_{g,n}$ records the conformal structure together with a marking by reference curves.  The mapping class group $\operatorname{Map}(\Sigma_{g,n})$ consists of orientation-preserving diffeomorphisms respecting the puncture labels, modulo those isotopic to the identity, and its quotient
\begin{equation}
 \mathcal M_{g,n}=\mathcal T_{g,n}/\operatorname{Map}(\Sigma_{g,n})
 \label{eq:s61moduli}
\end{equation}
is the moduli space.  Because of the factors in \eqref{eq:s61stateconstraints}, $\Psi$ is a section of a line bundle over $\mathcal M_{g,n}$.  For the four-punctured sphere the corresponding coordinate factors appear in the prefactor \eqref{eq:ellipticprefactor}, which is independent of the intermediate primary and which section~\ref{sec:inverse} removes before transforming the moduli dependence.}

\paragraph{{The physical pairing and the state choice.}}
The norm of this state is determined by the gravitational pairing.  The marked space $\widehat{\mathcal H}_{g,n}^{(b)}$ and the crossing-symmetric space $\mathcal H_{g,n}^{(b)}$ after the mapping-class quotient have different integration domains.  For matching external particle data, the leading pairings are \cite{CEM}
\begin{equation}
\begin{aligned}
 \langle\Psi_1|\Psi_2\rangle_{\widehat{\mathcal H}}
 &\propto\int_{\mathcal T_{g,n}}\dd\mu_{\rm gh}\,
              \Psi_1^*\Psi_2,\\
 \langle\Psi_1|\Psi_2\rangle_{\mathcal H}
 &\propto\int_{\mathcal M_{g,n}}\dd\mu_{\rm gh}\,
              \Psi_1^*\Psi_2.
\end{aligned}
\label{eq:s61statespaces}
\end{equation}
The ghost measure comes from fixing the redundant metric coordinates; fixed-topology and external-state normalizations are suppressed.  These pairings come from the cylinder topology.  Additional three-dimensional topologies may correct them.  A physical state must have finite norm in the chosen pairing, a condition beyond the Ward identities.

{Finite norm distinguishes the selected state from a no-boundary prescription.  For the torus, the continued and regulated AdS saddle sum \cite{CastroMaloney} has a constant Fourier mode growing as $e^{\pi\tau_2/6}$ at large $\tau_2$, so by Cauchy--Schwarz its norm diverges in the cusp of the torus moduli space.  The vacuum-block Hartle--Hawking candidate has the same divergence \cite{CEM}.  The Liouville continuum instead starts at $P=0$, where $\Delta_{\min}=Q^2/2$ and the effective central charge is $c-12\Delta_{\min}=1$.  This milder long-cylinder behaviour supports normalizability of the selected state in the leading pairing \cite{CEM}.}

\paragraph{{Sewing at fixed moduli.}}
For the four-punctured sphere, the selected state is
\begin{equation}
\Psi_{0,4}^{(b)}(z,\bar z)
 =\left.\big\langle V_{P_1}(0)V_{P_2}(z,\bar z)V_{P_3}(1)V_{P_4}(\infty)\big\rangle_b
 \right|_{b=re^{\ii\pi/4}}.
\label{eq:s2wavefunctioncorrelator}
\end{equation}
After fixing three punctures, the cross-ratio $z$ is the remaining complex modulus.  Cutting between $(12)$ and $(34)$ inserts the primary resolution \eqref{eq:s2primaryresolution}.  The resulting sum over representations and the gravitational integral over geometries are
\begin{equation}
\begin{aligned}
\Psi_{0,4}^{(b)}(z,\bar z)
 &=\int_{e^{\ii\pi/4}\mathbb R_+}\dd P\,
 \rho_s(P)\,\mathcal F_P^{(s)}(z)\mathcal F_P^{(s)}(\bar z),\\
 \langle\Psi_1|\Psi_2\rangle_{\rm grav}
 &=\int_{\mathcal M_{0,4}}\dd^2z\,
 \Psi_1(z,\bar z)^*\Psi_2(z,\bar z).
\end{aligned}
\label{eq:unknownstate}
\end{equation}
The first line expands one state at fixed $z$.  The second pairs two specified states and lets their common geometry fluctuate; equal states give a norm.  It uses the gauge-fixed four-puncture measure, with the normalization factors suppressed as in \eqref{eq:s61statespaces}.  The distinction between a wavefunction and a cosmological expectation also appears in \cite{HarlowStanford}; here the integrated final data are the boundary moduli.  

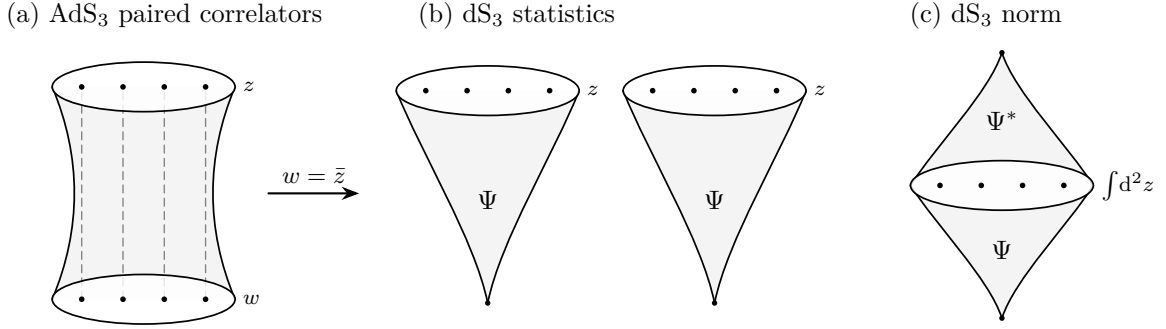
\begin{figure}[t]
\centering
\begin{tikzpicture}[x=1cm,y=1cm,line width=.65pt,line cap=round,line join=round,>=Stealth,font=\small,
  lab/.style={font=\scriptsize,inner sep=1.2pt}]
\node[anchor=west,paneltitle] at (-1.9,3.95) {(a) AdS$_3$ paired correlators};
\node[anchor=west,paneltitle] at (3.55,3.95) {(b) dS$_3$ statistics};
\node[anchor=west,paneltitle] at (10.05,3.95) {(c) dS$_3$ norm};
\begin{scope}[shift={(0,.2)},scale=.78]
 \figwormhole{3.6}
 \foreach \x in {-1.05,-.35,.35,1.05}{\draw[black!50,densely dashed,line width=.5pt] (\x,0)--(\x,3.6);}
 \figboundary{0}\figboundary{3.6}
 \node[anchor=west,lab] at (1.62,3.6) {$z$};
 \node[anchor=west,lab] at (1.62,0) {$w$};
\end{scope}
\draw[->,thick] (1.65,1.6)--(2.85,1.6) node[midway,above,font=\footnotesize]{$w=\bar z$};
\begin{scope}[shift={(4.55,.15)},scale=.78]
 \figfunnel{3.6}\figboundary{3.6}
 \node[anchor=west,lab] at (1.62,3.6) {$z$};
 \node at (0,1.75) {$\Psi$};
\end{scope}
\begin{scope}[shift={(7.55,.15)},scale=.78]
 \figfunnel{3.6}\figboundary{3.6}
 \node[anchor=west,lab] at (1.62,3.6) {$z$};
 \node at (0,1.75) {$\Psi$};
\end{scope}
\begin{scope}[shift={(11.35,-.05)},scale=.78]
 \figfunnel{2.25}
 \begin{scope}[shift={(0,4.5)},yscale=-1]\figfunnel{2.25}\end{scope}
 \figboundary{2.25}
 \node at (0,1.15) {$\Psi$};
 \node at (0,3.35) {$\Psi^{*}$};
 \node[anchor=west,lab] at (1.62,2.25) {$\int\!\dd^2z$};
\end{scope}
\end{tikzpicture}
\caption{{The two bilinears of the cosmological state.  (a) In AdS the paired contraction of two four-point functions is the wormhole $\Sigma_{0,4}\times I$; its two chiral factors join the moduli $z$ and $w$, \eqref{eq:statisticalstatepair}.  (b) At mirrored moduli $w=\bar z$ each chiral factor joins $z$ to $\bar z$ and folds at the topological end into one inflating universe.  The paired statistics is therefore $\Psi(z,\bar z)^2$, two universes at the same moduli and the same $b$.  (c) The gravitational norm instead glues the state to its conjugate along $\mathcal I^+$ and integrates over the moduli.  Conjugation sends $b$ to $-\ii b$, so the two halves carry central charges $c$ and $26-c$.  At real $b$ the conjugate half is the same theory, and the two bilinears are built from the same factors.}}
\label{fig:dsstatepairing}
\end{figure}

{Figure~\ref{fig:dsstatepairing} contrasts the two bilinears built from this state.  The same-$b$ product $\Psi^2$, which section~\ref{ssec:cylinder} identifies with paired OPE statistics, is the AdS wormhole at mirrored moduli.  The gravitational norm instead glues $\Psi$ to its conjugate along $\mathcal I^+$ and integrates over $z$.}  {At three points the state has no moduli and equals $C_0(A)$, so the leading second moment is $\Psi_{0,3}(A)^2$, as shown in section~\ref{ssec:cylinder}.}

The first-line density is the continued DOZZ density in our primary normalization \cite{CEM},
\begin{equation}
\rho_s(P)=\rho_0(P)\,
 C_0(P_1,P_2,P)C_0(P_3,P_4,P).
\label{eq:s55knownOPEdensity}
\end{equation}
As in \eqref{eq:s2liouvillesewing}, the inverse two-point pairing supplies $\rho_0$ and the two three-punctured spheres supply the $C_0$ factors.  This is known deterministic data of the chosen Liouville state.  Section~\ref{sec:boundaryamp} first recovers it from the state's modulus dependence, then applies the extraction to a specified crossed-channel component to obtain the coefficient used in the OPE-statistical prescription.

\paragraph{{The conjugate norm and matrix amplitudes.}}
To write that norm in the conventions of the complex Liouville string \cite{CLS}, set $p=-\ii P$ and $h_b(p)=Q^2/4-p^2$.  This is the same representation with the momentum convention used in the string and matrix literature.  On the physical rays, $b^*=-\ii b$ and $p^*=\ii p$, so
\begin{equation}
\begin{aligned}
 c(-\ii b)&=13-6(b^2+b^{-2})=26-c(b)=c(b)^*,\\
 h_b(p)+h_{-\ii b}(\ii p)&=\frac{(b+b^{-1})^2-(b-b^{-1})^2}{4}=1.
\end{aligned}
\label{eq:conjugationmap}
\end{equation}
The norm integrand is therefore a correlator of two Liouville theories at conjugate central charges.  Each puncture carries the product of one primary from each theory; \eqref{eq:conjugationmap} makes its holomorphic and antiholomorphic weights both one.  Such a product can be integrated over its insertion position.  The reparametrization ghosts cancel the combined conformal anomaly,
\begin{equation}
c+c^*=26,\qquad c_{\rm gh}=-26,
\qquad c+c^*+c_{\rm gh}=0.
\label{eq:s55ghostcancellation}
\end{equation}
Thus the two Liouville theories and the ghosts form the worldsheet theory of the complex Liouville string.  Its worldsheet is the boundary surface $\Sigma_{g,n}$ whose conformal structures are integrated in the gravitational pairing.  The genus-$g$, $n$-puncture string amplitude is therefore the moduli integral
\begin{equation}
\mathsf A_{g,n}^{(b)}
 \ \propto\ \int_{\mathcal M_{g,n}}\dd\mu_{\rm gh}(m)\,
 \Psi_{g,n}^{(b)}[m;P_i]^*\Psi_{g,n}^{(b)}[m;P_i].
\label{eq:s55integratedstate}
\end{equation}
The proportionality suppresses the external leg factors and the fixed-topology normalization \cite{CEM,CLS}. Our sphere-state convention is $\Psi_{0,n}=(-1)^{n+1}\Psi_{0,n}^{\rm CLS}$ and $p=-\ii P$, so the three-point coefficient agrees and the two- and four-point states have the opposite overall sign. This phase drops out of the norm. The state is evaluated at a fixed conformal structure, while the norm integrates over conformal structures and puncture positions.

A double-scaled two-matrix model computes these integrated string amplitudes \cite{CEM,Matrix}.

The distinction from an OPE moment can already be seen in \eqref{eq:unknownstate}. Expanding $\Psi^*\Psi$ before the moduli integral produces overlaps of blocks with different internal labels. Ordinary Virasoro blocks are nonorthogonal in this gravitational pairing \cite{CEM}. The norm therefore tests the full channel coefficients and their conjugates. Individual channel coefficients, in contrast, are extracted at fixed moduli.

For the marked extraction we choose a curve on the future boundary and retain its sewing channel. A collar of the curve is an annulus, conformally a cylinder; its radial modulus is the cylinder length and its twist is the relative rotation of the ends. Section~\ref{sec:boundaryamp} averages the twist, transforms the radial modulus and takes a discontinuity to isolate a primary coefficient. This is a functional of a specified boundary state. Observer-relative constructions instead define gravitationally dressed operators with reference to an observer worldline \cite{CLPW,WittenObs,AAIL,HUZ,BTV}; identifying our functional with such an operator requires an additional prescription.

\FloatBarrier
\section{The cosmological state and boundary OPE reconstruction}\label{sec:boundaryamp}\label{sec:liouvillestate}

{We begin with the Virasoro cylinder, which relates the selected Liouville state to paired CFT statistics. Its three-point specialization fixes the $C_0$ normalization and gives the theta second moment. We then extract the primary density of the four-point reference state, which calibrates the inverse map. Applying that map to an $s/t$ crossed preparation gives the $K_4$ tetrahedral coefficient, while the intersecting cycles of a punctured torus give the linked-handcuff coefficient. These cases have different graphs and compact fillings. The theta example is the degenerate core of figure~\ref{fig:v72wormhole}; the four-point example is the two-cell core of figure~\ref{fig:reviewcores}(a). Section~\ref{sec:gravity} develops the latter geometry, section~\ref{sec:bulkinverse} assembles the coefficients into OPE moments, with the six-vertex fillings in appendix~\ref{sec:sixth}. Section~\ref{sec:physicalinterpretation} explains the random Euclidean CFT interpretation and its relation to the selected cosmological state.}

The input consists of the standard Virasoro data $\rho_0$, $C_0$, $\mathbb F$ and $\mathbb S${, defined at real $b$ in section~\ref{sec:review} and continued by the conventions of section~\ref{ssec:dsproblem}}.  We first fix the extraction normalization using the known primary density of the reference state, then apply the same operation to a crossed preparation and remove the three-point and sewing factors.  The reference calculation has three steps,
\begin{equation}
\Psi\ \xrightarrow{\ \text{twist average}\ }\ Z_{\gamma_s}(y)
\ \xrightarrow{\ \text{Laplace in }y\ }\ \widehat Z_{\gamma_s}(\zeta)
\ \xrightarrow{\ \Disc\ \text{at }\zeta=P^2\ }\ \rho_s(P),
\label{eq:extractionchain}
\end{equation}
and the last two use the third and fourth rows of \eqref{eq:adsdscompare}.  At real momenta the primary and descendant cuts overlap.  On the dS ray they separate, so the primary jump isolates the reference coefficient under the convergence assumptions below.  A crossed preparation requires a different order.  Its twist projection is defined at real frequencies, and only the extracted coefficient is continued to dS.

\paragraph{What the continuation changes.}
We use the same meromorphic Virasoro kernels in the recoupling identities, CTV weights and inverse transform, continued with the specified integration contours.  The continuation nevertheless changes the real integration domains, the size of geometric contributions and the physical conjugation rule.  The following comparison anticipates the calculations that address these changes,
\begin{equation}
\renewcommand{\arraystretch}{1.3}
\begin{array}{l|c|c}
 & b>0\ \ (\text{AdS}_3) & b=re^{\ii\pi/4}\ \ (\text{dS}_3)\\\hline
 \text{external label} & \text{length }4\pi bP & \text{angle }\psi=-2\pi\ii bP\\
 \text{fixed-angle }\{6j\}^2 & e^{-cV_A/(3\pi)} & O(r^0)\times e^{2\ii V_A/(\pi r^2)}\\
 \text{level-}n\text{ cut in }\zeta & [n,\infty) & n+\ii\mathbb R_+\\
 \text{twist frequency }P_L^2-P_R^2 & \text{real} & \text{complex}\\
 \text{internal measure }\rho_0 & \text{one dominant exponential} & \text{two phases of equal modulus}\\
 \text{conjugate parameter} & b & -\ii b,\ \ c\mapsto26-c
\end{array}
\label{eq:adsdscompare}
\end{equation}
The first row compares the usual real-length labels on the AdS slice with the real-angle labels \eqref{eq:introanglespsi} on the dS ray.  The second holds the angles fixed within one nondegenerate geometric family.  At real $b$ these angles require imaginary $P$, hence below-threshold analytic labels; a conical-defect interpretation applies only in the admissible geometric range.  The fixed-angle exponential suppression in \eqref{eq:revsemicl} becomes a volume phase with order-one modulus on the complex-$b$ asymptotic branch \eqref{eq:MY}, as derived in \eqref{eq:dSsize}.

In the third row, $\zeta$ is the Laplace variable conjugate to collar length and $n$ the descendant level.  In the fourth, $P_L,P_R$ are the two chiral internal labels before the marked extraction.  Their cuts and twist dependence require the prescription of section~\ref{sec:boundaryamp}.  The fifth row concerns the internal-edge measure at fixed real length, whose two dS phases are retained in {appendix~\ref{ssec:generalinternal}}.  The last row separates the same-$b$ statistical bilinear from the physical conjugate pairing of section~\ref{sec:physicalinterpretation}, which gives the matrix-model norm.

\Needspace{7\baselineskip}
\subsection[The Virasoro cylinder and \texorpdfstring{$C_0$}{C0}-normalized statistics]{{The Virasoro cylinder and $C_0$-normalized statistics}}\label{ssec:cylinder}

We first identify the statistical correlation computed by the cylinder that prepares the Liouville state.  Fix four distinct scalar external primaries and work at real $b$, with real intermediate momenta and no discrete residue contributions.  {We evaluate the cylinder and paired contraction at real $b$ and continue that particular contribution; the general moment functional is defined in section~\ref{ssec:ensemblesample}.}  A Virasoro cylinder $\Sigma_{0,4}\times I$ has two ends with independent conformal moduli $z,w$.  The interval $I$ is the cylinder direction.  Sewing a primary and all its descendants between these ends gives \cite{CEZ}
\begin{equation}
 Z_{\rm Vir}(\Sigma_{0,4}\times I;z,w)
 =\int_0^\infty\dd P\,\rho_s(P)\,
 \mathcal F_P^{(s)}(z)\mathcal F_P^{(s)}(w),
 \label{eq:cylinderkernel}
\end{equation}
The integral sums the intermediate primary $P$, and each conformal block includes its descendants.  The sewing density $\rho_s(P)=\rho_0(P)C_0(P_1,P_2,P)C_0(P_3,P_4,P)$ combines the primary measure with the two three-point vertices, as in \eqref{eq:s2channelnormalizations}.  The diagonal topological boundary condition {\eqref{eq:s55topologicalstate}} joins the two TQFT copies.  Unfolding that boundary turns them into the two ends of a single-copy cylinder.  Setting $w=\bar z$ then gives the Liouville four-point correlator, whose continuation in $b$ is the selected dS cosmological wavefunction of section~\ref{ssec:dsproblem} \cite{CEM}.

For the paired statistical contribution, expand two boundary four-point functions in the same channel {and take the real-$b$ ensemble average \eqref{eq:s2ensembleaverage}}.  As in \eqref{eq:reviewsamechannelwick}, the Gaussian contractions identify their intermediate primaries, and for scalar external operators the OPE permutation phases cancel \cite{CCHM}, giving
\begin{equation}
\begin{aligned}
 \overline{G_{1234}(z,\bar z)G_{1234}(w,\bar w)}\Big|_{\rm paired}
 &=\sum_p\overline{|c_{12p}|^2}\;\overline{|c_{34p}|^2}\,
 \mathcal F_{P_p}^{(s)}(z)\mathcal F_{P_p}^{(s)}(w)\,
 \mathcal F_{\bar P_p}^{(s)}(\bar z)\mathcal F_{\bar P_p}^{(s)}(\bar w)\\
 &\simeq\int\dd P\,\dd\bar P\,\rho_s(P)\rho_s(\bar P)\,
 \mathcal F_P^{(s)}(z)\mathcal F_P^{(s)}(w)\,
 \mathcal F_{\bar P}^{(s)}(\bar z)\mathcal F_{\bar P}^{(s)}(\bar w).
\end{aligned}
\label{eq:pairedderivation}
\end{equation}
Each OPE variance contributes one $C_0$ factor to each chirality, while the heavy-state sum supplies $\rho_0(P)\rho_0(\bar P)$.  Consequently the two momentum integrals factorize into the cylinder of \eqref{eq:cylinderkernel}, one for each chirality \cite{CCHM}.

{We perform this contraction at real $b$ and momenta, then continue its coefficient to the dS rays \eqref{eq:physicalray}--\eqref{eq:physicalP}.  The notation $[\,\cdot\,]_{\rm cont}$ includes the specified contour deformations and crossed-pole residues.  Before continuation, conjugate amplitudes are written as reflected amplitudes $Z^\vee$, so both chiral factors continue to the same $b$.  We denote this continued contribution by $\Eds|_{\rm paired}$; section~\ref{ssec:ensemblesample} defines the general graph-moment assignment.}

{A bar on $P,h,z$ labels the right-moving sector; an overline on an observable denotes the real-$b$ ensemble average. A star denotes complex conjugation, which sends $b$ to $-\ii b$ on the dS ray. Thus the same-$b$ statistical pairing differs from the gravitational norm \eqref{eq:s55integratedstate}.}

Writing $G=G_{1234}$, this choice gives
\begin{equation}
\boxed{\begin{aligned}
 &\Eds[G(z,\bar z)G(w,\bar w)]\big|_{\rm paired}\\
 &\quad\simeq
 Z_{\rm Vir}(\Sigma_{0,4}\times I;z,w)\,
 Z_{\rm Vir}(\Sigma_{0,4}\times I;\bar w,\bar z),\\
 &\Eds[G(z,\bar z)G(\bar z,z)]\big|_{\rm paired}
 \simeq \Psi_{0,4}^{(b)}(z,\bar z)^2,
 \qquad b=re^{\ii\pi/4}.
\end{aligned}}
\label{eq:statisticalstatepair}
\end{equation}
The specialization $(w,\bar w)=(\bar z,z)$ makes each cylinder the Liouville state.  {This is the passage from figure~\ref{fig:dsstatepairing}(a) to figure~\ref{fig:dsstatepairing}(b).}  {The result is the square $\Psi^2$ at one value of $b$; section~\ref{ssec:ensemblesample} formalizes its moment assignment in \eqref{eq:Edsdefinition}.}  It is a two-copy statistical correlation and can be nonzero for distinct external labels even when the mean four-point amplitude vanishes.  More generally, the connected correlations of two or three four-point amplitudes involve the fourth and sixth OPE moments and their lower-moment products, as organized in {section~\ref{ssec:ensemblesample}}; the paired-correlator construction is that of \cite{CCHM}.

\paragraph{{The three-point $C_0$ state and the theta core.}}

{At three points the cylinder carries no moduli. The state is $\Psi_{0,3}=C_0(A)$, where $A$ denotes the three external labels and $C_0$ is the Liouville three-point coefficient in our state normalization. Pairing the two trivalent vertices along their three equal labels gives the theta graph.}  With $c_A=C_0(A)C_A$ as in {\eqref{eq:revnormalized}}, generic first moments vanish, and the theta contribution and the norm give
\begin{equation}
 \Eds[c_Ac_A]\big|_\theta=C_0(A)^2=\Psi_{0,3}(A)^2,
 \qquad
 |\Psi_{0,3}(A)|^2=C_0(A)\,C_0(A)^* .
 \label{eq:twoquadratics}
\end{equation}
{Dividing the first expression by its two $C_0(A)$ factors gives the normalized theta amplitude $Z_\theta=1$. Its compact filling has topology $M_\theta\cong\Pi\times I$ and genus-two boundary; the classical core in figure~\ref{fig:v72wormhole} is the zero-thickness, zero-volume limit. This is the three-point example. The four-point crossed coefficient below has four trivalent vertices and a different compact filling.} The second expression in \eqref{eq:twoquadratics} is the Hermitian norm, whose parameter conjugation is made explicit below.

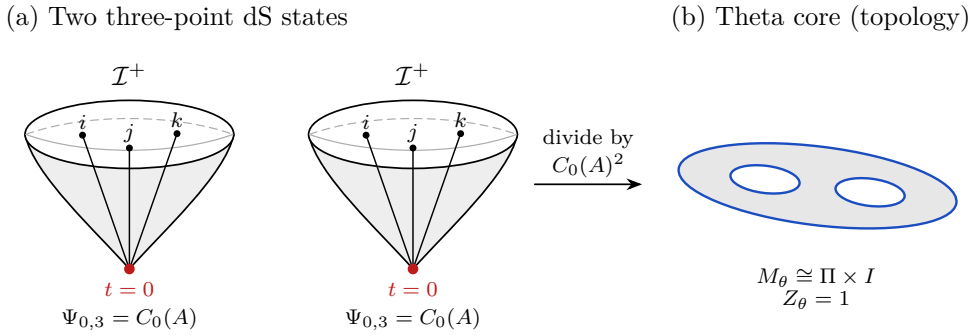
\begin{figure}[!htbp]
\centering
\begin{tikzpicture}[x=1cm,y=1cm,line width=.65pt,line cap=round,line join=round,
    >=Stealth,font=\small]
\node[anchor=west,paneltitle] at (-1.72,2.95) {(a) Two three-point dS states};
\node[anchor=west,paneltitle] at (7.08,2.95) {(b) Theta core (topology)};
\def\oneuniverse{%
  \path[fill=gray!13] (-1.38,1.42)
    .. controls (-1.27,.88) and (-.42,.12) .. (0,-.36)
    .. controls (.42,.12) and (1.27,.88) .. (1.38,1.42) --cycle;
  \draw (-1.38,1.42) .. controls (-1.27,.88) and (-.42,.12) .. (0,-.36);
  \draw (1.38,1.42) .. controls (1.27,.88) and (.42,.12) .. (0,-.36);
  \draw[fill=white] (0,1.42) ellipse (1.38 and .45);
  \draw[gray!65,densely dashed,line width=.45pt] (-1.37,1.42)
   .. controls (-.60,1.70) and (.60,1.70) .. (1.37,1.42);
  \draw[gray!65,line width=.45pt] (-1.37,1.42)
   .. controls (-.60,1.14) and (.60,1.14) .. (1.37,1.42);
  \foreach \x/\y/\lab in {-.62/1.41/i,0/1.24/j,.63/1.43/k}{
   \draw[line width=.55pt] (0,-.36)--(\x,\y);
   \fill (\x,\y) circle (1.3pt);
   \node[above,font=\scriptsize,inner sep=2pt] at (\x,\y) {$\lab$};}
  \fill[gs] (0,-.36) circle (2pt);
  \node[gs,below,font=\scriptsize] at (0,-.40) {$t=0$};
}
\begin{scope}\oneuniverse\end{scope}
\begin{scope}[shift={(3.75,0)}]\oneuniverse\end{scope}
\node[above] at (0,1.94) {$\mathcal I^+$};
\node[above] at (3.75,1.94) {$\mathcal I^+$};
\node[font=\scriptsize] at (0,-1.02) {$\Psi_{0,3}=C_0(A)$};
\node[font=\scriptsize] at (3.75,-1.02) {$\Psi_{0,3}=C_0(A)$};
\draw[->] (5.36,.76)--(6.78,.76);
\node[font=\scriptsize,align=center] at (6.07,1.17) {divide by\\$C_0(A)^2$};
\begin{scope}[shift={(9.10,.74)},rotate=-7]
 \path[fill=gray!19,even odd rule] (0,0) ellipse (1.85 and .52)
   (-.70,0) ellipse (.46 and .18) (.70,0) ellipse (.46 and .18);
 \draw[pt,line width=.9pt] (0,0) ellipse (1.85 and .52);
 \draw[pt,line width=.9pt] (-.70,0) ellipse (.46 and .18);
 \draw[pt,line width=.9pt] (.70,0) ellipse (.46 and .18);
\end{scope}
\node[align=center,font=\scriptsize] at (9.08,-.62) {$M_\theta\cong\Pi\times I$\\[-2pt] $Z_\theta=1$};
\end{tikzpicture}
\caption{{The theta second moment and its compact-core dictionary, compared with figure~\ref{fig:v72wormhole}. (a) Two three-point dS states at the same continued $b$ each contribute $\Psi_{0,3}(A)=C_0(A)$; their bilinear is $C_0(A)^2$, as in \eqref{eq:twoquadratics}. The marked $t=0$ tips belong to the cosmological preparations. (b) Dividing by $C_0(A)^2$ gives the theta-normalized amplitude $Z_\theta=1$. The right-hand pants schematic represents the topology of the graph-complement core $M_\theta=S^3\setminus N(\Gamma_\theta)\cong\Pi\times I$, whose boundary has genus two.}}
\label{fig:dsflares}
\end{figure}

\paragraph{{Four points and the tetrahedral contraction.}}

{Return to the four-point real-$b$ cylinder in \eqref{eq:cylinderkernel}. Each end now has two pairs of pants, so its channel tree has two trivalent vertices. Pairing an $s$ tree with a $t$ tree joins four vertices into $K_4$. The same $C_0$ normalization is applied locally at each vertex.}  Changing $(12)(34)$ to $(14)(23)$ uses the Virasoro fusion kernel $\mathbb F$, with sewing density $\rho_t(P_t)=\rho_0(P_t)C_0(P_1,P_4,P_t)C_0(P_2,P_3,P_t)$, and expands each $t$-channel block in $s$-channel blocks,
\begin{equation}
\mathcal F_{P_t}^{(t)}(z)=\int\dd P\,\mathbb F_{P_tP}\,\mathcal F_{P}^{(s)}(z).
\label{eq:fusionexpansion}
\end{equation}%
Multiply each conformal block by the square root of the corresponding sewing density.  In this sewing-normalized basis the \emph{same} fusion transformation has kernel
\begin{equation}
\begin{aligned}
 \widehat{\mathbb F}_{P_tP}
 &=\left[\frac{\rho_t(P_t)}{\rho_s(P)}\right]^{1/2}\mathbb F_{P_tP}\\
 &=\big[\rho_0(P_t)\rho_0(P)\big]^{1/2}
 \begin{Bmatrix}P_1&P_4&P_t\\P_3&P_2&P\end{Bmatrix}_{\!\rm Vir}.
\end{aligned}
\label{eq:Utransition}\eqalias{eq:F6j}
\end{equation}
{The second line is the Racah--Wigner normalization \eqref{eq:revfusion} applied to the $t\to s$ kernel, whose three-point factors are those of \eqref{eq:revfusion} with the two channels exchanged.  At real $b$ this relation expresses the fusion kernel through the Racah--Wigner coefficients of \cite{PT}, and \cite{TV} shows that the resulting Virasoro $6j$ symbol is tetrahedrally symmetric.  Both sides are meromorphic in the momenta and in $b$ for $\Re b>0$, being built from Barnes double-gamma functions and their contour integrals, so the identity continues to the ray \eqref{eq:physicalray}.}  The three-point factors cancel, leaving the two Plancherel densities multiplying the Virasoro $6j$.  {Thus $\widehat{\mathbb F}$ is the same change of channel, acting on sewing-normalized blocks.}  On the real starting contours it preserves the cylinder, which can be resolved in either channel,
\begin{equation}
\begin{aligned}
 Z_{\rm Vir}(\Sigma_{0,4}\times I;z,w)
 &=\int\dd P\,\rho_s(P)\,\mathcal F_P^{(s)}(z)\mathcal F_P^{(s)}(w)\\
 &=\int\dd P_t\,\rho_t(P_t)\,
  \mathcal F_{P_t}^{(t)}(z)\mathcal F_{P_t}^{(t)}(w).
\end{aligned}
\label{eq:twochannelcylinder}
\end{equation}
Substituting \eqref{eq:fusionexpansion} in the second line shows that their equality is the real-domain orthogonality of the sewing-normalized kernel,
\begin{equation}
\int\dd P_t\,\widehat{\mathbb F}_{P_tP}\widehat{\mathbb F}_{P_tQ}=\delta(P-Q).
\label{eq:Fhatorthogonality}
\end{equation}

The same kernel also appears when the two four-point amplitudes in a statistical contraction use different OPE channels.  Apply the fusion transformation to the second end of the cylinder, leaving the first in the $s$ channel.  The mixed resolution is
\begin{equation}
 Z_{\rm Vir}(\Sigma_{0,4}\times I;z,w)
 =\int\dd P\,\dd P_t\,\big[\rho_s(P)\rho_t(P_t)\big]^{1/2}\,
  \widehat{\mathbb F}_{P_tP}\,\mathcal F_P^{(s)}(z)\mathcal F_{P_t}^{(t)}(w).
 \label{eq:mixedcylinder}
\end{equation}
The corresponding statistical contraction contains the four OPE triples $(12p)$, $(34p)$, $(14q)$ and $(23q)$, with $P_p=P$ and $P_q=P_t$.  For generic distinct triples their Wick pairings vanish.  Each chiral factor instead supplies the change-of-channel kernel in \eqref{eq:mixedcylinder}.  Removing the Plancherel and three-point factors as in \eqref{eq:s2tetnormalization} gives the tetrahedral bilinear of \eqref{eq:revnongauss}.  This is the mixed-channel expression of the paired contribution just computed, as in \cite{Hartman}{, so the two expressions describe one contribution}.

\FloatBarrier

\subsection[The four-point reference density and its inverse extraction]{{The four-point reference density and its inverse extraction}}\label{sec:inverse}\label{sec:boundary}\label{sec:sewing}

{We first reconstruct the known $s$-channel density $\rho_s(P)$ of the complex Liouville four-point state, given in \eqref{eq:s55knownOPEdensity}. This reference calculation fixes the inverse map's normalization and analytic prescription. Its output is a coefficient in one channel of $\Psi_{0,4}$; the compact $K_4$ filling will be identified from the crossed-channel ratio in section~\ref{sec:crossed}. The construction averages the relative twist to match the two chiral descendant levels, then takes the discontinuity of a radial transform to isolate primary propagation. Both operations are linear in the specified state.}

{The state under discussion is the four-point wavefunction \eqref{eq:unknownstate}.  Its notation is the same as in \eqref{eq:s2wavefunctioncorrelator}.  The subscript $(0,4)$ means genus zero with four punctures, the superscript $(b)$ indicates the continued Liouville/Virasoro parameter, and the dependence on the external labels $P_1,\dots,P_4$ is suppressed.  Thus $\Psi_{0,4}^{(b)}(z,\bar z)$ is a function of the remaining modulus $z$ for fixed external primaries.}

{We use the same four-holed sphere and the same blue $s$-channel sewing circle $P$ as in figure~\ref{fig:v72sewing}.  Denote this marked circle by $\gamma_s$.  A neighbourhood of $\gamma_s$ on the two-dimensional surface is an annular collar.  It is not the three-dimensional Virasoro cylinder of section~\ref{ssec:cylinder}, nor does its radial parameter denote cosmological time.  The cross ratio $z$ parametrizes the four-punctured sphere, and the degeneration $z\to0$ pinches the $s$-channel curve separating $(12)$ from $(34)$.  To make propagation through its collar and the relative twist explicit, we replace $z$ by the elliptic modulus $\tau=u+\ii y$ of the double-cover torus,}
\begin{equation}
z=\frac{\vartheta_2(\tau)^4}{\vartheta_3(\tau)^4},\qquad q=e^{\ii\pi\tau}.
\label{eq:elliptic}
\end{equation}
{In these coordinates $q=e^{-\pi y}e^{\ii\pi u}$.  Thus $y$ controls the magnitude of the collar's plumbing parameter, while $u$ measures its relative twist.  From the standard theta-function expansions $\vartheta_2(\tau)=2q^{1/4}(1+q^2+\cdots)$ and $\vartheta_3(\tau)=1+2q+\cdots$, equation~\eqref{eq:elliptic} gives $z=16q+O(q^2)$.  Therefore the $s$-channel OPE degeneration $z\to0$ is the same limit as $q\to0$, namely $y\to\infty$.  In this limit the conformal collar becomes long even though its central geodesic becomes short.  The hyperbolic plumbing relation $\ell_\Sigma(\gamma_s)\simeq2\pi^2/(-\log|z|)$ \cite{Wolpert} gives}
\begin{equation}
\ell_\Sigma(\gamma_s)\simeq\frac{2\pi}{y}.
\label{eq:plumbinglength}
\end{equation}
{Here $\ell_\Sigma(\gamma_s)$ is a length on the two-dimensional boundary surface, not an edge length of the compact bulk core.  A shift $u\to u+2$ changes the relative twist by one full Dehn twist and leaves $q$ invariant.  This is the period used below in the first step of the inverse map.  The reason for introducing these coordinates is that they make each Virasoro descendant level a definite twist Fourier mode.  In Zamolodchikov's elliptic representation \cite{Zam},}
\begin{equation}
\mathcal F_P^{(s)}(z)=E(\tau)(16q)^{P^2}\sum_{n\ge0}a_n(P)q^n,\qquad a_0=1 .
\label{eq:block}
\end{equation}
{The power $(16q)^{P^2}$ carries the primary momentum, and $a_n(P)q^n$ collects the level-$n$ contribution with $a_0(P)=1$.  Since $q^n=e^{-\pi ny}e^{\ii\pi nu}$, the level is visible in both radial propagation and twist dependence.  The remaining factor $E(\tau)$ depends on the external weights and coordinates but not on the exchanged momentum $P$.  In the ordering $(1,2,3,4)$ at $(0,z,1,\infty)$ it is \cite{Zam}}
\begin{equation}
E(\tau)=z^{Q^2/4-h_1-h_2}(1-z)^{Q^2/4-h_2-h_3}
 \vartheta_3(\tau)^{3Q^2-4\sum_{i=1}^4h_i}.
\label{eq:ellipticprefactor}
\end{equation}
{We remove this known $P$-independent factor from each chiral block before projecting.  The second elliptic nome is $q(-\bar\tau)=\bar q=e^{-\pi y-\ii\pi u}$; evaluating at $\bar z$ changes the coordinate, not the continued parameter $b$ shared by the two factors.  The reduced wavefunction is therefore defined by}
\begin{equation}
\Psi(z,\bar z)=E(\tau)E(-\bar\tau)\,\Psi_{\rm red}(u,y).
\label{eq:Psired}
\end{equation}

{The first inverse operation is a Fourier projection in the twist $u$, with the radial modulus $y$ fixed.  For the reference state, the two chiral blocks have the same intermediate primary momentum, so only the difference of their descendant levels contributes to the twist phase.  We average a single reduced wavefunction over one full period,}
\begin{equation}
Z_{\gamma_s}[\Psi](y)
 =\frac12\int_0^2\dd u\,\Psi_{\rm red}(u,y),\qquad y>\frac12.
\label{eq:Zboundary}
\end{equation}
{We call $Z_{\gamma_s}[\Psi](y)$ the marked boundary amplitude.  It depends on the remaining modulus $y$, and the integral is normalized by the twist period rather than by a gravitational inner product.  Substituting the $s$-channel expansion \eqref{eq:unknownstate} and the two elliptic blocks \eqref{eq:block} shows explicitly why it is a Fourier projection.  With $q=e^{-\pi y+\ii\pi u}$ and $\bar q=e^{-\pi y-\ii\pi u}$, the factor from levels $(m,n)$ is $q^{P^2+m}\bar q^{P^2+n}=e^{-\pi y(2P^2+m+n)}e^{\ii\pi u(m-n)}$, leaving}
\begin{equation}
\Psi_{\rm red}(u,y)
 =\int\dd P\,\rho_s(P)16^{2P^2}\sum_{m,n\ge0}a_m(P)a_n(P)
 e^{-\pi y(2P^2+m+n)}e^{\ii\pi u(m-n)}.
\label{eq:scalarfull}
\end{equation}
{Here $m$ and $n$ are the two chiral descendant levels.  The phase $e^{\ii\pi u(m-n)}$ depends only on their integer difference, so integration over the full labelled-sphere twist period gives exactly}
\begin{equation}
\frac12\int_0^2\dd u\,e^{\ii\pi u(m-n)}=\delta_{mn}.
\label{eq:twistprojector}
\end{equation}
{Thus the reference-state average keeps $(m,n)=(0,0),(1,1),(2,2),\ldots$ and annihilates all terms with $m\neq n$.  It imposes equality of the two levels; it does \emph{not} remove the descendants.  In particular, the primary pair $(0,0)$ and all matched descendant pairs $(n,n)$ survive, giving}
\begin{equation}
Z_{\gamma_s}(y)
 =\int_{e^{\ii\pi/4}\mathbb R_+}\dd P\,\rho_s(P)16^{2P^2}
 \sum_{n\ge0}a_n(P)^2e^{-2\pi(P^2+n)y}.
\label{eq:scalarZ}
\end{equation}
{In the remaining propagation factor $e^{-2\pi(P^2+n)y}$, the primary has $n=0$ and each matched descendant shifts the frequency from $P^2$ to $P^2+n$.  A second inverse operation must separate these frequencies; otherwise \eqref{eq:scalarZ} still contains the full descendant tower.}\footnote{For an integer-spin primary the twist condition is $j+m-n=0$.  The reference state used here has scalar primaries.  A reconstruction including spinning primaries must keep their own level-matching and threshold conditions.}

{Before defining that second operation, we fix the measure and the range used in the twist average.  The ``observer'' specifies the marking $\gamma_s$, not an additional physical worldline.  The gravitational pairing supplies a natural measure on moduli space, but it is distinct from the single-state functional \eqref{eq:Zboundary}.  For the four-punctured sphere the gauge-fixed measure is $\dd^2z$ \cite{CEM,CLS}.  In the scalar principal-series domain $\Re h_i=1/2$ and $\Re c=13$, so $|E(\tau)E(-\bar\tau)|^2=|z(1-z)|^{-2}|\vartheta_3(\tau)|^{-8}$.  With $\dd z/\dd\tau=\ii\pi\vartheta_3(\tau)^4z(1-z)$, the Jacobian cancels the removed prefactors,}
\begin{equation}
\dd^2z\,|E(\tau)E(-\bar\tau)|^2
 =\left|\frac{\dd z}{\dd\tau}\right|^2
 \frac{\dd u\,\dd y}{|z(1-z)|^{2}|\vartheta_3(\tau)|^{8}}
 =\pi^2\,\dd u\,\dd y ,
\label{eq:flatmeasure}
\end{equation}
{Accordingly the physical pairing becomes a flat integral of \emph{two} reduced wavefunctions,}
\begin{equation}
\langle\Psi_1|\Psi_2\rangle_{\rm grav}
 =\pi^2\int\dd u\,\dd y\,\Psi_{1,\rm red}(u,y)^*\Psi_{2,\rm red}(u,y).
\label{eq:flatpairing}
\end{equation}

{The norm \eqref{eq:flatpairing} integrates over $u$ and $y$ and includes complex conjugation, whereas \eqref{eq:Zboundary} averages one state over $u$ at fixed $y$.  The flat measure supports the uniform twist average.  Its geometric domain is the cusp of $\mathcal M_{0,4}$, the upper half-plane modulo $\Gamma(2)$ (with the trivially acting centre understood).  The cusp stabilizer identifies $u$ with $u+2$.  For an element outside that stabilizer, write $\gamma\tau=(a\tau+b)/(c\tau+d)$, where the matrix entry $c$ is unrelated to central charge.  Since $|c|\ge2$,}
\begin{equation}
\Im(\gamma\tau)=\frac{y}{(cu+d)^2+c^2y^2}\le\frac1{4y}.
\label{eq:cusp}
\end{equation}
{If $y>1/2$, the right-hand side is smaller than $1/2$.  Consequently no other modular image of the patch $0\le u<2$, $y>1/2$ lies in the same patch; the only identification is the one-period twist.  This verifies the normalization and domain of \eqref{eq:Zboundary} without changing the level-matching result \eqref{eq:twistprojector}.}

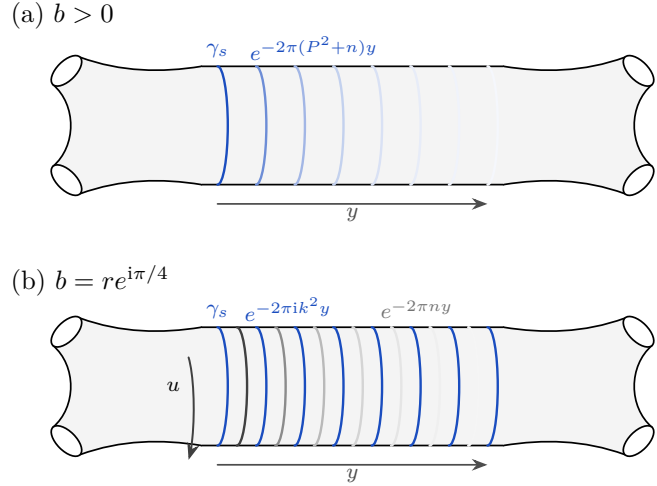
\begin{figure}[!htbp]
\centering
\begin{tikzpicture}[x=1cm,y=1cm,line width=.65pt,line cap=round,line join=round,>=Stealth,font=\small,
  lab/.style={font=\scriptsize,inner sep=1.2pt}]
\newcommand{\longcollar}{%
 \begin{scope}[shift={(-3.25,0)}]\figpants\figleftcuffs\end{scope}
 \begin{scope}[shift={(3.25,0)},xscale=-1]\figpants\figleftcuffs\end{scope}
 \fill[gray!9] (-2.35,.92) rectangle (2.35,-.92);
 \draw (-2.35,.92)--(2.35,.92);\draw (-2.35,-.92)--(2.35,-.92);}
\newcommand{\ring}[2]{\draw[#2,line width=.9pt] (#1,-.92) arc[start angle=270,end angle=450,x radius=.16,y radius=.92];}
\node[anchor=west,paneltitle] at (-4.6,1.45) {(a) $b>0$};
\begin{scope}[scale=.85]
 \longcollar
 \foreach \i in {0,...,7}{\pgfmathsetmacro{\op}{100*exp(-.45*\i)}\ring{-2.10+.60*\i}{pt!\op!white}}
 \draw[->,black!70] (-2.10,-1.22)--(2.10,-1.22) node[midway,below,lab]{$y$};
 \node[pt,lab] at (-.6,1.20) {$e^{-2\pi(P^2+n)y}$};
 \node[pt,lab] at (-2.10,1.16) {$\gamma_s$};
\end{scope}
\node[anchor=west,paneltitle] at (-4.6,-2.05) {(b) $b=re^{\ii\pi/4}$};
\begin{scope}[shift={(0,-3.45)},scale=.85]
 \longcollar
 \foreach \i in {0,...,6}{\pgfmathsetmacro{\op}{75*exp(-.50*\i)}\ring{-1.80+.60*\i}{black!\op!white}}
 \foreach \i in {0,...,7}{\ring{-2.10+.60*\i}{pt}}
 \draw[->,black!75,line width=.75pt] (-2.55,.45) arc[start angle=60,end angle=-60,x radius=.16,y radius=.92];
 \node[lab,anchor=east] at (-2.62,0) {$u$};
 \draw[->,black!70] (-2.10,-1.22)--(2.10,-1.22) node[midway,below,lab]{$y$};
 \node[pt,lab] at (-1.0,1.20) {$e^{-2\pi\ii k^2y}$};
 \node[pt,lab] at (-2.10,1.16) {$\gamma_s$};
 \node[black!55,lab] at (1.0,1.20) {$e^{-2\pi ny}$};
\end{scope}
\end{tikzpicture}
\caption{{The marked boundary collar, using the same blue sewing circle $\gamma_s$ and three-holed surfaces as figure~\ref{fig:v72sewing}.  The repeated cross-sectional rings indicate copies of the same cuff along the annulus, not different descendant states.  The coordinate $y$ runs along the conformal collar and $u$ is its relative twist.  (a) For real $b$ and real momentum, primary propagation decays as $e^{-2\pi P^2y}$ and level-$n$ descendants acquire the additional factor $e^{-2\pi ny}$.  (b) On the dS ray $P^2=\ii k^2$, primary propagation is oscillatory with unit modulus, while the descendant factors still decay.  The subsequent radial transform separates their spectral cuts and the primary discontinuity yields $\rho_s(P)$ through \eqref{eq:inverse}.}}
\label{fig:dsworkflow}
\end{figure}

{We now perform the second, dS-specific part of the inversion.  The twist average did not discard the terms with $n>0$.  What distinguishes them is their shifted radial frequency $P^2+n$.  On the physical ray $P=e^{\ii\pi/4}k$, $k>0$, we have $P^2=\ii k^2$, so the $n=0$ primary factor $e^{-2\pi\ii k^2y}$ is oscillatory, whereas a level-$n$ contribution has the additional decay $e^{-2\pi ny}$.  Decay by itself does not isolate the primary exactly.  Instead, a Laplace transform in $y$ turns the frequencies into singularities of a new complex variable $\zeta$, so the contribution with $n=0$ can be distinguished by its own cut.}

{Because the momentum continuum is oscillatory on this ray, we first shift $y$ by $-\ii\varepsilon/(2\pi)$.  A frequency $P^2+n$ then acquires $e^{\ii\varepsilon(P^2+n)}=e^{-\varepsilon k^2}e^{\ii\varepsilon n}$; the first factor regulates large momentum and the second leaves the descendant-level dependence explicit.  We work initially in a common convergence domain, with the regulated momentum integral and locally uniformly convergent level series, and define the radial Laplace transform there,}
\begin{equation}
\widehat Z_{\gamma_s}(\zeta;\varepsilon)
 =\int_Y^\infty\dd y\,e^{2\pi\zeta y}
 Z_{\gamma_s}\!\left(y-\frac{\ii\varepsilon}{2\pi}\right),\qquad \Re\zeta<0,
\label{eq:transform}
\end{equation}
{Here $Y>1/2$ fixes the starting point within the nonoverlapping cusp patch, and $\Re\zeta<0$ specifies the initial Laplace domain.  The defining calculation can be done for each frequency $\omega=P^2+n$.  Combining its propagation with the Laplace kernel gives}
\begin{equation}
\int_Y^\infty\dd y\,e^{2\pi\zeta y}
 e^{-2\pi\omega(y-\ii\varepsilon/2\pi)}
 =\frac{e^{\ii\varepsilon\omega}}{2\pi(\omega-\zeta)}\,e^{-2\pi(\omega-\zeta)Y}.
\label{eq:singlefrequency}
\end{equation}
{The denominator $1/(\omega-\zeta)$ is the crucial new feature, since radial propagation has become a Cauchy-type resolvent in the spectral variable.  Applying \eqref{eq:singlefrequency} to all terms in \eqref{eq:scalarZ}, using the stated interchange of the sum and integrals, yields}
\begin{equation}
\widehat Z_{\gamma_s}(\zeta;\varepsilon)
 =\sum_{n\ge0}\int\dd P\,
 \frac{\rho_s(P)16^{2P^2}a_n(P)^2e^{\ii\varepsilon(P^2+n)}}
 {2\pi(P^2+n-\zeta)}e^{-2\pi(P^2+n-\zeta)Y}.
\label{eq:transformed}
\end{equation}
{For each fixed $P$, the level-$n$ term has a potential pole at $\zeta=P^2+n$.  After integrating over continuous $P$, these poles produce a spectral cut.  The distinction between AdS starting data and dS data is now sharp.  For real $P$, the possible cuts $[n,\infty)$ overlap on the real axis.  For $P^2=\ii k^2$, however, they are translated vertical rays $n+\ii\mathbb R_+$.  The primary cut at $\zeta\in\ii\mathbb R_+$ is therefore isolated from every $n>0$ cut.  Figure~\ref{fig:dsworkflow} shows the original collar propagation responsible for these distinct frequencies.  For identical external weights, the reduced elliptic coefficients further obey \cite{Cardona2021}}
\begin{equation}
h_1=h_2=h_3=h_4\quad\Longrightarrow\quad
a_{2j+1}(P)=0,\qquad
\zeta\in2j+\ii\mathbb R_+,\quad j=0,1,\ldots .
\label{eq:evenellipticcuts}
\end{equation}
{Thus for this symmetric external configuration only $n=0,2,4,\ldots$ can contribute cuts of the reduced amplitude; the odd-level states have not been deleted from the Virasoro module.  Starting from $\Re\zeta<0$, continue beneath the origin to the east side of the primary cut.  Across a point $\zeta=P^2\neq0$ on this cut, the $n>0$ contributions are locally analytic.  Hence, provided the full descendant series converges locally uniformly there, only the $n=0$ term contributes to the discontinuity.}

{To evaluate that discontinuity, write the primary term of \eqref{eq:transformed} as a Cauchy transform.  Changing variables from $P$ to $\omega=P^2$ gives $\dd P=\dd\omega/(2P)$ on the oriented physical ray.  This Jacobian and the $1/(2\pi)$ from the Laplace integral combine to produce the spectral density in}
\begin{equation}
\begin{aligned}
 \widehat Z_{\gamma_s}^{(n=0)}(\zeta;\varepsilon)
 &=e^{2\pi\zeta Y}\int_{\ii\mathbb R_+}\dd\omega\,
       \frac{e^{-2\pi\omega Y}g_\varepsilon(\omega)}{\omega-\zeta},\\
 g_\varepsilon(\omega)
 &=\frac{\rho_s(\sqrt\omega)16^{2\omega}e^{\ii\varepsilon\omega}}
         {4\pi\sqrt\omega}.
\end{aligned}
\label{eq:omegaform}
\end{equation}
{The square root is taken along the chosen physical momentum ray.  With the contour $\ii\mathbb R_+$ oriented upward, define $\Disc$ as the west-minus-east boundary value of the Cauchy transform.  The Plemelj relation then gives $2\pi\ii$ times the density at $\omega=\zeta=P^2$.  At that point the factor $e^{2\pi\zeta Y}$ outside the integral cancels $e^{-2\pi\omega Y}$ inside, so the result is independent of the arbitrary lower endpoint $Y$,}
\begin{equation}
\begin{aligned}
 \Disc\widehat Z_{\gamma_s}(P^2;\varepsilon)
 &=2\pi\ii\,e^{2\pi P^2Y}e^{-2\pi P^2Y}g_\varepsilon(P^2)\\
 &=2\pi\ii\,\frac{\rho_s(P)16^{2P^2}e^{\ii\varepsilon P^2}}{4\pi P}.
\end{aligned}
\label{eq:Plemelj}
\end{equation}
{This equation has only the desired $\rho_s(P)$ and the known Jacobian, elliptic normalization and regulator factors.  Multiplying by their inverses gives the reconstruction formula}
\begin{equation}
\boxed{\rho_s(P)
 =\frac{2P}{\ii}\,16^{-2P^2}e^{-\ii\varepsilon P^2}
 \Disc\widehat Z_{\gamma_s}(P^2;\varepsilon).}
\label{eq:inverse}
\end{equation}
{The factor $2P/\ii$ in \eqref{eq:inverse} reverses the Cauchy jump and the Jacobian $\dd(P^2)=2P\,\dd P$; $16^{-2P^2}$ removes the primary normalization of the elliptic block, and $e^{-\ii\varepsilon P^2}$ cancels the regulator at the discontinuity.  The matched $n>0$ descendants were included throughout \eqref{eq:transformed}; their analytic continuation contributes no jump across the isolated primary cut under the local uniform convergence assumption.  The reconstruction thus uses the full wavefunction and is independent of the arbitrary lower endpoint $Y$.}

{The output must agree with the known Liouville sewing density if the inversion is correctly normalized.  It does, since $\rho_s(P)$ contains one Plancherel factor and the two three-point coefficients adjacent to the marked $s$-channel cuff.  The corresponding $t$-channel density is obtained by the other pants decomposition,}
\begin{equation}
\begin{aligned}
\rho_s(P)&=\rho_0(P)C_0(P_1,P_2,P)C_0(P_3,P_4,P),\\
 \rho_t(P)&=\rho_0(P)C_0(P_1,P_4,P)C_0(P_2,P_3,P).
\end{aligned}
\label{eq:rhos}
\end{equation}

{This completes the calibration of the inverse map, not an independent computation of the Liouville structure constants.  The reference input was diagonal in its $s$-channel primary momentum, making the twist projection a simple level-matching condition $m=n$.  We now keep the same observer curve $\gamma_s$ but prepare a fixed $t$-channel component.  Its $s$-channel expansion has independent chiral momenta, and the twist projection must be defined first in the real starting domain.  The resulting crossed coefficient can then be continued to dS and compared with the reference density to recover a nontrivial compact-core coefficient.}

\subsection[Crossed extraction for \texorpdfstring{$K_4$}{K4} and the linked handcuff]{{Crossed extraction for $K_4$ and the linked handcuff}}\label{sec:crossed}\label{ssec:channelcomponent}

{The inverse map now becomes a nontrivial channel calculation.  The observer continues to mark $\gamma_s$, but instead of the reference state's full $s$-channel sum we prepare a single unweighted $t$-channel block pair at fixed $P_t$.  Its $s$-channel primary coefficient $\rho_{st}(P;P_t)$ differs from the known density $\rho_s(P)$ because it contains the fusion transformation between distinct pants decompositions.  After extraction, the normalized ratio of these coefficients yields the tetrahedral $6j^2$ associated with figure~\ref{fig:reviewcores}(a).  The new difficulty is that the $s$-channel expansion of this crossed preparation has independent left and right intermediate momenta; the real-domain twist projection must therefore precede continuation to dS.  The prepared component is}
\begin{equation}
\Psi_t(z,\bar z;P_t)
 =\mathcal F_{P_t}^{(t)}(z)\mathcal F_{P_t}^{(t)}(\bar z).
\label{eq:crossedstate}
\end{equation}
{The fixed label $P_t$ identifies one component of the $t$-channel decomposition, not a second independent cosmological state.  The block $\mathcal F^{(t)}(z)$ is expressed in the original cross ratio using its natural local coordinate $1-z$, as in \eqref{eq:reviewblockfusion}.  The full Liouville state sums these components with the known density $\rho_t(P_t)$,}
\begin{equation}
 \Psi^{(b)}_{0,4}(z,\bar z)
 =\int\dd P_t\,\rho_t(P_t)\,\Psi_t(z,\bar z;P_t).
 \label{eq:statecomponents}
\end{equation}
{To extract a coefficient at the observer's $s$-channel cuff, expand each factor of \eqref{eq:crossedstate} using the fusion kernel \eqref{eq:fusionexpansion}.  This produces independent output momenta $P_L$ and $P_R$.  Their twist frequency is now $P_L^2+m-P_R^2-n$, whereas the reference state in \eqref{eq:scalarfull} had the simpler frequency $m-n$.  The two inverse problems therefore share the collar and radial transform but not the initial twist projection.}

\begin{figure}[!htbp]
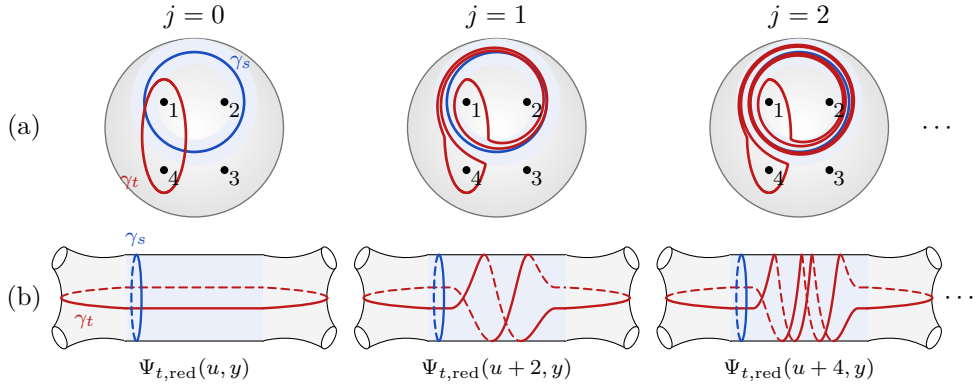

\centering

\caption{{The crossed preparation and the Dehn twist about the observer curve.  (a) On $\Sigma_{0,4}$ the prepared curve $\gamma_t$ separates $\{1,4\}$ from $\{2,3\}$ and meets the observer curve $\gamma_s$ twice.  The $j$-th power of the Dehn twist about $\gamma_s$ keeps $\gamma_s$ fixed and, inside its collar (shaded), winds each of the two crossings $j$ times around it.  (b) The same images on the collar of figure~\ref{fig:dsworkflow}, where both strands of $\gamma_t$ turn $j$ times around the collar.  The $j$-th image is the state at the twisted modulus, $\Psi_{t,\rm red}(u+2j,y)$, and summing the images and averaging over one period gives the marked lift \eqref{eq:crossedZdef}.  A term of twist frequency $\omega$ enters the $j$-th image with weight $e^{2\pi\ii j\omega}$.  At real $b$ this weight is a phase, and the sum implements the delta function \eqref{eq:realdelta}.  On the dS ray its modulus is $e^{-2\pi j\Im\omega}$, and the sum diverges in one direction.  The projection is therefore evaluated first at real $b$, and only the extracted coefficient is then continued.}}
\label{fig:v91crossed}
\end{figure}

A single $t$-channel block pair is not periodic under the observer's Dehn twist $u\to u+2$.  It belongs to the marked description before the mapping-class quotient, whose distinction from the quotient Hilbert space is discussed in \cite{CEM}.  Our prescription sums only its Dehn-twist images in the embedded cusp $y>1/2$.  On the marked cover, $u\in\mathbb R$ and the intervals $[2j,2j+2)$ are related by these twists.  Summing the images and averaging over one period unfolds to the full line.  We define the sum distributionally, with a common regulator and test wavepackets in the twist frequency,
\begin{equation}
Z_{\gamma_s}^{(t)}(y)
 =\frac12\int_0^2\dd u\sum_{j\in\mathbb Z}\Psi_{t,\rm red}(u+2j,y)
 =\frac12\int_{\mathbb R}\dd u\,\Psi_{t,\rm red}(u,y).
\label{eq:crossedZdef}
\end{equation}
We call the unfolded expression the \emph{marked lift}.  With $b$ and the output momenta initially real, it is a distributional projection onto zero total twist frequency, tested in the frequency variables.  On the dS ray the frequency $P_L^2+m-P_R^2-n$ is complex, and $\int_{\mathbb R}\dd u\,e^{\ii\pi u\omega}$ diverges when $\Im\omega\neq0$.  We therefore evaluate the projection where it is a delta function and continue only the extracted coefficient.  {Figure~\ref{fig:v91crossed} shows the twist images.}

A term at levels $(m,n)$ has twist dependence $e^{\ii\pi u(P_L^2+m-P_R^2-n)}$.  The distributional Fourier integral is therefore
\begin{equation}
\frac12\int_{\mathbb R}\dd u\,e^{\ii\pi u(P_L^2+m-P_R^2-n)}
 =\delta(P_L^2+m-P_R^2-n).
\label{eq:realdelta}
\end{equation}
For the reference state, equality of primary weights reduced this projection to $m=n$.  For the crossed preparation it instead equates the full frequencies $P_L^2+m=P_R^2+n=\omega$.  Since both output momenta are real at this stage, $\omega$ is at least $m$ and at least $n$.  The primary pair $(m,n)=(0,0)$ thus begins at zero; all other pairs begin at $\max(m,n)\ge1$.  Changing each output momentum to its frequency also supplies the two Jacobians used in the next formula.

The primary term makes both Jacobians explicit.  On the positive real momentum contours, integrating $\delta(P_L^2-P_R^2)$ and setting $\omega=P_L^2$ gives $\dd\omega/(4\omega)$.  Combining this with the radial integral \eqref{eq:singlefrequency}, in the same regulated convergence domain, gives
\[
 \left.\widehat Z_{\gamma_s}^{(t)}(\zeta;\varepsilon)\right|_{m=n=0}
 =\int_0^\infty\dd\omega\,
 \frac{16^{2\omega}e^{\ii\varepsilon\omega}
       \mathbb F_{P_t,\sqrt\omega}^{\,2}}
      {8\pi\omega(\omega-\zeta)}
 e^{-2\pi(\omega-\zeta)Y}.
\]
Near $\zeta=0$, only this primary pair reaches its threshold; every descendant threshold is at least one unit away.  For generic labels away from kernel poles, the factor $\rho_0(P)=O(P^2)$ in \eqref{eq:revfusion} makes $\mathbb F_{P_tP}=O(P^2)$, with an even analytic remainder.  Thus $\mathbb F_{P_t,\sqrt\omega}^{\,2}/\omega$ is analytic at $\omega=0$.  The Cauchy integral has a logarithm whose coefficient is minus its analytic numerator evaluated at $\omega=\zeta$; the cutoff factor is then one.  The descendant integrals are single-valued in a disk ending before their first threshold.  The local monodromy is therefore fixed by
\begin{equation}
\widehat Z_{\gamma_s}^{(t)}(\zeta;\varepsilon)
 =-\frac{16^{2\zeta}e^{\ii\varepsilon\zeta}}{8\pi\zeta}
 \mathbb F_{P_t,\sqrt\zeta}^{\,2}\log(-\zeta)+\text{holomorphic},
\label{eq:monodromyexact}
\end{equation}
and we define $\Disc_0$ as the change after one clockwise circuit around the origin.  Since $\log(-\zeta)$ changes by $-2\pi\ii$,
\begin{equation}
\Disc_0\widehat Z_{\gamma_s}^{(t)}(\zeta;\varepsilon)
 =\frac{\ii\,16^{2\zeta}e^{\ii\varepsilon\zeta}}{4\zeta}
 \mathbb F_{P_t,\sqrt\zeta}^{\,2}.
\label{eq:monodromyexactdisc}
\end{equation}
The descendant terms are single-valued on this circuit.  The clockwise convention gives the same positive $2\pi\ii$ density factor as the west-minus-east jump in \eqref{eq:Plemelj}.  We use this local monodromy in place of the reference cut discontinuity in \eqref{eq:inverse}.  After removing the primary normalization it defines the crossed coefficient,
\begin{equation}
\rho_{st}(P;P_t)
 =\frac{2P}{\ii}\,16^{-2P^2}e^{-\ii\varepsilon P^2}
 \Disc_0\widehat Z_{\gamma_s}^{(t)}(P^2;\varepsilon)
 =\frac{\mathbb F_{P_tP}^{\,2}}{2P}.
\label{eq:crossinverse}
\end{equation}

We continue the extracted meromorphic coefficient in $P$, $P_t$ and $b$ from the real starting domain to the dS ray.  In the continued expression, $\mathbb F_{P_tP}^{\,2}$ remains the bilinear product of the two chiral fusion kernels on the same branch.  Keeping the generic-label, pole-avoidance conditions just stated, for equal external labels we use the nondegenerate chamber
\begin{equation}
\begin{aligned}
 P_0&=\frac{\ii\psi_0}{2\pi b},\qquad
 P=\frac{\ii\psi}{2\pi b},\\
 &0<\psi_0<\frac\pi2,\qquad 0<\psi<\pi-2\psi_0 ,
\end{aligned}
\label{eq:externalchamber}
\end{equation}
with the coefficient followed on the analytic sheet fixed by the real starting domain.  This chamber avoids the fusion-kernel pole families \eqref{eq:fusionpoles} of appendix~\ref{app:contour}.

We now convert the extracted coefficient to the compact normalization.  The factor $2P$ removes the twist-delta Jacobian, $\rho_t(P_t)$ restores the weight of the selected component in \eqref{eq:statecomponents}, and division by $\rho_s(P)$ gives the sewing-normalized channel coefficient.  Thus \eqref{eq:Utransition} gives
\begin{equation}
 \frac{2P\,\rho_t(P_t)}{\rho_s(P)}\,\rho_{st}(P;P_t)
 =\frac{\rho_t(P_t)}{\rho_s(P)}\,\mathbb F_{P_tP}^{\,2}
 =\widehat{\mathbb F}_{P_tP}^{\,2}
 =\rho_0(P_t)\rho_0(P)
 \begin{Bmatrix}P_1&P_4&P_t\\P_3&P_2&P\end{Bmatrix}_{\!\rm Vir}^{\!2}.
 \label{eq:componentextraction}\eqalias{eq:ratioderivation}
\end{equation}
The observer-channel three-point factors cancel in this ratio.  The preparation factors are $\rho_t(P_t)/\rho_0(P_t)$, equal to $C_0(P_0,P_0,P_t)^2$ for equal external labels.  Below, $\Disc$ is the jump across the reference state's dS primary cut, whereas $\Disc_0$ is the real-domain threshold monodromy followed by continuation.  Using the common extraction normalization \eqref{eq:inverse} gives
\begin{equation}
\boxed{\begin{aligned}
 \frac{2P\,\rho_t(P_t)}{\rho_0(P_t)\rho_0(P)}\,
 \frac{\rho_{st}(P;P_t)}{\rho_s(P)}
 &=\begin{Bmatrix}P_1&P_4&P_t\\P_3&P_2&P\end{Bmatrix}_{\!\rm Vir}^{\!2},\\
 \frac{2P\,C_0(P_0,P_0,P_t)^2}{\rho_0(P)}
 \frac{\Disc_0\widehat Z_{\gamma_s}^{(t)}(P^2;\varepsilon)}
      {\Disc\widehat Z_{\gamma_s}(P^2;\varepsilon)}
 &=\begin{Bmatrix}P_0&P_0&P_t\\P_0&P_0&P\end{Bmatrix}_{\!\rm Vir}^{\!2}.
\end{aligned}}
\label{eq:mainratio}\eqalias{eq:generallabelratio}\eqalias{eq:statefunctional}
\end{equation}

The normalized ratio of crossed and reference primary discontinuities is the tetrahedral $6j^2$.  Equivalently, it is the square of the sewing-normalized fusion kernel divided by the two Plancherel densities.  This construction selects channel components algebraically.  A physical orthogonal projection would additionally require a basis for the gravitational pairing, since ordinary block products are nonorthogonal and complex Virasoro partial waves remain to be constructed \cite{CEM}.

\phantomsection\paragraph{{The punctured torus.}}\label{sec:torus}

{The punctured torus supplies the second application of the same inverse methodology.  The sphere calculation changed a pair-of-pants channel by the fusion kernel and produced the tetrahedral coefficient.  On the torus the corresponding change is modular.  The observer fixes the $a$ cycle, while the preparation uses the intersecting $b$ cycle, with $a\cdot b=1$.  Its crossed/reference extraction leads to the linked-handcuff coefficient of figure~\ref{fig:reviewcores}(b).  The nome is now $q=e^{2\pi\ii\tau}$ rather than the sphere's $e^{\ii\pi\tau}$, so the twist period and the normalization of the radial transform also change.  The torus block with external momentum $P_0$ is}
\begin{equation}
\mathcal F_P^{(1)}[P_0](\tau)
 =\frac{q^{P^2}}{\eta(\tau)}\sum_{n\ge0}a_n^{(1)}(P)q^n .
\label{eq:torusblock}
\end{equation}
Here $\eta$ is the Dedekind eta function, $a_0^{(1)}(P)=1$ fixes the primary normalization{, and $h_0=h(P_0)$ is the weight of the puncture}.  In the torus pairing of \cite{CLS}, the ghost contribution and the order-two automorphism give $\frac12|2\pi\eta(\tau)^2|^2$.  Removing the eta denominator from each chiral block cancels $|\eta|^4$ in this measure, leaving
\begin{equation}
\langle\Psi_1|\Psi_2\rangle_{1,1}
 =2\pi^2\int\dd u\,\dd y\,\Psi_{1,\rm red}^{*}\Psi_{2,\rm red}.
\label{eq:torusmeasure}
\end{equation}
The reference state has $a$-channel sewing density $\rho_0(P)C_0(P,P,P_0)$, since the torus is obtained by sewing two legs of one three-point vertex.  The cusp has period one, and its marked amplitude is the twist average
\begin{equation}
Z_a[\Psi](y)=\int_0^1\dd u\,\Psi_{\rm red}(u,y),\qquad y>1.
\label{eq:torusreference}
\end{equation}
The crossed preparation is a single block in the $b$-cycle channel.  It is expanded in $a$-cycle blocks by the punctured modular transformation,
\begin{equation}
\mathcal F^{(1)}_{P_t}[P_0](-1/\tau)
 =(-\ii\tau)^{h_0}\int_0^\infty\dd P\,e^{-\ii\pi h_0/2}
 \mathbb S_{P_tP}[P_0]\mathcal F^{(1)}_P[P_0](\tau).
\label{eq:torusmodular}
\end{equation}
Before averaging the crossed torus block, we remove the puncture factor $(-\ii\tau)^{h_0}$ from each chiral copy, as well as the eta prefactor.  The puncture factor depends on the twist and would otherwise prevent a Fourier projection.  Denoting the resulting state by $\Psi_{S,\rm red}$, the period-one marked lift is
\[
 Z_a^{(S)}(y)
 =\int_0^1\dd u\sum_{j\in\mathbb Z}\Psi_{S,\rm red}(u+j,y)
 =\int_{\mathbb R}\dd u\,\Psi_{S,\rm red}(u,y).
\]
As on the sphere, this identity is defined first at real $b$ and real output momenta, with a common regulator and frequency wavepackets, and only the extracted coefficient is continued.  The nome $q=e^{2\pi\ii\tau}$ gives radial propagation $e^{-4\pi(P^2+n)y}$.  Accordingly, for either torus preparation we define
\begin{equation}
 \widehat Z_a(\zeta;\varepsilon)
 =\int_Y^\infty\dd y\,e^{4\pi\zeta y}
 Z_a\!\left(y-\frac{\ii\varepsilon}{4\pi}\right),
 \qquad Y>1,\quad\Re\zeta<0 .
 \label{eq:torusLaplace}
\end{equation}
A single frequency contributes $e^{\ii\varepsilon\omega}e^{-4\pi(\omega-\zeta)Y}/[4\pi(\omega-\zeta)]$.  With the same regulated convergence and generic-label analyticity assumptions as on the sphere, the real-domain marked lift and primary monodromy give
\begin{equation}
\Disc_0\widehat Z_a^{(S)}(\zeta;\varepsilon)
 =\frac{\ii e^{-\ii\pi h_0}e^{\ii\varepsilon\zeta}}{8\zeta}
 \mathbb S_{P_t,\sqrt\zeta}[P_0]^2 .
\label{eq:torusmonodromy}
\end{equation}
The torus exponent halves the $y$ integral relative to the sphere, which gives the denominator $8$ in place of $4$ in \eqref{eq:monodromyexactdisc}.  To compare the extracted coefficients with graph amplitudes, we use the Racah--Wigner normalization \cite{CTV,PostTsiares}
\begin{equation}
\widehat{\mathbb S}_{P_tP}[P_0]
 =\frac{1}{\rho_0(P)}
 \sqrt{\frac{C_0(P_t,P_t,P_0)}{C_0(P,P,P_0)}}\,
 \mathbb S_{P_tP}[P_0],
\label{eq:Shat}
\end{equation}
{in the phase convention of \eqref{eq:torusmodular}.}  Dividing the crossed discontinuity by the reference one removes the observer flare exactly as in \eqref{eq:componentextraction},
\begin{equation}
\begin{aligned}
 &\frac{2P C_0(P_t,P_t,P_0)}{\rho_0(P)}
 \frac{\Disc_0\widehat Z_a^{(S)}(P^2;\varepsilon)}
      {\Disc\widehat Z_a(P^2;\varepsilon)}
 =e^{-\ii\pi h_0}\widehat{\mathbb S}_{P_tP}[P_0]^2
 =Z_{L,H}^{\rm cont}.
\end{aligned}
\label{eq:torusresult}
\end{equation}
The real-domain punctured modular kernel carries phase $e^{\ii\pi h_0/2}$ \cite{CTV}.  In the reversed framing, $\widehat{\mathbb S}^{\vee}=e^{-\ii\pi h_0}\widehat{\mathbb S}$, so the right side of \eqref{eq:torusresult} is the bilinear product $\widehat{\mathbb S}\widehat{\mathbb S}^{\vee}$.  Here $Z_{L,H}^{\rm cont}$ denotes one labelled compact-region amplitude, \vtwo{with the fixed-length normalization of \eqref{eq:revZL}. Its external transform is \eqref{eq:handcuffCTV} below.}  The normalized OPE moment for the linked handcuff also counts two inequivalent flare attachments to $C_{113}C_{223}$ \cite{Hartman}.  We keep that combinatorial factor separate from the kernel extraction and include it as $\nu_\Gamma=2$ in section~\ref{ssec:momentdictionary}.

In the boundary diagram \eqref{eq:boundarycurvesketch}, $(\gamma_1,\gamma_2,\gamma_3)$ carry the labels $(P_t,P,P_0)$ of this subsection.  In the one-cell realization of figure~\ref{fig:reviewcores}, four original tetrahedron edges join end to end to form $\gamma_3$, while the other two edges close separately to form $\gamma_1$ and $\gamma_2$ \cite{Hartman}.

Write $(P_1,P_2)=(P_t,P)$.  The fixed-angle amplitude of the same core is obtained by applying the external cosine kernel $S_{P'P}$ of section~\ref{sec:review} to all three edges.  This is the boundary-polarization transform, while $\mathbb S$ in \eqref{eq:torusresult} changes the punctured-torus channel.  The modular kernel and its external transform combine into a single $6j$ symbol, so the CTV amplitude has one cell,
\begin{equation}
\begin{aligned}
Z_{A,H}^{\rm cont}(P_1',P_2',P_0')
 &=\left[\int_0^\infty\prod_{j=0}^2\dd P_j\,S_{P_j'P_j}
 e^{-\ii\pi h_0}\widehat{\mathbb S}_{P_1P_2}[P_0]^2\right]_{\rm cont}\\
 &=\begin{Bmatrix}P_0'&P_0'&P_1'\\P_0'&P_0'&P_2'\end{Bmatrix}_{\!\rm Vir}.
\end{aligned}
\label{eq:handcuffCTV}
\end{equation}
In this example the fixed-length coefficient is a framed modular-kernel product, while the fixed-angle coefficient is a single $6j$.  They describe the same specified core in different boundary representations and are related by the external transform.  Section~\ref{sec:gravity} returns to their geometric interpretation.

\FloatBarrier

\section[Compact amplitudes and Lorentzian reconstruction]{\vtwo{Compact amplitudes and Lorentzian reconstruction}}\label{sec:gravity}
\begin{figure}[!t]
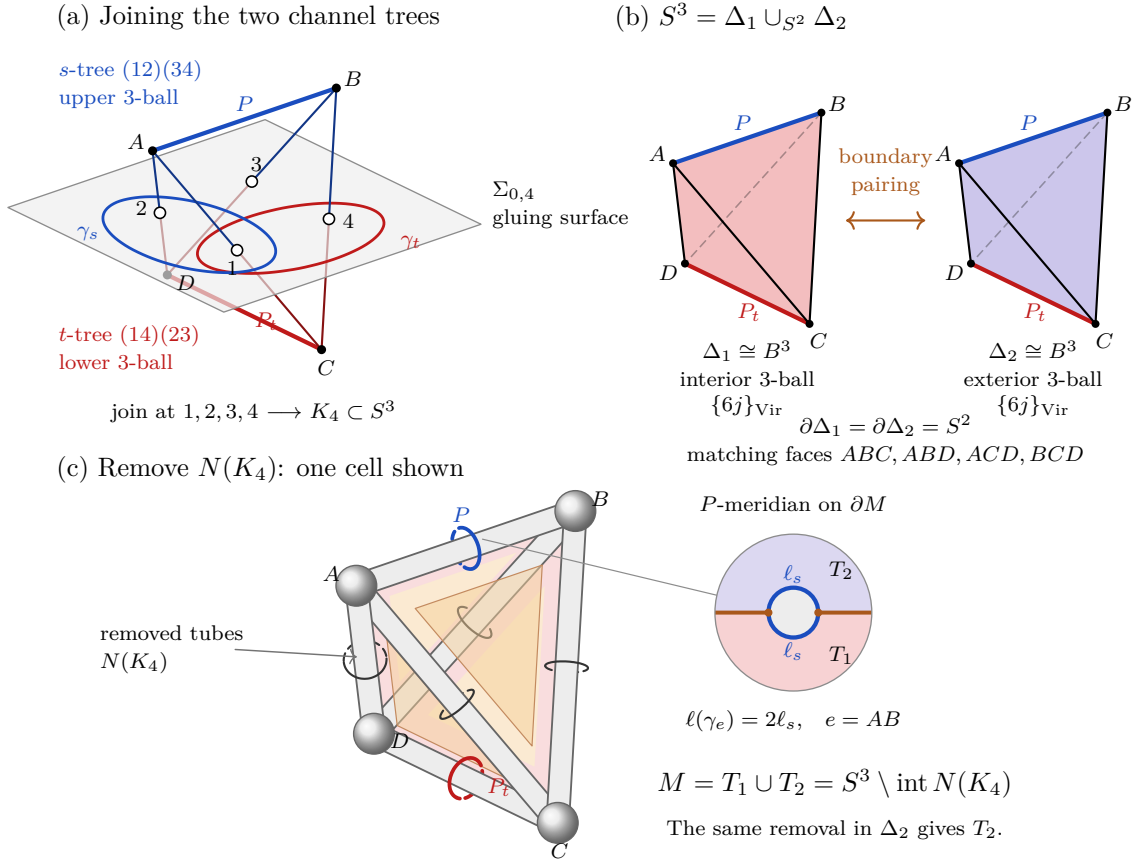

\centering
\definecolor{f4a}{RGB}{247,214,150}\definecolor{f4b}{RGB}{176,216,176}
\definecolor{f4c}{RGB}{242,180,182}\definecolor{f4d}{RGB}{196,190,232}

\caption{{The tetrahedral coefficient and its compact core.
(a) The two channel trees join at equal external labels on $\Sigma_{0,4}$
to form $K_4\subset S^3$.
(b) A different cut presents $S^3$ as two tetrahedral three-balls sharing
\vtwo{their entire $S^2$ boundaries.}
(c) Removing $N(K_4)$ produces $M$; one cell and the two-cell meridian are shown.}}
\label{fig:dsbranches}
\end{figure}

\begin{vtwoblock}
The four-point coefficient extracted in \eqref{eq:mainratio} has a compact gravitational filling. We first identify its two tetrahedra and continue their amplitude to the dS ray. The resulting volume phase determines hyperbolic edge lengths. We then recover the same cells from real Lorentzian spacetime data and compare their compact weight with the gravitational action.

Joining the observer's $s$-channel tree to the preparation's $t$-channel tree gives the graph in figure~\ref{fig:dsbranches}(a). Its vertices are $A=(1,2,s)$, $B=(3,4,s)$, $C=(1,4,t)$ and $D=(2,3,t)$. The six edges are $AB=P$, $CD=P_t$, $AC=P_1$, $AD=P_2$, $BD=P_3$ and $BC=P_4$. Thus every pair of vertices is joined once, giving $\Gamma_{\rm tet}=K_4$. The prescribed gluing of the two marked balls fixes its unknotted embedding in $S^3$ and its framing.

Panel~(b) gives a two-cell decomposition of this $S^3$. Each solid tetrahedron $\Delta_i$ is a three-ball. Identifying their entire boundary spheres joins an upper and a lower hemisphere of $S^3$, with four matching triangular faces and distinct interiors. Removing a regular neighbourhood $N(K_4)$ from both cells gives the compact complement $M=S^3\setminus\operatorname{int}N(K_4)$. The truncation and edge dualization of section~\ref{ssec:core} express it as two generalized tetrahedra \cite{Hartman,CTV}. In the hyperideal chamber,
\begin{equation}
 M=T_1\cup_{\mathrm I,\mathrm {II},\mathrm {III},\mathrm {IV}}T_2,
 \qquad \operatorname{Vol}(M)=2V_A(\psi).
\label{eq:doubletet}
\end{equation}
Their four pairs of triangular truncation faces are glued, leaving four pants on $\partial M$. Hence $\chi(\partial M)=4(2-3)=-4$ and $\partial M\simeq\Sigma_3$. This is the nonzero-volume $K_4$ core of figure~\ref{fig:reviewcores}(a), distinct from the zero-volume theta core used to normalize the three-point state. The sphere separating the two channel trees in panel~(a) and the tetrahedral boundary sphere in panel~(b) are different cuts of the same $S^3$.

The theta-normalized chiral graph is the Virasoro $6j$ in \eqref{eq:revtet}. The external $S$ transform of its square exchanges the fixed-length and fixed-angle polarizations and swaps the opposite-edge roles. With $P_1=P_2=P_3=P_4=P_0$, tetrahedral symmetry takes the row-swapped symbol back to the displayed one \cite{Hartman,CTV,Barrett2003}. Combining this evaluation with \eqref{eq:mainratio} gives
\begin{equation}
\begin{aligned}
 \frac{2P\,C_0(P_0,P_0,P_t)^2}{\rho_0(P)}
 \frac{\Disc_0\widehat Z_{\gamma_s}^{(t)}(P^2;\varepsilon)}
      {\Disc\widehat Z_{\gamma_s}(P^2;\varepsilon)}
 &=\left[\begin{Bmatrix}P_0&P_0&P_t\\P_0&P_0&P\end{Bmatrix}_{\!\rm Vir}^{\!2}\right]_{\rm cont}\\
 &=[Z_{\rm CTV}(S^3,\Gamma_{\rm tet})]_{\rm cont}
 =Z_A^{\rm cont}(M,\gamma).
\label{eq:coreexact}
\end{aligned}
\end{equation}
The right side is the continued state sum of \eqref{eq:revgeneralCTV}, with two cell weights, six fixed external labels and no internal momentum integral. We use the vertex normalization and continued Barnes contour specified in appendix~\ref{app:contour}, retaining any crossed-pole residues. The equality identifies the extracted coefficient with this marked compact filling. We next evaluate its classical action and geometric phase.
\end{vtwoblock}

\subsection[The dS continuation of the \texorpdfstring{$K_4$}{K4} amplitude]{{The dS continuation of the $K_4$ amplitude}}\label{ssec:labels}\label{ssec:polargeometry}

{We now continue the labels and action of the two-cell filling \eqref{eq:doubletet}, keeping its topology fixed.  The relevant angles and lengths belong to the compact core, while $y$ in section~\ref{sec:inverse} is a boundary collar modulus.}  Hold each real dihedral angle $\psi_e=-2\pi\ii bP_e\in(0,\pi)$ fixed, so that $P_e=\ii\psi_e/(2\pi b)=e^{\ii\pi/4}\psi_e/(2\pi r)$ on the dS ray.  {In the compact core, $\psi_e$ is the interior angle of a cell and $\pi-\psi_e$ is the bend of $\partial M$ along $\gamma_e$.  The same label has a separate meaning in the cosmological geometry \eqref{eq:cosmology_cone_domain_v46}, where a scalar particle has opening angle $2\psi_e$.}

The gluing in \eqref{eq:doubletet} gives $\chi(M)=1+1-4=-2$ and $\partial M=\Sigma_3$.  Each closed curve $\gamma_e$ consists of one edge from each cell, so $\ell(\gamma_e)=2\ell_e$.  Orthogonality at the truncation faces makes the gluing smooth away from these prescribed boundary bends.

\phantomsection\label{ssec:action}
The fixed-angle action \eqref{eq:revactions} has the on-shell value \eqref{eq:revonshell}, and \eqref{eq:s2conjugatevariables} gives its variation.  On the real AdS starting slice, in unit-radius conventions, $\ell(\gamma_e)=2\ell_e$ and $c=3/(2G_N)$ at classical order,
\begin{equation}
 I_A[g_{\rm cl};\psi]=\frac{c}{6\pi}\operatorname{Vol}(M)=\frac{c}{3\pi}V_A(\psi),\qquad
 \dd I_A[g_{\rm cl};\psi]=-\frac{c}{6\pi}\sum_e\ell_e\,\dd\psi_e ,
\label{eq:coreaction}
\end{equation}
Both equalities follow by substituting $\operatorname{Vol}(M)=2V_A$ and the Schl\"afli relation $\dd V_A=-\frac12\sum_e\ell_e\,\dd\psi_e$ into the fixed-angle action.  The Legendre term in \eqref{eq:revonshellL} gives the corresponding fixed-length action.

The Einstein representative behind this continuation can also be written
at the level of the metric.  Let $h_{\mu\nu}$ now denote the unit-curvature
hyperbolic \emph{three-dimensional} metric on the compact core.  Rescale
it and the cosmological constant together,
\begin{equation}
 \begin{aligned}
 g_{\mu\nu}(\theta)&=\ell_{\rm dS}^2e^{-2\ii\theta}h_{\mu\nu},\\
 \Lambda(\theta)&=-\frac{e^{2\ii\theta}}{\ell_{\rm dS}^2} .
 \end{aligned}
 \label{eq:complex_core_metric_v46}
\end{equation}
{The unit-curvature hyperbolic metric $h$ obeys $R_{\mu\nu}(h)=-2h_{\mu\nu}$.  The phase-dependent factor multiplying $h$ in \eqref{eq:complex_core_metric_v46} is constant on $M$, so it leaves the Christoffel symbols and the Ricci tensor with both indices lowered unchanged.  The product $2\Lambda(\theta)g(\theta)$ is also $-2h$; therefore the continued family satisfies Einstein's equation at every $\theta$,}
\begin{equation}
 R_{\mu\nu}\big(g(\theta)\big)=-2h_{\mu\nu}
 =2\Lambda(\theta)g_{\mu\nu}(\theta) .
 \label{eq:complex_core_einstein_v46}
\end{equation}
{At $\theta=0$ the rescaled metric is hyperbolic and $\Lambda<0$, as on the real AdS starting slice.  At $\theta=\pi/2$, $e^{2\ii\theta}=-1$, so $\Lambda=+1/\ell_{\rm dS}^2$ and $g=-\ell_{\rm dS}^2h$; this endpoint is negative definite.  \vtwo{Section~\ref{ssec:dsquantumtriangulation} connects this core to a real Lorentzian region through a complex time contour.}  The family gives an analytic Einstein-action continuation of the fixed compact filling.  Continuing the volume form with the branch inherited from $\theta=0$ gives}
\begin{equation}
 I_A[g(\theta)]
 =\frac{\ell_{\rm dS}}{4\pi G_N}e^{-\ii\theta}
   \operatorname{Vol}_h(M)
 \ \longrightarrow\
 -\frac{\ii\ell_{\rm dS}}{4\pi G_N}\operatorname{Vol}_h(M) .
 \label{eq:complex_core_action_v46}
\end{equation}
The corner cancellation is unchanged because constant rescaling preserves
the dihedral angles.  Thus the classical weight is
$\exp[\ii\ell_{\rm dS}\operatorname{Vol}_h(M)/(4\pi G_N)]$, which agrees
with the leading $r^{-2}$ phase below.  This computation establishes the
Einstein equation and the action of the continued core.
{Appendix~\ref{ssec:internaledge}} tests saddle selection in the exact
Barnes and length integrals, which is the contour statement available
in the continued CTV representation.  An admissible integration
cycle for the cosmological metric path integral is a further input
\cite{CEM,WittenPL}.

The same leading action can be expressed in the $b$ parametrization.  With $c=6b^{-2}+13+6b^2$,
\begin{equation}
I_A[g_{\rm cl};\psi]=\frac{c}{3\pi}V_A(\psi)
 =\frac{2}{\pi b^2}V_A(\psi)+O(b^0),
\label{eq:continuedaction1}
\end{equation}
and on the ray \eqref{eq:physicalray} $b^2=\ii r^2$, so $1/b^2=-\ii/r^2$ and
\begin{equation}
\boxed{\;-I_A^{\rm cont}[g_{\rm cl};\psi]
 =\frac{2\ii}{\pi r^2}V_A(\psi)+O(r^0).\;}
\label{eq:continuedaction}
\end{equation}
{Since $1/b^2=1/(\ii r^2)=-\ii/r^2$, the real leading exponent $-2V_A/(\pi b^2)$ becomes $+2\ii V_A/(\pi r^2)$.  The same real hyperbolic volume is being evaluated; only the complex coupling multiplying it has changed.  In particular, the leading weight is oscillatory rather than exponentially volume-suppressed, with its nontrivial finite prefactor treated in section~\ref{sec:geometry}.}  After restoring curvature radii, the sign of $\Im c$ chosen in \eqref{eq:physicalc} corresponds at leading order to $c\simeq-\ii\,3\ell_{\rm dS}/(2G_N)$, so that $r^2\simeq4G_N/\ell_{\rm dS}$.  The opposite continuation conjugates the phase.

Both factors of the CTV amplitude \vtwo{\eqref{eq:coreexact}} continue with the same $b$, so the amplitude is a bilinear square and its leading phase is twice the phase of one cell.  A Hermitian norm would pair $b$ with $\bar b$ and conjugate one factor, and the two phases would cancel.  The same-$b$ compact amplitude keeps both phases, so the two-cell core carries the phase $2\ii V_A(\psi)/(\pi r^2)$.  The transformed handcuff of section~\ref{sec:torus} is the one-cell counterpart,
\begin{equation}
\operatorname{Vol}(M_H)=V_A,\qquad
 I_A(M_H)=\frac{c}{6\pi}V_A=\frac{V_A}{\pi b^2}+O(1),\qquad
 e^{-I_A^{\rm cont}(M_H)}=
 \exp\left[\frac{\ii}{\pi r^2}V_A+O(1)\right].
\label{eq:handcuffaction}
\end{equation}
The corresponding real-angle inequalities are the hyperideal-vertex conditions for the one-cell weight, and its local kernel asymptotic is covered by \cite{MY}.  {The two-cell and one-cell topologies are thus unchanged, while their leading weights are oscillatory rather than exponentially suppressed.  We now extract geometric information from this phase.}

\subsection[Reconstructing the \texorpdfstring{$K_4$}{K4} geometry from its phase]{{Reconstructing the $K_4$ geometry from its phase}}\label{sec:geometry}

We now compare the edge lengths determined by the core angles with those read from the phase of the extracted boundary coefficient.  The Gram matrix gives the first answer, and the $6j$ asymptotic and Schl\"afli relation give the second.  Their agreement also determines the angular resolution needed to retain the oscillatory coefficient.  To obtain the edge lengths, represent the four original faces by their outward unit normals in Minkowski space.  Their inner products form the Gram matrix $G$.  In the all-angle hyperideal chamber, its diagonal entries are $1$ and an off-diagonal entry is minus the cosine of the interior dihedral angle between the corresponding faces.  With the edge ordering used here,
\begin{equation}
G=\begin{pmatrix}
 1&-\cos\psi_1&-\cos\psi_2&-\cos\psi_6\\
 -\cos\psi_1&1&-\cos\psi_3&-\cos\psi_5\\
 -\cos\psi_2&-\cos\psi_3&1&-\cos\psi_4\\
 -\cos\psi_6&-\cos\psi_5&-\cos\psi_4&1
 \end{pmatrix}.
\label{eq:generalGram}
\end{equation}
{For example, $G_{12}=-\cos\psi_1$ records the angle on the common edge of faces 1 and 2.  The inverse matrix instead pairs the vertices opposite those faces.  Its entry $(G^{-1})_{12}$ therefore determines the length of the complementary edge, between the truncation planes at vertices 1 and 2.}  The tetrahedron is hyperideal when $G$ has signature $(3,1)$ and the angles at each vertex sum to less than $\pi$.  Let $\ell_{ij}$ denote this segment joining the vertices opposite faces $i$ and $j$.  Its length is
\begin{equation}
\cosh\ell_{ij}
 =-\frac{(G^{-1})_{ij}}{\sqrt{(G^{-1})_{ii}(G^{-1})_{jj}}}.
\label{eq:Graminverse}
\end{equation}
On the equal-external slice, $\psi_1=\psi_2=\psi_4=\psi_5=\psi_0$, $\psi_6=\psi_s$ and $\psi_3=\psi_t$.  Thus $\ell_s=\ell_{23}$ is obtained from $(G^{-1})_{23}$, and $\ell_t=\ell_{14}$ from $(G^{-1})_{14}$.  The hyperideal vertex conditions are $2\psi_0+\psi_s<\pi$ and $2\psi_0+\psi_t<\pi$, with all angles positive.  Substituting this slice in \eqref{eq:Graminverse} gives
\[
 \cosh\ell_s
 =\frac{2\cos^2\psi_0+(1-\cos\psi_s)\cos\psi_t}
        {2\cos^2\psi_0-1+\cos\psi_s}.
\]
Using $\cosh\ell_s-1=2\sinh^2(\ell_s/2)$ gives the first formula below; exchanging $s$ and $t$ gives the second,
\begin{equation}
\ell_s
 =2\arsinh\frac{\cos(\psi_t/2)\sin(\psi_s/2)}
 {\sqrt{\cos^2\psi_0-\sin^2(\psi_s/2)}},\qquad
 \ell_t
 =2\arsinh\frac{\cos(\psi_s/2)\sin(\psi_t/2)}
 {\sqrt{\cos^2\psi_0-\sin^2(\psi_t/2)}}.
\label{eq:explicitlengths}
\end{equation}
For a nondegenerate all-angle cell we use the complex-$b$ asymptotic theorem of \cite{MY}.  Its entries are the Liouville charges $a_e=Q/2+\ii P_e=Q/2-\psi_e/(2\pi b)$, whereas our Virasoro symbol uses momenta $P_e$.  When the six fixed real angles belong to a truncated hyperideal tetrahedron and $\arg b=\pi/4$, the theorem gives
\begin{equation}
\begin{aligned}
&\begin{Bmatrix}
 \frac Q2-\frac{\psi_1}{2\pi b}&\frac Q2-\frac{\psi_2}{2\pi b}&\frac Q2-\frac{\psi_3}{2\pi b}\\
 \frac Q2-\frac{\psi_4}{2\pi b}&\frac Q2-\frac{\psi_5}{2\pi b}&\frac Q2-\frac{\psi_6}{2\pi b}
 \end{Bmatrix}_{\!b}
 =\frac{e^{-V_A(\psi)/(\pi b^2)}}{2(-\det G)^{1/4}}
 \big(1+O(b^2)\big).
\end{aligned}
\label{eq:MY}
\end{equation}
{The asymptotic theorem links the special-function coefficient to the geometry.  Its exponential is precisely the hyperbolic volume $V_A(\psi)$ of one cell, and its determinant prefactor measures the leading non-exponential magnitude.  Squaring the one-cell weight produces the two-cell exponent of \eqref{eq:continuedaction}, so the independent geometric Einstein-action continuation and the $6j$ asymptotics agree.}  The Gram prefactor combines the subleading Barnes amplitude with its Gaussian Hessian; appendix~\ref{app:generalmixed6j} states the variable map and continued normalization.  A gravitational one-loop interpretation additionally requires a metric-fluctuation analysis.  A ratio to a reference configuration, denoted by a star, removes a momentum-independent phase,
\begin{equation}
\begin{aligned}
&\frac{\left(\sixj\right)^2}
 {\left(\begin{Bmatrix}P_{0*}&P_{0*}&P_{t*}\\P_{0*}&P_{0*}&P_*\end{Bmatrix}_{\!\rm Vir}\right)^2}\\
 &\qquad=\sqrt{\frac{-\det G_*}{-\det G}}\,
 \exp\left[\frac{2\ii}{\pi r^2}\big(V_A(\psi)-V_A(\psi_*)\big)\right]
 \big(1+O(r^2)\big),
\end{aligned}
\label{eq:relativegeometry}
\end{equation}
and the relative amplitude is fixed entirely by geometric quantities.  The volume and its branch are those of the hyperideal saddle in \cite{MY}.

The modulus of the same asymptotic compares the two continuations of this fixed-angle family.  At fixed real angles, \eqref{eq:revsemicl} and \eqref{eq:MY}, with the normalization of appendix~\ref{app:generalmixed6j}, give
\begin{equation}
\left|\begin{Bmatrix}P_1&P_2&P_3\\P_4&P_5&P_6\end{Bmatrix}_{\!\rm Vir}^{\!2}\right|
\simeq
\begin{cases}
 e^{-cV_A(\psi)/(3\pi)}, & b>0,\\[4pt]
 \big[4\sqrt{-\det G}\,\big]^{-1}, & b=re^{\ii\pi/4},
\end{cases}
\label{eq:dSsize}
\end{equation}
where $\simeq$ keeps the leading exponential in the first line and the leading term in the second.  At real $b$, the real angles give imaginary momenta below the continuum threshold, with the semiclassical conical-defect interpretation described in section~\ref{ssec:core}.  On the dS ray the coefficient is of order $r^0$ at fixed angles, and its volume determines how fast it oscillates as an angle varies.  The conditional $K_{3,3}$ window contribution \eqref{eq:k33_closed_prefactor_v46} has the same $r^0$ scaling.  These are normalized coefficients, before the spectral and conformal-block integrals.  In a four-point correlation the same-channel Gaussian term and the mixed-channel tetrahedral term are two expansions of one contribution, related by the fusion transform of section~\ref{ssec:core}; their spectral densities and three-point normalizations must be included in either channel.

{Having determined that the same hyperbolic volume governs the continued coefficient, we can return to the boundary extraction of \eqref{eq:mainratio}.  A phase derivative is meaningful on a fixed nonzero analytic branch and, for the stated asymptotics, requires uniformity under one angle derivative.}  The phase of \eqref{eq:relativegeometry} also measures edge lengths.  The Schl\"afli relation and its restriction to the equal-external slice at fixed $\psi_0$ are
\begin{equation}
\dd V_A=-\frac12\sum_{e=1}^6\ell_e\,\dd\psi_e,\qquad
 \dd V_A\big|_{\psi_0\,\mathrm{fixed}}
 =-\frac12\ell_s\,\dd\psi_s-\frac12\ell_t\,\dd\psi_t .
\label{eq:Schlafli}
\end{equation}
\begin{vtwoblock}
Here $V_A$ is the volume of one generalized tetrahedron, with curvature radius one. In the Klein ball it has the explicit integral below. Integrating the Schl\"afli relation also computes its change from any reference cell in the same chamber,
{\color{black}
\begin{equation}
\begin{aligned}
 V_A(\boldsymbol\psi)
 &=\operatorname{Vol}_{H_3}(T_1(\boldsymbol\psi))
 =\int_{T_1(\boldsymbol\psi)}
       \frac{\dd^3\boldsymbol x}{(1-|\boldsymbol x|^2)^2},\\
 V_A(\boldsymbol\psi)-V_A(\boldsymbol\psi_*)
 &=-\frac12\int_{\boldsymbol\psi_*}^{\boldsymbol\psi}
       \sum_{e=1}^{6}\ell_e(\boldsymbol\psi')\,\dd\psi_e',\\
 V_A(\psi)-V_A(\psi_*)
 &=-3\int_{\psi_*}^{\psi}
       \operatorname{arcosh}\!\left(\frac{\cos u}{2\cos u-1}\right)\dd u
       \qquad(\psi_e=\psi).
\end{aligned}
\label{eq:lorentzVolumeFromAngles}
\end{equation}
}
The second line is an angle-space integral along a path in that chamber. The last line inserts the equal-angle length from \eqref{eq:explicitlengths}, with $0<\psi,\psi_*<\pi/3$. The reference cell fixes the additive constant.
\end{vtwoblock}

{The phase of \eqref{eq:relativegeometry} is $2V_A/(\pi r^2)$ to leading order, up to the constant reference phase. At fixed $\psi_0$ and $\psi_t$, \eqref{eq:Schlafli} therefore gives its derivative $-\ell_s/(\pi r^2)$. A longer edge produces a faster phase variation. Locally, a full $2\pi$ change corresponds to $\Delta\psi_s\simeq2\pi^2r^2/\ell_s$, provided $\ell_s$ varies negligibly across that interval. By \eqref{eq:coreexact}, the boundary-extracted ratio contains this same phase. On a continuous nonzero branch, with uniform control of the asymptotic and one angle derivative on the compact angle domain, its derivative gives}
\begin{equation}
-\pi r^2\,\partial_{\psi_s}\arg\left[
 \frac{2P C_0(P_0,P_0,P_t)^2}{\rho_0(P)}
 \frac{\Disc_0\widehat Z_{\gamma_s}^{(t)}(P^2)}
 {\Disc\widehat Z_{\gamma_s}(P^2)}\right]
 \longrightarrow\ell_s .
\label{eq:geometricreadout}
\end{equation}
{Equation~\eqref{eq:geometricreadout} contains the external $C_0$ and $\rho_0$ factors as well as the two discontinuities.  To express the answer using only the two extracted densities, we must ensure that removing the prefactors introduces no additional $\psi_s$-dependent phase.  The $C_0(P_0,P_0,P_t)$ factor does not depend on $P$, and the ratio $\rho_0(P)/(2P)$ has constant phase on $0<\psi_s<\pi$,}
\begin{equation}
\frac{\rho_0(e^{\ii\pi/4}k)}{2e^{\ii\pi/4}k}
 =\frac{2\sqrt2\,e^{\ii\pi/4}}{k}
 \sin(2\pi rk)\sinh(2\pi k/r).
\label{eq:constantphase}
\end{equation}
With $\psi_s=2\pi rk$ the edge length is therefore read from the phase of the ratio of the two spectral coefficients,
\begin{equation}
r\,\partial_k\arg\frac{\rho_{st}(e^{\ii\pi/4}k;P_t)}
 {\rho_s(e^{\ii\pi/4}k)}
 \longrightarrow-2\ell_s(\psi_0,\psi_s,\psi_t).
\label{eq:ratiophase}
\end{equation}

The Einstein action \eqref{eq:continuedaction} with the face gluings \eqref{eq:doubletet}, and the finite-$b$ identity \eqref{eq:coreexact} with the asymptotic \eqref{eq:MY}, therefore give the same leading phase.  The factor of two comes from the two cells.  Keeping only this exponential in the first line below, with the Gram prefactor given in \eqref{eq:relativegeometry}, we find
\begin{equation}
 \boxed{\;\begin{aligned}
 Z_A^{\rm cont}(M,\gamma)&\simeq\exp\Big[\frac{2\ii}{\pi r^2}V_A(\psi)\Big],\\
 r\,\partial_k\arg\frac{\rho_{st}(e^{\ii\pi/4}k;P_t)}{\rho_s(e^{\ii\pi/4}k)}
 &\longrightarrow-2\ell_s(\psi_0,\psi_s,\psi_t).
 \end{aligned}\;}
 \label{eq:boxgeometry}\eqalias{eq:boxcore}\eqalias{eq:twoevaluations}
\end{equation}
The first line is the dS form of the AdS weight $e^{-cV_A/(3\pi)}$ in \eqref{eq:revsemicl}, with the same volume entering as a phase.  The second line makes that geometry measurable from the boundary data of section~\ref{sec:boundaryamp}, since $\rho_{st}$ and $\rho_s$ are extracted primary coefficients.

{This phase variation also sets the angular resolution needed to retain the extracted coefficient under an average over labels.} Hold five angles fixed, vary the sixth as $\psi=\psi_*+x$ inside the hyperideal chamber, and write $\{6j\}^2_{\rm Vir}(\psi)$ for the continued tetrahedral coefficient as a function of $\psi$.  By \eqref{eq:relativegeometry} and \eqref{eq:Schlafli},
\begin{equation}
\begin{aligned}
\frac{\{6j\}^2_{\rm Vir}(\psi_*+x)}{\{6j\}^2_{\rm Vir}(\psi_*)}
&=\sqrt{\frac{\det G(\psi_*)}{\det G(\psi_*+x)}}\,
 \exp\Big[\frac{2\ii}{\pi r^2}\big(V_A(\psi_*+x)-V_A(\psi_*)\big)\Big]\big(1+O(r^2)\big)\\
&=\exp\Big[-\frac{\ii\ell_*x}{\pi r^2}\Big]\Big(1+O(x)+O\big(x^2r^{-2}\big)+O(r^2)\Big),
\end{aligned}
\label{eq:anglevariation}
\end{equation}
where $\ell_*$ is the cell-edge length conjugate to $\psi$ at $\psi_*$.  Choose a fixed $x_0>0$ so that $|x|\leq x_0$ stays inside the nondegenerate hyperideal chamber, and assume the asymptotic is uniform there.  Average over this interval with a normalized Gaussian of width $\lambda r^2$, with $\lambda>0$ fixed.  Its typical $x$ is $O(r^2)$, so each error term contributes $O(r^2)$ after averaging.  The Gaussian Fourier transform then gives
\begin{equation}
\frac{\displaystyle\int_{-x_0}^{x_0}\dd x\,e^{-x^2/(2\lambda^2r^4)}
\frac{\{6j\}^2_{\rm Vir}(\psi_*+x)}{\{6j\}^2_{\rm Vir}(\psi_*)}}
{\displaystyle\int_{-x_0}^{x_0}\dd x\,e^{-x^2/(2\lambda^2r^4)}}
=\exp\Big[-\frac{\ell_*^2\lambda^2}{2\pi^2}\Big]+O(r^2).
\label{eq:angleresolution}
\end{equation}
Extending the Gaussian transform of the linearized phase to the real line changes the answer only by an exponentially small tail.  The coefficient itself is sampled only inside the chosen interval.  {An angle resolution of order $\pi r^2/\ell_*$, corresponding to an order-one change in phase, therefore retains a finite fraction of the coefficient.} The length measured in \eqref{eq:ratiophase} sets this resolution.  A boundary correlator also includes spectral densities, three-point normalizations and conformal blocks, whose phases enter its own stationary-phase analysis.

\subsection[Reconstructing the marked compact core from Lorentzian \texorpdfstring{$dS_3$}{dS3} data]{\vtwo{Reconstructing the marked compact core from Lorentzian $dS_3$ data}}\label{ssec:inverseLorentzian}
\begin{vtwoblock}
Can the two-cell compact core be recovered from a real Lorentzian de Sitter spacetime? We answer this for a marked tetrahedral family. First we remove the flat collars from its future boundary and recover four hyperbolic pants joined along geodesic cuffs. The initial arc lengths then determine the shape of one tetrahedron. The boost holonomies and the identified pants seams determine whether two copies of that tetrahedron fit the recovered boundary and how they are joined.

\paragraph{The Lorentzian input.}
We start with the pants--wedge solutions of the hyperbolic-end/de Sitter correspondence \cite{Mess,ABBGLSS,BenedettiBonsante}. On a spatial slice, four pants are joined by six cylindrical strips. The spacetime region swept out by each strip is a wedge. The pants and cuffs carry the $K_4$ labels. We retain the seam endpoints and their identifications, with no additional shift along a cuff. This is the marked, seam-aligned family considered here. We also retain the timelike curves obtained by varying $t$ at fixed spatial coordinates, together with their local embeddings in $dS_3$. In a tetrahedral fundamental region, the four initial vertices are represented by $U_1,\ldots,U_4$. Their Gram matrix is assumed to lie in the nondegenerate hyperideal chamber used above.

On each pants region the metric is
\begin{equation}
 \dd s^2=\ell_{\rm dS}^2\big[-\dd t^2+\sinh^2t\,h_{\Pi_i}\big],
 \qquad i=1,\ldots,4,
\label{eq:lorentzInverseFuchsian}
\end{equation}
where $h_{\Pi_i}$ is a curvature-$-1$ pants metric. To see how a wedge joins two such regions, use the unit-curvature embedding models
\begin{equation}
 H_3=\{Y:\eta(Y,Y)=-1,\ Y^0>0\},\qquad
 dS_3=\{X:\eta(X,X)=1\},\qquad
 \eta=\operatorname{diag}(-1,1,1,1).
\label{eq:lorentzInverseModels}
\end{equation}
Choose an orthonormal frame $E_0,E_1,U_i,E_3$, with $E_0$ timelike and $U_j=\cos d_e\,U_i+\sin d_e\,E_3$. The coordinate $x$ runs along the cuff and $\varphi$ across the wedge. These directions are described by a hyperbolic geodesic $Y(x)$ and an arc of normals $n(\varphi)$,
\begin{equation}
\begin{aligned}
 Y(x)&=\cosh x\,E_0+\sinh x\,E_1,&
 n(\varphi)&=\cos\varphi\,U_i+\sin\varphi\,E_3,\\
 X(t,x,\varphi)&=\sinh t\,Y(x)+\cosh t\,n(\varphi),&
 0&\leq\varphi\leq d_e .
\end{aligned}
\label{eq:lorentzWedgeEmbedding}
\end{equation}
Both curves have unit speed and obey $\eta(Y,Y)=-1$, $\eta(n,n)=1$ and $\eta(Y,n)=0$. Hence $\eta(X,X)=-\sinh^2t+\cosh^2t=1$, so $X$ lies on $dS_3$. Its derivatives are
\[
 \partial_tX=\cosh t\,Y+\sinh t\,n,\qquad
 \partial_xX=\sinh t\,Y',\qquad
 \partial_\varphi X=\cosh t\,n'.
\]
They are mutually orthogonal, with squared norms $-1$, $\sinh^2t$ and $\cosh^2t$. Restoring the de Sitter radius therefore gives
\begin{equation}
 \dd s^2=\ell_{\rm dS}^2\big[-\dd t^2+\cosh^2t\,\dd\varphi^2+\sinh^2t\,\dd x^2\big],
 \qquad 0\leq\varphi\leq d_e,\quad x\sim x+\lambda_e .
\label{eq:lorentzWedge}
\end{equation}
\finalpolish{At $\varphi=0$ and $\varphi=d_e$, the wedge embedding agrees with the expanding cuffs of the two adjacent pants. Both sides induce the metric $\ell_{\rm dS}^2(-\dd t^2+\sinh^2t\,\dd x^2)$ on each joining surface and have vanishing extrinsic curvature there, so the pieces join smoothly. The width $d_e$ measures the turn between the pants normals, while $\lambda_e$ is the period around their common cuff.}

What do these two parameters measure in the Lorentzian spacetime? Near $t=0$, the $x$ direction shrinks but the $\varphi$ direction remains. The spatial line elements along these directions are $e^1=\ell_{\rm dS}\cosh t\,\dd\varphi$ and $e^2=\ell_{\rm dS}\sinh t\,\dd x$. Compared with the pants regions, their limit is
\begin{equation}
 \lim_{t\to0}(e^1,e^2)\big|_t=
 \begin{cases}
 (0,0),&\Pi_i,\\
 (\ell_{\rm dS}\,\dd\varphi,0),&\text{wedge of }\gamma_e,
 \end{cases}
\label{eq:lorentzInitialFrame}
\end{equation}
Each pants contracts to one vertex $U_i$, while each wedge leaves a spacelike arc of physical length $\ell_{\rm dS}d_e$. In the embedding this is $X(0,x,\varphi)=n(\varphi)$, independent of $x$. The four vertices and six arcs form the initial $K_4$ graph. At $t>0$, the pants and strips instead form a closed genus-three spatial slice.

The second measurement comes from going once around the periodic $x$ direction. The embedding is defined before this periodic identification. In \eqref{eq:lorentzWedgeEmbedding}, the shift $x\mapsto x+\lambda_e$ is implemented by
\begin{equation}
 E_0\pm E_1\longmapsto e^{\pm\lambda_e}(E_0\pm E_1),
 \qquad U_i\longmapsto U_i,\quad E_3\longmapsto E_3 ,
\label{eq:lorentzBoost}
\end{equation}
This is a Lorentz boost of rapidity $\lambda_e$ in the $E_0,E_1$ plane. It identifies the two ends of one period in $x$ and leaves the initial normal arc fixed. In $SL(2,\mathbb C)$, the corresponding holonomy is
\begin{equation}
 H_{\gamma_e}\sim\pm
 \begin{pmatrix}e^{\lambda_e/2}&0\\0&e^{-\lambda_e/2}\end{pmatrix},
 \qquad |\operatorname{tr}H_{\gamma_e}|=2\cosh\frac{\lambda_e}{2}.
\label{eq:lorentzBoostTrace}
\end{equation}
Its trace is independent of the chosen frame. We have therefore measured two different quantities for each edge, the initial arc length $\ell_{\rm dS}d_e$ and the boost rapidity $\lambda_e$.

\paragraph{Removing the geometric collars.}
At late times, $\sinh^2t$ and $\cosh^2t$ both approach $e^{2t}/4$. Dividing the spatial metric by $\ell_{\rm dS}^2e^{2t}/4$ leaves $h_{\Pi_i}$ on each pants and $\dd\varphi^2+\dd x^2$ on each wedge. Each strip therefore becomes a flat cylinder of width $d_e$ and circumference $\lambda_e$. These are the six collars at the future boundary. Inserting such cylinders between hyperbolic pants is called grafting \cite{McMullen}. To recover the pleated surface, we must collapse each cylinder to a geodesic and retain its width as the bend between the adjacent pants.

\finalpolish{Invert $X=\sinh t\,Y+\cosh t\,n$ together with its $t$ derivative to obtain}
\begin{finalpolishblock}
\begin{equation}
\begin{aligned}
 Y&=\cosh t\,\partial_tX-\sinh t\,X,\\
 n&=\cosh t\,X-\sinh t\,\partial_tX.
\end{aligned}
\label{eq:lorentzInverseNormalFlow}
\end{equation}
\end{finalpolishblock}
\finalpolish{These formulas recover the hyperbolic point $Y\in H_3$ and its unit normal $n$. They satisfy $\eta(Y,Y)=-1$, $\eta(n,n)=1$ and $\eta(Y,n)=0$. Along the chosen timelike geodesics, $\partial_t^2X=X$, so differentiating either formula gives zero. The recovered data are therefore independent of the time $t>0$ at which we perform the inversion.}

\finalpolish{On a pants region, $n=U_i$ is constant and the inverse recovers the hyperbolic pants in the plane $\eta(Y,U_i)=0$. On a wedge, the terms proportional to $n(\varphi)$ cancel, leaving}
\begin{finalpolishblock}
\begin{equation}
 \cosh t\,\partial_tX-\sinh t\,X
 =Y(x),\qquad \partial_\varphi Y=0 .
\label{eq:collarungrafting}
\end{equation}
\end{finalpolishblock}
The result no longer depends on $\varphi$. Thus every point across the collar at the same $x$ is sent to the same point $Y(x)$. Keeping the periodic identification in $x$, the cylinder becomes a closed geodesic of length $\lambda_e$,
\begin{equation}
 [0,d_e]\times(\mathbb R/\lambda_e\mathbb Z)\longrightarrow\gamma_e,
 \qquad (\varphi,x)\longmapsto [Y(x)],
 \qquad \ell(\gamma_e)=\lambda_e .
\label{eq:lorentzCollarCollapse}
\end{equation}
The brackets mean that points related by the boost are identified. The collar has lost its transverse direction but retained its circumference.

\finalpolish{The width $d_e$ survives as the bending angle between the two adjacent pants. The second inverse formula recovers the normals $n(\varphi)$, with endpoints $U_i,U_j$ satisfying $\eta(U_i,U_j)=\cos d_e$. Thus collar removal retains both the bend $d_e$ and the cuff length $\lambda_e$. Figure~\ref{fig:lorentzcore}(a)--(b) shows this collapse. Applying it to all six collars recovers the pleated surface $\partial M$, with its four hyperbolic pants and marked seam identifications.}
\end{vtwoblock}

\begin{figure}[!htbp]
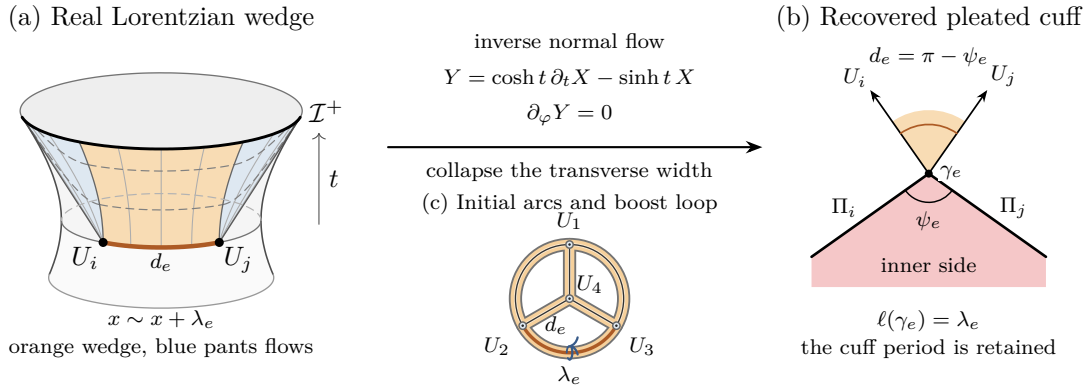

\centering
\definecolor{f4c}{RGB}{242,180,182}

\caption{\vtwo{Local inverse normal flow. (a) Real Lorentzian section at fixed $x$, with constant-$t$ curves dashed. (b) The recovered cuff retains period $\lambda_e$ and bend $d_e$, while $\partial_\varphi Y=0$ collapses the wedge width. (c) Initial arcs and a boost loop supply the two Lorentzian measurements.}}
\label{fig:lorentzcore}
\end{figure}
\begin{figure}[!htbp]
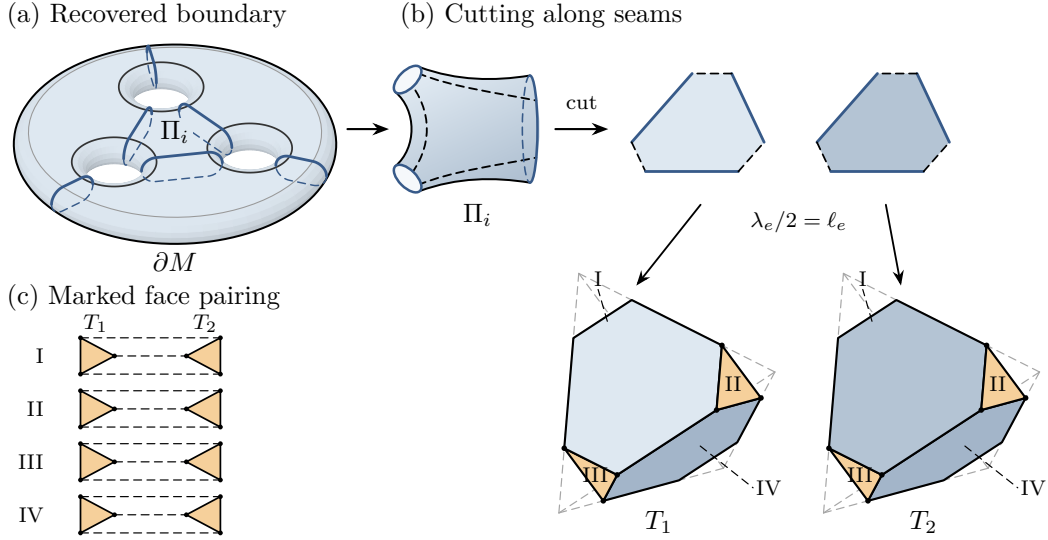

\centering
\begingroup\color{black}

\endgroup
\caption{\finalpolish{Reconstruction of the marked compact core. (a) Four hyperbolic pants. (b) One pants cut along its three seams into two hexagons with alternating sides $\lambda_e/2=\ell_e$. (c) The eight blue hexagons form the boundary faces of $T_1,T_2$. The four pairs of orange truncation faces I--IV are glued at the marked seam endpoints.}}
\label{fig:lorentzPantsAssembly}
\end{figure}

\FloatBarrier
\begin{vtwoblock}

\paragraph{Recovering the tetrahedra and their gluing.}
The initial arc has physical length $\ell_{\rm dS}d_e$. After collar removal, the same $d_e$ is the angle between two outward hyperbolic face normals. The interior dihedral angle is its supplement, $\psi_e=\pi-d_e$. Thus the six arc measurements give the angle Gram matrix
\begin{equation}
 G_{ii}=1,\qquad
 G_{ij}=\eta(U_i,U_j)=\cos d_e=-\cos\psi_e,\qquad
 \psi_e=\pi-d_e,\qquad e=(i,j).
\label{eq:lorentzInverseGram}
\end{equation}
This turns an initial arc length into an angle, not into a tetrahedral edge length. To determine the cell itself, recall that $U_i$ specifies the hyperbolic face plane $\eta(Y,U_i)=0$ \cite{Hartman}. The nondegenerate Gram matrix fixes the relative positions of the four normals up to a Lorentz isometry. With the chosen half-spaces and future sheet, it therefore fixes the four face planes. The truncation planes follow from the dual basis,
\begin{equation}
 V_i=\sum_{j=1}^{4}(G^{-1})_{ij}U_j,\qquad
 \eta(V_i,U_j)=\delta_{ij},\qquad
 W_i=\frac{V_i}{\sqrt{(G^{-1})_{ii}}}.
\label{eq:lorentzInverseDual}
\end{equation}
The vector $W_i$ is the unit normal to the polar plane that cuts off the vertex opposite face $i$. The hyperideal condition makes $(G^{-1})_{ii}>0$, and $\eta(W_i,U_j)=0$ for $j\ne i$ makes this cut orthogonal to its three neighbouring faces. The six initial arc lengths have therefore fixed both the original faces and the four truncations. In the assumed chamber, this determines one generalized tetrahedron up to an isometry.

The edge between original faces $i,j$ ends at the two opposite truncation planes. If $\{k,l\}$ is the complementary pair of indices, their separation is
\begin{equation}
 \cosh\ell_e=-\eta(W_k,W_l)
 =-\frac{(G^{-1})_{kl}}{\sqrt{(G^{-1})_{kk}(G^{-1})_{ll}}},
 \qquad \{k,l\}=\{1,2,3,4\}\setminus\{i,j\}.
\label{eq:lorentzInverseLengths}
\end{equation}
We have obtained the cell-edge lengths $\ell_e$ from the initial arcs alone. The boost rapidities $\lambda_e$ were measured independently. We next check whether they agree with the cuff lengths produced by gluing two copies of this cell.

\finalpolish{The two-cell structure follows from the hexagon decomposition of the recovered boundary. Cutting each pants along its three orthogeodesic seams gives two congruent right-angled hexagons, hence eight hexagons in total. Each generalized tetrahedron has four original hexagonal faces. The eight boundary hexagons can therefore supply the faces of two tetrahedra once their shapes and marked seams match, as in figure~\ref{fig:lorentzPantsAssembly}.}

In a pants hexagon, three alternating sides are half-cuffs of length $\lambda_e/2$, and the remaining sides are seams. In a cell face, the corresponding alternating sides are tetrahedral edges of length $\ell_e$. The right angles follow from the orthogonal polar cuts. The required length match is therefore
\begin{equation}
 \lambda_e=\ell(\gamma_e)=2\ell_e,\qquad
 \cosh\frac{\lambda_e}{2}
 =-\frac{(G^{-1})_{kl}}{\sqrt{(G^{-1})_{kk}(G^{-1})_{ll}}}.
\label{eq:lorentzRecoveredMatching}
\end{equation}
\finalpolish{This compares the independently measured boosts with the edge lengths determined by $G$. A right-angled hexagon is fixed by its three alternating side lengths \cite{SchlenkerHyperideal}. Thus $\lambda_e/2=\ell_e$ makes each pants hexagon isometric to the corresponding tetrahedral face, including its three seam sides. One hexagon from each pants supplies a face of $T_1$, and the other supplies the corresponding face of $T_2$.}

\finalpolish{The marked seams fix how the two cells are joined. Each triangular truncation face borders three hexagons. Glue it to the triangle with the same label in the other cell, identifying the marked seam endpoints. After all four triangle pairs are glued, each corresponding hexagon pair has been joined along its three seams and becomes one of the original pants. Each cuff contains one edge from $T_1$ and one from $T_2$, giving length $2\ell_e$. The specified seam endpoints fix the relative twist. The reflection calculation in appendix~\ref{app:markedfacepairings} checks that these same four face pairings reproduce all six boost holonomies.}
\end{vtwoblock}

\FloatBarrier
\begin{vtwoblock}

For these compatible marked data, the recovered filling and its boundary are
\begin{equation}
 M=T_1\cup_{\mathrm I,\mathrm {II},\mathrm {III},\mathrm {IV}}T_2,\qquad
 \partial M=\bigcup_{i=1}^{4}\Pi_i,\qquad
 \chi(\partial M)=4(2-3)=-4,\qquad g=3 .
\label{eq:lorentzBoundaryTopology}
\end{equation}
\finalpolish{The six initial arc lengths determine the dihedral angles and hence the geometry of each tetrahedron, with labels $P_e=\ii\psi_e/(2\pi b)$. The six boost holonomies test the length matching $\lambda_e=2\ell_e$, and the marked seams specify the four truncation-face pairings. Together these reconstruct the two-cell compact filling within the marked $K_4$ family. This is the filling of \eqref{eq:doubletet}, and the given spacetime is its associated Lorentzian region $\mathcal D=\mathcal D_M$.}
\end{vtwoblock}

\FloatBarrier
\subsection[The compact amplitude and the punctured-sphere limit]{\vtwo{The compact amplitude and the punctured-sphere limit}}\label{ssec:dsquantumtriangulation}\label{ssec:lorentztetrahedra}
\begin{vtwoblock}
\finalpolish{The recovered compact filling $M=T_1\cup T_2$ also gives a geometric test of the boundary coefficient. The coefficient extracted in section~\ref{sec:boundaryamp} equals the compact amplitude in \eqref{eq:coreexact}. We now compare its semiclassical phase with the gravitational action of the reconstructed geometry.}

\finalpolish{Three operations select this same marked compact core. OPE normalization removes four $C_0$ flare factors and leaves $Z_L(M,\gamma)$, whose external $S$ transform gives $Z_A$. Inverse normal flow collapses six geometric collars and recovers the pleated boundary $\partial M$. At the level of the action, we must subtract the contribution of the gravitational end $E$ to isolate $I_A(M)$. The first two steps were established above. The third is the calculation below.}

\paragraph{Subtracting the gravitational end.}
The complete hyperbolic filling $\widehat M=M\cup E$ consists of the compact core and an outward end $E$. The Lorentzian region $\mathcal D_M$ is related to this end by the time continuation below. We will subtract its contribution to isolate the compact action. Substituting the contour of \eqref{eq:cosmology_wormhole_contour_v46} into the pants and wedge metrics gives
{\color{black}
\begin{equation}
 t=\rho+\frac{\ii\pi}2\quad\Longrightarrow\quad
 \begin{aligned}
 \dd s^2&=-\ell_{\rm dS}^2\big[\dd\rho^2+\cosh^2\rho\,h_{\Pi_i}\big],\\
 \dd s^2&=-\ell_{\rm dS}^2\big[\dd\rho^2+\sinh^2\rho\,\dd\varphi^2+\cosh^2\rho\,\dd x^2\big],
 \end{aligned}
\label{eq:lorentzToEnd}
\end{equation}
}
Apart from the overall minus sign, these are the hyperbolic metrics outside $M$, with $\rho$ the distance from $\partial M$. At $\rho=0$ the continued embedding is $X=\ii Y$, where the end joins the negative-definite core of section~\ref{ssec:labels}. Here we continue time to compare actions. The preceding inverse normal flow instead kept time real and recovered $Y$ from the Lorentzian geometry.

We evaluate the Einstein action with its initial and final boundary terms and the area counterterm. The area integrals are given in appendix~\ref{app:endactions}. At the initial boundary, each arc contributes its length times the boost rapidity around it,
{\color{black}
\begin{equation}
 \frac1{8\pi G_N}\lim_{t\to0}\int\dd^2x\sqrt h\,K
 =\frac{\ell_{\rm dS}}{8\pi G_N}\sum_e(\pi-\psi_e)\,\ell(\gamma_e)
 =\frac{c_{\rm asym}}{12\pi}\sum_e(\pi-\psi_e)\,\ell(\gamma_e).
\label{eq:lorentzRegge}
\end{equation}
}
The product in each term is $d_e\lambda_e$. In Lorentzian units, $d_e$ is the initial arc length divided by $\ell_{\rm dS}$ and $\lambda_e$ is the boost rapidity. On the recovered core, the same product is the exterior bending angle times the cuff length. Including the bulk and final-boundary terms, with the counterterm cancelling the leading area divergence, gives
{\color{black}
\begin{equation}
 S_L[\mathcal D_M]=\frac{\ell_{\rm dS}}{4\pi G_N}\Big[\frac14\sum_e(\pi-\psi_e)\,\ell(\gamma_e)
 +\pi\chi(\partial M)\Big(t_c+\frac12\Big)\Big]+O\big(e^{-2t_c}\big).
\label{eq:lorentzAction}
\end{equation}
}
Here $t_c$ is the late-time cutoff and $\chi(\partial M)=-4$. A shift of $t_c$ rescales the boundary metric. The remaining linear term therefore gives the classical Weyl anomaly. With the chosen branch $c\simeq-\ii\,3\ell_{\rm dS}/(2G_N)$ and phase $e^{-\ii S_L}$, its coefficient is
{\color{black}
\begin{equation}
 \partial_{t_c}\log e^{-\ii S_L}
 \longrightarrow-\frac{\ii\ell_{\rm dS}}{4G_N}\chi(\partial M)
 \simeq\frac{c}{6}\chi(\partial M).
\label{eq:lorentzAnomalyBranch}
\end{equation}
}
Thus a cutoff shift $t_c\mapsto t_c+\sigma$ changes the logarithm of the phase by $c\,\chi(\partial M)\sigma/6$. Now compare the action of $\widehat M$ at radial cutoff $\rho_c$. It contains a compact volume term and the same bending and cutoff terms with opposite signs,
{\color{black}
\begin{equation}
 I_E[\widehat M]=\frac{c}{6\pi}\Big[\operatorname{Vol}(M)-\frac14\sum_e(\pi-\psi_e)\,\ell(\gamma_e)
 -\pi\chi(\partial M)\Big(\rho_c+\frac12\Big)\Big]+O\big(e^{-2\rho_c}\big).
\label{eq:lorentzEuclid}
\end{equation}
}
The finite part inside the brackets is the half-sum of the volume and dual volume of the convex core \cite{KrasnovSchlenker,SchlenkerRV}. Set $\rho_c=t_c$ so that the two actions use the same cutoff. After removing their overall constants, the bending and linear cutoff terms cancel in the sum,
{\color{black}
\begin{equation}
 \frac{6\pi}{c}\,I_E[\widehat M]+\frac{4\pi G_N}{\ell_{\rm dS}}\,S_L[\mathcal D_M]
 =\operatorname{Vol}(M)+\Big(4\pi-\frac12\sum_e(\pi-\psi_e)\,\ell(\gamma_e)\Big)e^{-2t_c},
\label{eq:lorentzSumExact}
\end{equation}
}
Only $\operatorname{Vol}(M)=2V_A$ survives as the cutoff is removed. Restore the constants and continue $c/(6\pi)\to1/(\pi b^2)=-\ii/(\pi r^2)$, with $\ell_{\rm dS}/(4\pi G_N)=1/(\pi r^2)$ at classical order. The cancellation then reads
{\color{black}
\begin{equation}
 -I_E^{\rm cont}[\widehat M]+\ii S_L[\mathcal D_M]
 =-I_A^{\rm cont}(M)+O\!\left(r^{-2}e^{-2t_c}\right),
 \qquad \rho_c=t_c .
\label{eq:lorentzBox}
\end{equation}
}
Subtracting the end action thus leaves the compact action. In the exponential weights, subtraction becomes division by $e^{-\ii S_L[\mathcal D_M]}$. Taking $t_c\to\infty$ at fixed coupling before the semiclassical limit gives
\begin{equation}
\begin{aligned}
 \lim_{\substack{t_c\to\infty\\ \rho_c=t_c}}
 \frac{e^{-I_E^{\rm cont}[\widehat M]}}{e^{-\ii S_L[\mathcal D_M]}}
 &\overset{\rm cl}{=}e^{-I_A^{\rm cont}(M)}\\
 &=\exp\!\left[\frac{\ii\,\operatorname{Vol}(M)}{\pi r^2}+O(r^0)\right]
 =\exp\!\left[\frac{2\ii V_A(\boldsymbol\psi)}{\pi r^2}+O(r^0)\right].
\end{aligned}
\label{eq:compactendremoval}
\end{equation}
The remaining phase is $2V_A/(\pi r^2)$, exactly the leading phase of the boundary coefficient. This checks the classical action, while the exact coefficient also contains a quantum prefactor. We can make the geometric comparison more explicit by differentiating the phase. The Schl\"afli relation turns the volume derivative into a cell-edge length, and the reconstruction identifies twice that length with the measured boost rapidity,
\begin{equation}
\begin{aligned}
 \arg Z_A^{\rm cont}(M,\gamma)
 &=\frac{2V_A(\boldsymbol\psi)}{\pi r^2}+O(1),\\
 -\pi r^2\,\partial_{\psi_e}\arg Z_A^{\rm cont}(M,\gamma)
 &\longrightarrow -2\,\partial_{\psi_e}V_A
 =\ell_e=\frac{\lambda_e}{2}.
\end{aligned}
\label{eq:lorentzInverseQuantumCheck}
\end{equation}
As in \eqref{eq:geometricreadout}, we use a continuous nonzero branch on which the asymptotics can be differentiated once. The relation $\psi_s=2\pi rk$ then makes the phase extracted from the wavefunction in \eqref{eq:ratiophase} yield the same $\ell_s=\lambda_s/2$. Thus the boundary coefficient and the compact gravitational action reproduce the boost measured in the real Lorentzian spacetime.

\paragraph{From the closed future boundary to punctured spheres.}
There is still a difference between the two boundaries. The real spacetime has a closed genus-three future boundary, whereas the four-point wavefunction is defined on a four-punctured sphere. Their connection can be seen by cutting and then pinching four necks of the closed surface.

Cut the cuffs labelled $1,2,3,4$ in figure~\ref{fig:dsbranches}. The pants at $A,B$ remain joined along $s$, and those at $C,D$ along $t$. Each pair forms a sphere with four circular holes, denoted $F_s$ and $F_t$. Joining their corresponding boundary circles restores $\Sigma_3$. The same construction applies to two $n$-leaf pants trees $\tau,\sigma$,
\begin{equation}
\partial M_{\Gamma_{\tau\sigma}}
 =F_\tau\cup_{\gamma_1,\ldots,\gamma_n}F_\sigma,
\qquad
F_\tau\simeq F_\sigma\simeq\Sigma_{0,n}^{\rm holed},
\qquad
g(\partial M)=n-1.
\label{eq:spheretreeboundarysplit}
\end{equation}
For the canonical boundary-parallel embedding, the compact manifold is the product $M\simeq F\times I$, where $F=\Sigma_{0,n}^{\rm holed}$. Its boundary consists of two copies of $F$ and one connecting annulus for each hole,
\begin{equation}
\partial(F\times I)
 =(F\times\{0\})\cup(\partial F\times I)\cup(F\times\{1\}),
\qquad
\chi(\partial M)=2\chi(F)=4-2n.
\label{eq:spheretreeproductboundary}
\end{equation}
Each copy of $F$ contains $n-2$ pants and $n-3$ internal cuffs. The connecting annuli add no Euler characteristic, so $2-2g=4-2n$ gives $g=n-1$. The examples are
\begin{center}
\begin{tabular}{@{}lccc@{}}
\toprule
Sphere component $F$ & Closed boundary & Joining annuli & Graph example\\
\midrule
$\Sigma_{0,3}^{\rm holed}$ & $\Sigma_2$ & $3$ & theta\\
$\Sigma_{0,4}^{\rm holed}$ & $\Sigma_3$ & $4$ & $K_4$\\
$\Sigma_{0,5}^{\rm holed}$ & $\Sigma_4$ & $5$ & prism, $K_{3,3}$\\
$\Sigma_{0,n}^{\rm holed}$ & $\Sigma_{n-1}$ & $n$ & $\Gamma_{\tau\sigma}$\\
\bottomrule
\end{tabular}
\end{center}
For $K_{3,3}$ the two trees are $((12)3(45))$ and $((14)3(25))$. The $n$ joining annuli are distinct from the $n-3$ OPE collars within either sphere.

Cutting produces circular holes, not yet punctures. To obtain punctures, make the connecting cylinders long compared with their circumference. For $K_4$, set the four external angles to the common value $\psi_0$ and retain the two internal angles $\psi_s,\psi_t$. The inverse Gram matrix gives the external half-cuff length
\begin{equation}
\begin{aligned}
\cosh\ell_{\rm ext}
 &=\frac{\cos\psi_0\cos(\psi_s/2)\cos(\psi_t/2)}
 {\sqrt{[\cos^2\psi_0-\sin^2(\psi_s/2)]
 [\cos^2\psi_0-\sin^2(\psi_t/2)]}},\\
\ell_{\rm ext}
 &=\psi_0\sqrt{\frac{1}{\cos^2(\psi_s/2)}
  +\frac{1}{\cos^2(\psi_t/2)}-1}+O(\psi_0^3),
 \qquad \frac{\pi-\psi_0}{2\ell_{\rm ext}}\longrightarrow\infty .
\end{aligned}
\label{eq:externalcuffpinch}
\end{equation}
Here $\psi_0\to0$ with $\psi_s,\psi_t$ fixed in the hyperideal chamber. The cylinder circumference $2\ell_{\rm ext}$ tends to zero while its width $\pi-\psi_0$ stays finite, so their ratio diverges. A flat cylinder of width $d_i$ and circumference $\lambda_i$ has sewing parameter $|q_i|=e^{-2\pi d_i/\lambda_i}$, with phase fixed by the twist. Substituting $d_i=\pi-\psi_i$ and $\lambda_i=2\ell_i$ gives
\begin{equation}
\begin{aligned}
 |q_i|&=\exp\!\left[-\frac{2\pi d_i}{\lambda_i}\right]
       =\exp\!\left[-\frac{\pi(\pi-\psi_i)}{\ell_i}\right]
       \longrightarrow0,\qquad i=1,\ldots,4,\\
 \Sigma_3&\longrightarrow
 \Sigma_{0,4}^{(s)}\cup_{\text{four nodes}}\Sigma_{0,4}^{(t)} .
\end{aligned}
\label{eq:futureannuluspinch}
\end{equation}
Each long neck pinches to a node. Separating the two branches at all four nodes leaves two four-punctured spheres. The parameters $q_i$ belong to these connecting necks, not to the OPE collars inside either sphere. We reach the limit by varying the spacetime parameters, not by a topology change during time evolution.

The four external necks have now pinched, but each component still contains its internal flat collar. To obtain the hyperbolic four-punctured sphere used in the inflating metric, shrink these two remaining widths as well. Take the external angles to zero faster than the internal bending angles,
\begin{equation}
\psi_0=\varepsilon,\qquad
\psi_s=\pi-\delta,\qquad \psi_t=\pi-\kappa\delta,
\qquad \varepsilon=o(\delta),\qquad
\delta\longrightarrow0,\quad 0<\kappa<\infty.
\label{eq:matchedspherepath}
\end{equation}
The first length in \eqref{eq:explicitlengths} then gives
\begin{equation}
\begin{aligned}
\sinh\frac{\ell_s}{2}
 &=\frac{\sin(\kappa\delta/2)\cos(\delta/2)}
 {\sqrt{\sin^2(\delta/2)-\sin^2\varepsilon}}
 \longrightarrow\kappa,\\
\ell_s&\longrightarrow2\arsinh\kappa,\qquad
\ell_t\longrightarrow2\arsinh(\kappa^{-1}),\\
\ell_{\rm ext}
 &\sim\frac{2\varepsilon}{\delta}\sqrt{1+\kappa^{-2}}
 \longrightarrow0 .
\end{aligned}
\label{eq:matchedspherelengths}
\end{equation}
Exchanging $s,t$ gives the second internal length. The exact formula for $\ell_{\rm ext}$ gives the last line. Thus the external cuffs still pinch, while the internal cuff lengths approach finite values.

The marked face pairings keep the corresponding seam endpoints aligned as the two internal widths $\delta,\kappa\delta$ vanish. The resulting components are oppositely oriented copies of the same four-punctured sphere, obtained by doubling an ideal quadrilateral. The limiting half-cuff lengths obey $\sinh(\ell_s/2)\sinh(\ell_t/2)=1$, as shown above. With no internal flat band left, the metric on compact sets away from the cusps approaches
\begin{equation}
\dd s^2=\ell_{\rm dS}^{\,2}
\big[-\dd t^2+\sinh^2t\,h_{0,4}\big],
\label{eq:matchedsphereinflation}
\end{equation}
where $h_{0,4}$ is the complete hyperbolic metric on the four-punctured sphere. This is the inflating form used in \cite{CEM}. The tetrahedron itself becomes degenerate in this limit. Indeed, substituting the same path in its Gram determinant gives
\begin{equation}
\begin{aligned}
\det G
 &=16\cos^2\frac{\psi_s}{2}\cos^2\frac{\psi_t}{2}
 \left[\sin^2\frac{\psi_s}{2}\sin^2\frac{\psi_t}{2}
       -\cos^2\psi_0\right]\\
 &\sim-\frac{\kappa^2(1+\kappa^2)}4\,\delta^6 .
\end{aligned}
\label{eq:spherelimitGram}
\end{equation}

The vanishing determinant explains why the corresponding quantum limit must be taken in the exact kernel, not by substituting into the nondegenerate fixed-angle asymptotics.

For general sphere trees, the same two steps are a cusp limit at the external necks and matched small bending at the internal cuffs. They give mirrored punctured components when the two pants decompositions fill the sphere, meaning that no essential closed curve avoids both sets of cuffs. The prism and $K_{3,3}$ pairs obey this condition. Appendix~\ref{ssec:sphere-tree-extension} specifies the bending conditions and limiting metric \cite{BonahonOtal,LecuireBending,SeriesBending}.
\end{vtwoblock}

\section[OPE statistics and statistical sewing]{{OPE statistics and statistical sewing}}\label{sec:bulkinverse}\label{sec:ensemble}

{The compact filling amplitudes of a marked graph are complex numbers on the dS ray.  This section defines from them the moments of normalized OPE coefficients, assembles the moments into correlations of boundary four-point functions, and imposes the sphere and torus sewing conditions on these correlations.}

\subsection{The moment functional and boundary correlations}\label{ssec:ensemblesample}\label{ssec:bulkstatisticalinput}\label{ssec:statisticalobservables}

{We hold fixed the central charge, a labelled, locally finite primary spectrum with multiplicities, a unique identity, integer spins $h_i-\bar h_i\in\mathbb Z$, and a self-dual basis in which $c_{\mathbf1ij}=\delta_{ij}$. These specify the proposed CFT sample data, not the normalizable spectrum of the collective cosmological state. The statistical variables are the nonidentity OPE coefficients, normalized by the leading real-$b$ variance,}
\begin{equation}
 C_A=\frac{c_{ijk}}{\big[C_0(P_i,P_j,P_k)\,C_0(\bar P_i,\bar P_j,\bar P_k)\big]^{1/2}},\qquad A=(ijk).
 \label{eq:s5variables}
\end{equation}
{For scalar primaries the triple $A$ is unordered, and distinct entries may be correlated.}

{At real $b>0$ the overline is the ensemble average \eqref{eq:s2ensembleaverage}, an integral against a positive weight.}
We first give the filling assignment for scalar labels, $\bar P_e=P_e$, so the two chiral graphs have the same momenta.  The spinning handle sector below keeps its left and right labels independent.
{Fix a connected trivalent graph $\Gamma$, embedded with a framing in a closed three-manifold $M_E$, with edge labels $\boldsymbol P=(P_e)$, and write}
\begin{equation}
 Z_\Gamma(b,\boldsymbol P):=Z_{\rm Vir}\big(M_E,\Gamma(\boldsymbol P)\big)
 \label{eq:s5ZGamma}
\end{equation}
{for its chiral Virasoro TQFT amplitude at $c=1+6(b+b^{-1})^2$, with every vertex normalized by the theta graph.  Its compact core is $M=M_E\setminus N(\Gamma)$.  One flare attachment gives the factorization \eqref{eq:s2flarefactorization}, and summing the $\nu_\Gamma$ inequivalent attachments that realize the same index contraction gives the contribution of this filling to the AdS moments,}
\begin{equation}
\begin{aligned}
 \left.\overline{\prod_{v\in V(\Gamma)}c_v}\,\right|_{(M_E,\Gamma)}
 &=\nu_\Gamma\prod_{v\in V(\Gamma)}\big[C_0(P_{i_v},P_{j_v},P_{k_v})\,C_0(\bar P_{i_v},\bar P_{j_v},\bar P_{k_v})\big]^{1/2}\,
   Z_L\big(M,\gamma(\boldsymbol P)\big),\\
 \left.\overline{\prod_{v\in V(\Gamma)}C_v}\,\right|_{(M_E,\Gamma)}
 &=\nu_\Gamma\,Z_\Gamma(b,\boldsymbol P)\,Z_\Gamma(b,\boldsymbol P)^*,\qquad b>0,\quad P_e>0 .
\end{aligned}
\label{eq:s5AdSfilling}
\end{equation}
\Needspace{10\baselineskip}
{The second line divides by the flare factors as in \eqref{eq:s5variables} and uses $Z_L=|Z_\Gamma|^2$ from \eqref{eq:revZL}. The restriction $|_{(M_E,\Gamma)}$ selects one connected filling contribution. The compact amplitude $Z_L$ itself contains no information about which asymptotic three-point boundaries were attached before their flares were stripped off. The integer $\nu_\Gamma$ counts the inequivalent ways of reattaching one $C_0$ flare per graph vertex that give the \emph{same prescribed OPE index contraction}. Every such attachment carries the same normalized compact contribution. For generic distinct scalar labels, the theta, linked-handcuff and unbraided-tetrahedral examples give \cite{Hartman}}
\[
 {\nu_\theta=1,\qquad \nu_H=1+1=2,\qquad \nu_{\mathrm{tet}}=1,\qquad
 \left.\overline{C_{113}C_{223}}\right|_H=(1+1)Z_{L,H}=2Z_{L,H}.}
\]
{The two terms in $1+1$ count two attachments of the same labelled handcuff core, not two additional topologies or a factor chosen to make an amplitude integer-valued. In particular $\nu_\Gamma$ is a combinatorial multiplicity, whereas $Z_L$ and the resulting moment need not be integers. The corresponding spin-weighted exchange factor is distinguished below. For scalar operators in a real basis the normalized coefficients are real, so the stars on individual coefficients can be dropped.}

{Complex conjugation is antiholomorphic.  To analytically continue \eqref{eq:s5AdSfilling} off the real axis we therefore write its conjugate factor as a holomorphic function, the reflected amplitude}
\begin{equation}
 Z_\Gamma^\vee(b,\boldsymbol P):=Z_\Gamma(b^*,\boldsymbol P^*)^* .
 \label{eq:veedefinition}
\end{equation}
{It is holomorphic wherever $Z_\Gamma$ is, and for real $(b,\boldsymbol P)$ it equals $Z_\Gamma(b,\boldsymbol P)^*$.  Geometrically, $Z_\Gamma^\vee$ is the amplitude of the mirror image of the framed graph, in which each framing and braiding phase $e^{2\pi\ii h}$ of $Z_\Gamma$ appears as $e^{-2\pi\ii h}$ \cite{CTV}.  The two cases used below are}
\begin{equation}
 \begin{Bmatrix}P_1&P_2&P_3\\P_4&P_5&P_6\end{Bmatrix}_{\!\rm Vir}^{\!\vee}
 =\begin{Bmatrix}P_1&P_2&P_3\\P_4&P_5&P_6\end{Bmatrix}_{\!\rm Vir},\qquad
 \mathbb S^\vee=e^{-\ii\pi h_0}\,\mathbb S .
 \label{eq:s5veeexamples}
\end{equation}
{The unbraided tetrahedral amplitude carries no framing phase and is real for real arguments, so it equals its own reflection.  The punctured-torus kernel is written in the convention of \cite{CTV}.}

\begin{figure}[!htbp]
\centering
\begin{tikzpicture}[x=1.22cm,y=1.22cm,line cap=round,line join=round,font=\small,
  lab/.style={font=\scriptsize,inner sep=1.1pt},
  axis/.style={black,line width=.5pt},
  ray/.style={black,densely dashed,line width=.5pt}]
\newcommand{\bplane}{%
  \draw[axis] (-.25,0) -- (2.55,0); \draw[axis] (0,-1.75) -- (0,1.75);
  \node[lab,black] at (2.55,.17) {$\Re b$};
  \node[lab,black] at (.30,1.70) {$\Im b$};
  \draw[ray] (0,0) -- (45:2.35); \draw[ray] (0,0) -- (-45:2.35);
  \draw[black,line width=2.2pt] (.12,0) -- (2.30,0);}
\node[anchor=west,paneltitle] at (-.35,2.25) {(a) $\Eds$};
\begin{scope}
 \bplane
 \draw[pt,line width=1.3pt] (1.80,0) arc[start angle=0,end angle=45,radius=1.80];
 \draw[paircolor,line width=1.3pt] (1.80,0) arc[start angle=0,end angle=-45,radius=1.80];
 \draw[black,densely dotted,line width=.7pt] (45:1.80) -- (-45:1.80);
 \fill[black] (1.80,0) circle (1.8pt);
 \fill[pt] (45:1.80) circle (2.1pt);
 \fill[paircolor] (-45:1.80) circle (2.1pt);
 \node[lab] at (2.02,-.20) {$b$};
 \node[lab,pt] at ($(45:1.80)+(.38,.12)$) {$b_{\rm dS}$};
 \node[lab,paircolor] at ($(-45:1.80)+(.40,-.12)$) {$b_{\rm dS}^*$};
 \node[lab,pt] at ($(45:1.80)+(-.30,.24)$) {$Z_\Gamma$};
 \node[lab,paircolor] at ($(-45:1.80)+(-.30,-.26)$) {$Z_\Gamma^*$};
\end{scope}
\node[anchor=west,paneltitle] at (3.65,2.25) {(b) $|Z_\Gamma(b_{\rm dS})|^2$};
\begin{scope}[shift={(4.0,0)}]
 \bplane
 \draw[pt,line width=1.3pt] (1.80,0) arc[start angle=0,end angle=45,radius=1.80];
 \fill[black] (1.80,0) circle (1.8pt);
 \fill[pt] (45:1.80) circle (2.1pt);
 \draw[paircolor,line width=.9pt] (45:1.80) circle (3.6pt);
 \node[lab] at (2.02,-.20) {$b$};
 \node[lab,pt] at ($(45:1.80)+(.38,.12)$) {$b_{\rm dS}$};
 \node[lab,pt] at ($(45:1.80)+(-.30,.24)$) {$Z_\Gamma$};
 \node[lab,paircolor] at ($(45:1.80)+(-.10,-.32)$) {$Z_\Gamma^*$};
\end{scope}
\node[anchor=west,paneltitle] at (7.75,2.25) {(c) Tetrahedral values};
\begin{scope}[shift={(8.75,0)}]
 \draw[axis] (-.25,0) -- (2.30,0); \draw[axis] (0,-1.55) -- (0,1.95);
 \node[lab,black] at (2.30,.17) {$\Re$};
 \node[lab,black] at (.22,1.90) {$\Im$};
 \def\ph{32}
 \draw[black,densely dashed,line width=.45pt] (-\ph-12:1.1) arc[start angle=-\ph-12,end angle=\ph+12,radius=1.1];
 \draw[black,densely dashed,line width=.45pt] (-12:1.75) arc[start angle=-12,end angle=2*\ph+14,radius=1.75];
 \fill[pt] (\ph:1.1) circle (2.0pt);
 \fill[black] (-\ph:1.1) circle (1.7pt);
 \node[lab,pt] at ($(\ph:1.1)+(.32,.08)$) {$\{6j\}$};
 \node[lab,black] at ($(-\ph:1.1)+(.32,-.14)$) {$\{6j\}^*$};
 \draw[pt,line width=.6pt] (.45,0) arc[start angle=0,end angle=\ph,radius=.45];
 \node[lab,pt] at (15:.64) {$\varphi$};
 \fill[pt] (2*\ph:1.75) circle (2.4pt);
 \draw[paircolor,line width=.9pt] (2*\ph:1.75) circle (4pt);
 \node[lab,anchor=west] at ($(2*\ph:1.75)+(.14,.06)$) {$\{6j\}^2$};
 \draw[black,line width=.6pt] (.80,0) arc[start angle=0,end angle=2*\ph,radius=.80];
 \node[lab] at (52:.98) {$2\varphi$};
 \fill[black] (1.75,0) circle (2.4pt);
 \node[lab,black] at (1.80,-.30) {$|\{6j\}|^2$};
\end{scope}
\end{tikzpicture}
\caption{{Analytic continuation of one graph contribution.  (a) The AdS contribution $Z_\Gamma(b)Z_\Gamma(b)^*$ evaluates both factors at the same real $b$.  Its analytic continuation \eqref{eq:Edsdefinition} carries $Z_\Gamma$ along the upper path to $b_{\rm dS}$ and the reflected factor $Z_\Gamma^\vee(b)=Z_\Gamma(b^*)^*$ along the mirror path, so the second factor is $Z_\Gamma$ at $b_{\rm dS}^*$, conjugated.  The edge momenta move with $b$, to $\boldsymbol P_{\rm dS}$ and $\boldsymbol P_{\rm dS}^*$.  (b) The modulus at the dS point conjugates after the continuation and evaluates both factors at $b_{\rm dS}$.  (c) For the unbraided tetrahedron the reflected factor equals $\{6j\}$ itself, so the dS moment is $\{6j\}^2$, with twice the phase $\varphi$ of $\{6j\}$, while the modulus $|\{6j\}|^2$ is real.  Radii are schematic.}}
\label{fig:dsstatcontinuation}
\end{figure}
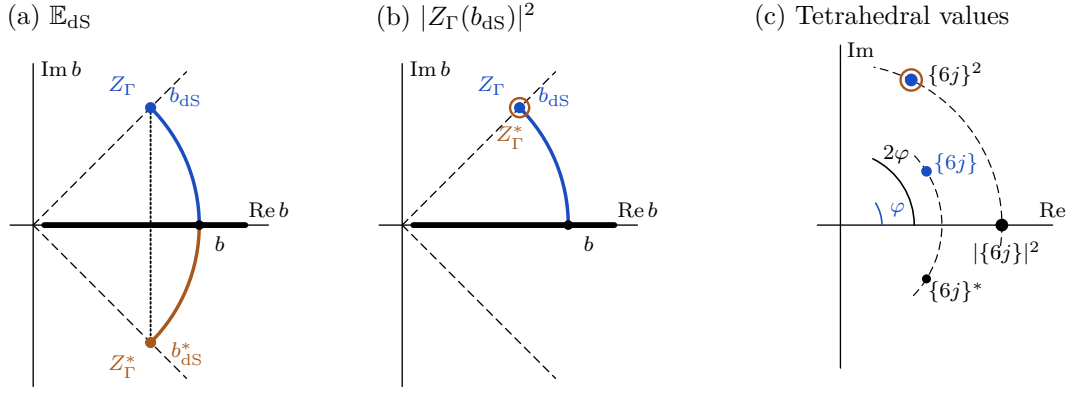

{The dS point is $b_{\rm dS}=re^{\ii\pi/4}$ with edge labels $P_{e,\rm dS}=e^{\ii\pi/4}k_e$, $k_e>0$.  For a function $F(b,\boldsymbol P)$ holomorphic near the real domain $b>0$, $P_e>0$, the bracket is its value at the dS point after analytic continuation,}
\begin{equation}
 \begin{aligned}
 [F]_{\rm cont}&:=F(b_{\rm dS},\boldsymbol P_{\rm dS}),\\
 (b,P_e)&=(re^{\ii\theta},e^{\ii\theta}k_e),\qquad 0\le\theta\le\tfrac\pi4 .
 \end{aligned}
 \label{eq:s5contdefinition}
\end{equation}
{Here $F$ on the right is analytically continued from the real domain along the path in the second line. The path is deformed where needed to keep the prescribed side of the pole families of appendix~\ref{app:contour}, and the analytic continuation of a product is the product of the analytic continuations of its factors. Fixing $c$ and the spectrum in the statistical problem means they are not random variables in any one moment; the path above relates amplitudes at different parameter values and does not average over $b$ or $P_e$. At the end point $4\pi bP_e=2\ii\psi_e$ with real angle $\psi_e=2\pi rk_e$. Along this path the reflected factor \eqref{eq:veedefinition} evaluates $Z_\Gamma$ on the mirror path $(re^{-\ii\theta},e^{-\ii\theta}k_e)$, as drawn in figure~\ref{fig:dsstatcontinuation}(a). This joint rotation is the scalar-label continuation; left and right momenta in spinning sectors require the fixed-spin prescription \eqref{eq:fixedspintransport}. The dS moment of the filling is the analytic continuation of \eqref{eq:s5AdSfilling},}
\begin{equation}
\boxed{\begin{aligned}
 \left.\Eds\Big[\prod_{v\in V(\Gamma)}C_v\Big]_c\right|_{(M_E,\Gamma)}
 &:=\nu_\Gamma\big[Z_\Gamma Z_\Gamma^\vee\big]_{\rm cont}\\
 &=\nu_\Gamma\,Z_\Gamma(b_{\rm dS},\boldsymbol P_{\rm dS})\;Z_\Gamma(b_{\rm dS}^*,\boldsymbol P_{\rm dS}^*)^* .
\end{aligned}}
\label{eq:Edsdefinition}
\end{equation}
{\vtwo{The bracket is the continued fixed-length amplitude $Z_L^{\rm cont}(M,\gamma)$ of \eqref{eq:revZL}. The external transform \eqref{eq:revZA} relates it to the fixed-angle CTV weight.}  Placing the three bilinears built from one graph amplitude side by side,}
\begin{equation}
\begin{aligned}
 \left.\overline{\prod_vC_v}\,\right|_{(M_E,\Gamma)}
 &=\nu_\Gamma\,Z_\Gamma(b,\boldsymbol P)\;Z_\Gamma(b,\boldsymbol P)^*,
 &&b>0,\\
 \left.\Eds\Big[\prod_vC_v\Big]_c\right|_{(M_E,\Gamma)}
 &=\nu_\Gamma\,Z_\Gamma(b_{\rm dS},\boldsymbol P_{\rm dS})\;Z_\Gamma(b_{\rm dS}^*,\boldsymbol P_{\rm dS}^*)^*,\\
 \nu_\Gamma\,\big|Z_\Gamma(b_{\rm dS},\boldsymbol P_{\rm dS})\big|^2
 &=\nu_\Gamma\,Z_\Gamma(b_{\rm dS},\boldsymbol P_{\rm dS})\;Z_\Gamma(b_{\rm dS},\boldsymbol P_{\rm dS})^* .
\end{aligned}
\label{eq:s5threebilinears}
\end{equation}
{At real $b$ the two arguments in the first line coincide, and the AdS average is real and non-negative.  The dS moment evaluates its second factor at the mirror point $b_{\rm dS}^*=re^{-\ii\pi/4}$, so it is in general a complex number.  The third line conjugates after the analytic continuation and evaluates both factors at $b_{\rm dS}$, as in figure~\ref{fig:dsstatcontinuation}(b), a different, real number.  By \eqref{eq:veedefinition} it equals $\nu_\Gamma Z_\Gamma(b_{\rm dS},\boldsymbol P_{\rm dS})Z_\Gamma^\vee(b_{\rm dS}^*,\boldsymbol P_{\rm dS}^*)$, the reflected factor at the conjugate parameter $b_{\rm dS}^*=-\ii b_{\rm dS}$ with $c(b_{\rm dS}^*)=26-c(b_{\rm dS})$, which is the pairing of the physical norm $\Psi^*\Psi$.  The dS moment keeps the reflected factor at the same $b_{\rm dS}$, which is the pairing of $\Psi\Psi$. The third line of \eqref{eq:s5threebilinears} only illustrates complex conjugation of a single graph coefficient. The physical gravitational norm also pairs full states and integrates their boundary moduli, and is not numerically identified with this fixed-graph modulus.}

{The unbraided tetrahedron makes the difference explicit.  With}
\[
 \{6j\}(b,\boldsymbol P):=\begin{Bmatrix}P_1&P_2&P\\P_3&P_4&R\end{Bmatrix}_{\!\rm Vir},
\]
{\eqref{eq:revnongauss} and \eqref{eq:Edsdefinition} give}
\begin{equation}
\begin{aligned}
\left.\overline{C_{12P}C_{34P}C^*_{14R}C^*_{23R}}\right|_{\rm tet}
 &=\{6j\}(b,\boldsymbol P)\;\{6j\}(b,\boldsymbol P)^*
  =\{6j\}(b,\boldsymbol P)^2,
 &&b>0,\\
\left.\Eds\big[C_{12P}C_{34P}C_{14R}C_{23R}\big]_c\right|_{\rm tet}
 &=\{6j\}(b_{\rm dS},\boldsymbol P_{\rm dS})\;\{6j\}(b_{\rm dS}^*,\boldsymbol P_{\rm dS}^*)^*
  =\{6j\}(b_{\rm dS},\boldsymbol P_{\rm dS})^2 .
\end{aligned}
\label{eq:tetAdSdS}
\end{equation}
{The first line uses that $\{6j\}$ is real for real arguments, and the second uses its reflection invariance \eqref{eq:s5veeexamples}.  On the dS ray $\{6j\}$ has a phase $\varphi$, equal to $V_A/(\pi r^2)$ at leading order by \eqref{eq:MY}, so the moment has the phase $2\varphi$ of figure~\ref{fig:dsstatcontinuation}(c), while $|\{6j\}(b_{\rm dS},\boldsymbol P_{\rm dS})|^2$ is real.  For graphs with framing phases the reflected factor carries the reversed phases, as for $\mathbb S^\vee$ in \eqref{eq:s5veeexamples}.}

{The dS moment functional $\Eds$ is the complex linear functional on polynomials in the $C_A$ fixed by $\Eds[1]=1$ and by its moments.  For a finite polynomial,}
\begin{equation}
 \Eds\Big[\sum_n f^{A_1\cdots A_n}C_{A_1}\cdots C_{A_n}\Big]
 =\sum_n f^{A_1\cdots A_n}\,m_{A_1\cdots A_n},
 \label{eq:s5linear}
\end{equation}
{with the raw moments $m_{A_1\cdots A_n}$ assembled below.  These are moments of the coefficients $C_A$ alone, and moments containing $C_A^*$ are separate data.}

{Full moments sum the filling contributions. For an ordered set $(A_1,\ldots,A_n)$, a contraction graph $\Gamma$ identifies equal primary labels along its edges. Its tensor $\delta^\Gamma_{A_1\cdots A_n}$ contains both the Kronecker deltas that enforce these index identifications and the reattachment multiplicity $\nu_\Gamma$. For instance, for one specified matching of the displayed OPE slots, the theta and linked-handcuff tensors read \cite{Hartman}}
\[
 {\delta^\theta_{(ijk),(lmn)}=\delta_{il}\delta_{jm}\delta_{kn},\qquad
 \delta^H_{(iik),(jjl)}=2\,\delta_{kl}\quad(i\ne j).}
\]
{In the handcuff the two copies of $i$ and the two copies of $j$ close into loops, while the $k$ and $l$ legs join the vertices and must agree. The factor $2$ counts the two flare attachments, not a second equality of primary indices. Other assignments of the ordered OPE entries are included in the graph sum. A marked embedding $\iota$ of $\Gamma$ in $M_E$ specifies a filling $M_E-N(\iota\Gamma)$ and its framing, with relative weight $w_{M_E,\iota}$ among the fillings of the same contraction. The raw moment is}
\begin{equation}
 m_{A_1\cdots A_n}
 :=\sum_\Gamma\delta^\Gamma_{A_1\cdots A_n}
 \sum_{[M_E,\iota]}w_{M_E,\iota}
 Z_L^{\rm cont}\big(M_E-N(\iota\Gamma),\gamma(\boldsymbol P)\big).
\label{eq:allgraphinverse}
\end{equation}
{The first sum runs over contraction graphs, including disconnected ones, and the second over inequivalent marked, framed fillings of each graph, each counted once with any residual symmetry factor in $w_{M_E,\iota}$. The factor $\nu_\Gamma$ is already included \emph{once} in $\delta^\Gamma$; it must not be multiplied into $Z_L^{\rm cont}$ a second time in this graph sum. Thus the fixed-filling formula \eqref{eq:Edsdefinition} and the full graph sum above use the same attachment count. The fixed-filling coefficient without that count is $Z_L^{\rm cont}=[Z_\Gamma Z_\Gamma^\vee]_{\rm cont}$, and the weights $w_{M_E,\iota}$ are constrained by statistical sewing in section~\ref{ssec:momentdictionary}. Spinning entries use separately prescribed analytic continuations of their chiral graph factors. Different triangulations of the same marked filling give the same TQFT coefficient wherever the state sum is defined.}

For a realization by genuine CFT tables, the fixed spectral data must admit a modular-invariant torus partition function, and the OPE entries must obey sphere crossing, braiding and punctured-torus sewing. The graph assignment supplies moment contributions to be tested against these conditions. The smooth density $\rho_0(P)\rho_0(\bar P)$ approximates heavy state sums in the real-$b$ reference ensemble, as in \eqref{eq:s2densitysum}. We continue the resulting sewn amplitudes with their contours; a complex continued $\rho_0$ is not itself a positive state-counting density. The central charge, labelled spectrum and light-sector data remain fixed in $\Eds$.

For disconnected fillings carrying OPE insertions we choose
\begin{equation}
 w_\varnothing=1,\qquad
 w_{F_1\sqcup F_2}=w_{F_1}w_{F_2}.
 \label{eq:commonfillingweights}
\end{equation}
Here $F_1,F_2$ are fillings carrying OPE insertions. Disconnected components carrying no OPE insertions cancel against the empty-insertion normalization. The tensor $\delta^\Gamma$ counts assignments of the ordered OPE insertions to graph vertices and includes the flare attachments exactly once; for example, disconnected theta graphs give the three Wick pairings at fourth order and fifteen at sixth. The connected subtraction at these orders is written explicitly in \eqref{eq:s4connectedfour}.

Introduce an auxiliary source $J_A$ for each coefficient $C_A$.  For a finite retained set of heavy coefficients, with distributional momentum labels integrated against test functions, the moments are collected in one formal generating function,
\begin{equation}
 Z_{\rm ens}[J]\equiv\Eds[e^{J_AC_A}]
 =1+\sum_{n\geq1}\frac{J_{A_1}\cdots J_{A_n}}{n!}
                 m_{A_1\cdots A_n}.
 \label{eq:commonmomentprescription}
\end{equation}
Repeated component indices are summed, and source derivatives insert the corresponding OPE coefficients.  A connected moment, or cumulant, removes products of lower moments, as illustrated at fourth order in \eqref{eq:s2connectedfour}.  Write $B_{AB}=\Eds[C_AC_B]$, $\kappa_{ABCD}=\Eds[C_AC_BC_CC_D]_c$ and $\kappa_{ABCDEF}=\Eds[C_AC_BC_CC_DC_EC_F]_c$.  The number of indices distinguishes the fourth and sixth cumulants.

These connected moments are generated by $W[J]=\log Z_{\rm ens}[J]$.  We assume that all odd moments of the fluctuating heavy coefficients vanish, including their means; identity and light coefficients remain fixed.  This is an assumption about the heavy-sector statistics.  Thus
\begin{equation}
\begin{aligned}
W[J]={}&\frac12B_{AB}J_AJ_B
+\frac1{4!}\kappa_{ABCD}J_AJ_BJ_CJ_D\\
&+\frac1{6!}\kappa_{ABCDEF}J_AJ_BJ_CJ_DJ_EJ_F+O(J^8),
\end{aligned}
\label{eq:s4cumulantexpansion}
\end{equation}
Exponentiating $W[J]$ restores the products of lower moments.  At fourth order these are the three pairings; at sixth order they are the $2+4$ and $2+2+2$ partitions,
\begin{equation}
\begin{aligned}
 \Eds[C_AC_BC_CC_D]
 &=\kappa_{ABCD}+B_{AB}B_{CD}+B_{AC}B_{BD}+B_{AD}B_{BC},\\
 \Eds[C_AC_BC_CC_DC_EC_F]
 &=\kappa_{ABCDEF}+15B_{(AB}\kappa_{CDEF)}+15B_{(AB}B_{CD}B_{EF)} .
\end{aligned}
\label{eq:s4connectedfour}\eqalias{eq:commonmomentsix}\eqalias{eq:appendixBK}
\end{equation}
Parentheses denote normalized symmetrization over the six free indices, and the two factors of fifteen count the $2+4$ and three-pair partitions.  A Gaussian assignment has $\kappa_{ABCD}=0$, and the tetrahedral filling supplies a connected fourth-moment contribution.

The moments predict correlations between boundary CFT amplitudes.  For example, the heavy-channel part of a four-point function is
\begin{equation}
\begin{aligned}
 G_{1234}^{\rm heavy}(z,\bar z)
 &=\sum_{i\,\mathrm{heavy}}c_{12i}c_{34i}
       \mathcal F^{(s)}_{P_i}(z)\mathcal F^{(s)}_{\bar P_i}(\bar z),\\
 G_a[C]&=T_a^{AB}C_AC_B,\qquad T_a^{AB}=T_a^{BA}.
\end{aligned}
 \label{eq:quadraticboundaryamplitude}
\end{equation}
In the second line we have substituted $c_{ijk}=[C_0(P_i,P_j,P_k)C_0(\bar P_i,\bar P_j,\bar P_k)]^{1/2}C_{ijk}$ and collected the known conformal blocks, three-point factors and state sum into $T_a^{AB}$.  The label $a$ fixes a particular four-point configuration (external operators, positions and channel).  Thus each amplitude is quadratic in the fluctuating OPE coefficients $C_A$.

More generally, a regulated amplitude $G_a[C]$ is a polynomial in the retained OPE coefficients.  Its correlations follow by replacing each $C_A$ with a source derivative,
\begin{equation}
 \Eds\!\left[\prod_{a=1}^rG_a[C]\right]
 =\left.\prod_{a=1}^rG_a(\partial_J)\,Z_{\rm ens}[J]\right|_{J=0}.
 \label{eq:statisticalobservables}
\end{equation}
The mean amplitude is $\Eds[G_a]=T_a^{AB}B_{AB}$.  Subtracting the products of means, two and three amplitudes have the connected correlations
\begin{equation}
\boxed{\begin{aligned}
 \Eds[G_aG_b]_c
 &=T_a^{AB}T_b^{CD}
       \big(\kappa_{ABCD}+B_{AC}B_{BD}+B_{AD}B_{BC}\big),\\
 \Eds[G_aG_bG_c]_c
 &=T_a^{AB}T_b^{CD}T_c^{EF}\big[\kappa_{ABCDEF}+8B_{AC}B_{BE}B_{DF}\\
 &\qquad+4\big(B_{AC}\kappa_{BDEF}+B_{AE}\kappa_{BCDF}+B_{CE}\kappa_{ABDF}\big)\big].
\end{aligned}}
\label{eq:boundaryfluctuations}\eqalias{eq:threeboundaryfluctuations}
\end{equation}
{The subscript $c$ on the left denotes connected correlations of the composite observables $G_a$. Each $G_a$ has exactly two OPE slots and all odd heavy-sector cumulants vanish by assumption. A connected partition of their six slots can therefore have only block sizes $6$, $4+2$, or $2+2+2$. The single size-six block gives $\kappa_6$. In the $4+2$ case, the size-two block must join two different observables, and the remaining size-four block must meet the third. There are three choices of the joined pair and two choices of slot in each, giving $3\times2^2=12$ partitions. Symmetry of the three $T$ tensors groups them into the three displayed terms with coefficient four. With only size-two blocks, every observable must connect to both others; the three-cycle has $2^3=8$ pairings. Disconnected partitions cancel in the connected correlator. Hence \eqref{eq:boundaryfluctuations} exhausts the allowed partition types at this order, including their multiplicities. This is the moment--cumulant decomposition under the stated assumptions, and the values of $B$, $\kappa_4$ and $\kappa_6$ come from the filling amplitudes. The $K_4$, prism and $K_{3,3}$ amplitudes supply selected non-Gaussian cumulant contributions; their consistency is tested below.}

\subsection{Statistical sewing and the low-order moments}\label{ssec:momentdictionary}\label{ssec:lowbootstrap}\label{ssec:ensembleclosure}\label{ssec:sourceequations}

To describe consistent CFT data, the proposed moments must give equivalent boundary correlations in different OPE channels and respect the relevant torus modular relations.  We call the corresponding conditions \emph{statistical sewing}.  They test whether the second and fourth moments fit together and constrain the higher connected contributions that accompany them.  Each compact amplitude is computed separately and carries no information about the relative weights of different fillings, so statistical sewing is the condition that relates them.

For scalar labels, the normalization is $c_{ijk}=C_0(P_i,P_j,P_k)C_{ijk}$, with both chiral factors analytically continued at the same $b$.  The theta contribution is one, and the tetrahedral term is \vtwo{\eqref{eq:tetAdSdS}}.  The linked handcuff instead gives a framed product of punctured modular kernels.  Its two inequivalent flare attachments give the factor of two below, and the kernels are normalized as in {\eqref{eq:Shat}}.  The restrictions $\theta$, $H$ and ${\rm tet}$ select the indicated graph contribution at unit filling weight, with the attachment multiplicity included once,
\begin{equation}
\begin{aligned}
\left.\Eds[C_{123}C_{123}]\right|_\theta&=1,&&\qquad\thetagraph\\[4pt]
 \left.\Eds[C_{113}C_{223}]\right|_H
 &=2e^{-\ii\pi h_3}\widehat{\mathbb S}_{P_1P_2}[P_3]^2,&&\qquad\handcuffgraph\\[4pt]
 \left.\Eds[C_{123}C_{156}C_{246}C_{345}]_c\right|_{\rm tet}
 &=\begin{Bmatrix}P_1&P_2&P_3\\P_4&P_5&P_6\end{Bmatrix}_{\!\rm Vir}^{\!2},&&\qquad\tetgraph
\end{aligned}
\label{eq:bootstrapagreement}
\end{equation}
The labels in the first line are generic and distinct.  {In the second line, reversing the handcuff framing contributes $e^{-\ii\pi h_3}$.}  The unbraided tetrahedron has no framing phase, so its two-copy coefficient is the displayed square.

{Its phase measures the boost holonomies of \vtwo{the Lorentzian spacetime from which section~\ref{ssec:inverseLorentzian} recovers the same core}, \eqref{eq:s6statisticalgeometry}.}

{These three coefficients enter different consistency tests, rather than three independent proofs of full crossing. The theta graph fixes the normalized second moment used in Gaussian pairings. The tetrahedral graph supplies the non-Gaussian fourth cumulant needed in the sphere $s/t$-channel relation, whose homogeneous heavy-sector hierarchy is derived below. The handcuff tests the repeated-label torus sector through the punctured modular $S$ kernel. Its finite $S$-closed pair is constructed in \eqref{eq:repeatedfiniteS}; the separate $T$ constraint and the full modular orbit are not settled by that pair. All three are specified graph contributions at unit filling weight, while the complete functional also sums over other allowed fillings.}

In one CFT the $s$- and $t$-channel expansions of a four-point function agree, with blocks in the convention of \eqref{eq:reviewblockfusion}.  Each channel contains two three-point vertices, so their difference is quadratic in the normalized coefficients, while the modular relation of a torus one-point function is linear \cite{CMMT,Hartman}.

Expand both OPE channels in a common Virasoro-block basis and project their difference against a test function labelled by $\lambda$. {Concretely, the sphere residual is the test-function projection of $G_{1234}^{(s)}(C)-G_{1234}^{(t)}(C)$, with each channel expanded as in \eqref{eq:quadraticboundaryamplitude}; it is not an independently chosen quadratic polynomial.} For the torus, similarly subtract the two modular-related one-point expansions. The coefficients of the heavy OPE variables in these differences define the full CFT crossing and torus residuals,
\begin{equation}
\begin{aligned}
 R_\lambda(C)&=r^\lambda_{AB}C_AC_B+r^\lambda_A C_A+r^\lambda_0,\\
 L_\lambda(C)&=\ell^\lambda_A C_A+\ell^\lambda_0 .
\end{aligned}
\label{eq:residualdef}
\end{equation}
The indices $A,B$ label fluctuating heavy OPE coefficients.  The coefficients $r^\lambda_A,r^\lambda_0,\ell^\lambda_0$ are fixed by terms containing light or identity OPE coefficients.  All $r^\lambda,\ell^\lambda$ also contain the conformal-block kernels, external labels, normalization factors and test function.  For the chosen fixed spectrum, exact CFT crossing requires $R_\lambda=L_\lambda=0$.

For any statistical functional supported on CFT tables satisfying these equations, multiplication by a heavy-coefficient polynomial $\mathcal P(C)$ and averaging gives
\begin{equation}
\Eds[\mathcal P(C)R_\lambda(C)]=0,
\qquad
\Eds[\mathcal P(C)L_\lambda(C)]=0.
\label{eq:s4polynomialcrossing}
\end{equation}
With $\partial_A=\partial/\partial J_A$, these equations are equivalent to the full source equations
\begin{equation}
\begin{aligned}
 (r^\lambda_{AB}\partial_A\partial_B+r^\lambda_A\partial_A+r^\lambda_0)Z_{\rm ens}[J]&=0,\\
 (\ell^\lambda_A\partial_A+\ell^\lambda_0)Z_{\rm ens}[J]&=0
 \qquad\text{for all }J.
\end{aligned}
\label{eq:CFTsourceequations}
\end{equation}
For the homogeneous sphere equations below, choose a channel projection that removes the fixed vacuum and light exchanges while retaining a heavy exchange. Appendix~\ref{app:lightprojection} gives an explicit construction and the condition for this surviving coefficient to be nonzero.

For a heavy-sector projection satisfying the homogeneous equations, the torus relation gives $\ell^\lambda_A B_{AB}=0$ and corresponding higher-moment relations.  Dividing the homogeneous quadratic source equation by $Z_{\rm ens}$ and using $\partial_A\partial_B Z_{\rm ens}/Z_{\rm ens}=W_{AB}+W_AW_B$ gives
\begin{equation}
r^\lambda_{AB}\big(W_{AB}+W_AW_B\big)=0 .
\label{eq:Wcrossing}
\end{equation}
Subscripts on $W$ denote source derivatives.  Writing $Z_{\rm ens}=e^W$ separates connected cumulants $W_{AB}$ from the products $W_AW_B$ generated by lower moments.

Substituting the cumulant expansion \eqref{eq:s4cumulantexpansion} into \eqref{eq:Wcrossing} and comparing the orders $J^0,J^2,J^4$ yields
\begin{equation}
\begin{aligned}
r^\lambda_{AB}B_{AB}&=0,\\
 r^\lambda_{AB}\big(\kappa_{ABCD}+B_{AC}B_{BD}\big)
 &=-r^\lambda_{AB}B_{AD}B_{BC},\\
 r^\lambda_{AB}\Big[\kappa_{ABCDEF}
 &+B_{AC}\kappa_{BDEF}+B_{AD}\kappa_{BCEF}+B_{AE}\kappa_{BCDF}+B_{AF}\kappa_{BCDE}\\
 &+B_{BC}\kappa_{ADEF}+B_{BD}\kappa_{ACEF}\\
 &+B_{BE}\kappa_{ACDF}+B_{BF}\kappa_{ACDE}\Big]=0.
\end{aligned}
\label{eq:sourcehierarchy}
\end{equation}
The first line is the homogeneous heavy-sector crossing equation with no extra OPE insertion.  For the second line, insert \eqref{eq:s4connectedfour} into the homogeneous test $r^\lambda_{AB}\Eds[C_AC_BC_CC_D]=0$.  The term $r^\lambda_{AB}B_{AB}B_{CD}$ drops out by the first line, leaving the two cross-pairings and the fourth cumulant.  Within this projected sector, this is the algebraic form of the mixed-channel comparison, in which a connected tetrahedral term compensates the change in Gaussian pairings.  {In the third line, only the contraction $r^\lambda_{AB}\kappa_{ABCDEF}$ of the connected sixth moment is required to cancel the eight displayed products of second and connected fourth moments in the chosen heavy-sector test. This projected equation does not determine the entire sixth cumulant, or the weights of all fillings, uniquely.}

{Appendix~\ref{sec:sixth} solves the third line of \eqref{eq:sourcehierarchy} for the prism joining-edge component.  The tree contraction $\kappa^{\rm tet}B_\theta^{-1}\kappa^{\rm tet}$ of two tetrahedral cumulants is the prism, because removing the sewn vertex from each tetrahedron leaves one of its two triangles, figure~\ref{fig:prismmoves}(a,b).  The fusion move on a joining edge opens both triangles into quadrilaterals and produces $K_{3,3}$, figure~\ref{fig:prismmoves}(c).  This graph contains no triangle and receives no tree contribution, so the independently evaluated $K_{3,3}$ coefficient cancels the joining-edge residual, \eqref{eq:localprismcompletion}, and the channel difference fixes the relative filling weights, \eqref{eq:localweightconstraint}.  The two-collar extraction of appendix~\ref{ssec:fiveboundary} realizes both coefficients on the boundary through \eqref{eq:fivenormalized}, and the internal edge of $K_{3,3}$ has the zero-deficit window contribution \eqref{eq:k33_closed_prefactor_v46}.  With $Z_{\rm pr}$ and $Z_{K_{3,3}}$ the two-copy amplitudes of \eqref{eq:localprismtransport}, the sixth-order results are}
\begin{equation}
\boxed{\begin{gathered}
 (\mathbb F_e^{(\theta)}\otimes\mathbb F_e^{(\theta),\vee})\,Z_{\rm pr}=Z_{K_{3,3}},\qquad
 w_K=w_{\rm pr}=\frac{w_{\rm tet}^{\,2}}{w_\theta},\\
 \left.\Eds\Big[\prod_{v\in V(\Gamma)}C_v\Big]_c\right|_{(S^3,\Gamma)}
 =Z_{\rm Vir}(S^3,\Gamma)\,Z_{\rm Vir}^{\vee}(S^3,\Gamma),\qquad \Gamma=\Gamma_{\rm pr},\ K_{3,3}.
\end{gathered}}
\label{eq:sixthsummary}
\end{equation}

\begin{figure}[!htbp]
\centering
\begin{tikzpicture}[x=1cm,y=1cm,line cap=round,line join=round,
 >=Stealth,font=\small,
 vlab/.style={font=\small,inner sep=2pt,fill=white},
 elab/.style={font=\scriptsize,inner sep=1.5pt,fill=white}]
\node[paneltitle,anchor=west] at(0,2.05){(a) Sew the two tetrahedral cumulants};
\begin{scope}[shift={(7,0)}]
 \coordinate(A)at(-3.1,1.15);\coordinate(D)at(-3.1,-1.15);
 \coordinate(B)at(-1.80,0);\coordinate(G)at(-.65,0);
 \coordinate(F)at(3.1,1.15);\coordinate(E)at(3.1,-1.15);
 \coordinate(C)at(1.80,0);\coordinate(H)at(.65,0);
 \fill[faceblue] (A)--(B)--(D)--cycle;
 \fill[caporange!55] (F)--(C)--(E)--cycle;
 \draw[black!75] (A)--(D)--(B) (C)--(F) (C)--(E);
 \draw[pt,line width=1pt] (A)--(B) (B)--(G);
 \draw[gs,line width=1pt] (F)--(E) (D)--(G);
 \draw[black!75] (A)--(G)node[elab,pos=.55,above]{$2$}
  (F)--(H)node[elab,pos=.55,above]{$2$};
 \draw[pt,line width=1pt] (C)--(H);
 \draw[gs,line width=1pt] (E)--(H);
 \node[elab,above] at(-1.22,0){$b$};
 \node[elab,below] at(-1.80,-.60){$x$};
 \node[elab,above] at(1.22,0){$b$};
 \node[elab,below] at(1.80,-.60){$x$};
 \draw[black!65,densely dotted,line width=.9pt] (G)--(H);
 \node[above=7pt] at(0,0){$B_\theta^{-1}$};
 \foreach \v in {A,B,C,D,E,F}{\fill(\v)circle(1.8pt);}
 \node[vlab,above left] at(A){$A$};\node[vlab,below left] at(D){$D$};
 \node[vlab,left=3pt] at(B){$B$};
 \node[vlab,above right] at(F){$F$};\node[vlab,below right] at(E){$E$};
 \node[vlab,right=3pt] at(C){$C$};
 \foreach \v in {G,H}{\draw[fill=white] (\v)circle(2pt);}
 \node[vlab,below=4pt] at(G){$G$};\node[vlab,below=4pt] at(H){$H$};
\end{scope}
\node at(7,-1.85){$\kappa^{\rm tet}_{ABDG}(B_\theta^{-1})^{GH}
                         \kappa^{\rm tet}_{HCEF}=Z_{\rm pr}$};
\node[paneltitle,anchor=west] at(0,-2.70){(b) Prism};
\node[paneltitle,anchor=west] at(9.15,-2.70){(c) Crossed graph $K_{3,3}$};
\begin{scope}[shift={(2.70,-4.35)}]
 \coordinate(A)at(-1.60,1.05);\coordinate(D)at(-1.60,-1.05);
 \coordinate(B)at(-.65,0);\coordinate(F)at(1.60,1.05);
 \coordinate(E)at(1.60,-1.05);\coordinate(C)at(.65,0);
 \fill[faceblue] (A)--(B)--(D)--cycle;
 \fill[caporange!55] (F)--(C)--(E)--cycle;
 \draw[black!75] (A)--(D)--(B) (C)--(F);
 \draw[black!75] (C)--(E) (A)--(F)node[elab,midway,above]{$2$};
 \draw[pt,line width=1pt] (A)--(B) (B)--(C)node[elab,midway,above]{$b$};
 \draw[gs,line width=1pt] (F)--(E);
 \draw[gs,line width=1.5pt] (D)--(E)node[elab,midway,below]{$x$};
 \foreach \v in {A,B,C,D,E,F}{\fill(\v)circle(1.8pt);}
 \node[vlab,above left] at(A){$A$};\node[vlab,below left] at(D){$D$};
 \node[vlab,left=3pt] at(B){$B$};\node[vlab,right=3pt] at(C){$C$};
 \node[vlab,above right] at(F){$F$};\node[vlab,below right] at(E){$E$};
 \node at(0,-1.75){$Z_{\rm pr}$};
\end{scope}
\draw[->,line width=.8pt] (5.25,-4.35)--(8.40,-4.35);
\node[align=center] at(6.82,-3.80){$\mathbb F_x\otimes\mathbb F_x^\vee$};
\begin{scope}[shift={(11.20,-4.35)}]
 \coordinate(A)at(-1.65,1.05);\coordinate(F)at(1.65,1.05);
 \coordinate(B)at(-.85,0);\coordinate(C)at(.85,0);
 \coordinate(Dp)at(-.70,-1.10);\coordinate(Ep)at(.70,-1.10);
 \draw[pt,line width=1pt] (A)--(B) (B)--(C);
 \draw[black!75] (A)--(F) (C)--(F) (A)--(Dp);
 \draw[black!75] (B)--(Ep);
 \draw[white,line width=3.5pt] (C)--(Dp);
 \draw[black!75] (C)--(Dp);
 \draw[gs,line width=1pt] (F)--(Ep);
 \draw[copper,line width=1.5pt] (Dp)--(Ep)
  node[elab,midway,below=2pt]{$P_t$};
 \foreach \v in {A,B,C,F,Dp,Ep}{\fill(\v)circle(1.8pt);}
 \node[vlab,above left] at(A){$A$};\node[vlab,above right] at(F){$F$};
 \node[vlab,left=3pt] at(B){$B$};\node[vlab,right=3pt] at(C){$C$};
 \node[vlab,below left] at(Dp){$D'$};\node[vlab,below right] at(Ep){$E'$};
 \node at(0,-1.75){$Z_{K_{3,3}}$};
\end{scope}
\end{tikzpicture}
\caption{{Sixth-order sewing on a joining edge.  (a) Contracting the vertices $G,H$ with $B_\theta^{-1}$ joins the three equally labelled legs.  (b) The remaining triangles $ABD$ and $CEF$ form the prism.  (c) The fusion move on $x$ gives $K_{3,3}$; the crossing without a dot is not a vertex.  This graph has no triangle, so its coefficient is supplied by the sixth-order completion, \eqref{eq:localprismcompletion}.}}
\label{fig:prismmoves}
\end{figure}

The low-order checks use the AdS fusion and modular identities \cite{Hartman,CTV}, analytically continued with the same contour and residue prescriptions as the coefficients in \eqref{eq:bootstrapagreement}.  Other moment components involve additional contraction patterns and fillings.

{The sphere calculation above concerns the quadratic crossing residual of four-point functions. It constrains the hierarchy of moments, but it does not test the independent modular relation of a torus one-point function, where two legs of the same three-point vertex are identified. That identification produces the repeated-label coefficients $C_{iis}$ appearing in the handcuff contribution of \eqref{eq:bootstrapagreement}. We therefore turn from the sphere hierarchy to the punctured-torus handle. The purpose is to test how the handcuff filling pairs with its modular $S$ image and to show how a local handle relation persists when extra graph insertions are attached outside the handle.}

\paragraph{Repeated labels and the handle equation.}
The one-point function on a torus contains a three-point coefficient $c_{iis}$ because two operator legs are sewn together.  Exchanging the torus cycles relates its repeated-label correlations under the punctured modular kernel.  We cut the compact filling so that the handle can be reglued without changing the graph insertions in an attached three-ball.  The modular transformation then acts on the handle, while the state prepared by those insertions is held fixed.  We first calculate a triangle in that ball, then express the same sewing operation for an arbitrary spectator state \cite{UniversalTopologies}\cite{CrossingOPE}.

Work at $b>0$ with independent real left- and right-moving momenta. {The continuous two-copy measure used for this kernel calculation is an auxiliary real-domain sewing measure. It is not already the measure of an integer-spin spectrum. The condition $J_I=P_i^2-\bar P_i^2\in\mathbb Z$ is imposed separately through the modular $T$ constraint and its projection.} The non-vacuum primary $s$ has integer spin $J_s=h_s-\bar h_s$. Write $I=(P_i,\bar P_i)$, $K=(P_k,\bar P_k)$, and
\begin{equation}
 \dd\nu(I)=\rho_0(P_i)\rho_0(\bar P_i)\dd P_i\dd\bar P_i,
 \qquad
 D(I,K)=\frac{\delta(P_i-P_k)\delta(\bar P_i-\bar P_k)}
 {\rho_0(P_k)\rho_0(\bar P_k)}.
 \label{eq:repeatedpairing}
\end{equation}
Thus $D$ acts as the identity on a test function $f$ with respect to the sewing measure,
\[
 \int\dd\nu(K)\,D(I,K)f(K)=f(I).
\]
For each marked handle filling $\gamma$, let $B_\gamma(I,K;s)$ be its contribution to the second OPE moment $B_{AB}$ for $A=(iis)$ and $B=(kks)$.  The subscripts $I,K$ carry the left- and right-moving momentum labels of the repeated indices.  The seed has two parallel labelled strands in $S^2\times S^1$; cutting the two-punctured sphere gives $D(I,K)$.  Its two Wick contractions differ by exchanging the identical legs and have relative sign $(-1)^{J_s}$, hence total {spin-weighted attachment factor} $\nu_s=1+(-1)^{J_s}$. {For a scalar $s$, both contractions add and $\nu_s=2$, reproducing the handcuff count $\nu_H$ above. For odd spin they cancel and $\nu_s=0$; this sign-weighted sum is not an extra factor to multiply onto the scalar $\nu_H=2$. Neither $\nu_H$ nor $\nu_s$ imposes integer spin. The latter depends on that separate condition.}

Now replace the $(k,k,s)$ vertex inside a three-ball by the triangle with vertices $(k,y,q)$, $(k,z,q)$, and $(s,y,z)$.  The Wilson-triangle identity gives
\begin{equation}
\begin{aligned}
 R_k&=\begin{Bmatrix}P_k&P_k&P_s\\P_z&P_y&P_q\end{Bmatrix}_{\rm Vir},\\
 \mathcal G_\triangle^\gamma(I,K)
 &:=\left.\Eds[C_{iis}C_{kyq}C_{kzq}C_{syz}]\right|_\gamma
   =R_kR_{\bar k}^{\vee}B_\gamma(I,K;s).
\end{aligned}
\label{eq:repeatedtriangle}
\end{equation}
The second factor uses all right-moving labels.  The normalization factors cancel as shown in \eqref{eq:repeatedtriangleproof}.  Changing the handle gluing leaves this spectator ball unchanged, so the same factor multiplies every handle filling.

{For this concrete triangle attachment, the modular transformation acts only on the handle label $I$ and leaves the local factor $R_kR_{\bar k}^{\vee}$ fixed.  Let $U_S$ be the punctured-torus modular operator of \eqref{eq:repeatedoperators}; it obeys $U_S^2=(-1)^{J_s}$ on the real starting contour.  Summing the identity seed $D$ and its $S$ image with the already counted spin-weighted multiplicity $\nu_s$ gives}
\begin{equation}
\begin{aligned}
 B_{\{1,S\}}&=\nu_s(1+U_S)D,\qquad (1-U_S)B_{\{1,S\}}=\nu_s(1-U_S^2)D=0,\\
 (1-U_S)\mathcal G_\triangle^{\{1,S\}}(I,K)
 &=R_kR_{\bar k}^{\vee}\,(1-U_S)B_{\{1,S\}}(I,K;s)=0.
\end{aligned}
\label{eq:repeatedfiniteS}
\end{equation}
{For odd $J_s$, $\nu_s=0$; for even $J_s$, $U_S^2=1$.  The second line uses the triangle factorization \eqref{eq:repeatedtriangle}, and shows directly that this specific fourth-moment contribution inherits the handle's finite $S$ closure.  More generally, the same argument works for a spectator whose attachment is unchanged by the handle move.  Appendix~\ref{ssec:repeatedproof}, \eqref{eq:repeatedreduction}, defines that spectator sewing and states its conditions;}
the sphere fusion moves are the separate calculation of appendix~\ref{ssec:heegaard}.
{The finite pair does not solve the independent modular $T$ equation.}

{The linked-handcuff coefficient is explicit because the real-$b$ Hopf-link evaluation is $Z_S^H=\widehat{\mathbb S}_{ik}[s]/\sqrt{C_0(i,i,s)C_0(k,k,s)}$, as recalled in \eqref{eq:repeatedHopfproof}. Normalizing the two OPE vertices removes these $C_0$ factors. For scalar labels the two inequivalent flare attachments give $\nu_s=2$, and the reversed chiral framing contributes $e^{-\ii\pi h_s}$. The local Wilson-triangle identity \eqref{eq:repeatedtriangleproof} then multiplies the chiral Hopf result by $R_k$ and its reversed copy by $R_{\bar k}^{\vee}$. On the analytically continued scalar slice their product is $R_k^2$. Thus the following coefficients are evaluated in terms of known Virasoro kernels, rather than left as unevaluated topology sums \cite{UniversalTopologies},}
\begin{equation}
\begin{aligned}
 B_H^{\rm cont}(i,k;s)&=2e^{-\ii\pi h_s}
                    \widehat{\mathbb S}_{P_iP_k}[P_s]^2,\\
 \mathcal G_{\triangle,H}^{\rm cont}(i;k,s,y,z,q)
   &=2e^{-\ii\pi h_s}\widehat{\mathbb S}_{P_iP_k}[P_s]^2R_k^2.
\end{aligned}
\label{eq:repeatedHopf}
\end{equation}
These are contributions to the raw moments of the normalized coefficients $C_A$; connected subtraction is given by \eqref{eq:s4connectedfour}.  {The calculation is explicit at the level of the meromorphic $\widehat{\mathbb S}$ and Virasoro $6j$ kernels and their normalized product; it relies on the previously established Hopf-link and Wilson-triangle evaluations, not a new direct bulk integral for each set of labels.} Only the smooth Hopf and triangle coefficients have been analytically continued here.  The identity distribution $D$ is still defined on the real reference contours.  The scalar boundary state {of section~\ref{sec:boundaryamp}} uses $P\in e^{\ii\pi/4}\mathbb R_+$; spinning sectors instead require a joint analytic continuation satisfying \eqref{eq:fixedspintransport}.  The $T$ constraint and transport of the marked modular orbit are described in appendix~\ref{ssec:repeatedproof} and discussed further in section~\ref{sec:discussion}.
\FloatBarrier

\section[A random Euclidean CFT interpretation of dS\texorpdfstring{$_3$}{3} gravity]{A random Euclidean CFT interpretation of dS$_3$ gravity}\label{sec:physicalinterpretation}\label{sec:matrixsources}

The preceding sections extract a coefficient from the cosmological state, identify it with a compact-filling amplitude and assign it to an OPE moment. This section assembles these results into a boundary description. We show why the boundary theory is a Euclidean CFT, what its OPE statistics measure in the Lorentzian spacetime, and how the statistical dictionary meets the vacuum condition of \cite{CEM}.

\subsection[Euclidean conformal data at future infinity]{Euclidean conformal data at future infinity}\label{ssec:euclideanboundary}

The Euclidean signature follows directly from the Lorentzian solution. Multiply the pants and wedge metrics of section~\ref{ssec:inverseLorentzian} by the conformal factor $4e^{-2t}\ell_{\rm dS}^{-2}$. They become
\begin{equation}
 \frac{4e^{-2t}}{\ell_{\rm dS}^2}\,\dd s^2
 =-4\big(\dd e^{-t}\big)^2+
 \begin{cases}
 (1-e^{-2t})^2h_{\Pi_i},&\text{pants},\\
 (1+e^{-2t})^2\dd\varphi^2+(1-e^{-2t})^2\dd x^2,&\text{wedges}.
 \end{cases}
\label{eq:s6conformalmetric}
\end{equation}
The normal coordinate $e^{-t}$ vanishes at future infinity. Pulling back to that surface removes its timelike differential and gives
\begin{equation}
\boxed{\left.\lim_{t\to\infty}\frac{4e^{-2t}}{\ell_{\rm dS}^2}\dd s^2
       \right|_{\dd(e^{-t})=0}
 =\begin{cases}
 h_{\Pi_i},&\text{pants},\\
 \dd\varphi^2+\dd x^2,&\text{grafted bands}.
 \end{cases}}
\label{eq:s6inducedmetric}
\end{equation}
Both pieces are positive definite and join to the \vnew{conformal boundary of this $K_4$ geometry}. The same pull-back for the reference-state metric \eqref{eq:cosmology_metric_v46} gives the Euclidean metric $h$ on $\Sigma_{g,n}$. Locally, future infinity is the sphere of null directions in $\mathbb R^{3,1}$. A complex coordinate $z$ on this sphere describes two real spatial directions, with $\bar z=z^*$, and the isometry group $SO^+(3,1)\simeq PSL(2,\mathbb C)$ acts on it by M\"obius transformations \cite{Strominger,Mess,BenedettiBonsante}.

The gravitational constraints supply the conformal transformation law. Linearizing the Weyl identity \eqref{eq:s61stateconstraints} gives \cite{CEM}
\begin{equation}
 \delta_\sigma\Psi[h]
 =\left[\frac{c}{24\pi}\int\dd^2x\sqrt h\,R_h\sigma
              -\sum_i\Delta_i\sigma(x_i)\right]\Psi[h].
\label{eq:s6weylvariation}
\end{equation}
The two terms are the conformal anomaly and the Weyl weights of the punctures. For constant $\sigma$ on an unpunctured surface, $\delta_\sigma\log\Psi=c\chi\sigma/6$, whose semiclassical sign agrees with the Lorentzian end phase in \eqref{eq:lorentzAnomalyBranch}. Together with diffeomorphism invariance, \eqref{eq:s6weylvariation} gives the conformal Ward identities of two Virasoro algebras with the complex charge $c$ of \eqref{eq:physicalc}, distinct from the real asymptotic charge $c_{\rm asym}$ \cite{Strominger,CEM}. In a single OPE term,
\begin{equation}
 z^{h_k-h_i-h_j}\bar z^{\bar h_k-\bar h_i-\bar h_j}
 =|z|^{\Delta_k-\Delta_i-\Delta_j}
   e^{\ii(J_k-J_i-J_j)\arg z},
 \qquad \Delta=h+\bar h,\quad J=h-\bar h,
\label{eq:s6euclideanope}
\end{equation}
the bar on $\bar h$ denotes the right-moving weight and $\bar z$ is the conjugate coordinate. Integer spins make the angular factor single valued, while complex dimensions produce oscillations in $\log|z|$ on Euclidean coordinates. \nw{The proposed boundary description is therefore Euclidean and nonunitary.}

The random variables are the normalized nonidentity OPE coefficients $C_A$ of \eqref{eq:s5variables}. The central charge, labelled spectrum, identity and two-point normalization are fixed. Each correlator therefore has fixed conformal blocks and fluctuating coefficients, and the Ward identities act separately on each correlator in a product. The moments of the $C_A$ are those assigned in \eqref{eq:inversedictionary} and tested by statistical sewing in section~\ref{sec:bulkinverse}. \nw{The paired correlation obeys an independent conformal Ward identity in each of its two boundary arguments; simultaneous Weyl variation therefore acts on two factors, not on one Liouville wavefunction.}

\subsection[OPE correlations and Lorentzian geometry]{OPE correlations and Lorentzian geometry}\label{ssec:randomEuclidean}

Two four-point functions of the same random CFT are correlated through their shared OPE coefficients. Their paired covariance is the square $\Psi_{0,4}^{(b)}(z,\bar z)^2$ of the cosmological state, \eqref{eq:statisticalstatepair}. Its crossed channel isolates the connected tetrahedral moment, whose normalization is fixed by the extraction \eqref{eq:mainratio} and whose value is the compact amplitude of the $K_4$ core, \eqref{eq:bootstrapagreement}. Together with the theta normalization, this moment carries the Lorentzian content of the filling. \hnew{Its phase derivative measures the Lorentzian boost conjugacy class associated with the filling.} Table~\ref{tab:s6dictionary} collects how each of its features is realized in the spacetime $\mathcal D_M$ of section~\ref{ssec:inverseLorentzian}.

\begin{table}[H]
\centering
\begingroup\small
\renewcommand{\arraystretch}{1.25}
\begin{tabular}{@{}p{.40\linewidth}p{.45\linewidth}l@{}}
\toprule
Euclidean data at future infinity & Lorentzian spacetime $\mathcal D_M$ & \\
\midrule
Label $P_e=e^{\ii\pi/4}k_e$ of a shared primary & Arc of the initial $K_4$ graph, of length $\pi-\psi_e$ with $\psi_e=2\pi r|P_e|$ & \eqref{eq:lorentzInitialFrame}\\
Modulus of the tetrahedral moment & Face Gram determinant of the cell, $1/(4\sqrt{-\det G})$ & \eqref{eq:MY}\\
Phase of the tetrahedral moment & \nw{Volume-dependent phase $2V_A/(\pi r^2)$ of the continued core, reached from $\mathcal D_M$ at $t=\ii\pi/2$} & \eqref{eq:lorentzBox}\\
Angle derivative of the phase & Boost rapidity $\lambda_e=\ell(\gamma_e)$ around the arc & \eqref{eq:lorentzInverseQuantumCheck}\\
Label resolution $\pi r^2/\ell_e$ & \nw{Angular resolution of an order-one phase change} & \eqref{eq:angleresolution}\\
Weyl anomaly $c\chi\sigma/6$ & Cutoff term of the Lorentzian action & \eqref{eq:lorentzAnomalyBranch}\\
Prism and $K_{3,3}$ sixth moments & \nw{Zero-deficit internal-edge gluing for the selected $K_{3,3}$ saddle} & \eqref{eq:oneedge3nj}\\
\bottomrule
\end{tabular}
\endgroup
\caption{The tetrahedral OPE statistics and the Lorentzian spacetime of the same compact core.}
\label{tab:s6dictionary}
\end{table}

The central entry is the rapidity. At fixed hyperideal angles the complex-$b$ asymptotic \eqref{eq:MY} gives the tetrahedral moment up to a label-independent phase, and the Schl\"afli relation \eqref{eq:Schlafli}, $\partial_{\psi_e}V_A=-\ell_e/2$, turns the angle derivative of its phase into an edge length. With the one-derivative uniformity assumed below \eqref{eq:relativegeometry},
\begin{equation}
\boxed{\begin{aligned}
 \left.\Eds[C_{123}C_{156}C_{246}C_{345}]_c\right|_{\rm tet}
 &\simeq\frac{e^{2\ii V_A(\psi)/(\pi r^2)}}{4\sqrt{-\det G}},\\
 -2\pi r^2\,\partial_{\psi_e}
 \arg\Big(\left.\Eds[C_{123}C_{156}C_{246}C_{345}]_c\right|_{\rm tet}\Big)
 &=\hnew{-4\,\partial_{\psi_e}V_A+O(r^2)}
 \longrightarrow\ell(\gamma_e)=2\ell_e .
\end{aligned}}
\label{eq:s6statisticalgeometry}\eqalias{eq:cumulantholonomy}
\end{equation}
Here $V_A$ is the volume of one tetrahedral cell and $G$ its face Gram matrix. The length $\ell(\gamma_e)$ is the rapidity of the boost around the arc $e$ of the initial singularity, \eqref{eq:lorentzBoost}, so this is \eqref{eq:lorentzInverseQuantumCheck} written as a statement about Euclidean OPE statistics. \hnew{Equivalently, the phase derivative fixes the frame-independent conjugacy invariant $|\operatorname{tr}H_{\gamma_e}|=2\cosh(\ell(\gamma_e)/2)>2$, as calculated from the Lorentzian identification in \eqref{eq:lorentzBoostTrace}.}

The two members of each angle--length pair therefore enter the statistics in different ways. The angle is a label. A hyperideal polyhedron is determined by its dihedral angles \cite{BaoBonahon,SchlenkerHyperideal}, \vtwo{so the six labels fix the tetrahedral cell. The marked seams then specify the doubled core recovered from $\mathcal D_M$.} The length is a phase derivative and measures a holonomy of that spacetime. Reading it requires OPE data resolved in angle at the scale $\pi r^2/\ell_e$. A Gaussian average of width $\lambda r^2$ multiplies the moment by $\exp[-\ell_e^2\lambda^2/(2\pi^2)]$, \eqref{eq:angleresolution}.

\subsection[Vacuum compatibility of the collective correlation]{Vacuum compatibility of the collective correlation}\label{ssec:vacuumcompletion}\label{ssec:positiveensembles}

When allowed by the external labels, vacuum exchange has a fixed coefficient across the CFT ensemble. We first check its cancellation from centered correlations and then examine the Liouville continuum in a degenerating cylinder. The relevant normalizability criterion for a cosmological wavefunction is \cite{CEM}
\begin{equation}
 c_{\rm eff}=c-12\Delta_{\min},\qquad
 \Re c_{\rm eff}>1\ \Longrightarrow\ \|\Psi\|^2=\infty,
\label{eq:s6ceff}
\end{equation}
where $\Delta_{\min}$ is the lowest scaling dimension in the normalizable spectrum. Liouville theory has $\Re c_{\rm eff}=1$ because its identity is absent from that spectrum. By contrast, an individual compact CFT includes the identity with $\Delta_{\min}=0$, so $c_{\rm eff}=c$ and $\Re c_{\rm eff}=13$ on the continued branch \cite{CLM,CEM}.

For a four-point configuration $a$, separate the exchanges with fixed OPE coefficients from the fluctuating ones,
\begin{equation}
\begin{aligned}
 G_a^{\rm full}[C]&=G_a^{\rm fixed}+T_a^{AB}C_AC_B,\\
 \delta G_a[C]&=G_a^{\rm full}[C]-\Eds\big[G_a^{\rm full}\big]
 =T_a^{AB}(C_AC_B-B_{AB}),
 \qquad B_{AB}=\Eds[C_AC_B].
\end{aligned}
\label{eq:vacuumcenteredcorrelator}
\end{equation}
The kernel $T_a^{AB}$ contains the conformal blocks and the $C_0$ normalization, and $G_a^{\rm fixed}$ contains every exchange whose coefficient is fixed. In a chosen channel the vacuum family contributes
\begin{equation}
 G_a^{\rm vac}
 =c_{12\mathbf1}c_{34\mathbf1}\,
   \mathcal F_{\mathbf1}^{(s)}(z)\mathcal F_{\mathbf1}^{(s)}(\bar z),
 \qquad G_a^{\rm vac}-\Eds\big[G_a^{\rm vac}\big]=0 .
\label{eq:s6vacuumfamily}
\end{equation}
Its prefactor is fixed by the two-point normalization and its descendants by the Virasoro Ward identities, so the entire block product is the same in every sample. The vacuum therefore remains in each sample correlator and cancels from every fluctuation, together with all of its descendants. The covariance of two fluctuations contains the connected contractions,
\begin{equation}
\begin{aligned}
 \Eds[\delta G_a\delta G_b]
 &=T_a^{AB}T_b^{CD}
   \big(\Eds[C_AC_BC_CC_D]-B_{AB}B_{CD}\big)\\
 &=T_a^{AB}T_b^{CD}
   \big(\kappa_{ABCD}+B_{AC}B_{BD}+B_{AD}B_{BC}\big),
\end{aligned}
\label{eq:s6centeredcovariance}
\end{equation}
where the fourth moment has been split into its cumulant and three pairings. The pairing within one correlator cancels. The pairings across two correlators give \eqref{eq:statisticalstatepair}, with two Liouville cylinder factors; the cumulant contains the tetrahedral filling of section~\ref{ssec:randomEuclidean}. \nw{This algebraic cancellation does not determine the spectrum of either cylinder. Its Liouville resolution is the independent input of section~\ref{sec:boundaryamp}.}

Crossing survives the subtraction. The averaged residual $\Eds[R_\lambda]$ of \eqref{eq:residualdef} retains the fixed vacuum in $r^\lambda_0$, and subtracting it from $R_\lambda$ removes that term,
\begin{equation}
 R_\lambda-\Eds[R_\lambda]
   =r^\lambda_{AB}(C_AC_B-B_{AB})+r^\lambda_A\big(C_A-\Eds[C_A]\big).
\label{eq:centeredfullcrossing}
\end{equation}
Multiplying by $C_CC_D-B_{CD}$ and averaging with the vanishing odd moments of section~\ref{ssec:ensemblesample} gives
\begin{equation}
 r^\lambda_{AB}\big(\kappa_{ABCD}+B_{AC}B_{BD}+B_{AD}B_{BC}\big)=0 ,
\label{eq:centeredfourthcrossing}\eqalias{eq:s6fourthsewing}
\end{equation}
the fourth-order relation of \eqref{eq:Wcrossing} with the fixed terms subtracted. \nw{The uncentered mean equation $r^\lambda_{AB}B_{AB}+r^\lambda_0=0$ (for zero odd moments) remains an independent crossing condition: subtracting a common vacuum block does not impose it.}

After the fixed terms cancel in \eqref{eq:s6vacuumfamily}, the paired term of \eqref{eq:s6centeredcovariance} is resolved into two Liouville cylinder factors. Their propagating scalar momenta lie on the dS ray $P=e^{\ii\pi/4}k$, with $k\geq0$. Using $Q^2=2+\ii(r^2-r^{-2})$ and $P^2=\ii k^2$ in $c=1+6Q^2$ and $\Delta=Q^2/2+2P^2$ gives
\begin{equation}
\begin{aligned}
 c&=13+6\ii(r^2-r^{-2}),\\
 \Delta(k)&=1+\ii\left(2k^2+\frac{r^2-r^{-2}}2\right),\\
 c-12\Delta(0)&=1.
\end{aligned}
\label{eq:vacuumthreshold}
\end{equation}
Thus the continuum starts at $\Re\Delta=1$ and each Liouville cylinder has $c_{\rm eff}=1$. To evaluate its contribution near a degenerating collar, write $q=e^{\ii\pi(u+\ii y)}$ with $y>0$. The long-collar limit is $y\to\infty$.

We restrict $k$ to a compact interval containing $k=0$ and avoiding poles. After extracting the universal conformal-block prefactors, $F(k)$ denotes the remaining smooth channel coefficient, including an even spectral cutoff. In particular, $F(k)=F(0)+O(k^2)$ near the threshold. The reduced cylinder contribution is
\begin{equation}
\begin{aligned}
 I(y)&=e^{\ii\pi/4}\int_0^\infty\dd k\,
       \rho_0(e^{\ii\pi/4}k)F(k)e^{-2\pi\ii yk^2},\\
 \rho_0(e^{\ii\pi/4}k)
 &=4\sqrt2\,\ii\sin(2\pi rk)\sinh(2\pi k/r)
   =16\sqrt2\,\ii\pi^2k^2+O(k^4).
\end{aligned}
\label{eq:randomcollarintegral}
\end{equation}
The factor $e^{\ii\pi/4}$ comes from $\dd P=e^{\ii\pi/4}\dd k$. The double zero of $\rho_0$ determines the long-collar decay. With a positive regulator $\epsilon$, the leading integral is
\begin{equation}
\begin{aligned}
 \int_0^\infty\dd k\,k^2e^{-2\pi\ii yk^2}
 &=\lim_{\epsilon\downarrow0}
   \frac{\sqrt\pi}{4(\epsilon+2\pi\ii y)^{3/2}}
 =\frac{\sqrt\pi e^{-3\pi\ii/4}}{4(2\pi y)^{3/2}},\\
 I(y)&=2\pi F(0)y^{-3/2}+O(y^{-5/2}).
\end{aligned}
\label{eq:randomcollartail}
\end{equation}

\Needspace{11\baselineskip}\noindent Let $Y>0$ be the lower limit of the long-collar region $y\geq Y$. Integrating the large-$y$ behavior gives the first line below. For the second, set $x=k^2$ and apply Plancherel to the Fourier transform in $y$.
\begin{equation}
\begin{aligned}
 \int_Y^\infty\dd y\,|I(y)|^2
 &=\frac{2\pi^2|F(0)|^2}{Y^2}+O(Y^{-3})
 \qquad(Y\to\infty),\\
 \int_{-\infty}^\infty\dd y\,|I(y)|^2
 &=\int_0^\infty\dd x\,
 \left|\frac{\rho_0(e^{\ii\pi/4}\sqrt{x})F(\sqrt{x})}{2\sqrt{x}}\right|^2\\
 &=\frac12\int_0^\infty\frac{\dd k}{k}
 \big|\rho_0(e^{\ii\pi/4}k)F(k)\big|^2<\infty.
\end{aligned}
\label{eq:randomcollarnorm}
\end{equation}
The last integrand behaves as $k^3$ at $k=0$ and has compact support, so the integral is finite.

At mirrored moduli, the paired contribution contains two such cylinder factors. With the same localization, its reduced amplitude is $I_{\rm pair}(y)=I(y)^2$. It follows directly that
\begin{equation}
\begin{aligned}
 I_{\rm pair}(y)&=4\pi^2F(0)^2y^{-3}+O(y^{-4}),\\
 \int_Y^\infty\dd y\,|I_{\rm pair}(y)|^2
 &=\frac{16\pi^4|F(0)|^4}{5Y^5}+O(Y^{-6}).
\end{aligned}
\label{eq:s6pairedcollar}
\end{equation}
The connected correlation therefore has an integrable long-collar contribution after spectral localization. The cancellation of its fixed vacuum family and the decay of the Liouville continuum follow from different parts of the calculation.

\subsection{A finite Virasoro OPE ensemble}\label{ssec:positivetensor}

The theta filling fixes a unit second moment and the tetrahedral filling gives the Virasoro $6j$ bilinear $K_{pq}$. These data do not fix individual OPE coefficients or their joint distribution. We construct a positive probability measure $\mu$ on the normalized OPE variables $C_A$ of \eqref{eq:s5variables}. A realization drawn from $\mu$ assigns values to all these coefficients together. The aim is to reproduce the theta and tetrahedral moments while satisfying the retained crossing equations for every realization. We fix four external primaries and retain finitely many intermediate labels $p$ and $q$ in the $s$ and $t$ channels, with generic distinct OPE triples.

\paragraph{From the two OPE channels to the crossing equations.}
Define the products of normalized OPE coefficients in the two channels,
\begin{equation}
 A_p=C_{12p}C_{34p},\qquad B_q=C_{14q}C_{23q}.
\label{eq:s6channelFactors}
\end{equation}
The labels in these products are fixed, but the values of the OPE coefficients remain random. In particular, $\mathbb E_\mu[A_pB_q]$ is unknown until the measure $\mu$ is constructed. The conformal blocks and the Virasoro fusion kernel, in contrast, are fixed by the labels.

The normalization factors multiplying the products in an unnormalized CFT four-point function are
\[
\begin{aligned}
 s_p^2&=C_0(12p)C_0(34p)C_0(\bar1\bar2\bar p)C_0(\bar3\bar4\bar p),\\
 t_q^2&=C_0(14q)C_0(23q)C_0(\bar1\bar4\bar q)C_0(\bar2\bar3\bar q).
\end{aligned}
\]
Here $C_0(12p)$ abbreviates $C_0(P_1,P_2,P_p)$. Barred labels denote right-moving momenta, not complex conjugates. Thus $s_pA_p$ and $t_qB_q$ are the corresponding products of unnormalized OPE coefficients. Omitting only the common external kinematic factor, crossing starts from the equality of the two expansions,
\begin{equation}
\begin{aligned}
 G_{1234}^{(s)}(z,\bar z)
 &=\sum_p s_pA_p\,\mathcal F_{P_p}^{(s)}(z)\mathcal F_{\bar P_p}^{(s)}(\bar z),\\
 G_{1234}^{(t)}(z,\bar z)
 &=\sum_q t_qB_q\,\mathcal F_{P_q}^{(t)}(z)\mathcal F_{\bar P_q}^{(t)}(\bar z),
 \qquad G_{1234}^{(s)}=G_{1234}^{(t)}.
\end{aligned}
\label{eq:s6crossingChannels}
\end{equation}
The fusion move \eqref{eq:fusionexpansion} expresses each $t$ block in the $s$ basis. Comparing coefficients of the resulting $s$ blocks gives, in the homogeneous heavy-sector projection,
\begin{equation}
\sum_i s_i A_i\,\delta_\Gamma\big((P,\bar P)-(P_i,\bar P_i)\big)
 =\sum_q t_q B_q\,\mathbb F_{P_qP}\mathbb F_{\bar P_q\bar P}.
\label{eq:s6continuumCoefficient}
\end{equation}
The contour delta distribution $\delta_\Gamma$ is normalized against $\dd P\,\dd\bar P$. This homogeneous projection retains the heavy OPE products. Fixed identity and light terms are included in the full residual \eqref{eq:residualdef}.

The fusion identity in the $C_0$ normalization of \eqref{eq:Utransition} reads
\begin{equation}
\begin{aligned}
 \mathbb F_{P_qP_p}\mathbb F_{\bar P_q\bar P_p}
 &=\rho_0(P_p)\rho_0(\bar P_p)\frac{s_p}{t_q}\,K_{pq},\\
 K_{pq}&=
 \begin{Bmatrix}P_1&P_4&P_q\\P_3&P_2&P_p\end{Bmatrix}_{\!\rm Vir}
 \begin{Bmatrix}\bar P_1&\bar P_4&\bar P_q\\\bar P_3&\bar P_2&\bar P_p\end{Bmatrix}_{\!\rm Vir}.
\end{aligned}
\label{eq:s6fullKernelFactors}
\end{equation}
Both factors use the same continued $b$, so $K_{pq}$ carries the complex tetrahedral phase rather than an absolute square.

To retain finitely many $s$-channel output nodes, we choose quadrature weights $\omega_p$. In \eqref{eq:s6continuumCoefficient} the contour delta is then represented by $\delta_{pi}/\omega_p$. Evaluating at the node $p$ and multiplying the equation by $\omega_p/s_p$ gives the finite crossing residual,
\begin{equation}
\boxed{\begin{aligned}
 R_p(C)&=A_p-\sum_q U_{pq}B_q,\\
 U_{pq}&=\omega_p\frac{t_q}{s_p}
             \mathbb F_{P_qP_p}\mathbb F_{\bar P_q\bar P_p}=D_pK_{pq},\\
 D_p&=\omega_p\rho_0(P_p)\rho_0(\bar P_p).
\end{aligned}}
\label{eq:s6finiteVirasoroResidual}
\end{equation}
The three-point factors $s_p,t_q$ cancel in $U_{pq}$. The Plancherel density $\rho_0$ is fixed by the Virasoro theory. We choose the quadrature weights $\omega_p$, which determine $D_p$ in this finite calculation. The weight $\omega_p$ belongs to the output label $p$, while the sum over $q$ runs over the retained $t$-channel primaries.

\paragraph{Which joint moments do we need?}
The theta normalization requires $\mathbb E_\mu[C_AC_B]=\delta_{AB}$. For four distinct OPE triples in the tetrahedral contraction, its disconnected pairings then vanish, and the gravitational coefficient requires
\[
 \kappa^\mu_{12p,34p,14q,23q}=\mathbb E_\mu[A_pB_q]=K_{pq}.
\]
We now determine what this target means for the still-unknown law of the products $B_q$. If $R_p(C)=0$ for every sample, multiplication of $A_p=\sum_qU_{pq}B_q$ by $B_r$ and averaging gives
\begin{equation}
 \mathbb E_\mu[A_pB_r]
 =\sum_qU_{pq}\mathbb E_\mu[B_qB_r]
 =D_p\sum_qK_{pq}\mathbb E_\mu[B_qB_r].
\label{eq:s6finiteCompatibility}
\end{equation}
The gravitational tetrahedral coefficient requires the left-hand side of \eqref{eq:s6finiteCompatibility} to equal $K_{pr}$. Choose the quadrature weights so that $D_p=D_0\ne0$ for all $p$. Then the simple choice $\mathbb E_\mu[B_qB_r]=D_0^{-1}\delta_{qr}$ gives
\[
 D_0\sum_qK_{pq}D_0^{-1}\delta_{qr}=K_{pr}.
\]
We will construct a positive measure realizing this choice. The moment $\mathbb E_\mu[B_qB_r]$ is a fourth moment of the original OPE variables and is not fixed by their theta normalization. Because the crossing equations hold for every realization, they also determine the remaining product moments,
\begin{equation}
\boxed{\begin{aligned}
 \mathbb E_\mu[B_qB_r]&=D_0^{-1}\delta_{qr},\\
 \mathbb E_\mu[A_pB_r]&=K_{pr},\\
 \mathbb E_\mu[A_pA_r]&=D_0\sum_qK_{pq}K_{rq}.
\end{aligned}}
\label{eq:s6targetBlocks}
\end{equation}
Thus we have identified the required moment table before defining the ensemble.

\paragraph{Choosing the quadrature weights $\omega_p$.}
On the dS ray $P=e^{\ii\pi/4}k$, the known Liouville sewing density takes the form
\begin{equation}
 \rho_0(e^{\ii\pi/4}k)=\ii f(k),\qquad
 f(k):=4\sqrt2\sin(2\pi rk)\sinh(2\pi k/r)>0
 \quad\left(0<k<\frac1{2r}\right).
\label{eq:s6sineDensity}
\end{equation}
Here $r>0$ is the dS continuation parameter and $b=re^{\ii\pi/4}$. The function $f(k)$ is explicitly determined by $\rho_0$. Take a compact pole-free region $\Omega$ in the first sine chamber for the two output variables $(k,\bar k)$. Define the positive weight per cell by
\[
 d=\frac{1}{n_s}\int_\Omega f(k)f(\bar k)\,\dd k\,\dd\bar k>0.
\]
Divide $\Omega$ into $n_s$ cells $\Omega_p$ such that each has integral $d$, and choose one node $(k_p,\bar k_p)$ in each cell. Since $\dd P\,\dd\bar P=\ii\dd k\,\dd\bar k$, choose
\[
 \omega_p=\frac{\ii d}{f(k_p)f(\bar k_p)}.
\]
Then the factor appearing in the crossing residual becomes
\begin{equation}
 \boxed{D_p=\omega_p\rho_0(P_p)\rho_0(\bar P_p)
 =-\ii d=:D_0,\qquad U_{pq}=D_0K_{pq}.}
\label{eq:s6commonSewing}
\end{equation}
The quadrature weights are now fixed, and the Virasoro density $\rho_0$ retains its original form. The left- and right-moving labels are independent arguments of this fusion calculation. The physical primary spin is $J=h-\bar h=P^2-\bar P^2\in\mathbb Z$, as in \eqref{eq:s6euclideanope}. For the scalar restriction $P=\bar P$,
\[
 K_{pq}\big|_{\rm scalar}
 =\begin{Bmatrix}P_1&P_4&P_q\\P_3&P_2&P_p\end{Bmatrix}_{\!\rm Vir}^{\,2}.
\]
The integer-spin spectral support in spinning channels is specified in \eqref{eq:fixedspintransport}. The present finite model realizes the corresponding selected two-chiral fusion coefficients.

\paragraph{Constructing a positive probability measure.}
We now choose the values of the OPE variables in ways that make the desired moments easy to compute. Let $n_s,n_t$ be the numbers of retained $p,q$ channels and set $S=2n_s+3n_t$. There are $2n_s+2n_t$ OPE coefficients, each of which will have its own axis component, and there is one channel component for each of the $n_t$ values of $q$. As part of the construction, we choose each of these $S$ components with probability $1/S$.

The axis components are introduced to obtain $\mathbb E_\mu[C_A^2]=1$ without generating unwanted pair correlations. On the axis component labelled by $A$, set $C_A=\Lambda e^{\ii\theta}$ and all other $C_B=0$. Every $A_p=C_{12p}C_{34p}$ and $B_q=C_{14q}C_{23q}$ contains two distinct OPE coefficients. Because at most one coefficient is nonzero on an axis component, $A_p=B_q=0$ there and the crossing equations hold.

The channel components are designed to produce the tetrahedral correlations. On the component labelled by $q$, use a common angle $\theta$ and independent uniform pair phases $\phi_p,\varphi_r$. Set
\begin{equation}
\begin{aligned}
 C_{12p}&=\Lambda e^{\ii\theta+\ii\phi_p}\sqrt{U_{pq}},&
 C_{34p}&=\Lambda e^{\ii\theta-\ii\phi_p}\sqrt{U_{pq}},\\
 C_{14r}&=\Lambda e^{\ii\theta+\ii\varphi_r}\delta_{rq},&
 C_{23r}&=\Lambda e^{\ii\theta-\ii\varphi_r}\delta_{rq}.
\end{aligned}
\label{eq:s6finiteSample}
\end{equation}
For such a sample the pair phases cancel in each product,
\[
 A_p=\Lambda^2 e^{2\ii\theta}U_{pq},\qquad
 B_r=\Lambda^2 e^{2\ii\theta}\delta_{rq},\qquad
 A_p-\sum_r U_{pr}B_r=0.
\]
Crossing therefore holds for every axis or channel sample, independently of their probabilities. The remaining task is to choose those probabilities so that the theta and tetrahedral moments have the target values.

The angle $\theta$ has different probability distributions on axis and channel components. On each component we specify an infinitesimal probability measure for $\theta$. We choose
\begin{equation}
\begin{aligned}
 \dd\Pr_{\rm ax}(\theta)
 &=\frac{\dd\theta}{2\pi}
   \left(1+\frac{2S}{\Lambda^2}\cos 2\theta\right),\\
 \dd\Pr_{\rm ch}(\theta)
 &=\frac{\dd\theta}{2\pi}
   \left(1+\frac{2S}{d\Lambda^4}\sin 4\theta\right).
\end{aligned}
\label{eq:s6angularLaw}
\end{equation}
Each expression is the probability measure for $\theta$ conditioned on the corresponding component having been selected. Both integrate to one. If $\Lambda$ is a sufficiently large finite number such that
\begin{equation}
 \frac{2S}{\Lambda^2}<1,\qquad
 \frac{2S}{d\Lambda^4}<1,
\label{eq:s6positivityCondition}
\end{equation}
then both measures are strictly positive for every angle. Together with the probability $1/S$ assigned to each component, they define the positive joint measure $\mu$ on the OPE coefficients.

\paragraph{Checking the moments and crossing.}
An axis component contributes to $C_A^2$ only when its nonzero coordinate is $C_A$. Its contribution has probability $1/S$, and the chosen angular measure gives
\begin{equation}
\begin{aligned}
 \mathbb E_\mu[C_A^2]
 &=\frac{\Lambda^2}{S}\int_0^{2\pi}
      e^{2\ii\theta}\,\dd\Pr_{\rm ax}(\theta)\\
 &=\frac{\Lambda^2}{S}\frac{2S}{\Lambda^2}
   \underbrace{\frac1{2\pi}\int_0^{2\pi}
      e^{2\ii\theta}\cos2\theta\,\dd\theta}_{1/2}
 =1.
\end{aligned}
\label{eq:s6thetaDirect}
\end{equation}
The constant term in $\dd\Pr_{\rm ax}$ integrates to zero in this expression. The uniform pair phases on the channel components give zero for their two-point contractions, so $\mathbb E_\mu[C_AC_B]=0$ whenever $A\ne B$.

For fixed $p,q$, only the channel component labelled by $q$ contributes to $A_pB_q$. Using $D_0=-\ii d$ and the channel angular measure, we find
\begin{equation}
\begin{aligned}
 \mathbb E_\mu[A_pB_q]
 &=\frac{\Lambda^4U_{pq}}{S}
   \int_0^{2\pi} e^{4\ii\theta}\,\dd\Pr_{\rm ch}(\theta)\\
 &=\frac{\Lambda^4U_{pq}}{S}\frac{2S}{d\Lambda^4}
   \underbrace{\frac1{2\pi}\int_0^{2\pi}
     e^{4\ii\theta}\sin4\theta\,\dd\theta}_{\ii/2}\\
 &=\frac{\ii}{d}U_{pq}=D_0^{-1}U_{pq}=K_{pq}.
\end{aligned}
\label{eq:s6finiteMoments}
\end{equation}
In the same way, $\mathbb E_\mu[B_qB_r]=D_0^{-1}\delta_{qr}$ and $\mathbb E_\mu[A_pA_r]=D_0\sum_qK_{pq}K_{rq}$. These are the product moments required by \eqref{eq:s6targetBlocks}.
All one-point functions vanish. Because the four OPE coefficients in the selected tetrahedral contraction are distinct and their pairwise second moments vanish, this fourth moment equals the connected cumulant. Together with the crossing relation verified for every realization,
\begin{equation}
\boxed{\begin{aligned}
 \mathbb E_\mu[C_AC_B]&=\delta_{AB},\\
 \kappa^\mu_{12p,34p,14q,23q}&=K_{pq},\qquad
 K_{pq}\big|_{\rm scalar}=\begin{Bmatrix}P_1&P_4&P_q\\P_3&P_2&P_p\end{Bmatrix}_{\!\rm Vir}^{\,2},\\
 R_p(C)&=0\quad\text{for every realization and retained }p.
\end{aligned}}
\label{eq:s6exactFiniteModel}\eqalias{eq:s6tensorTarget}\eqalias{eq:s6finiteModelResults}
\end{equation}
The unit tetrahedral coefficient follows from the explicit angular integral. The measure also determines further fourth and higher moments that can be compared with additional gravitational fillings.

A probability measure $\mu$ describes exact CFT data when each realization satisfies the sphere and torus crossing equations. The relation $R_\lambda(C)=0$ holds $\mu$-almost surely if and only if
\begin{equation}
 \mathbb E_\mu\big|R_\lambda(C)\big|^2
 =r^\lambda_{AB}\,r^{\lambda*}_{CD}\,\mathbb E_\mu\big[C_AC_BC_C^*C_D^*\big]
  +\cdots=0 ,
\label{eq:s6samplewisecrossing}
\end{equation}
where the omitted terms contain fewer coefficients. The holomorphic conditions \eqref{eq:s4polynomialcrossing} impose moment relations. Crossing for every realization is controlled by the mixed moments, which pair the $b$ and $b^*$ sectors as in the Hermitian pairing of \cite{CEM}.

\section{Discussion}\label{sec:discussion}

\subsubsection*{OPE statistics of dS$_3$ gravity}

The central physical proposal of this paper is that pure dS$_3$ gravity, at fixed conformal geometry of future infinity, assigns statistics to the OPE data of a two-dimensional Euclidean CFT. The boundary is Euclidean because future infinity is spacelike, with the positive-definite induced metric \eqref{eq:s6inducedmetric}, and the cosmological state obeys conformal Ward identities on it. The complex central charge and conformal weights make the proposed boundary theory nonunitary. The fluctuating data are the OPE coefficients. \nw{The theta filling fixes the reference second moment, and the tetrahedral filling supplies a connected fourth-moment contribution.} The same tetrahedral coefficient is extracted from the crossed channel of the selected Liouville state at fixed moduli, \eqref{eq:mainratio}. This agreement supports its interpretation as a contribution to a collective CFT correlation. \revnew{Physical primary data obey the integer-spin condition $J=h-\bar h\in\mathbb Z$, and the two-chiral kernel used in the finite ensemble reduces to $\{6j\}_{\rm Vir}^{\,2}$ on its scalar diagonal.}

The fixed-angle tetrahedral coefficient changes from exponential suppression to an oscillatory volume phase under the dS continuation, \eqref{eq:dSsize}. The real-$b$ comparison uses imaginary, conical-defect momenta; it is not a comparison of complete correlators in the ordinary heavy spectrum. The phase varies on the angular scale $\pi r^2/\ell_e$, \eqref{eq:angleresolution}, so its geometric information requires probes that resolve the OPE labels at this accuracy. Densities, conformal blocks and the channel choice still determine the size of its contribution to an integrated correlation.

Sewing tests how the statistics fit together.  Sphere crossing balances the connected tetrahedral cumulant against the change of Gaussian pairings between channels.  At sixth order the independently evaluated $K_{3,3}$ filling cancels the channel difference of the prism contraction, \eqref{eq:localprismcompletion}, and this fixes the weights of the prism and $K_{3,3}$ fillings in terms of the theta and tetrahedral weights, \eqref{eq:localweightconstraint}.  Crossing of the boundary statistics thereby fixes the relative contributions of bulk fillings in this sector, a statistical counterpart of the sum over geometries.  The torus calculation tests the same coefficients against modular invariance in the repeated-label sector, \eqref{eq:repeatedreduction}.

\nw{The selected cosmological state enters each Liouville cylinder factor of the paired correlation, \eqref{eq:statisticalstatepair}. The fixed vacuum family cancels from the connected correlator, \eqref{eq:vacuumcenteredcorrelator}. Each Liouville factor has the threshold $c_{\rm eff}=1$, while the full two-factor covariance has the separately computed, integrable reduced collar tail \eqref{eq:s6pairedcollar}. Section~\ref{ssec:positivetensor} provides an explicit positive measure for a finite homogeneous sphere sector: the \aidrev{discretized fusion equations} hold for every realization, their theta pairing is exactly normalized, and their selected tetrahedral cumulant equals $K_{pq}$ with unit coefficient, \eqref{eq:s6exactFiniteModel}.}

\vone{The five-point state independently reproduces the normalized prism and $K_{3,3}$ coefficients \eqref{eq:fivenormalized}. Applying the common external $S$ transform gives their fixed-angle CTV amplitudes, as derived in appendix~\ref{ssec:sphere-tree-extension}.}

\subsubsection*{Lorentzian geometry from OPE phases}

The phase of the tetrahedral moment has a direct Lorentzian meaning. \vtwo{The real de Sitter geometry $\mathcal D_M$ of \eqref{eq:lorentzInverseFuchsian} and \eqref{eq:lorentzWedge} supplies the data needed to recover the compact core.} \vnew{For this tetrahedral $K_4$ filling,} as $t\to0$ the spatial slices collapse onto a metric graph with the incidences of $K_4$, the graph that labels the OPE contraction. The four OPE triples become the vertices of this initial singularity and the six shared primaries become its arcs. Each label is measured twice in $\mathcal D_M$, as the \aidrev{arc length $\ell_{\rm dS}(\pi-\psi_e)$} and as the rapidity $\ell(\gamma_e)$ of the boost around it, and the angle derivative of the OPE phase, multiplied by $-2\pi r^2$, returns that rapidity, \eqref{eq:s6statisticalgeometry}. \hnew{Its trace $|\operatorname{tr}H_{\gamma_e}|=2\cosh(\ell(\gamma_e)/2)$ identifies the hyperbolic holonomy conjugacy class; the connected tetrahedral OPE moment thereby probes an invariant global identification of the Lorentzian spacetime.} The de Sitter points that appear in the embedding geometry of generalized tetrahedra \cite{Hartman} acquire in this way a dynamical role, as the vertices of that singularity.

\vtwo{For the chosen marked $K_4$ double, the initial arcs reconstruct the hyperideal Gram matrix, the boost rapidities \aidrev{test $\lambda_e=2\ell_e$}, and the common cuff seam identifications reproduce the four truncation-face reflections,} \eqref{eq:pleatedfourreflections}--\eqref{eq:pleatedboostpairing}. The recovered pleated boundary is therefore the boundary of the same two-cell compact filling. \aidrev{The four-point coefficient extracted from the selected cosmological state equals the continued CTV tetrahedral coefficient}, \eqref{eq:coreexact}. \aidrev{Its leading phase measures the Lorentzian holonomies. Together these identify the compact triangulation underlying the boundary amplitude and the marked real spacetime.}

The gravitational action shows where the volume phase resides.  The real Lorentzian evolution contributes the bending term, which at the initial singularity is a sum of arc length times boost angle, \eqref{eq:lorentzRegge}, and a cutoff term that reproduces the Weyl anomaly of the boundary theory, \eqref{eq:lorentzAnomalyBranch}.  The compact volume is carried by the continued core, a negative-definite cap reached from $\mathcal D_M$ at $t=\ii\pi/2$.  With matched cutoffs the bending and anomaly terms cancel between the two contributions, \eqref{eq:lorentzBox}, and the compact amplitude retains the volume phase.  The OPE phase thus records a complex saddle attached to a real Lorentzian spacetime, and the Lorentzian spacetime supplies the arcs and holonomies through which that phase is measured.

The internal edge of the $K_{3,3}$ filling extends this picture beyond a single tetrahedron. In the symmetric length window of \eqref{eq:oneedge3nj}, with its uniform Barnes estimates, the oscillatory Plancherel measure selects a stationary point at which the cells fill a full $2\pi$ angle around the edge. The selected branch thus carries the smooth-gluing condition of the hyperbolic state sum into the dS integral.

The continued Cardy density gives a complementary check of the angle variable.  On the scalar ray $P=e^{\ii\pi/4}k$, with $\psi=2\pi rk$, the leading behaviour at fixed angle is
\begin{equation}
\begin{aligned}
\rho_0(P)^2&=-32\sin^2\psi\,\sinh^2\frac{\psi}{r^2},\\
 \frac{2\psi}{r^2}&=\frac{\psi}{\pi}\,
 \frac{\pi\ell_{\rm dS}}{2G_N}+O(r^0).
\end{aligned}
\label{eq:SdSentropy}
\end{equation}
For fixed $0<\psi<\pi$ the growth $2\psi/r^2$ equals the Gibbons--Hawking entropy of a static patch containing a particle of opening angle $2\psi$.  In the conventions of \cite{BTV} this patch has horizon radius $\psi/\pi$ in units of $\ell_{\rm dS}$.  The same label therefore controls the entropy of a static patch and the geometric phase of the OPE statistics \cite{GibbonsHawking,MaldacenaStrominger,CEM}.

\begin{vtwoblock}
\subsubsection*{The sphere limit of the quantum amplitude}

The sphere-tree construction organizes higher-genus fillings through lower-genus CFT data. Pinching the external cuffs exposes two sphere components, whose channel coefficients reconstruct their states through \eqref{eq:spheretreereassembly}. The higher-genus state combines these components with neck propagators and an integral over their puncture labels, as in \eqref{eq:joinedspheresprimary}. Component moduli and neck parameters specify different parts of the boundary geometry. Matched small bending gives mirrored component moduli, $\boldsymbol m_\sigma=\bar{\boldsymbol m}_\tau$, while the puncture labels still have to be specified.

The external polarization kernel makes this distinction explicit. For a smooth test function $f(P)$ of compact support,
\begin{equation}
S_{0P}=2\sqrt2,\qquad
\lim_{P'\to0}\int_0^\infty\dd P\,S_{P'P}f(P)
 =2\sqrt2\int_0^\infty\dd P\,f(P).
\label{eq:zeroanglenotprojection}
\end{equation}
Sending the angle label $P'$ to zero integrates over representations. A fixed-puncture wavefunction is recovered by keeping its momenta fixed during reconstruction. The unprojected sewing instead sums the two component states.

Identifying the full quantum degeneration requires control of the exact finite-$b$ transform. The vanishing Gram determinant \eqref{eq:spherelimitGram} places this family outside the uniform regime of the nondegenerate estimate \eqref{eq:MY}. For $K_{3,3}$, all nine external transforms, both internal recoupling integrals and their crossed-pole residues must have a common limit. The symmetric internal-edge estimate leaves this convergence problem open.
\end{vtwoblock}

\subsubsection*{Future directions}

\emph{A microscopic ensemble.} \revnew{Section~\ref{ssec:positivetensor} constructs a single positive measure realizing exact theta normalization, the unit-weight Virasoro tetrahedral entry and \aidrev{samplewise satisfaction of the discretized fusion equations}. Its two-chiral recoupling coefficient contains the scalar $\{6j\}_{\rm Vir}^{\,2}$. \aidrev{Extension to a physical spectrum requires the integer-spin condition} \eqref{eq:fixedspintransport}. The next stage is a shared random OPE table for overlapping external quartets with identity and light exchanges, modular $S,T$ relations, and filling weights constrained by the higher moments of section~\ref{sec:bulkinverse}. The same construction supplies concrete finite targets for studying the continuum limit and global cosmological norms. Virasoro random tensor dynamics \cite{BdBJNS}, statistical crossing \cite{UniversalTopologies,CrossingOPE}, boundary ensembles \cite{CHJL,HJL,WWW} and open--closed triangulations \cite{JafferisWang} offer complementary methods for developing this microscopic description.}

\emph{Spectral statistics and higher points.} At fixed spectrum the torus partition function carries no OPE data,
\begin{equation}
 Z_{\rm CFT}(\tau,\bar\tau)
 =\operatorname{Tr}_{\mathcal H}q^{L_0-c/24}\bar q^{\bar L_0-c/24},
 \qquad \delta Z_{\rm CFT}=0,\qquad
 \Eds[\delta Z_{\rm CFT}\delta Z_{\rm CFT}]=0.
\label{eq:fixedspectrumtorus}
\end{equation}
Here $q=e^{2\pi\ii\tau}$ and $\bar q=e^{-2\pi\ii\bar\tau}$ are the torus variables. A nonzero torus cosmological state therefore requires spectral statistics or a different collective observable. Higher-point vacuum degenerations can carry random lower-point amplitudes, so four-point centering need not remove their residues. The repeated-label sector requires the modular-covariant handle sum of appendix~\ref{ssec:repeatedproof}, analogous to the marked sums over geometries in \cite{MaloneyWitten,CLM}.

\emph{Global cosmological states.} The collar estimate treats a marked, spectrally localized degeneration. Extending it to a full state requires control of large momenta, simultaneous degenerations and the mapping-class identifications. The norm of the selected Liouville state and its integrated-observable description \cite{CEM,CLS,Matrix} pair amplitudes at $b$ and $b^*$, and are the natural starting point for computing the mixed OPE moments.

\emph{General fillings and their Lorentzian spacetimes.} For a compact core whose boundary consists of totally geodesic pants bent along the meridians of a graph, \vtwo{the hyperbolic-end/de Sitter normal flow produces} an initial singularity with the incidences of that graph and arcs of length \aidrev{$\ell_{\rm dS}(\pi-\psi_e)$}, in line with the dual-tree description of initial singularities \cite{Mess,BenedettiBonsante}. The reverse reconstruction for a general marked filling additionally requires its global pants twists and face-pairing holonomies; the six lengths and local boost traces used in the tetrahedral example do not by themselves select an arbitrary filling. Internal edges also require the coupled zero-deficit equations of unequal-angle cells and the selection among their oscillatory branches. Whether the continued core belongs to the integration cycle of the de Sitter path integral, and how it fares under the criteria for allowable complex metrics \cite{KontsevichSegal,WittenComplex}, is part of the same contour problem.

\emph{Quantum corrections and higher dimensions.}  \nw{The geometric phase and holonomy relations above are leading semiclassical statements at fixed angles.}  The $O(r^2)$ corrections to the $6j$ asymptotics, the one-loop determinants around the continued saddles and the descendant contributions would determine how sharply the boundary statistics resolve the initial geometry.  In higher dimensions one may ask whether late-time correlators of de Sitter gravity admit an OPE-statistical description whose phases measure holonomies around an initial singularity, a question for which the dS/CFT and cosmological bootstrap frameworks provide the natural setting \cite{AHS,ABLP}.

These directions share one aim, to determine how much of de Sitter quantum gravity is captured by the statistics of boundary OPE data and how that information is reorganized when the data are integrated into cosmological observables.

\section*{Acknowledgments}
The author used AI assistants for support with parts of the calculations, literature searches, writing and schematic figures. The author directed the research, selected the key literature and ideas, refined the conceptual framework, checked the calculations and conclusions, and edited the figures for geometric accuracy and presentation.

\appendix
\FloatBarrier
\section{Sphere crossing and six-vertex fillings}\label{sec:sixth}\label{sec:higher}
The homogeneous crossing equations of section~\ref{ssec:momentdictionary} relate the theta and tetrahedral moments to the six-OPE contractions. We first remove the fixed conformal families, then evaluate the prism and $K_{3,3}$ graphs and compare their sixth-order contributions. The five-point extraction identifies the same graph coefficients in the boundary state.

\subsection{Removing fixed conformal families}\label{app:lightprojection}

A sphere projection can be constructed directly from the Virasoro-channel expansions. For fixed nonidentity scalar external primaries, take both OPE coefficients associated with each heavy exchanged primary as fluctuating. Before projection, the difference of the two channels is
\[
 R(x;C)=L(x)+T_{AB}(x)C_AC_B,
 \qquad x=(z,\bar z),
\]
where $L(x)$ contains the fixed vacuum and light exchanges in both channels, including their Virasoro descendants, and $T_{AB}(x)$ is the difference of the $s$- and $t$-channel coefficients $T_a^{AB}$ of \eqref{eq:quadraticboundaryamplitude}, with the same $C_0$ normalization. Given two Euclidean cross ratios $x_1,x_2$, define
\begin{equation}
\begin{aligned}
 \lambda[f]&=L(x_2)f(x_1)-L(x_1)f(x_2),\\
 \lambda[L]&=0,\qquad R_\lambda(C)=r^\lambda_{AB}C_AC_B,\\
 r^\lambda_{AB}&=L(x_2)T_{AB}(x_1)-L(x_1)T_{AB}(x_2).
\end{aligned}
\label{eq:explicitlightprojection}
\end{equation}
The fixed contribution cancels identically. The heavy coefficient survives whenever $T_{AB}(x)/L(x)$ is nonconstant on a domain where $L(x)\ne0$, since the two evaluation points can then be chosen with $r^\lambda_{AB}\ne0$.

Inserting $\mathcal P=1$, $C_CC_D$ and $C_CC_DC_EC_F$ into $\Eds[\mathcal P(C)R_\lambda(C)]=0$ then gives the second-, fourth- and sixth-order moment relations in \eqref{eq:sourcehierarchy}.

\subsection[Prism and K33 amplitudes]{Prism and $K_{3,3}$ amplitudes}\label{ssec:heegaard}\label{ssec:sourceaudit}

\begin{figure}[t]
\centering
\begin{tikzpicture}[x=1cm,y=1cm,line width=.65pt,line cap=round,line join=round,>=Stealth,font=\small,
  lab/.style={font=\scriptsize,inner sep=1.5pt},
  vlab/.style={font=\scriptsize,inner sep=1pt}]
\newcommand{\fivetree}[9]{%
 \coordinate (X1) at (0,0);\coordinate (X2) at (1.15,0);\coordinate (X3) at (2.30,0);
 \draw (X1)--(-.55,.62) node[left,lab]{$#1$};
 \draw (X1)--(-.55,-.62) node[left,lab]{$#2$};
 \draw (X2)--(1.15,.78) node[above,lab]{$#3$};
 \draw (X3)--(2.85,.62) node[right,lab]{$#4$};
 \draw (X3)--(2.85,-.62) node[right,lab]{$#5$};
 \draw[#8,line width=1pt] (X1)--(X2) node[midway,above,lab]{$#6$};
 \draw[#8,line width=1pt] (X2)--(X3) node[midway,above,lab]{$#7$};
 \foreach \p in {X1,X2,X3}{\fill (\p) circle (1.5pt);}
 \foreach \p/\n in {#9}{\node[below=2pt,vlab] at (\p) {$\n$};}}
\node[anchor=west,paneltitle] at (-1.1,1.55) {(a) Prism};
\begin{scope}[shift={(0,0)}]
 \fivetree{P_1}{P_2}{P_3}{P_4}{P_5}{P_a}{P_b}{blue!65!black}{X1/A,X2/B,X3/C}
 \node at (1.15,-1.15) {$\tau_1$};
\end{scope}
\begin{scope}[shift={(4.95,0)}]
 \fivetree{P_1}{P_3}{P_5}{P_2}{P_4}{P_a'}{P_b'}{red!70!black}{X1/D,X2/E,X3/F}
 \node at (1.15,-1.15) {$\tau_2^{\rm pr}$};
\end{scope}
\draw[->,thick] (8.55,0)--(9.65,0) node[midway,above,align=center,font=\scriptsize]{join equal\\labels};
\begin{scope}[shift={(11.55,-.05)},scale=.78]
 \coordinate(A)at(90:2.0);\coordinate(B)at(-30:2.0);\coordinate(D)at(210:2.0);
 \coordinate(F)at(90:.72);\coordinate(C)at(-30:.72);\coordinate(E)at(210:.72);
 \draw[blue!65!black,line width=1pt](A)--(B);\draw[blue!65!black,line width=1pt](B)--(C);
 \draw[red!70!black,line width=1pt](D)--(E);\draw[red!70!black,line width=1pt](E)--(F);
 \draw(B)--(D);\draw(D)--(A);\draw(F)--(C);\draw(C)--(E);\draw(A)--(F);
 \foreach \p in {A,B,C,D,E,F}{\fill(\p)circle(1.9pt);}
 \node[vlab,above]at(A){$A$};\node[vlab,right]at(B){$B$};\node[vlab,left]at(D){$D$};
 \node[vlab,above left]at(F){$F$};\node[vlab,above right]at(C){$C$};\node[vlab,below]at(E){$E$};
\end{scope}
\node[anchor=west,paneltitle] at (-1.1,-1.95) {(b) $K_{3,3}$};
\begin{scope}[shift={(0,-3.5)}]
 \fivetree{P_1}{P_2}{P_3}{P_4}{P_5}{P_a}{P_b}{blue!65!black}{X1/A,X2/B,X3/C}
 \node at (1.15,-1.15) {$\tau_1$};
\end{scope}
\begin{scope}[shift={(4.95,-3.5)}]
 \fivetree{P_1}{P_4}{P_3}{P_2}{P_5}{P_a'}{P_b'}{red!70!black}{X1/D,X2/E,X3/F}
 \node at (1.15,-1.15) {$\tau_2^{K}$};
\end{scope}
\draw[->,thick] (8.55,-3.5)--(9.65,-3.5);
\begin{scope}[shift={(11.55,-3.5)},scale=.82]
 \coordinate(A)at(-1.65,1.05);\coordinate(D)at(1.65,1.05);
 \coordinate(B)at(-.85,0);\coordinate(C)at(.85,0);
 \coordinate(F)at(-.70,-1.10);\coordinate(E)at(.70,-1.10);
 \draw[blue!65!black,line width=1pt](A)--(B)--(C);
 \draw[red!70!black,line width=1pt](D)--(E)--(F);
 \draw(A)--(D) (A)--(F) (C)--(D) (B)--(E);
 \draw[white,line width=3.5pt](C)--(F);
 \draw(C)--(F);
 \foreach \p in {A,B,C,D,E,F}{\fill(\p)circle(1.8pt);}
 \node[vlab,above left]at(A){$A$};\node[vlab,above right]at(D){$D$};
 \node[vlab,left=3pt]at(B){$B$};\node[vlab,right=3pt]at(C){$C$};
 \node[vlab,below left]at(F){$F$};\node[vlab,below right]at(E){$E$};
\end{scope}
\end{tikzpicture}
\caption{Five-point channel trees and their six-vertex contractions. Matching equal external labels gives (a) the prism and (b) $K_{3,3}$. Blue and red edges retain the observer and preparation internal labels. The bridge in (b) marks a crossing of edges. The graph amplitudes use the Heegaard pairing and recoupling convention specified in the text.}
\label{fig:heegaard}
\end{figure}
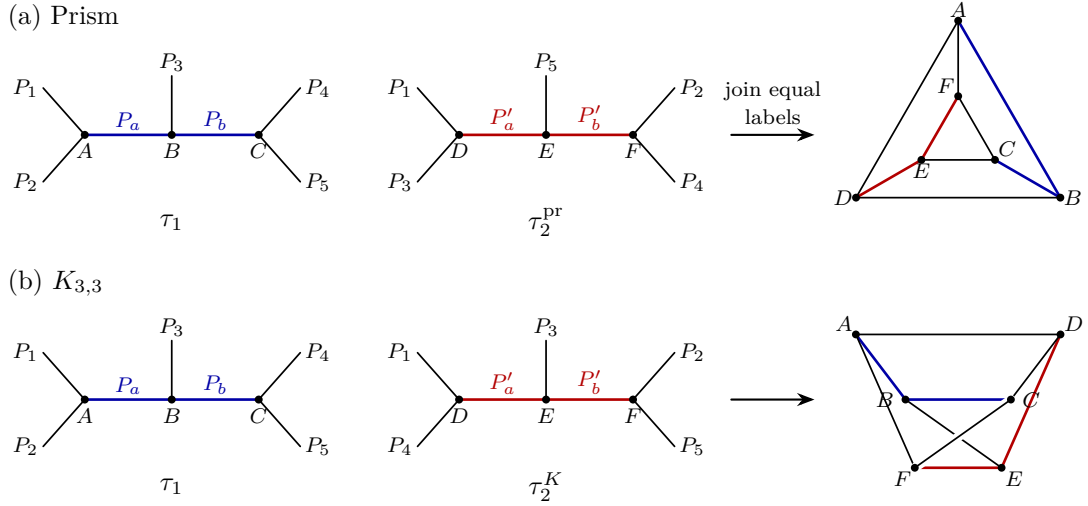

A five-punctured-sphere channel contains three OPE vertices, so pairing two such channels probes six coefficients. The lower theta and tetrahedral moments generate the prism contraction. A fusion move on a joining edge relates it to $K_{3,3}$ and determines their relative coefficients. Both sixth-moment contractions were evaluated in AdS in \cite{CEZ2}. Here the coefficients use the graph gluing of figure~\ref{fig:heegaard} and the recoupling convention below.

The two chiral copies use the same continued $b$.  The superscript $\vee$ denotes the reversed framing of \eqref{eq:veedefinition}.

A pants decomposition of $\Sigma_{0,5}$ uses two curves and three pairs of pants.  It is encoded by a trivalent tree with five leaves, one for each puncture, three vertices, one for each pair of pants, and two internal edges, one for each curve.  We write a tree by the order in which punctures fuse.  The observer uses $\tau_1=((12)\,3\,(45))$, in which punctures $1,2$ fuse into the internal label $P_a$, this fuses with $3$ into $P_b$, and $P_b$ fuses with $4,5$.  Its vertices are $A=(P_1,P_2,P_a)$, $B=(P_a,P_3,P_b)$ and $C=(P_b,P_4,P_5)$.  A preparation uses a second tree $\tau_2$ with internal labels $(P_a',P_b')$.  We abbreviate $P=(P_a,P_b)$ and $P'=(P_a',P_b')$.  The prime distinguishes the preparation labels.  The observer labels will be the spectral arguments recovered by two collar extractions, while the preparation labels are prescribed inputs.

To obtain a closed graph, place the observer and preparation trees in two three-balls and glue their boundary spheres so that equal puncture labels match, as in figure~\ref{fig:heegaard}.  This Heegaard pairing produces $S^3$ with six trivalent vertices and nine edges.  Five edges join corresponding leaves, and four are the internal edges of the trees.  We use the theta-normalized recoupling convention specified by \eqref{eq:prismZ} and \eqref{eq:K33Z} for the embedded graph amplitudes.  The two connected simple cubic graphs with six vertices are the triangular prism and $K_{3,3}$.  The preparations used here are
\begin{equation}
\tau_2^{\rm pr}=((13)\,5\,(24)),\qquad
 \tau_2^K=((14)\,3\,(25)),\qquad
 \tau_2'=((15)\,3\,(24)).
\label{eq:threepreparations}
\end{equation}
For $\tau_2^{\rm pr}$ the vertices are $D=(P_1,P_3,P_a')$, $E=(P_a',P_5,P_b')$ and $F=(P_2,P_4,P_b')$.  The edges $AB$, $BD$ and $DA$ form a triangle, and so do $CE$, $EF$ and $FC$, so the gluing is the triangular prism of figure~\ref{fig:heegaard}.  For $\tau_2^K$ every edge joins a vertex of $\{A,C,E\}$ to a vertex of $\{B,D,F\}$, and the gluing is $K_{3,3}$.  The preparation $\tau_2'$ is obtained from $\tau_2^{\rm pr}$ by the single fusion move $\{13\}\to\{15\}$ on its first internal edge.  Its gluing is again a $K_{3,3}$ graph, with a different assignment of external and internal labels, which enters the joining-edge cut below.

To read this as an OPE statistic, assign one coefficient $C_v$ to each vertex $v$, using its three incident labels as indices.  For the prism this gives
\begin{equation}
\prod_{v\in V(\Gamma_{\rm pr})}C_v
=C_{12a}C_{a3b}C_{b45}C_{13a'}C_{a'5b'}C_{24b'}.
\label{eq:s4prismproduct}
\end{equation}
Here an index such as $a$ denotes the primary labelled by $P_a$.  Each of the nine edge labels occurs at two vertices, just as each tetrahedral edge did in section~\ref{ssec:core}.  All nine labels are fixed arguments of this graph coefficient, and a label is integrated only when a sewing or recoupling operation introduces that integral.

To evaluate the prism, replace its triangle $ABD$ by a single trivalent vertex. Cutting out a three-ball around the triangle leaves three labelled strands on its boundary. The Virasoro block space on that three-punctured sphere is one-dimensional, so the triangle prepares a multiple of the single-vertex state \cite{CEZ2,UniversalTopologies}. In the theta normalization the multiplier is a $6j$ \cite{CTV}. The replacement gives the vertex $(P_2,P_b,P_a')$. Together with $C,E,F$, it forms a tetrahedral graph and supplies a second $6j$. Thus
\begin{equation}
Z_{\rm Vir}(S^3,\Gamma_{\rm pr})
 =\begin{Bmatrix}P_2&P_4&P_b'\\P_5&P_a'&P_b\end{Bmatrix}_{\!\rm Vir}
   \begin{Bmatrix}P_1&P_3&P_a'\\P_b&P_2&P_a\end{Bmatrix}_{\!\rm Vir}.
 \label{eq:prismZ}
\end{equation}
The second factor has triads $(P_1,P_3,P_a')$, $(P_1,P_2,P_a)$, $(P_b,P_3,P_a)$ and $(P_b,P_2,P_a')$, which are $D$, $A$, $B$ and the new vertex.  The graph $K_{3,3}$ contains no triangle.  One fusion move on an edge turns it into a prism, and the move carries an integral over the new label $X$,
\begin{equation}
\begin{aligned}
Z_{\rm Vir}(S^3,K_{3,3})
 &=\int_0^\infty\dd X\,\rho_0(X)
 \begin{Bmatrix}P_2&P_5&P_b'\\P_3&P_a'&X\end{Bmatrix}_{\!\rm Vir}
 \begin{Bmatrix}P_1&P_4&P_a'\\X&P_2&P_a\end{Bmatrix}_{\!\rm Vir}\\
 &\hspace{35mm}\times
 \begin{Bmatrix}P_3&P_5&X\\P_4&P_a&P_b\end{Bmatrix}_{\!\rm Vir}.
\end{aligned}
 \label{eq:K33Z}
\end{equation}
In \eqref{eq:K33Z}, the last $6j$ changes the channel and the first two evaluate the resulting prism. The nine graph labels remain fixed, while the new label $X$ is integrated with the inverse pairing $\rho_0(X)\dd X$ of \eqref{eq:twopoint}. This sewing integral joins the fixed-label kernels. The reversed-framing copy has its own intermediate label.

\subsection{The sixth-order crossing equation}\label{ssec:sixthtest}

The theta pairing and tetrahedral cumulant generate a prism term whose joining-edge equation fixes $w_{\rm pr}=w_{K_{3,3}}=w_{\rm tet}^2/w_\theta$ in the tree sector below. We derive this relation by comparing a cut through a triangle with a cut on an edge joining the two triangles.

First cut a triangle edge. This applies crossing to the local four-point subgraph around that edge. The two tetrahedral fourth cumulants, joined by the inverse theta pairing, reproduce the prism amplitude \eqref{eq:prismZ} in the corresponding two-copy contraction.
Crossing one edge of a triangle in the sixth-order equation of \eqref{eq:sourcehierarchy} contracts one tetrahedral factor with a fusion kernel.  If $R,R'$ label the channels paired across this cut, the contraction collapses by Racah orthogonality,
\begin{equation}
\int_0^\infty\dd P\,\rho_0(P)
 \begin{Bmatrix}P_1&P_2&P\\P_b&P_3&R\end{Bmatrix}_{\!\rm Vir}
 \begin{Bmatrix}P_1&P_2&P\\P_b&P_3&R'\end{Bmatrix}_{\!\rm Vir}
 =\frac{\delta(R-R')}{\rho_0(R)}.
\label{eq:prismorthogonality}
\end{equation}
The delta function and $1/\rho_0(R)$ form the reference two-point pairing. Applying the identity in both chiral copies leaves a theta pairing times a tetrahedral fourth cumulant. Denote these selected contributions by $B_\theta$ and $\kappa^{\rm tet}$. Their product is a $2+4$ partition of the raw sixth moment. Subtracting it gives the connected-moment equation. The three edges $AF$, $BC$ and $DE$ join the two triangles.

For a joining-edge cut, first contract two known fourth cumulants and substitute the result into \eqref{eq:sourcehierarchy}.

Recall that $\kappa_{ABCD}$ is the connected average of four OPE coefficients, whereas $\kappa_{ABCDEF}$ is the connected average of six.  Joining one coefficient slot from each of two fourth cumulants leaves six slots.  The partition $ABC|DEF$, for example, gives $\kappa_{ABCG}(B^{-1})^{GH}\kappa_{HDEF}$, with the joining normalized by $B_{AG}(B^{-1})^{GH}=\delta_A{}^H$.  Each index labels an entire OPE triple.

This tree contribution joins two cumulants by one inverse pairing. With tetrahedral inputs it produces a prism Wilson graph. Take the inverse on the independent coefficient directions obeying the homogeneous linear constraints, where $B$ is assumed nondegenerate. Summing the ten ways to split the six slots into two unordered triples gives
\begin{equation}
\kappa^{\rm tree}_{i_1\cdots i_6}
 =\sum_{\substack{\{S,S^c\}\\ |S|=3}}
 \kappa_{(i_j)_{j\in S}G}(B^{-1})^{GH}\kappa_{H(i_j)_{j\in S^c}}.
\label{eq:v12treecompletion}
\end{equation}
Here $S$ is a three-element subset of the six positions and $S^c$ its complement, giving $\tfrac12\binom63=10$ unordered partitions. The contracted indices $A,B$ can lie on the same or on different fourth cumulants. Four partitions put them on the same cumulant, with third external index $i=C,D,E,F$. The fourth-order equation then gives
\begin{equation}
\begin{aligned}
r^\lambda_{AB}\kappa_{ABiG}
 &=-r^\lambda_{AB}(B_{Ai}B_{BG}+B_{AG}B_{Bi}),\\
 r^\lambda_{AB}\kappa_{ABiG}(B^{-1})^{GH}\kappa_{Hjkl}
 &=-r^\lambda_{AB}(B_{Ai}\kappa_{Bjkl}+B_{Bi}\kappa_{Ajkl}).
\end{aligned}
\label{eq:v12samevertex}
\end{equation}
Each of these four terms produces two second--fourth contractions, cancelling all eight in the sixth-order equation. Six partitions remain, with $A$ and $B$ on opposite quartic vertices. Let $I$ contain two of $C,D,E,F$ and $I^c$ the other two. Then $\kappa_{AIG}$ denotes the quartic tensor with those four entries. Parentheses denote normalized symmetrization over $C,D,E,F$, so $4B_{A(C}\kappa_{DEF)B}$ is a sum of four terms. The uncancelled part is
\begin{equation}
\begin{aligned}
&r^\lambda_{AB}\left[
 \kappa^{\rm tree}_{ABCDEF}
 +4B_{A(C}\kappa_{DEF)B}+4B_{B(C}\kappa_{DEF)A}\right]\\
 &\qquad=\sum_{\substack{I\subset\{C,D,E,F\}\\ |I|=2}}
 r^\lambda_{AB}\kappa_{AIG}(B^{-1})^{GH}\kappa_{HBI^c}.
\end{aligned}
\label{eq:v12oppositevertex}
\end{equation}
The tree therefore leaves six uncancelled terms.  Write the sixth cumulant as $\kappa_{ABCDEF}=\kappa^{\rm tree}_{ABCDEF}+\Delta\kappa_{ABCDEF}$.  The additional connected contribution must cancel precisely these terms,
\begin{equation}
r^\lambda_{AB}\Delta\kappa_{ABCDEF}
 =-\sum_{\substack{I\subset\{C,D,E,F\}\\ |I|=2}}
 r^\lambda_{AB}\kappa_{AIG}(B^{-1})^{GH}\kappa_{HBI^c}.
\label{eq:v12neededcompletion}
\end{equation}
We now solve this requirement for the prism joining-edge component, keeping the theta pairing and tetrahedral fourth cumulants.  The missing crossed-channel contribution will be the independently evaluated $K_{3,3}$ amplitude.

Crossing a joining edge changes the preparation by one fusion move.  A Whitehead move replaces two adjacent trivalent vertices by the other pairing of their four outer strands.  On the joining edge $P_a'$ the move $\{13\}\to\{15\}$ replaces $\tau_2^{\rm pr}$ by $\tau_2'$.  Gluing the latter to the same observer tree gives $K_{3,3}(P_t)$, with $P_t$ joining the vertices $(P_1,P_5,P_t)$ and $(P_t,P_3,P_b')$.

The old joining label is integrated over and the new label $P_t$ is held fixed.  In the theta normalization the chiral move has kernel $\rho_0(P)$ times a $6j$ symbol.  Composing it with the two factors of the prism amplitude \eqref{eq:prismZ} gives an integral of three $6j$ symbols, conventionally called a $9j$-type expression,
\begin{equation}
\begin{aligned}
&\int_0^\infty\dd P\,\rho_0(P)
 \begin{Bmatrix}P_1&P_3&P\\P_b'&P_5&P_t\end{Bmatrix}_{\!\rm Vir}
 \begin{Bmatrix}P_2&P_4&P_b'\\P_5&P&P_b\end{Bmatrix}_{\!\rm Vir}
 \begin{Bmatrix}P_1&P_3&P\\P_b&P_2&P_a\end{Bmatrix}_{\!\rm Vir}\\
 &\qquad=Z_{\rm Vir}\big(S^3,K_{3,3}(P_t)\big).
\end{aligned}
\label{eq:rungidentity}
\end{equation}
In \eqref{eq:rungidentity}, the last two symbols are \eqref{eq:prismZ} at $P_a'=P$, and the first implements the joining-edge move. Apply this identity in both chiral copies, with reversed framing in the second. Keep the theta inverse pairing and tetrahedral cumulants, and take generic distinct external labels. If $e$ is the joining edge, its unobserved momenta $P,\bar P$ are integrated independently. The two-copy graph amplitudes and the theta-normalized fusion kernel are
\begin{equation}
\begin{aligned}
 \mathbb F_e^{(\theta)}(P_t,P)&=\rho_0(P)
 \begin{Bmatrix}P_1&P_3&P\\P_b'&P_5&P_t\end{Bmatrix}_{\!\rm Vir},\\
 Z_{\rm pr}(P,\bar P)&=Z_{\rm Vir}(S^3,\Gamma_{\rm pr}(P))
                  Z_{\rm Vir}^{\vee}(S^3,\Gamma_{\rm pr}(\bar P)),\\
 Z_{K_{3,3}}(P_t,\bar P_t)
 &=\int\dd P\,\dd \bar P\,\mathbb F_e^{(\theta)}(P_t,P)\mathbb F_e^{(\theta),\vee}(\bar P_t,\bar P)Z_{\rm pr}(P,\bar P)\\
 &=Z_{\rm Vir}(S^3,K_{3,3}(P_t))
                  Z_{\rm Vir}^{\vee}(S^3,K_{3,3}(\bar P_t)).
\end{aligned}
\label{eq:localprismtransport}
\end{equation}
The last equality applies \eqref{eq:rungidentity} in each chiral copy. Each move kernel includes its sewing density $\rho_0$. After the integrations we restrict to scalar labels $\bar P_t=P_t$.

For the prism component, $A,\ldots,F$ now denote the entire OPE triples assigned to the named graph vertices.  Only the partition $\{A,B,D\}\,|\,\{C,E,F\}$ in \eqref{eq:v12treecompletion} contributes, because removing the sewn vertex from either tetrahedron leaves one of the prism's triangles.  The contracted triple is $(2,b,x)$, with momenta $(P,\bar P)$ on $x$.  Writing $K$ for the crossed $K_{3,3}(P_t)$ component, the tree therefore gives
\[
\begin{aligned}
 \kappa^{\rm tree}\big|_{\rm pr}
 &=\kappa^{\rm tet}_{ABDG}(B_\theta^{-1})^{GH}\kappa^{\rm tet}_{HCEF}=Z_{\rm pr},\\
 \kappa^{\rm tree}\big|_K&=0.
\end{aligned}
\]
The second line follows because $K_{3,3}$ contains no triangle.  Thus the tree supplies the prism but no crossed graph, although the prism transforms to $Z_{K_{3,3}}$ by \eqref{eq:localprismtransport}.  Figure~\ref{fig:prismmoves} shows the mechanism.  Sewing two tetrahedra at a vertex produces the two triangles of the prism, and the move on a joining edge opens both triangles into quadrilaterals.  Retain the prism tree coefficient and add the independently computed crossed graph,
\[
 \Delta\kappa\big|_{\rm pr}=0,\qquad
 \Delta\kappa\big|_K=Z_{K_{3,3}}.
\]
Crossing compares the coefficient in the new channel with the transformed coefficient in the old channel.  Substituting the two assignments gives the cancellation explicitly,
\begin{equation}
\begin{aligned}
 &\left.(\kappa^{\rm tree}+\Delta\kappa)\right|_K
 -(\mathbb F_e^{(\theta)}\otimes\mathbb F_e^{(\theta),\vee})
       \left.(\kappa^{\rm tree}+\Delta\kappa)\right|_{\rm pr}\\
 &\quad=\underbrace{0-(\mathbb F_e^{(\theta)}\otimes\mathbb F_e^{(\theta),\vee})Z_{\rm pr}}_{\rm tree}
       +\underbrace{Z_{K_{3,3}}-0}_{\rm added\ graph}\\
 &\quad=-Z_{K_{3,3}}+Z_{K_{3,3}}=0.
\end{aligned}
\label{eq:localprismcompletion}
\end{equation}
Here $B_\theta^{-1}$ inverts only the theta pairing, which pairs identical OPE triples.  For generic distinct edge labels, no external triples coincide, so the second--fourth terms vanish in this component.  The displayed difference is then precisely the joining-edge sixth-order source residual.  Each labelled partition occurs once.  Triangle-edge cuts retain the $B_\theta\kappa^{\rm tet}$ contact term described above.

The same calculation relates the weights of these fillings.  Assign label-independent weights $w_\theta\ne0$ and $w_{\rm tet}$ to the theta and tetrahedral inputs.  The tree contraction $(w_{\rm tet}\kappa^{\rm tet})(w_\theta B_\theta)^{-1}(w_{\rm tet}\kappa^{\rm tet})$ carries $w_{\rm tet}^2/w_\theta$.  In the displayed sector with $\Delta\kappa|_{\rm pr}=0$, let $w_{\rm pr}$ and $w_K$ multiply the prism and crossed graph.  Requiring their channel difference to vanish gives
\begin{equation}
\boxed{\begin{aligned}
 w_{\rm pr}&=\frac{w_{\rm tet}^{\,2}}{w_\theta},\\
 0&=w_KZ_{K_{3,3}}-(\mathbb F_e^{(\theta)}\otimes\mathbb F_e^{(\theta),\vee})(w_{\rm pr}Z_{\rm pr})\\
  &=(w_K-w_{\rm pr})Z_{K_{3,3}},\\
 &\Longrightarrow\quad w_K=w_{\rm pr}=\frac{w_{\rm tet}^{\,2}}{w_\theta}
 \qquad(Z_{K_{3,3}}\not\equiv0).
\end{aligned}}
\label{eq:localweightconstraint}\eqalias{eq:treeweight}
\end{equation}
Unit theta and tetrahedral weights give unit weights for this prism tree term and its crossed partner. Additional fillings enter the same source equation. The $K_{3,3}$ coefficient retains the internal-label integral \eqref{eq:K33Z}, evaluated on the dS cycle in appendix~\ref{ssec:internaledge}.

The remaining joining edges obey the same fusion identity.  On triangle edges the result reduces to a theta pairing and a tetrahedral cumulant, as in \eqref{eq:prismorthogonality}.

For independent left- and right-moving labels, the vertex coefficients are related by $c_{ijk}=[C_0(P_i,P_j,P_k)C_0(\bar P_i,\bar P_j,\bar P_k)]^{1/2}C_{ijk}$, in agreement with \eqref{eq:revnormalized}.

Appendix~\ref{ssec:fiveboundary} realizes these coefficients on the boundary, and appendix~\ref{ssec:generalinternal} evaluates the internal edge of $K_{3,3}$.

\subsection{Five-point boundary extraction}\label{ssec:fiveboundary}

We now extract the prism and $K_{3,3}$ coefficients from the two marked collars of the five-point state. The observer and preparation trees are those of appendix~\ref{ssec:heegaard}. Their two commuting twists give two primary monodromies. Removing the momentum Jacobians and $C_0$ factors yields the graph coefficient entering the sixth-order sewing equation.

The two marked pants curves are disjoint, so their Dehn twists commute.  Near their simultaneous degeneration there is a product plumbing chart $q_j=e^{\ii\pi u_j-\pi y_j}$, $j=a,b$, local to that boundary stratum.  These are local sewing coordinates, and we choose blocks whose leading term is exactly $q_a^{P_a^2}q_b^{P_b^2}$ after the common kinematic prefactors are removed.  The fusion kernel and its Racah conversion use this same normalization, which omits the factors $16^{P^2}$ of the global elliptic representation \eqref{eq:block}.  Write $P=(P_a,P_b)$ and $P'=(P_a',P_b')$ for the observer and preparation labels.  The composed fusion kernel $\mathbb F_{\tau_2\tau_1}(P',P)$, including its prescribed internal integration, is evaluated before applying the two observer projections.

As in \eqref{eq:realdelta}, the marked lift is defined initially at real $b$ and real output momenta.  For left and right descendant levels $m_j,n_j$, the two unfolded twists give
\begin{equation}
 \frac14\int_{\mathbb R^2}\dd u_a\,\dd u_b\,
 e^{\ii\pi\sum_{j=a,b}u_j(P_{Lj}^2+m_j-P_{Rj}^2-n_j)}
 =\prod_{j=a,b}\delta(P_{Lj}^2+m_j-P_{Rj}^2-n_j).
 \label{eq:doublecollartwist}
\end{equation}
The two delta functions are independent and are tested against spectral wavepackets, with Gaussian twist regulators removed in the weak limit.  Each common frequency $\omega_j$ has threshold $\max(m_j,n_j)$.  The primary sector has both thresholds zero, and every other sector has a threshold at least one in at least one variable.

Let $\widehat Z$ be the double radial transform of the $\tau_2$ preparation at fixed $P'$, with observer channels $\tau_1$.  Its primary term in both collars is
\begin{equation}
\begin{aligned}
 \widehat Z^{(0,0)}(\zeta_a,\zeta_b)
 =\int_0^\infty\dd\omega_a\,\dd\omega_b\,
 \frac{\mathbb F_{\tau_2\tau_1}
       (P',\sqrt{\boldsymbol\omega})^2}
      {64\pi^2\omega_a\omega_b}\,
 \frac{e^{-2\pi\sum_{j=a,b}(\omega_j-\zeta_j)Y_j}}
 {(\omega_a-\zeta_a)(\omega_b-\zeta_b)} .
\end{aligned}
\label{eq:doublecollarCauchy}
\end{equation}
Here $\sqrt{\boldsymbol\omega}=(\sqrt{\omega_a},\sqrt{\omega_b})$.
The two chiral momentum Jacobians give $1/(4\omega_j)$ per collar,
and each radial Laplace transform gives $1/[2\pi(\omega_j-\zeta_j)]$.
The cutoffs are $y_j>Y_j$.  The equation is understood with common
spectral regulators where needed, initially with
$\operatorname{Re}\zeta_j<0$.

Work in a generic chamber where the primary frequency density is holomorphic near $(0,0)$ and the regulated descendant expansion converges locally uniformly and absolutely on the product cusp.  Choose a small disk around each $\zeta_j=0$, with radius less than one and away from input-kernel poles.  The double local monodromy isolates the coefficient of $\log(-\zeta_a)\log(-\zeta_b)$ in \eqref{eq:doublecollarCauchy}, because every descendant term is single-valued in at least one of these disks.  With the convention of \eqref{eq:monodromyexactdisc},
\begin{equation}
\begin{aligned}
 \Disc_{0,a}\Disc_{0,b}\widehat Z
 &=\Disc_{0,b}\Disc_{0,a}\widehat Z\\
 &=-\frac{\mathbb F_{\tau_2\tau_1}(P',P)^2}
          {16P_a^2P_b^2},
 \qquad \zeta_j=P_j^2 .
\end{aligned}
\label{eq:doublecollarmonodromy}
\end{equation}
The equality holds for $\zeta_j=P_j^2$ in the chosen polydisk, and the isolated coefficient then continues meromorphically. At finite regulator the product contour permits either monodromy order. Uniform convergence on these contours permits removal of the regulators. Applying $(2P_a/\ii)(2P_b/\ii)$ gives
\begin{equation}
\rho_{\tau_1\tau_2}(P;P')
 =\frac{\mathbb F_{\tau_2\tau_1}(P',P)^2}{4P_aP_b},
\label{eq:fivecrossedcoef}
\end{equation}
The square pairs the two chiral changes of channel.  Only this extracted meromorphic coefficient is then continued to the dS ray.  If poles cross an internal contour during continuation, its transported residues must be included before comparing kernels.  The pentagon identity equates different sequences implementing the same framed change of decomposition with these prescriptions \cite{CMMT,CTV}, under the local analyticity assumptions of \eqref{eq:doublecollarmonodromy}.

The reference density is the five-point version of \eqref{eq:rhos}.  There is one Plancherel factor for each sewn curve and one three-point factor for each pair of pants,
\begin{equation}
\begin{aligned}
\rho_{\tau_1}(P)
={}&\rho_0(P_a)\rho_0(P_b)C_0(P_1,P_2,P_a)\\
&\times C_0(P_a,P_3,P_b)C_0(P_b,P_4,P_5).
\end{aligned}
\label{eq:s4fivereferencedensity}
\end{equation}
Converting each fusion kernel by \eqref{eq:revfusion} cancels the three-point factors on intermediate decompositions. The factors on the observer and preparation trees remain,
\begin{equation}
 \mathbb F_{\tau_2\tau_1}(P',P)
 =\rho_0(P_a)\rho_0(P_b)
 \left[\frac{\prod_{v\in V(\tau_1)}C_0(v)}
 {\prod_{w\in V(\tau_2)}C_0(w)}\right]^{1/2}
 Z_{\rm Vir}(S^3,\Gamma_{\tau_1\tau_2}).
 \label{eq:fiveRacahconversion}
\end{equation}
Here $C_0(v)$ is evaluated on the three labels meeting at $v$. Substitute this expression in \eqref{eq:fivecrossedcoef} and divide by \eqref{eq:s4fivereferencedensity}. Removing the preparation flares gives
\begin{equation}
\boxed{\frac{2P_a}{\rho_0(P_a)}\frac{2P_b}{\rho_0(P_b)}
 \prod_{w\in V(\tau_2)}C_0(w)
 \frac{\rho_{\tau_1\tau_2}(P;P')}{\rho_{\tau_1}(P)}
 =Z_{\rm Vir}(S^3,\Gamma_{\tau_1\tau_2})^2 .}
\label{eq:fivenormalized}
\end{equation}
The factors $2P_a$ and $2P_b$ undo the Jacobians in \eqref{eq:fivecrossedcoef}. The two observer densities cancel against $1/\rho_0$, and the three preparation factors cancel against $\prod_w C_0(w)$. This leaves the two-copy graph coefficient assigned to the six-OPE contraction in \eqref{eq:s4prismproduct}.

The marked extraction has now reproduced the graph coefficients assigned to the six-OPE contractions.  We next evaluate the internal-edge integral that distinguishes the $K_{3,3}$ coefficient.

\Needspace{9\baselineskip}
\subsection{Continued CTV at sixth order and a sphere-tree extension}\label{ssec:sphere-tree-extension}
\begin{voneblock}
\paragraph{From five-point extraction to CTV.}
The five-point extraction \eqref{eq:fivenormalized} gives $Z_L=Z_{\rm Vir}^2$ for the prism and $K_{3,3}$ in our scalar, braid-free convention. The same external $S$ transform as for the tetrahedron converts their nine edge labels $P_e$ to $R_e$ \cite{Hartman,CTV},
\begin{equation}
\begin{aligned}
 Z_A(M,\gamma(R_1,\ldots,R_9))
 &=\int_{\mathbb R_+^9}\prod_{e=1}^{9}
 [\dd P_e\,S_{R_eP_e}]\,Z_L(M,\gamma(P_1,\ldots,P_9))\\
 &=Z_{\rm CTV}(S^3,\Gamma(R_1,\ldots,R_9)),
 \qquad \Gamma=\Gamma_{\rm pr},K_{3,3},\quad b>0 .
\end{aligned}
\label{eq:sixthCTVcorollary}\eqalias{eq:sixthCTVcont}
\end{equation}
Here $M=S^3\setminus\operatorname{int}N(\Gamma)$. Continuing the whole transform to the dS ray gives the fixed angles $\psi_e=-2\pi\ii bR_e$.

\paragraph{The general sphere-tree coefficient.}
The same normalization works for an $n$-point sphere. Choose observer and preparation trees $\tau,\sigma$, with $n\ge4$, fixed external primaries and no braiding. Number their internal cuffs by $j=1,\ldots,n-3$ and collect the momenta $P_j,P'_j$ in $\boldsymbol P,\boldsymbol P'$. Joining corresponding leaves gives $\Gamma_{\tau\sigma}$. There is one Plancherel factor per internal cuff and one $C_0$ per pants,
\begin{equation}
 \rho_\tau(\boldsymbol P)
 =\prod_{j=1}^{n-3}\rho_0(P_j)
  \prod_{v\in V(\tau)}C_0(v).
 \label{eq:allntreereference}
\end{equation}
Here $V(\tau)$ contains the $n-2$ pants vertices, and $C_0(v)$ uses their incident labels. Use blocks with leading term $\prod_{j=1}^{n-3}q_j^{P_j^2}$ and the regulated twists and local uniform convergence of appendix~\ref{ssec:fiveboundary} on all collars and monodromy contours.

Each twist integral equates the two chiral frequencies. Taking the primary monodromy in every collar removes descendants and leaves one momentum Jacobian per cuff,
\begin{equation}
 \rho_{\tau\sigma}(\boldsymbol P;\boldsymbol P')
 =\frac{\mathbb F_{\sigma\tau}(\boldsymbol P',\boldsymbol P)^2}
 {\prod_{j=1}^{n-3}2P_j}.
 \label{eq:allnprimaryprojection}
\end{equation}
Applying \eqref{eq:revfusion} to each fusion move cancels the $C_0$ factors of intermediate decompositions. Only the observer densities and the two endpoint trees remain,
\begin{equation}
 \mathbb F_{\sigma\tau}(\boldsymbol P',\boldsymbol P)
 =\left(\prod_{j=1}^{n-3}\rho_0(P_j)\right)
 \left[\frac{\prod_{v\in V(\tau)}C_0(v)}
 {\prod_{w\in V(\sigma)}C_0(w)}\right]^{1/2}
 Z_{\rm Vir}(S^3,\Gamma_{\tau\sigma}).
 \label{eq:allnRacahconversion}
\end{equation}
Squaring this expression, substituting it in \eqref{eq:allnprimaryprojection}, and dividing by \eqref{eq:allntreereference} gives
\begin{equation}
\boxed{
 \left(\prod_{j=1}^{n-3}\frac{2P_j}{\rho_0(P_j)}\right)
 \left(\prod_{w\in V(\sigma)}C_0(w)\right)
 \frac{\rho_{\tau\sigma}(\boldsymbol P;\boldsymbol P')}
 {\rho_\tau(\boldsymbol P)}
 =Z_{\rm Vir}(S^3,\Gamma_{\tau\sigma})^2 .}
\label{eq:allntreecoefficient}
\end{equation}
The $2P_j$ factors cancel the Jacobians, while $1/\rho_0(P_j)$ and the preparation $C_0$ factors remove the remaining normalizations. For $n=4,5$ this is \eqref{eq:mainratio} and \eqref{eq:fivenormalized}. The pentagon identity makes the result independent of the fusion sequence at fixed framing. With general framing the square is replaced by $Z_{\rm Vir}Z_{\rm Vir}^{\vee}$.

The joined graph has $3n-6$ edges. Transforming all their labels to $\boldsymbol R=(R_1,\ldots,R_{3n-6})$ as above gives
\begin{equation}
 Z_A(M,\gamma(\boldsymbol R))
 =Z_{\rm CTV}(S^3,\Gamma_{\tau\sigma}(\boldsymbol R)),
 \qquad b>0 .
 \label{eq:allntreeCTV}
\end{equation}
with $M=S^3\setminus\operatorname{int}N(\Gamma_{\tau\sigma})$.
\end{voneblock}

\begin{vtwoblock}
\paragraph{Reassembling the sphere state.}
To recover the sphere wavefunction, restore its conformal blocks and integrate the channel momenta. At real $b$, all these integrals run over positive momenta. Fix the puncture labels and the $n-3$ moduli $\boldsymbol m$. Changing the first block from $\sigma$ to $\tau$ gives
\begin{equation}
\begin{aligned}
\Psi_{0,n}^{(b)}(\boldsymbol m,\bar{\boldsymbol m})
 &=\int\dd^{\,n-3}\boldsymbol P'\,
   \rho_\sigma(\boldsymbol P')\,
   \mathcal F_{\sigma,\boldsymbol P'}(\boldsymbol m)
   \mathcal F_{\sigma,\boldsymbol P'}(\bar{\boldsymbol m})\\
 &=\int\dd^{\,n-3}\boldsymbol P\,\dd^{\,n-3}\boldsymbol P'\,
   \sqrt{\rho_\tau(\boldsymbol P)\rho_\sigma(\boldsymbol P')}\,
   \widehat{\mathbb F}_{\sigma\tau}(\boldsymbol P',\boldsymbol P)\,
   \mathcal F_{\tau,\boldsymbol P}(\boldsymbol m)
   \mathcal F_{\sigma,\boldsymbol P'}(\bar{\boldsymbol m}).
\end{aligned}
\label{eq:spheretreereassembly}
\end{equation}
As for the four-point cylinder, $\widehat{\mathbb F}_{\sigma\tau}=\sqrt{\rho_\sigma/\rho_\tau}\,\mathbb F_{\sigma\tau}$. The arguments are those of \eqref{eq:spheretreereassembly}. Substituting \eqref{eq:allnRacahconversion} gives
\[
 \widehat{\mathbb F}_{\sigma\tau}(\boldsymbol P',\boldsymbol P)
 =\prod_{j=1}^{n-3}\sqrt{\rho_0(P_j)\rho_0(P'_j)}\,
 Z_{\rm Vir}(S^3,\Gamma_{\tau\sigma}).
\]
Thus the graph coefficient supplies the change of channel, while $\sqrt{\rho_\tau\rho_\sigma}$ restores the $C_0$ factors needed for the state. Pairing with the reflected cylinder gives $Z_{\rm Vir}Z_{\rm Vir}^{\vee}$ in the statistic.

For $K_{3,3}$ the two copies retain independent $X$ integrals from \eqref{eq:K33Z}. The puncture labels remain fixed here and are integrated only when the sphere components are sewn together below.

\paragraph{The geometric sphere limit.}
The two sphere metrics agree after pinching the joining annuli and then flattening the internal bends. For $M\simeq F\times I$ in \eqref{eq:spheretreeproductboundary}, assign bending $\pi-\varepsilon$ to the joining curves and $\theta w_e$ to the internal cuffs, with $w_e>0$ fixed. The pants curves must fill the sphere, so every essential closed curve intersects at least one of them.

The total bend around an essential compressing disk must exceed $2\pi$ \cite{BonahonOtal}. A disk crossing the joining curves twice gives matching essential arcs on the two spheres. An arc avoiding both pants decompositions would yield an essential closed curve avoiding both, by taking a neighbourhood of the arc and its endpoint circles. Filling excludes this, giving at least
\begin{equation}
2(\pi-\varepsilon)+\theta\min_e w_e>2\pi,
\qquad 2\varepsilon<\theta\min_e w_e.
\label{eq:spheretreediskbound}
\end{equation}
Four or more crossings give $4(\pi-\varepsilon)>2\pi$ for $\varepsilon<\pi/2$. Essential annuli avoiding the joining curves meet the filling pants curves. Together with $\theta\max_e w_e<\pi$, this supplies the remaining bending conditions \cite{BonahonOtal}.

At fixed $\theta$, the limit $\varepsilon\to0$ pinches the joining curves to cusps \cite{BonahonOtal,LecuireBending}. Next $\theta\to0$ makes the two components approach the same complete hyperbolic metric. It is the unique minimum of the weighted cuff length \cite{SeriesBending},
\begin{equation}
 \sum_{\text{internal cuffs }e}
 w_e\,\ell_h(\gamma_e).
 \label{eq:spheretreelengthminimum}
\end{equation}
The sum includes the $n-3$ cuffs of each tree, and $h$ ranges over marked complete finite-area metrics of curvature $-1$ on $\Sigma_{0,n}$. For $K_4$, \eqref{eq:matchedspherepath} also controls a simultaneous limit. The prism and $K_{3,3}$ pairs in \eqref{eq:threepreparations} fill the five-punctured sphere because their pants decompositions have no common curve.

\paragraph{The marked tetrahedral face pairings.}\label{app:markedfacepairings}
The $K_4$ reconstruction also requires the cells to meet at the same seam endpoints. Use the normals $U_i,W_k$ of \eqref{eq:lorentzInverseDual}. An edge between original faces $i,j$ ends on the two remaining truncation planes $k,l$. Its endpoints are
\begin{equation}
Y_k=\pm\frac{W_l+\cosh\ell_e\,W_k}{\sinh\ell_e},\qquad
Y_l=\pm\frac{W_k+\cosh\ell_e\,W_l}{\sinh\ell_e},
\label{eq:pleatedseamfeet}
\end{equation}
with the same sign chosen so that $Y_k^0,Y_l^0>0$. Inserting $\eta(W_k,W_l)=-\cosh\ell_e$ gives $\eta(Y_k,W_k)=\eta(Y_l,W_l)=0$ and $-\eta(Y_k,Y_l)=\cosh\ell_e$. Both points also lie on the faces normal to $U_i,U_j$. Thus the recovered faces share the same endpoints and edge length.

Reflection in the $k$th truncation plane is
\begin{equation}
R_k(Z)=Z-2\eta(Z,W_k)W_k,\qquad R_kU_i=U_i\quad(i\ne k).
\label{eq:pleatedfourreflections}
\end{equation}
Since $\eta(W_k,U_i)=0$ for $i\ne k$, this reflection preserves the three incident faces. Gluing the two truncation triangles joins their hexagons along matching seams. The four reflections therefore align the endpoints of all six cuffs.

On each incident pants plane, $R_kX_i(t,Y)=X_i(t,R_kY)$. The product $R_lR_k$ fixes $U_i,U_j$ and translates their common axis by $2\ell_e$. Choosing the positive $x$ direction accordingly gives
\begin{equation}
R_lR_kX_e(t,x,\varphi)=X_e(t,x+2\ell_e,\varphi),\qquad
\operatorname{rapidity}(R_lR_k)=2\ell_e=\lambda_e .
\label{eq:pleatedboostpairing}
\end{equation}
Because $R_l^2=1$, successive identifications obey $(R_mR_l)(R_lR_k)=R_mR_k$. The same four face pairings therefore recover both the doubled cell and its Lorentzian boost identifications.

\paragraph{Sewing the two sphere components.}
Sewing the two spheres integrates their $n$ puncture labels with $\rho_0(P_i)\dd P_i$ from \eqref{eq:twopoint}. At real $b$, let $q_i,\bar q_i$ describe the joining cylinders. A scalar primary propagates with
\begin{equation}
q_i^{h(P_i)-c/24}\bar q_i^{h(P_i)-c/24}
 =(q_i\bar q_i)^{P_i^2-1/24},
\label{eq:joinedprimarypropagator}
\end{equation}
since $h(P)-c/24=P^2-1/24$. Keeping primary propagation in all joining cylinders gives
\begin{equation}
\begin{aligned}
\left.\Psi_{n-1,0}^{(b)}\right|_{\rm primary}
={}&\int_{\mathbb R_+^n}
\prod_{i=1}^n\big[
 \dd P_i\,\rho_0(P_i)
 (q_i\bar q_i)^{P_i^2-1/24}\big]\\
&\times\Psi_{0,n}^{(b)}
 (P_1,\ldots,P_n;\boldsymbol m_\tau,\bar{\boldsymbol m}_\tau)\,
 \Psi_{0,n}^{(b)}
 (P_1,\ldots,P_n;\boldsymbol m_\sigma,\bar{\boldsymbol m}_\sigma).
\end{aligned}
\label{eq:joinedspheresprimary}
\end{equation}
Here $P_i$ are the puncture momenta, and $\boldsymbol m_\tau,\boldsymbol m_\sigma$ are the component moduli. The other sewing terms propagate descendants \cite{CEZ,CEM}. At fixed $P_i$, removing the displayed propagators and densities leaves the two sphere states, including their $C_0$ factors. Mirrored moduli $\boldsymbol m_\sigma=\bar{\boldsymbol m}_\tau$ give the same-$b$ square of \eqref{eq:statisticalstatepair}.

Continuation transports all external and internal contours and includes crossed-pole residues. The geometric pinching limit can be taken through these full integrals only with uniform tail bounds and convergent residue terms. This remaining quantum limit is discussed in section~\ref{sec:discussion}.
\end{vtwoblock}

\FloatBarrier
\section[The internal edge of K33]{The internal edge of $K_{3,3}$}\label{ssec:generalinternal}\label{app:contour}
The internal momentum integral in \eqref{eq:K33Z} selects a smooth gluing when the three cell angles around the edge sum to $2\pi$. We compute the saddle length and its prefactor for equal external angles. The exact outer contour rotation is proved in appendix~\ref{app:contourC1}, and the uniform cell estimates used in the saddle expansion are stated in appendix~\ref{app:generalmixed6j}.

\subsection{Mixed-cell geometry and the gluing equation}

The three cells in \eqref{eq:K33Z} share one integrated label. Write its length as $L=4\pi b\widetilde P$. On the real-length contour obtained below, the external and internal labels are
\begin{equation}
P_i=\frac{\ii\psi_i}{2\pi b},\qquad
 \widetilde P=\frac{L}{4\pi b}.
\label{eq:generalmixedlabels}
\end{equation}
Here $L$ has the cuff normalization of \eqref{eq:revlabels}, so the geometric segment in each cell has length $L/2$. A cell with both angle and length labels is a mixed tetrahedron. We use a real, deeply truncated realization with the face incidences of \cite{BellettiYang}.

Let $\vartheta_e^{(\Delta)}$ be the cell angle conjugate to $L$, and retain $\ell_i^{\rm cell}$ for the geometric lengths on its angle edges, as in \eqref{eq:revcellschlafli}. Schläfli's formula and its Legendre transform give
\begin{equation}
\begin{aligned}
\dd V_\Delta
 &=-\frac12\sum_i\ell_i^{\rm cell}\,\dd\psi_i
   -\frac L4\,\dd\vartheta_e^{(\Delta)},\\
\dd\left(V_\Delta+\frac L4\vartheta_e^{(\Delta)}\right)
 &=-\frac12\sum_i\ell_i^{\rm cell}\,\dd\psi_i
   +\frac14\vartheta_e^{(\Delta)}\,\dd L .
\end{aligned}
\label{eq:mixedgeneratingdiff}
\end{equation}
The factor $1/4$ on the length edge follows from its segment length $L/2$. With the uniform Barnes estimates of appendix~\ref{app:generalmixed6j}, this mixed potential gives
\begin{equation}
\begin{Bmatrix}P_1&P_2&P_3\\P_4&P_5&P_6\end{Bmatrix}_{\!\rm Vir}
 =\frac{1}{2[-\det G]^{1/4}}
   \exp\!\left[\frac{\ii}{\pi r^2}
   \left(V_\Delta+\frac L4\vartheta_e^{(\Delta)}\right)\right]
   \bigl(1+O(r^2)\bigr),
 \qquad b=re^{\ii\pi/4}.
\label{eq:generalmixed6j}
\end{equation}
A pair of intersecting original faces gives the Gram entry $-\cos\psi_i$. The length entry is $-\cosh(L/2)$ for two disjoint faces separated by a common perpendicular of length $L/2$. The other two truncation planes meet along this perpendicular and determine its dihedral angle.

The angle along this new edge is computed using the normals to the two truncation planes.  Write $C_{rs}$ for the cofactors of $G$.  If $n_s$ are the original unit face normals, the vector normal to the truncation plane opposite face $r$ is
\begin{equation}
v_r=\sum_s(G^{-1})_{rs}\,n_s,\qquad
 \langle v_r,n_t\rangle=\delta_{rt},\qquad
 \langle v_r,v_s\rangle=\frac{C_{rs}}{\det G} .
\label{eq:dualvectors}
\end{equation}
For the length edge $e$, choose $r,s$ complementary to the indices of the Gram entry containing $L$. The normalized polar vectors give the cosine of its dihedral angle. Jacobi's identity, $C_{rr}C_{ss}-C_{rs}^2=\sinh^2(L/2)(-\det G)$, fixes the positive sine on $0<\vartheta_e<\pi$,
\begin{equation}
\cos\vartheta_e=\frac{C_{rs}}{\sqrt{C_{rr}C_{ss}}},\qquad
 \sin\vartheta_e=\frac{\sinh(L/2)\sqrt{-\det G}}{\sqrt{C_{rr}C_{ss}}} .
\label{eq:mixedangle}
\end{equation}
The rotated Plancherel measure supplies two phases $e^{\mp\ii L/(2r^2)}$, derived in \eqref{eq:v12densitybranches}. Combine the first with the three cell weights. The function multiplying $\ii/(\pi r^2)$ in the exponent is
\begin{align}
\sum_{a=1}^{3}V_a
 +\frac L4\left(\sum_{a=1}^{3}\vartheta_e^{(a)}-2\pi\right).
\label{eq:generalPhi}
\end{align}
At fixed external angles, \eqref{eq:mixedgeneratingdiff} gives
\begin{equation}
\frac{\dd}{\dd L}\left[\sum_{a=1}^{3}
 \left(V_a+\frac L4\vartheta_e^{(a)}\right)-\frac\pi2L\right]
 =\frac14\left(\sum_{a=1}^{3}\vartheta_e^{(a)}-2\pi\right).
\label{eq:Phiderivative}
\end{equation}
Setting this to zero gives
\begin{equation}
\sum_{a=1}^{3}\vartheta_e^{(a)}=2\pi .
\label{eq:generalsmoothness}
\end{equation}
This is the zero-deficit condition that the cells fill a full angle around the internal edge \cite{Hartman}. The other Plancherel phase replaces $-2\pi$ by $+2\pi$. Positive cell angles then exclude a stationary point on the same real branch. Thus the two equal-modulus dS phases select different stationary equations, and only the first admits this smooth gluing.
At the saddle the length terms cancel,
\begin{equation}
\left.\left[
  \sum_{a=1}^{3}V_a
  +\frac L4\left(\sum_{a=1}^{3}\vartheta_e^{(a)}-2\pi\right)
 \right]\right|_{L=L_*}
  =\sum_{a=1}^{3}V_a
 =\operatorname{Vol}(M),
\label{eq:generalvolumeonshell}
\end{equation}
For the stated geometric gluing, the stationary value is the volume of the three-cell core. We next solve for its internal length on the equal-angle slice.

\subsection[The symmetric K33 saddle]{The symmetric $K_{3,3}$ saddle}\label{ssec:internaledge}\label{ssec:internalsaddle}

The equal-angle coefficient has one outer representation integral and three Barnes integrals defining its cell weights. We first rotate the outer contour to real length, then compute its geometric stationary point. The cell saddle is selected by the transported Barnes contours \cite{WittenPL} under the estimates stated in appendix~\ref{app:generalmixed6j}.
Set all external labels equal, $P_0=\ii\psi/(2\pi b)$ with $0<\psi<\pi/3$.  One chiral copy of the $K_{3,3}$ amplitude \eqref{eq:K33Z} then has three identical cells around one internal edge of label $X$, and its integrand is
\begin{equation}
\mathcal W(X)=\begin{Bmatrix}P_0&P_0&P_0\\P_0&P_0&X\end{Bmatrix}_{\!\rm Vir}^{\!3}.
\label{eq:v12symmetricK}
\end{equation}
For the tetrahedral graph, the external $S$ transform returns the same $6j^2$ with relabelled opposite edges, giving the two-cell CTV core of \eqref{eq:doubletet}.  Here we analyze one chiral factor of the fixed-length $K_{3,3}$ coefficient.  Its fixed-angle CTV amplitude is obtained by pairing with the reversed-framing factor and then applying the external $S$ transform.  The calculation below concerns the chiral integral before that transform.
The outer contour can be rotated to the ray on which $L=4\pi bX$ is real.  Section~\ref{app:v12arc} places all possible $X$ poles outside the swept wedge and bounds the large arc by a polynomial times $e^{-\pi rR/\sqrt2}$.  Hence, at fixed $b=re^{\ii\pi/4}$,
\begin{equation}
\begin{aligned}
&\int_0^\infty\dd X\,\rho_0(X)\mathcal W(X)\\
&\qquad=e^{-\ii\pi/4}\int_0^\infty\dd x\,
 \rho_0(e^{-\ii\pi/4}x)\mathcal W(e^{-\ii\pi/4}x).
\end{aligned}
\label{eq:v12rotationproved}
\end{equation}
On the original contour $X>0$ the length $L=4\pi bX$ runs along $e^{\ii\pi/4}\mathbb R_+$.  The pole-free rotation puts the outer integral on the real-length ray, where the candidate zero-deficit length can be a stationary point.  On this ray $L=4\pi bX>0$, and the Plancherel measure becomes
\begin{equation}
\rho_0(X)
 =4\sqrt2\sinh\frac{L}{2}\,\sinh\frac{L}{2b^2}
 =2\sqrt2\sinh\frac L2\left(e^{-\ii L/(2r^2)}-e^{\ii L/(2r^2)}\right).
\label{eq:v12densitybranches}
\end{equation}
The two exponentials have equal modulus. On the real geometric branch, $e^{-\ii L/(2r^2)}$ gives the zero-deficit equation \eqref{eq:generalPhi}. The other sign has no positive-length geometric saddle.

For the symmetric cell, the cofactor formula \eqref{eq:mixedangle} gives
\begin{equation}
 \cos\vartheta_X(L)
 =\frac{\cos\psi\,[2\cos\psi+1-\cosh(L/2)]}
        {2\cos^2\psi+\cosh(L/2)-1}.
 \label{eq:symmetrictheta}
\end{equation}
Differentiating $-\sin\vartheta_X\,\vartheta_X'=(\cos\vartheta_X)'$ yields
\begin{equation}
 \vartheta_X'(L)
 =\frac{\cos^2\psi(1+\cos\psi)\sinh(L/2)}
        {\sin\vartheta_X(L)\,[2\cos^2\psi+\cosh(L/2)-1]^2}>0
 \qquad(L>0).
 \label{eq:symmetricthetaderivative}
\end{equation}
The endpoint values $\vartheta_X(0)=0$ and $\vartheta_X(\infty)=\pi-\psi$ follow from $\cosh(L/2)=1$ and $\cosh(L/2)\to\infty$.
For three cells there is one finite real saddle exactly when $3(\pi-\psi)>2\pi$, that is $0<\psi<\pi/3$, and it is characterized by
\begin{equation}
3\vartheta_X(L_*)=2\pi .
\label{eq:v12smoothness}
\end{equation}
Since $\cos(2\pi/3)=-\frac12$, inserting \eqref{eq:symmetrictheta} gives a linear equation for $\cosh(L_*/2)$,
\begin{equation}
\begin{aligned}
2\cos\psi\,\big[2\cos\psi+1-\cosh(L_*/2)\big]
 &=-\big[2\cos^2\psi+\cosh(L_*/2)-1\big],\\
 \cosh\frac{L_*}{2}
 &=1+\frac{6\cos^2\psi}{2\cos\psi-1},
\end{aligned}
\label{eq:Lstar}
\end{equation}
which is finite and positive for $0<\psi<\pi/3$. The three cells leave an angular deficit for $L<L_*$, meet smoothly at $L=L_*$, and have total angle greater than $2\pi$ for $L>L_*$. The length Hessian is $H_*=3\vartheta_X'(L_*)/4>0$.

\phantomsection\label{ssec:reduction}

Choose a smooth length cutoff supported in a closed window $W$ containing $L_*$ and equal to one near the saddle. On $W$, assume the cell asymptotics and derivative bounds of appendix~\ref{app:generalmixed6j} hold uniformly, with the strict contour condition \eqref{eq:certificate-minimum}. Denote this cutoff contribution by $Z_{\rm Vir}(S^3,K_{3,3})|_W$. With $G_*=G(L_*)$, stationary phase gives
\begin{equation}
\left.Z_{\rm Vir}(S^3,K_{3,3})\right|_W
 =\frac{\sinh(L_*/2)}{\sqrt{H_*}}
 \frac{1}{8[-\det G_*]^{3/4}}
 \exp\!\left[\frac{\ii}{\pi r^2}\operatorname{Vol}(M)\right]
 \bigl(1+O(r^2)\bigr).
\label{eq:oneedge3nj}
\end{equation}
The stationary phase is \eqref{eq:generalvolumeonshell}, and each cell supplies its Gram prefactor.  The remaining normalization combines $\dd X=\dd L/(4\pi b)$, the selected Plancherel term and the length Gaussian,
\begin{equation}
\frac{1}{4\pi b}\cdot2\sqrt2\sinh\frac{L_*}{2}\cdot\frac{\pi r\sqrt2\,e^{\ii\pi/4}}{\sqrt{H_*}}
 =\frac{\sinh(L_*/2)}{\sqrt{H_*}},
\label{eq:prefactorcancel}
\end{equation}
since $b=re^{\ii\pi/4}$. The expansion evaluates the contribution of $W$ to the exact rotated integral.

For the symmetric $K_{3,3}$ integral the determinant and Hessian in this
local contribution can be simplified completely.  Writing
$x=\cosh(L/2)$ only in the following algebra, the Gram determinant is
\begin{equation}
 \det G=(1+\cos\psi)(x+1)
 \big[(\cos\psi-1)(x-1)-4\cos^2\psi\big] .
 \label{eq:k33_det_polynomial_v46}
\end{equation}
Substituting \eqref{eq:Lstar} in the determinant and in the explicit derivative \eqref{eq:symmetricthetaderivative} gives
\begin{align}
 -\det G(L_*)&=
 \frac{4\cos^2\psi(1+\cos\psi)^3(3\cos\psi-1)}{(2\cos\psi-1)^2},
 \label{eq:k33_saddle_det_v46}\\
 H_*&=\frac34\vartheta_X'(L_*)
 =\frac{3(2\cos\psi-1)}{16\cos\psi}
 \sqrt{\frac{3\cos\psi-1}{1+\cos\psi}} .
 \label{eq:k33_saddle_hessian_v46}
\end{align}
Both are strictly positive for $0<\psi<\pi/3$.  At the saddle the length
term cancels between the three cells and the Plancherel measure,
\begin{equation}
 3\left(V_\Delta+\frac14L_*\vartheta_X(L_*)\right)
 -\frac\pi2L_*=3V_\Delta,
 \qquad \vartheta_X(L_*)=\frac{2\pi}{3} .
 \label{eq:k33_onshell_cancel_v46}
\end{equation}
Using these expressions in \eqref{eq:oneedge3nj} gives
\begin{equation}
 \boxed{\left.Z_{\rm Vir}(S^3,K_{3,3})\right|_W
 =\frac{\exp\!\big[3\ii V_\Delta(\psi,L_*)/(\pi r^2)\big]}
 {2\sqrt2(1+\cos\psi)^{3/2}\sqrt{3\cos\psi-1}}
 \bigl(1+O(r^2)\bigr) .}
 \label{eq:k33_closed_prefactor_v46}
\end{equation}
With the contour and window hypotheses of \eqref{eq:oneedge3nj}, this formula evaluates the selected saddle contribution to the single chiral integral \eqref{eq:K33Z}.  Its phase is the three-cell volume, and its finite prefactor combines the cell fluctuations with the fluctuation of the internal length.  The zero-deficit equation makes that internal edge smooth.

\subsection{Barnes kernels and the outer contour}\label{app:contourC1}

The Virasoro $6j$ is meromorphic in its labels and is represented by a Barnes integral \cite{Hartman,TV}.  With $Q=b+b^{-1}$ and $S_b(z)=\Gamma_b(z)/\Gamma_b(Q-z)$, the poles of $\Gamma_b$ occur at $z=-mb-nb^{-1}$, $m,n\ge0$,
\begin{equation}
 S_b(z)S_b(Q-z)=1,\qquad
 \operatorname{Poles}(\Gamma_b)=\{-mb-nb^{-1}:m,n\ge0\}.
\label{eq:Barnes}
\end{equation}
Our normalization is
\begin{align}
\begin{Bmatrix}P_1&P_2&P_3\\P_4&P_5&P_6\end{Bmatrix}_{\!\rm Vir}
 &=V_{123}V_{156}V_{246}V_{345}
 \int_{-\ii Q/4+\mathbb R}\dd u\,
 \frac{S_b(\ii u-\ii P_{123})S_b(\ii u-\ii P_{156})}
 {S_b(Q/2+\ii u-\ii P_{1245})S_b(Q/2+\ii u-\ii P_{1346})}
 \nonumber\\[-2pt]
 &\quad\times
 \frac{S_b(\ii u-\ii P_{246})S_b(\ii u-\ii P_{345})}
 {S_b(Q/2+\ii u-\ii P_{2356})S_b(Q/2+\ii u)},
 \label{eq:exact6j}\\
 V_{ijk}&=\left[\prod_{\sigma=\pm1}
 S_b\!\left(\tfrac Q2+\ii P_k+\ii\sigma(P_i+P_j)\right)
 S_b\!\left(\tfrac Q2-\ii P_k+\ii\sigma(P_i-P_j)\right)
 \right]^{1/2}.
 \label{eq:vertexfactors}
\end{align}
Here $P_{ijk}$ and $P_{ijkl}$ denote sums of the indicated labels.  The square roots are fixed in the real-$b$ starting normalization and transported with the contour.  The contour separates the upward and downward pole lattices and is transported with residues when a pole crosses it.

The label-dependent singularities relevant to the equal-external-label sphere fusion are contained in
\begin{equation}
 P_{\rm pole}=\pm\ii(Q/2+mb+nb^{-1})+2\sigma P_0,
 \qquad m,n\ge0,\quad\sigma=0,\pm1.
\label{eq:fusionpoles}
\end{equation}
For $P_0=\ii\psi_0/(2\pi b)$ with $0<\psi_0<\pi/2$ and $b=re^{\ii\beta}$, $0\le\beta\le\pi/4$, these families stay off the rotated positive output ray.  To compare them, multiply both the pole locations and the ray by $b$.  In the $bP$ plane the upper family has negative real part for $\beta>0$, while the lower family has negative imaginary part.  The real-angle chamber of \eqref{eq:externalchamber} avoids the nearest pole.  These estimates apply to the extracted primary coefficient, following the order of operations in section~\ref{sec:crossed}.

\phantomsection\label{app:v12arc}
We now verify the pole separation and large-arc estimate used in \eqref{eq:v12rotationproved}.
Fix $b=re^{\ii\pi/4}$.  For the equal-label $K_{3,3}$ integrand \eqref{eq:v12symmetricK}, the chiral cell factor is
\begin{equation}
\begin{aligned}
\begin{Bmatrix}P_0&P_0&P_0\\P_0&P_0&X\end{Bmatrix}_{\!\rm Vir}
 &=V_{000}^{2}V_{00X}^{2}\int\dd u\,B(u,X),\\
 B(u,X)&=\frac{S_b(\ii u-3\ii P_0)^2S_b(\ii u-2\ii P_0-\ii X)^2}
 {S_b(Q/2+\ii u-4\ii P_0)
 S_b(Q/2+\ii u-3\ii P_0-\ii X)^2S_b(Q/2+\ii u)}.
\end{aligned}
\label{eq:v12symmetricBarnes}
\end{equation}
The numerator poles start at $3P_0$ and $2P_0+X$ and extend upward by $\ii(mb+nb^{-1})$. The downward poles start at $4P_0-\ii Q/2$, $3P_0+X-\ii Q/2$ and $-\ii Q/2$. Their pinches and the vertex factors lie in the families
\begin{equation}
\begin{aligned}
 X_{\rm pole}&=\pm\ii(Q/2+mb+nb^{-1})+2\sigma P_0,\\
 bX_{\rm pole}&=\mp(m+\tfrac12)r^2
 +\ii\big[\pm(n+\tfrac12)+\sigma\psi/\pi\big].
\end{aligned}
\label{eq:v12Xpole}
\end{equation}
Here $m,n\ge0$ and $\sigma=0,\pm1$. For $0<\psi<\pi/3$, the upper-sign family has negative real part and positive imaginary part in the $bX$ plane. The lower-sign family lies below the real axis. The rotation $X=Re^{-\ii\theta}$, $0\le\theta\le\pi/4$, sweeps the wedge $0\le\arg(bX)\le\pi/4$, which is therefore pole free.

For the large arc choose a Barnes line separating the two families,
\begin{equation}
 u=s e^{-\ii\theta}+\frac{\ii\chi}{b},\qquad s\in\mathbb R,
 \qquad \frac{2\psi}{\pi}-\frac12<\chi<\frac{\psi}{\pi}.
\label{eq:v12kcontour}
\end{equation}
Use the double-sine product formula obtained from \cite{MY}, with $(x;q)_\infty=\prod_{n\ge0}(1-xq^n)$,
\[
 S_b(z)=e^{-\frac{\ii\pi}{2}(z^2-Qz+(Q^2+1)/6)}
 \frac{(e^{2\pi\ii(z/b-b^{-2})};e^{-2\pi\ii/b^2})_\infty}
 {(e^{2\pi\ii bz};e^{2\pi\ii b^2})_\infty}.
\]
Each argument of $B(u,X)$ has the form $z=z_0+\ii s e^{-\ii\theta}$ or $z=z_0+\ii(s-R)e^{-\ii\theta}$.  The changes of sign at $s=0,R$ give the three integration regions below.  The five constant offsets $\Re(bz_0)$ are
\[
3\psi/(2\pi)-\chi,\quad 2\psi/(2\pi)-\chi,\quad
1/2+2\psi/\pi-\chi,\quad1/2+3\psi/(2\pi)-\chi,\quad1/2-\chi.
\]
They lie strictly between $0$ and $1$, and $\Im(bz_0)\ge0$. Reflect the negative rays by $S_b(Q-z)=S_b(z)^{-1}$. The reflected offsets have the same properties. On every resulting positive ray $z=z_0+\ii t e^{-\ii\theta}$, $t\ge0$,
\[
\begin{aligned}
 |e^{2\pi\ii bz}|
 &=e^{-2\pi[\Im(bz_0)+rt\cos(\pi/4-\theta)]},\\
 |e^{2\pi\ii(z/b-b^{-2})}|
 &=e^{-2\pi[(1-\Re(bz_0))/r^2+(t/r)\cos(\pi/4+\theta)]}.
\end{aligned}
\]
The second modulus is strictly below one even at $\theta=\pi/4$.  The first can reach one only at $t=0$, where its phase is $e^{2\pi\ii\Re(bz_0)}\ne1$.  Since both product bases have modulus below one, the quotients and their inverses are uniformly bounded on $s<0$, $0<s<R$ and $s>R$.  For $a=P_0$, the growing part of $\log B(u,X)$ is therefore obtained solely from the quadratic exponential,
\[
\begin{array}{ll}
  s<0&-\pi QX+2\pi Qu+2\pi\ii aX+O_{b,\psi}(1),\\
  0<s<R&-\pi QX+2\pi\ii aX-4\pi\ii au+O_{b,\psi}(1),\\
  s>R&\pi QX-2\pi Qu-2\pi\ii aX+O_{b,\psi}(1).
\end{array}
\]
The vertex factor is also uniformly bounded.  Indeed, with $z=Q/2+\ii X$ and $a=P_0$, \eqref{eq:vertexfactors} and reflection give
\[
 V_{00X}^2=\frac{S_b(z+2\ii a)S_b(z-2\ii a)}{S_b(z)^2},
 \qquad
 -\frac{\ii\pi}{2}\big[(z+2\ii a)^2+(z-2\ii a)^2-2z^2\big]
 =4\pi\ii a^2.
\]
The linear terms cancel as well, and the remaining product quotients obey the same bound, with offsets $1/2$ and $1/2\pm\psi/\pi$.  Thus $|V_{00X}^2|\le C_{b,\psi}$.  Uniformly for $0\le\theta\le\pi/4$, the three Barnes estimates now give
\begin{equation}
 \left|\begin{Bmatrix}P_0&P_0&P_0\\P_0&P_0&X\end{Bmatrix}_{\!\rm Vir}\right|
 \le C_{b,\psi}(1+R)
 e^{-\pi\Re(QX)+\psi\Re(X/b)},\qquad X=Re^{-\ii\theta}.
\label{eq:v12arcbound}
\end{equation}
The middle exponent is affine in $u=tX+\ii\chi/b$, so its maximum occurs at an endpoint. The two tails decay with rate $\Re(Qe^{-\ii\theta})>0$. Combining this estimate with the three-cell weight and the exact Plancherel density gives
\[
|\rho_0(X)\mathcal W(X)|
\le C'_{b,\psi}(1+R)^3
 e^{-\pi\Re(bX)-(\pi-3\psi)\Re(X/b)}.
\]
Since $0<\psi<\pi/3$ and $\Re(bX)\ge rR/\sqrt2$, the integral on the outer arc tends to zero.  This proves the deformation \eqref{eq:v12rotationproved} at fixed continued $b$ without assuming the asymptotics of a mixed cell.

\subsection{The mixed-cell saddle approximation}\label{app:generalmixed6j}

First match the representation labels to the MY convention, $a_e=Q/2+\ii P_e$ and $\alpha_e=\pi b a_e-\pi b^2/2$.  Angle edges have $\alpha_e=(\pi-\psi_e)/2$, and length edges have $\alpha_e=\pi/2+\ii L_e/4$.

The change of Barnes variable $u_{\rm MY}=3Q/2+\ii u$, $\xi=\pi b u_{\rm MY}$ maps \eqref{eq:exact6j} to the MY integrand, with measure $\dd\xi/(\ii\pi b)$.  Contour orientation, logarithms and square roots are transported together from this real-$b$ normalization.

The potential below uses $\tau_i$, the sums of $\alpha_e$ over $(123),(156),(246),(345)$, and $\eta_{1,2,3}$, the sums over $(1245),(1346),(2356)$, with $\eta_4=2\pi$ \cite{MY}.  It is written in terms of the auxiliary dilogarithm function $L_{\rm MY}(v)$,
\begin{equation}
\begin{aligned}
 L_{\rm MY}(v)&=v^2-\pi v+\frac{\pi^2}{6}-\operatorname{Li}_2(e^{2\ii v}),\\
 U(\xi)&=-\frac12\sum_{i,j=1}^4L_{\rm MY}(\eta_j-\tau_i)
 +\sum_{i=1}^4L_{\rm MY}(\xi-\tau_i)+\sum_{j=1}^4L_{\rm MY}(\eta_j-\xi),\\
 \prod_i\sin(\xi_*-\tau_i)&=\prod_j\sin(\eta_j-\xi_*),
 \qquad U'(\xi_*)=0.
\end{aligned}
\label{eq:general-potential}
\end{equation}
On the real mixed branch with the stated face incidences, the continued hyperideal critical value is the Legendre transform over the cell's length edges \cite{MY,BellettiYang},
\begin{equation}
 \frac{\ii}{2}U(\xi_*)
 =V_\Delta+\frac14\sum_eL_e\vartheta_e^{(\Delta)}.
\label{eq:generalcriticalvalue}
\end{equation}
The derivative with respect to $L_e$ is $\vartheta_e^{(\Delta)}/4$, as in \eqref{eq:mixedgeneratingdiff}. Write the subleading coefficient $\kappa_{\boldsymbol\alpha}$ of \cite{MY} as $a_{\rm MY}$ here to distinguish it from the OPE cumulants. On a fixed saddle neighborhood $\Gamma_*$, the Barnes expansion and Hessian identity give
\[
\begin{aligned}
 \left.\{6j\}_{\rm Vir}\right|_{\Gamma_*}
 &=\frac1{\ii\pi b}\int_{\Gamma_*}\dd\xi\,
 e^{U(\xi)/(2\pi\ii b^2)}e^{a_{\rm MY}(\xi)/(2\pi\ii)}
 \bigl(1+O(b^2)\bigr),\\
 a_{\rm MY}(\xi)
 &=8\pi^2+14\pi\sum_e\alpha_e-28\pi\xi
 -4\pi\ii\sum_i\log(1-e^{2\ii(\xi-\tau_i)})\\
 &\qquad+3\pi\ii\sum_j\log(1-e^{2\ii(\eta_j-\xi)}),\\
 -U''(\xi_*)e^{-a_{\rm MY}(\xi_*)/(\pi\ii)}
 &=16\sqrt{\det G}.
\end{aligned}
\]
The Hessian identity is continued on the same branch, away from $\det G=0$ and the logarithmic singularities.

Assume the continued coefficient has an exact contour representation that deforms without crossing poles to the line $\xi=\xi_*+\ii t$, $t\in\mathbb R$, through a unique nondegenerate saddle. Require uniform admissible-domain bounds for the double-sine expansion near the saddle and uniformly exponentially smaller exact inner-contour tails on compact parameter sets. On $b^2=\ii r^2$ the exponential is $\exp[-U/(2\pi r^2)]$, so the required minimum is
\begin{equation}
 \Re[U(\xi_*+\ii t)-U(\xi_*)]>0\quad(t\ne0),\qquad
 -\Re U''(\xi_*)>0,
\label{eq:certificate-minimum}
\end{equation}
with uniform positive gaps off the saddle neighborhood. On the ascending line $\dd\xi=\ii\dd t$, the Jacobian cancels the $1/\ii$ in the Virasoro measure, and the Gaussian gives
\[
 \frac{2r}{b}\,
 \frac{e^{a_{\rm MY}(\xi_*)/(2\pi\ii)}}{\sqrt{-U''(\xi_*)}}
 =\frac{1}{2[-\det G]^{1/4}},\qquad b=re^{\ii\pi/4}.
\]
Here $\sqrt{\det G}=-\ii\sqrt{-\det G}$ is fixed by continuation from $\alpha_e=\pi/2$, $\xi_*=7\pi/4$, where $U''=16\ii$ and $e^{a_{\rm MY}/(2\pi\ii)}=1/2$.  The normalized local result is therefore
\begin{equation}
 \begin{Bmatrix}P_1&P_2&P_3\\P_4&P_5&P_6\end{Bmatrix}_{\!\rm Vir}
 =\frac{\exp\!\left[\frac{\ii}{\pi r^2}
 \left(V_\Delta+\frac14\sum_eL_e\vartheta_e^{(\Delta)}\right)\right]}
 {2[-\det G]^{1/4}}
 (1+O(r^2)),\qquad b=re^{\ii\pi/4}.
\label{eq:certificate-result}
\end{equation}
These estimates, including their remainder and derivative bounds, are required uniformly on the length window of \eqref{eq:oneedge3nj}. The result determines that window contribution. Estimates at the endpoints of the full length cycle and at degenerate saddles complete the remaining contour analysis.

\FloatBarrier
\section{Finite torus sewing and integer-spin support}\label{app:torussewing}\label{ssec:repeatedproof}
The theta seed and linked handcuff form an $S$-closed pair with equal weights. We derive its normalization, apply a $T$ transformation and then attach spectator insertions. The twist sum supplies the integer-spin condition.

Work first at real $b$ and real momenta, with puncture spin $J_s\in\mathbb Z$. In the modular kernels, $i=P_i$, $k=P_k$ and $x=P_x$ denote chiral momenta. The sewing measure is $\dd\nu(x)=\rho_0(x)\dd x$, with two-chiral product $\dd\nu(I)$ for $I=(i,\bar i)$ as in \eqref{eq:repeatedpairing}. The normalized modular kernel satisfies \cite{UniversalTopologies}
\begin{equation}
 \int\dd\nu(x)\,
 \widehat{\mathbb S}_{ix}[s]\widehat{\mathbb S}^{\vee}_{xk}[s]
 =\frac{\delta(i-k)}{\rho_0(k)},\qquad
 \widehat{\mathbb S}^{\vee}_{ik}[s]=e^{-\ii\pi h_s}\widehat{\mathbb S}_{ik}[s].
\label{eq:repeatedSinverse}
\end{equation}
With $I=(i,\bar i)$ and spin $J_I=h_i-\bar h_i$, the two-chiral modular operators act by
\begin{equation}
\begin{aligned}
 (U_Sf)(I)&=e^{-\ii\pi J_s}\!\int\dd\nu(E)\,
 \widehat{\mathbb S}_{ie}[s]
 \widehat{\mathbb S}_{\bar i\bar e}^{\vee}[\bar s]f(E),\\
 (U_Tf)(I)&=e^{2\pi\ii J_I}f(I),\qquad
 U_S^2=(U_SU_T)^3=(-1)^{J_s}.
\end{aligned}
\label{eq:repeatedoperators}
\end{equation}
For the twist sign in \eqref{eq:repeatedoperators}, the chiral inverse and $ST$ relations contribute $e^{\ii\pi h_s}$ and $e^{2\pi\ii h_s}$ \cite{UniversalTopologies}. Pairing with the reversed-framing copy and including the external phase in $U_S$ gives
\begin{align*}
 U_S^2&=e^{-2\pi\ii J_s}e^{\pi\ii J_s}=(-1)^{J_s},\\
 (U_SU_T)^3&=e^{-3\pi\ii J_s}e^{2\pi\ii J_s}=(-1)^{J_s}.
\end{align*}
For even $J_s$ these operators therefore define a $PSL(2,\mathbb Z)$ action.

Let $D(I,K)$ be the identity pairing of \eqref{eq:repeatedpairing}. The two repeated-index Wick contractions have total weight $\nu_s=1+(-1)^{J_s}$. Adding the seed and its $S$ image gives
\begin{equation}
 B_{\{1,S\}}=\nu_s(1+U_S)D,\qquad
 (1-U_S)B_{\{1,S\}}=\nu_s(1-U_S^2)D=0.
\label{eq:repeatedSproof}
\end{equation}
For even $J_s$, independent weights give $w_0D+w_HU_SD$. Its $S$ residual is $(w_0-w_H)(D-U_SD)$, so the point-supported seed and smooth Hopf term have equal weights. Fix $K$ to have integer spin, for which $U_TD(I,K)=D(I,K)$. The remaining twist residual is
\begin{equation}
 (1-U_T)B_{\{1,S\}}(I,K;s)
 =\nu_s(1-e^{2\pi\ii J_I})(U_SD)(I,K).
\label{eq:repeatedTresidual}
\end{equation}
Summing integer twists of the Hopf term projects onto integer spin. On the real starting contour, Poisson resummation gives
\begin{equation}
 \sum_{n\in\mathbb Z}e^{-\epsilon n^2+2\pi\ii nJ}
 =\sqrt{\frac\pi\epsilon}\sum_{m\in\mathbb Z}
 e^{-\pi^2(J-m)^2/\epsilon}
 \xrightarrow[\epsilon\to0^+]{\mathcal D'}\sum_{m\in\mathbb Z}\delta(J-m).
\label{eq:handlecomb}
\end{equation}

Let $Z_\gamma^H(i,k;s)$ denote the chiral handcuff amplitude of filling $\gamma$. For the linked filling its value is \cite{UniversalTopologies}
\begin{equation}
 Z_S^H=\frac{\widehat{\mathbb S}_{ik}[s]}{\sqrt{C_0(i,i,s)C_0(k,k,s)}}.
\label{eq:repeatedHopfproof}
\end{equation}
To check its statistical source normalization directly, insert \eqref{eq:Shat} and the symmetry $\widehat{\mathbb S}_{xi}[s]=\widehat{\mathbb S}_{ix}[s]$,
\begin{equation}
\begin{aligned}
 &\int_0^\infty\dd x\,\mathbb S_{xi}[s]\rho_0(x)
 \sqrt{C_0(x,x,s)C_0(k,k,s)}\,Z_\gamma^H(x,k;s)\\
 &\quad=\rho_0(i)\sqrt{C_0(i,i,s)C_0(k,k,s)}
 \int\dd\nu(x)\,\widehat{\mathbb S}_{ix}[s]Z_\gamma^H(x,k;s).
\end{aligned}
 \label{eq:repeatedsourceproof}
\end{equation}
Thus the modular transformation of the density-weighted OPE source acts by $\widehat{\mathbb S}$ on the graph amplitude.

Now replace the $(k,k,s)$ vertex by the Wilson triangle with vertices $(k,y,q)$, $(k,z,q)$ and $(s,y,z)$. The triangle identity gives
\begin{equation}
\begin{aligned}
 Z_\gamma^\triangle
 &=\sqrt{\frac{C_0(k,k,s)}{C_0(k,y,q)C_0(k,z,q)C_0(s,y,z)}}\,
 R_k\,Z_\gamma^H,\\
 R_k&=\begin{Bmatrix}k&k&s\\z&y&q\end{Bmatrix}_{\!\rm Vir}.
\end{aligned}
\label{eq:repeatedtriangleproof}
\end{equation}
Multiplying the linked triangle by $\sqrt{C_0(i,i,s)C_0(k,y,q)C_0(k,z,q)C_0(s,y,z)}$ leaves $\widehat{\mathbb S}_{ik}[s]R_k$. The right-moving copy supplies $R_{\bar k}^{\vee}$, giving the factor in \eqref{eq:repeatedtriangle}.

For a general spectator attachment, cut the handle along its sewing surface. Let $E=(P_e,\bar P_e)$ be the intermediate pair at the cut, and let the spectator insertions $X$ prepare the coefficient $\psi_X(E)$, including the vertex factors. Gluing this state to a marked handle filling $\gamma$ gives
\begin{equation}
\begin{aligned}
 \mathcal G_X^\gamma(I)&=\int\dd\nu(E)\,\psi_X(E)B_\gamma(I,E;s),\\
 (1-U_g)\mathcal G_X^\gamma(I)
 &=\int\dd\nu(E)\,\psi_X(E)\big[(1-U_g)B_\gamma\big](I,E;s),\qquad g=S,T .
\end{aligned}
\label{eq:repeatedreduction}
\end{equation}
Here the handle move acts on $I$ and preserves the spectator state. The sewing pairing and its prescribed distributional continuation must admit interchange with $U_g$. For the triangle, $\psi_X(E)=R_kR_{\bar k}^{\vee}D(E,K)$, which gives \eqref{eq:repeatedfiniteS}. A complete family of spectator tests and a nondegenerate pairing make this implication reversible.

For the extracted torus coefficient, the double-sine representation of \cite{PostTsiares} gives $\mathbb S_{P_tP}[P_0]=O(P^2)$ at generic labels as $P\to0$. We continue that coefficient with its kernel contour and residues, using the same prescription in the Moore--Seiberg identities \cite{CTV,PT}.

The full modular orbit requires a regulator compatible with both $S$ and $T$. Its dS continuation must preserve integer spin and $\Re\Delta_i=1$. Since $\Delta_i=Q^2/2+P_i^2+\bar P_i^2$ and $\Re Q^2=2$, these conditions give
\begin{equation}
 P_i^2=\frac{\ii\lambda+m}{2},\qquad
 \bar P_i^2=\frac{\ii\lambda-m}{2},\qquad
 J_I=m\in\mathbb Z,\quad\lambda\in\mathbb R.
\label{eq:fixedspintransport}
\end{equation}
This is the support to which the regulated modular orbit must be transported.

\section[The Lorentzian and hyperbolic end actions]{\vtwo{The Lorentzian and hyperbolic end actions}}\label{app:endactions}
\begin{vtwoblock}
This appendix evaluates the end contributions used in section~\ref{ssec:dsquantumtriangulation}. We use the same boundary terms and area counterterm for the matched Lorentzian and hyperbolic cutoffs.

Each of the four curvature-$-1$ pants has area $2\pi$. A Lorentzian wedge has width $\pi-\psi_e$ and period $\ell(\gamma_e)$. The slice area in units of $\ell_{\rm dS}^2$ is therefore
\begin{equation}
 A(t)=8\pi\sinh^2t+\sinh t\cosh t\,\sum_e(\pi-\psi_e)\,\ell(\gamma_e).
\label{eq:lorentzArea}
\end{equation}
On shell, $R=6\ell_{\rm dS}^{-2}$ and $\Lambda=\ell_{\rm dS}^{-2}$. With $K$ defined using the future-pointing unit normal, $\int\sqrt h\,K=\ell_{\rm dS}A'(t)$. The action reduces to
\begin{equation}
\begin{aligned}
 S_L[\mathcal D_M]
 &=\frac1{16\pi G_N}\int\dd^3x\sqrt{-g}\,(R-2\Lambda)
 -\frac1{8\pi G_N}\Big[\int\dd^2x\sqrt h\,K\Big]_{t\to0}^{t=t_c}
 +\frac{1}{8\pi G_N\ell_{\rm dS}}\int_{t=t_c}\dd^2x\sqrt h\\
 &=\frac{\ell_{\rm dS}}{4\pi G_N}\Big[\int_0^{t_c}\dd t\,A(t)-\frac12A'(t_c)+\frac12A'(0)+\frac12A(t_c)\Big].
\end{aligned}
\label{eq:lorentzActionDef}
\end{equation}
The initial term gives \eqref{eq:lorentzRegge}. The remaining integrals are
\begin{equation}
\begin{aligned}
 \int_0^{t_c}\dd t\,A(t)&=2\pi\big(\sinh2t_c-2t_c\big)+\frac12\sinh^2t_c\,\sum_e(\pi-\psi_e)\,\ell(\gamma_e),\\
 A'(t)&=8\pi\sinh2t+\cosh2t\,\sum_e(\pi-\psi_e)\,\ell(\gamma_e).
\end{aligned}
\label{eq:lorentzIntegrals}
\end{equation}
Substitution cancels the terms proportional to $e^{2t_c}$ and gives \eqref{eq:lorentzAction}.

The complete hyperbolic filling contains the compact volume as well as its end. With the normalization of \eqref{eq:s2actionpieces},
\begin{equation}
\begin{aligned}
 I_E[\widehat M]&=\frac{c}{6\pi}\Big[\operatorname{Vol}(M)+\int_0^{\rho_c}\dd\rho\,A_E(\rho)-\frac12A_E'(\rho_c)+\frac12A_E(\rho_c)\Big],\\
 A_E(\rho)&=8\pi\cosh^2\rho+\sinh\rho\cosh\rho\,\sum_e(\pi-\psi_e)\,\ell(\gamma_e),
\end{aligned}
\label{eq:lorentzEuclidDef}
\end{equation}
The primitives are
\begin{equation}
\begin{aligned}
 \int_0^{\rho_c}\dd\rho\,A_E(\rho)&=2\pi\big(\sinh2\rho_c+2\rho_c\big)+\frac12\sinh^2\rho_c\,\sum_e(\pi-\psi_e)\,\ell(\gamma_e),\\
 A_E'(\rho)&=8\pi\sinh2\rho+\cosh2\rho\,\sum_e(\pi-\psi_e)\,\ell(\gamma_e).
\end{aligned}
\label{eq:lorentzEuclidIntegrals}
\end{equation}
Substitution gives \eqref{eq:lorentzEuclid}. Keeping the exponentially small terms in both actions yields the finite-cutoff cancellation \eqref{eq:lorentzSumExact}.
\end{vtwoblock}

\end{document}